\documentclass[]{aa}
\usepackage{graphicx}
\usepackage[varg]{txfonts}
\usepackage{natbib,twoopt}
\usepackage{booktabs}    
\usepackage{float}

\newcommand{\kms}{km\,s$^{-1}$}

\newcommand{\rstar}{R$_{\star}$}
\newcommand{\rsun}{R$_{\sun}$}
\begin{document}
\title{An observational study of dust nucleation in Mira ($o$\,Ceti):}
\subtitle{I. Variable features of AlO and other Al-bearing species}

\author{T. Kami\'nski\inst{\ref{inst1}}, K.T. Wong\inst{\ref{inst2}}, M. R. Schmidt\inst{\ref{inst3}}, 
        H.S.P. M\"uller\inst{\ref{inst4}}, C.A. Gottlieb\inst{\ref{inst5}}, I. Cherchneff\inst{\ref{inst6}},\\
        K.M. Menten\inst{\ref{inst2}}, D. Keller\inst{\ref{inst2}}, S. Br{\"u}nken\inst{\ref{inst4}}, 
        J.M. Winters\inst{\ref{inst7}}, N.A. Patel\inst{\ref{inst5}} 
       }
\institute{\centering 
            ESO, Alonso de Cordova 3107, Vitacura, Casilla 19001, Santiago, Chile, \email{tkaminsk@eso.org}\label{inst1},
       \and Max-Planck-Institut f\"ur Radioastronomie, Auf dem H\"ugel 69, 53121 Bonn, Germany\label{inst2} 
       \and Nicolaus Copernicus Astronomical Center, Polish Academy of Sciences, Rabia{\'n}ska 8, 87-100 Toru\'n\label{inst3}
       \and I. Physikalisches Institut, Z\"ulpicher Strasse 77, 50937 K\"oln, Germany\label{inst4} 
       \and Harvard-Smithsonian Center for Astrophysics, 60 Garden Street, Cambridge, MA, USA\label{inst5}
       \and Departement Physik, Universit\"at Basel, Klingelbergstrasse 82, 4056, Basel, Switzerland\label{inst6}
       \and IRAM, 300 rue de la Piscine, Domaine Universitaire de Grenoble, 38406, St. Martin d'H{\'e}res, France\label{inst7}
       }

  \offprints{T. Kami{\'n}ski}

\date{Received; accepted}
\abstract
{Dust is efficiently produced by cool giant stars, but the condensation of inorganic dust is poorly understood.
Observations of key aluminum bearing molecules around evolved stars has allowed us to investigate the nucleation of alumina (Al$_2$O$_3$) dust in the gas.}
{Identify and characterize aluminum bearing species in the circumstellar gas of Mira ($o$\,Ceti) in order to elucidate their role in the production of Al$_2$O$_3$ dust.}
{Multiepoch spectral line observations at (sub-)millimeter, far-infrared, and optical wavelengths 
including: 
maps with ALMA which probe the gas distribution in the immediate vicinity of the star at $\sim$30\,mas; 
observations with ALMA, APEX, and {\it Herschel} in 2013--2015 for studying cycle and inter-cycle variability of the rotational lines of Al bearing molecules; 
optical records as far back as 1965 to examine variations in electronic transitions over time spans of days to decades; 
and velocity measurements and excitation analysis of the spectral features which constrain the physical parameters of the gas.}
{Three diatomic molecules AlO, AlOH, and AlH, and atomic \ion{Al}{I} are the main observable aluminum species in Mira, although a significant fraction of aluminum might reside in other species that have not yet been identified. 
Strong irregular variability in the (sub-)millimeter and optical features of AlO (possibly the direct precursor of Al$_2$O$_3$) indicates substantial changes in the excitation conditions, or varying abundance that is likely related to shocks in the star. The inhomogeneous distribution of AlO might influence the spatial and temporal characteristics of dust production.}
{We are unable to quantitatively trace aluminum depletion from the gas, but the rich observational material constrains time dependent chemical networks. 
Future improvements should include spectroscopic characterization of higher aluminum oxides, coordinated observations of dust and gas species at different variability phases, and tools to derive abundances in shock excited gas.}

        
\titlerunning{Draft}
\authorrunning{Kami\'nski et al.}
\maketitle
\section{Introduction}\label{sect-intro}
Evolved stars are primary producers of dust in galaxies. They are sources of both carbonaceous dust, originating mainly from carbon-rich asymptotic giant branch (AGB) stars, and inorganic dust, formed in oxygen-rich environments of M-type AGB stars and their massive analogs -- red supergiants. Despite the important role of dust in a broad range of astrophysical phenomena, the formation of stardust is poorly understood. In general, dust formation proceeds along a chain of chemical reactions starting from small gas-phase molecules which form successively larger species \citep{cherchneff2013}. These large molecules grow, form clusters, and end as macroscopic complexes. The formation of inorganic (silicate and alumina) dust is likely to start from oxides. Because the nucleation occurs at rather high temperatures ($\sim$1100--1700\,K), these oxides must be refractory. After ruling out more abundant elements (Si, Fe, and Mg) and under the assumption of thermodynamic equilibrium (TE), it was proposed that oxides of titanium (TiO, TiO$_2$) and of aluminum (AlO) are the gas-phase species that can potentially initiate the formation of critical clusters (seeds) in O-rich environments \citep{GS98,jeong}. The widely known ``silicates'', which determine the observed properties of warm and cold dust in circumstellar shells, are important for grain growth at lower temperatures (a few hundred K) and condense on the seeds formed at higher temperatures. This TE prediction, however, is challenged by strong non-equilibrium effects characterizing the dynamic atmospheres of AGB stars where the condensation takes place \citep{cherchneff2006}. The importance of Ti and Al oxides in grain formation is reinforced by meteorite studies, which show that these oxides are present in presolar grains \citep[i.e. grains thought to represent pristine stardust,][]{Nittler2008}. The meteoritic studies demonstrate that aluminum oxides are at the core of a few silicate grains originating from AGB stars, but it is definitely not a common feature among those grains \cite[e.g.][]{Alcore,presolar2016}. 

Most theoretical studies of the dust condensation in circumstellar material assumed chemical equilibrium in the gas \citep[e.g.][]{sharp}. Some alternative chemical studies of the circumstellar envelopes went beyond this assumption and more realistic non-equilibrium models paved the way toward a better understanding of circumstellar chemistry. They include the presence of shocks which substantially change our view of the chemistry in the envelopes of pulsating giants \citep[e.g.][]{WC,cherchneff2006}. Recent models highlight the chief role of Al oxides as a separate form of warm dust in AGB stars \citep{gobrecht}. An observational verification of the nucleation and chemical reactions leading to the formation of critical clusters is still very much desired. With modern instruments and by combining data from different wavelength regimes, such a verification is now possible \citep[e.g.][]{kami_alo,kami_tio}.

Here, we make an attempt to trace Al-bearing species in the gas phase around Mira in different phases and different pulsation cycles. Using techniques of mainly optical and submillimeter-wave (submm) spectroscopy, we investigate the link between these species and the formation of metal-oxide grains. Mira was chosen because of ({\it i}) the wealth of optical data that was collected over a long period of time and ({\it ii}) relatively high brightness of spectral lines at millimeter (mm) and submm wavelengths. Because in many respects Mira is the prototypical M-type AGB star, understanding Mira has direct implications for understanding the entire class it represents. The numerous studies of dust formation of Mira provide a basis and a rich context for our study, thereby allowing more conclusive interpretations. In this paper -- being the first of a series -- we focus on aluminum-bearing species and their role in the formation of dust in innermost parts of the Mira's envelope. The forthcoming papers will investigate the impact of titanium-, silicon-, and iron-bearing species on the nucleation process using similar methods as those presented here.


The paper is organized as follows. In the remainder of Sect.\,\ref{sect-intro}, we provide further introductory material on $o$\,Ceti,  structure, chemistry, and dust formation in the envelopes of Mira variables. Next, we present and analyze the observational material in two parts. Sections \ref{sec-obs-submm}--\ref{sect-mm-results} focus on mm/submm and far-infrared (FIR) data, mainly of AlO and AlOH. Section\,\ref{opt} presents visual spectroscopy of $o$\,Ceti which yields information primarily on \ion{Al}{I}, AlO, and AlH. In Sect.\,\ref{discussion}, results are discussed within the context of the circumstellar chemistry and dust formation. 

\subsection{$o$\,Ceti}\label{sect-intro-mira}
Mira has a well documented history of light variations. The visual magnitude changes by up to 8$^{\rm mag}$, i.e. in some maxima Mira is $\sim$1600 times brighter than the minima. These light variations are partially caused by changes in the effective temperature of the pulsating photosphere but, even more importantly, the fluxes change owing to variable opacity caused by metal oxides which have strong bands at optical wavelengths \citep{reid02}. The variable temperature and dynamical processes lead to non-equilibrium chemistry and variable abundances of these oxides \citep[e.g.][]{cherchneff2006}. Complex dynamics and chemistry are heavily contributing to the spectacular light changes in Miras. 


Mira's companion, Mira\,B, is likely a white dwarf \citep[but a main-sequence star was also postulated by several authors; see][and references therein]{disk,SokoloskiBildsten}. At an orbit of radius $\sim$90\,AU (projected distance 0\farcs5) and a period of $\sim$600\,yr \citep{disk}, the companion is accreting material from the wind of Mira\,A \citep{mohamed}. The accretion process gives rise to ultraviolet (UV) and X-ray radiation, including emission in highly-ionized species, e.g. of \ion{N}{V} \citep{Reimers1985}. Because Mira\,B is embedded in the dusty wind which absorbs most of the energetic photons in the direction of the cool giant \citep{disk}, the harsh radiation does not affect the inner wind of Mira\,A (within, say, 10\,AU) but may be responsible for dissociation of molecules in the immediate vicinity of Mira\,B \citep[cf.][]{nhung}. Also, Mira\,B is not expected to have a significant dynamic influence on the inner wind of Mira\,A \citep{lynn}.

An important parameter for our analysis is the radius of $o$\,Ceti. The size of the photosphere of a Mira variable varies considerably with wavelength owing to different dominant sources of opacity \citep[cf.][]{stewart2013}. Also, the size changes with phase mainly owing to pulsations in the fundamental mode which change the physical extent of the atmosphere \citep[e.g.][]{woodruff,ireland2011}. Furthermore, the temperature changes as a combined result of expansion/contraction and the propagation of shock waves, influencing the gas excitation and thus the opacity. Photospheric sizes have been derived by different authors using a myriad of observational techniques. The smallest radius we found in the literature, 12.19$\pm$0.02\,mas, was determined near the maximum visual light at $\sim$2\,$\mu$m \citep{perrin}. From $K$-band IR data, a radius of 14.8$\pm$0.6\,mas was found \citep{stewart2013} which is similar to 14.3--14.4\,mas observed at 0.94\,$\mu$m \citep{woodruff,wittkowski2015}; at shorter wavelengths, near 0.701\,$\mu$m and 0.45\,$\mu$m, the continuum source was measured to be a disk of a radius of 26--28\,mas \citep{haniff} and 35$\pm$10\,mas \citep{specle}, respectively. All these values are considerably smaller than the radius of 47$\pm$7\,mas determined at 4.93\,$\mu$m and at $\varphi$=0.2--0.3 \citep{stewart2013}. Most recently, also the millimeter photosphere of Mira has been resolved giving a radius of R$_{\rm mm,ph}$=23\,mas near visual phase 0.5 (i.e. when the atmosphere is more extended than at maximum) \citep{vlemmings,lynn,wong}. \citet{RM07} measured a radius of the radio photosphere at 43\,GHz of 26\,mas at phase 0.05. Throughout this paper, we adopt $R_{\star}\!\equiv$14.4\,mas as the stellar radius and the characteristic distance scale. This is equivalent to 2.4$\times$10$^{13}$\,cm or 331\,\rsun\ at the distance of 107\,pc to the star \citep{distance}. The temperature of the star changes within about\footnote{\citet{arkharov} suggest a range 2200--3000\,K.} 2900--3200\,K and its bolometric luminosity is of about 9000\,L$_{\sun}$ \citep{woodruff,perrin}.

\subsection{The atmosphere and circumstellar envelope of Mira}\label{sect-intro-structure}
The complex kinematic and dynamical structure of the atmosphere and circumstellar envelope of Mira has a crucial bearing on the analysis and discussion in this work. The complex velocity fields of Mira variables have been studied observationally at optical and infrared wavelengths but the greatest successes were achieved by observations of the infrared ro-vibrational lines of CO  \citep[e.g.][]{Hinkle-COI,IRveloCO,Nowotny2010}. Theoretical effort has also been done to understand the structure of Mira variables and a number of models have satisfactorily reproduced some observational data \citep[e.g.][]{ireland2011,hofner2008,gobrecht}. The deepest parts of the atmosphere are dynamically most influenced by the stellar pulsation and the associated shock wave that is created in each cycle. The shock is radiative and ionizing, so that in its wake intense emission lines of high excitation, such as hydrogen recombination lines, are formed. The temperatures within the shock are elevated to 35\,000\,K \citep{fox}. The shock propagates outward and eventually reaches the upper atmosphere. Because a large part of the shock's energy is radiated away and because of the expansion, the shock slows down in progressively higher parts of the atmosphere. Its influence on the velocity field can be traced up to approximately 3--4\,\rstar, where its velocity is a few \kms. This region, which we call hereafter the {\it extended  atmosphere}, shows the highest temporal variations in the kinematics of gas. The radial velocity of high-excitation lines, such as those of the  $\Delta\varv = 3$ band of CO, vary with a full amplitude of 20--30\,\kms\ in $o$\,Ceti. The motions include infall with velocities of up to 13\,\kms. Alumina dust is observed in Mira stars at $\sim$2\,\rstar, i.e. within this dynamically active region. The part of the extended atmosphere up to $\sim$2\,R$_{\star}$ is known for a high-opacity molecular layer of CO and H$_2$O which has been resolved in NIR interferometric observations \citep[e.g.][]{perrin,wittkowski2015}. It is also the region where the variable SiO masers are observed. In Mira, the maser rings are of radii of 2.0--2.8\,R$_{\star}$ \citep{cotton,RM07}.  

Above the extended atmosphere, starting at about 4\,R$_{\star}$, is the zone of silicate-dust formation that is probably somewhat affected by the outward and inward movements. Once (silicate) dust of high opacity is formed, the radiation pressure from the star can accelerate the envelope to form a steady wind. It is often assumed that above 10\,R$_{\star}$ the outflow is fully accelerated. No short-term velocity changes are expected for features arising in the wind.



These three regions are characterized by kinetic gas temperatures of about 2000--4500\,K (photosphere and just above it), 800--1200\,K (the silicate dust formation zone), and $\lesssim$500\,K in the extended wind \citep[cf.][]{Nowotny2010}. At these temperatures, the two inner regions are warm enough to cool down radiatively through electronic transitions of neutral atoms and ions and through electronic and rovibrational bands of refractory molecules. The material in the wind is expected to cool down through low-excitation emission lines in the optical, especially in resonance lines of alkali metals and pure rotational lines of molecules.

\emph{Absorption} lines in $o$\,Ceti change their radial velocity by up to 12\,\kms\ in the optical \citep{joy2}\footnote{\citet{joy} first suggested an amplitude of 12\,\kms\ which however was revoked in his later paper in favor of an amplitude of 4.5\,\kms.  Measurement presented later in our work, however, support amplitudes above 4.5\,\kms.}, and are displaced symmetrically around the stellar (center-of-mass) velocity by 24\,\kms\ in the infrared \citep{IRveloCO}. Optical lines are thought to generally trace higher parts of the stellar atmosphere compared to infrared lines of similar excitation, an effect caused by increased continuum opacity at optical wavelengths \citep[through the Rayleigh scattering;][]{willson1982}. \emph{Emission} lines, mainly optical ones, vary in position as well but they are always shifted toward the blue, by as much as 16--18\,\kms\ \citep{joy,joy2}. They are thought to be excited by and located close to a shock front \citep{willson1982,richter2001,richter2003}. Because the front always moves outwards, the lines appear only at negative (or zero) velocities with respect to the star, i.e. they are always blueshifted. The shock is directly responsible for a temperature inversion which allows us to see emission features even when the gas is seen against the stellar disk.

The wind of Mira has been extensively observed in the classical pure rotational lines of CO \citep[e.g.][]{mira_co,ramstedt,nhung}. The terminal velocity of the wind, $\varv_{\infty}$, is usually quoted as about 5\,\kms, but velocities as low as 2.5\,\kms\ have been suggested in the literature \citep{mira_co}. From observations of multiple lines of CO up to $J_{\rm up}$=16, we constructed a model of the Mira's wind which strongly disfavors the low terminal velocity and here we adopt $\varv_{\infty}$=5\,\kms, equivalent to 1.5\,R$_{\star}$\,yr$^{-1}$. On the basis of the same model and literature data \citep[e.g.][]{GerardBeougois1993}, we also adopt the stellar center-of-mass velocity of  $V_{\rm LSR,sys}$=46.8$\pm$0.5\,\kms\ in the local standard of rest, which is equivalent to $V_{h,\rm sys}$=57.0\,\kms\ in the heliocentric rest frame.

The mass-loss rate of $o$\,Ceti, $\sim$2$\times$10$^{-7}$\,M$_{\sun}$\,yr$^{-1}$ \citep{mira_co,taco}, is typical for Mira stars but the distribution of rates has a large scatter so that some sources can differ by more than one order of magnitude above and below the value derived for $o$\,Ceti \citep{taco}.

\subsection{Dust formation in $o$\,Ceti and other Mira variables}

A relation between dust formation and pulsation cycle of Mira stars is not well established -- it is unclear whether dust formation occurs within certain phases of a given cycle or takes place independent of it. In general, individual cycles are characterized by somewhat different amplitude and shape of the light curve and there is a possibility that the rate of dust production may change from cycle to cycle. Temporal variations in the emission of the innermost dust shells have been reported in the infrared for some objects, including $o$\,Ceti \citep[see e.g.][]{lopez,lobel}. They suggest that dust production itself may indeed be variable. For the Mira variable IK\,Tau, theoretical models show that dust formation is time-dependent and occurs at specific pulsation phases in the shocked upper atmosphere, i.e. $\varphi$=0.8--1.0 for alumina and $\varphi$=0.5--1.0 for silicates \citep{gobrecht}.

Mira variables have been categorized depending on the type of spectral features observed at mid-infrared (MIR) wavelengths. These include: (I) a broad feature of Al$_2$O$_3$ (corundum), (II) a mix of alumina and silicate features, or (III) a dominant silicate feature \citep{lorenz,little}. These three types are thought to be an evolutionary sequence, where objects with alumina-dominated dust are less evolved  stars with dust production initiated recently, and group III has a long history of dust production. [More recent studies suggest that the three groups reflect different regimes of mass-loss rates, with group I representing very low rates \citep[e.g.][]{karovicova}.] $o$-Ceti's MIR spectrum is dominated by the silicate feature \citep[e.g.][]{lobel}, placing it in the third group, i.e. among the most evolved objects in the sequence. It is thought that this group of stars, including Mira, forms alumina dust at high temperatures close to the star \citep[2--3.5\,R$_{\star}$, $\sim$30--50\,mas in $o$\,Ceti,][]{bester,degiacomi,lopez} but most of the mass of dust is built up farther from the photosphere, in the form of silicate grains ($\geq$200\,mas in $o$\,Ceti); the silicate dust may partially form on the seeds provided by the Al$_2$O$_3$ clusters. Therefore, although their MIR spectra are dominated by the silicate feature, Mira variables are still thought to be efficient producers of alumina-based solids. This is also consistent with popular hypotheses on the dust-nucleation sequence in evolved stars. It has been confirmed observationally by infrared interferometric techniques that the alumina dust forms closer to the photosphere than silicates \citep[e.g.][]{zhao,karovicova}. Recent theoretical models agree with those observational findings and show that alumina dust forms at radii $\lesssim$2\,R$_{\star}$ and the silicate dust is present at radii $\geq$4\,R$_{\star}$ \citep{gobrecht}. 


\subsection{Which Al-bearing species are important?}
In order to trace the species containing Al in the envelope of Mira, we first identify the most likely carriers on the basis of previous observations of cool circumstellar envelopes and chemical models. \citet{gobrecht} consider Al, AlH, AlO, AlOH, AlO$_2$, AlCl, Al$_2$, and Al$_2$O to be important species for Al gas-phase chemistry. In addition, AlF and AlNC were observed in the carbon star IRC+10216 \citep{AlF,AlNC}, and AlS and AlCN are potentially interesting carriers as well \citep[cf.][]{ref-AlOH}. Of these, homonuclear Al$_2$ is not easily observable. Similarly, the main isomers of AlO$_2$ and Al$_2$O are linear and their rotational lines are not observable. We found no high-resolution spectroscopic studies of gas-phase forms   of higher Al oxides that would allow us to identify their spectral features. This poses a serious shortcoming for our current study. Other molecules listed here have been observed at visual (AlH, AlO) or mm/submm (AlO, AlOH, AlCl, AlF, AlNC) wavelengths around cool evolved stars. From these, only the optical electronic bands of AlO have been reported to date in Mira. 


Atomic aluminum is present mainly in the stellar photosphere where thermal equilibrium holds. For an effective temperature of 2200--3000\,K of Mira, equilibrium calculations indicate that atomic aluminum is mainly in the neutral and singly-ionized forms \citep[cf.][]{ref-AlOH}. 

Because Mira is thought to be a low-mass star \citep[$\sim$1\,M$_{\sun}$,][]{ireland2011}, one does not expect the rare unstable isotope of $^{26}$Al to be present in its atmosphere and circumstellar environment \citep{26Al}. Therefore, no rare isotopologues were targeted in this study.

\section{Millimeter to  FIR observations}\label{sec-obs-submm}
In search for the Al-bearing species, we observed Mira in a broad wavelength range, from mm to FIR wavelengths. Additionally, optical observations are presented in Sect.\,\ref{opt}.


\subsection{APEX/FLASH}
Mira was observed in 13--14 August 2013 and 30 June--8 July 2014  using the dual-band FLASH+ receiver at APEX. Multiple frequency setups were used with the primary aim to detect rotational transitions of AlO (and lines of CO in the first excited vibrational state). FLASH+ has two frequency units (covering the 345 and 460\,GHz atmospheric bands) and separates the upper and lower side bands (USB and LSB), what produces four spectra in a single observation. This allowed us to observe ten different frequency ranges within the 345\,GHz and 460\,GHz atmospheric windows, each 4\,GHz wide, at the spectral resolution of 38\,kHz (in the 345\,GHz band) and 76\,kHz (in the 460\,GHz band). Observational details including central frequencies, integration times, rms-noise levels, and exact dates of observations are given in Table\,\ref{tab-log} (Appendix\,\ref{appendix-log}). 

All observations were performed with a wobbler which was switched by 1\arcmin. The data were calibrated with the default APEX pipeline \citep{APEXcalib} and are here presented in the antenna brightness temperature ($T_A^*$) or converted to main-beam brightness temperature ($T_{\rm mb}$), as indicated in each case. Baselines of low order were subtracted from the spectra. All spectra presented here were smoothed to resolutions which allow more readily presentation. 


\subsection{APEX/HET230}
Spectra were also acquired in the 1-mm band with the SHeFI (HET230) instrument at APEX. These observations were obtained between 3 and 23 December 2013 in two setups and were intended to detect ({\it i}) two lines of TiO centered at 222.7\,GHz (in LSB), and ({\it ii}) lines of AlO and vibrationally excited CO, with the band centered at 229.8\,GHz (in USB). The second setup was re-observed in 9--12 June 2015. The HET230 instrument produces one single-sideband spectrum. The XFFTS spectrometer produced spectra at a resolution of 76\,kHz.


Two frequency ranges, centered on 229.800\,GHz and 344.454\,GHz, were each observed twice, 20 and 11 months apart with APEX (cf. Table\,\ref{tab-log}). A detailed technical analysis of the data, summarized in Appendix\,\ref{sect-var-apex}, have shown that some of the AlO lines covered are variable.

\paragraph{Herschel/HIFI}
In the search for the Al-bearing species, we browsed the archives of the Herschel Space Observatory ({\it Herschel}) for observations of Mira obtained with the HIFI instrument. Publicly available observations were mostly performed within the {\small HIFISTARS} project (PI: V. Bujarrabal), but some additional data were acquired within the Performance-Verification (PV) of HIFI. We used the processed data from the {\small HIFISTARS}' User Provided Data Products Release\footnote{\url{http://herschel.esac.esa.int/UserProvidedDataProducts.shtml}} or -- in the case of the PV data -- the pipeline-processed spectra which were corrected for baseline and converted to $T_{\rm mb}$ units. Here, we analyze only one spectrum which covers the $N$=29--28 line of AlO.



\begin{table*}[!ht]
\caption{Transitions of AlO and AlOH covered by APEX, {\it Herschel}, and ALMA observations.}\label{tab-alo}
\centering
\begin{tabular}{cc cc cc cc cc}
\hline\hline
Transition&$\langle\nu\rangle$\tablefootmark{a}&$\Sigma S_i\mu^2$&$E_u$&Date of&Vis.&Telesc.&Beam\tablefootmark{b}& $I(T_{\rm mb})$\tablefootmark{c}&Flux$\pm$1$\sigma$\\ 
$N_{\rm up}\!-\!N_{\rm low}$ & (GHz)     &(D$^2$)&(K)  &observations& phase&        &(\arcsec)&(K\,\kms)&(Jy\,\kms)\\
\hline\hline\\[-7pt]
\multicolumn{10}{c}{AlO}\\
\hline\hline
6--5   & 229.67025503& 1522.89&  38.65& 6--23 Dec. 2013    & 0.5   & APEX &27.2 & $<$0.035                   &$<$1.02\\
6--5   & 229.67025503& 1522.89&  38.65&29\,Oct./1\,Nov.\,2014 & 0.5   & ALMA &0.029& 24591.6\tablefootmark{d}   &0.89$\pm$0.02\\
6--5   & 229.67025503& 1522.89&  38.65& 9--12 Jun. 2015    & 0.1   & APEX &27.2 &    0.076                   &2.21$\pm$0.72\\[3pt]
8--7   & 306.19734234& 2030.60&  66.20&  7--8 Jul. 2014    & 0.1   & APEX &20.4 &    0.079                   &2.34$\pm$0.65\\[3pt]
9--8   & 344.45177512& 2284.46&  82.73&       Feb. 2014    &  0.7  & ALMA &0.88 &    16.79\tablefootmark{d}  &1.25$\pm$0.09\\  
9--8   & 344.45177512& 2284.46&  82.73&        May 2014    &  0.9  & ALMA &0.43 &   119.28\tablefootmark{d}  &2.16$\pm$0.29\\  
9--8   & 344.45177512& 2284.46&  82.73&12--15 June 2014    &  1.0  & ALMA &0.33 &   291.31\tablefootmark{d}  &3.05$\pm$0.03\\  
9--8   & 344.45177512& 2284.46&  82.73& Aug.2013/Jul.2014&0.1/0.1& APEX &18.1 &    0.114                   &3.41$\pm$0.81\\
9--8   & 344.45177512& 2284.46&  82.73&   21 July 2015     &  0.2  & ALMA &0.15 &  1838.58\tablefootmark{d}  &3.84$\pm$0.02\\[3pt]
11--10 & 420.93832496& 2792.05& 121.30& 13--14 Aug. 2013   & 0.1   & APEX &14.8 &    0.408                   &11.8$\pm$3.0\\ 
12--11 & 459.16920150& 3040.87& 143.33&    14 Aug. 2013    & 0.1   & APEX &13.6 & $<$0.140                   &$<$4.03\\
13--12 & 497.38987615& 3300.05& 167.20&  1--8 Jul. 2014    & 0.1   & APEX &12.5 &    0.413                   &11.9$\pm$2.3\\
29--28 &1106.98018793& 7362.47& 798.08&    20 Jul. 2010    & 0.7   & HSO  &19.2 &    0.160                   &57.1$\pm$19.0\\
\hline\hline\\[-7pt]
\multicolumn{10}{c}{AlOH}\\
\hline\hline
11--10 & 346.15555028 &  71.39 &   99.69 & Aug. 2013/Jul. 2014&0.1/0.1&APEX& 18.0 &  0.059 &1.77$\pm$0.49 \\  
13--12 & 409.03103805 &  84.36 &  137.45 & 13--14 Aug. 2013   & 0.1&APEX& 15.3 &     0.105 &3.02$\pm$0.97 \\ 
15--14 & 471.87557840 &  97.35 &  181.23 & 14 Aug. 2013       & 0.1&APEX& 13.2 &	 0.180 &5.18$\pm$1.54 \\ 
18--17 & 566.07395655 & 116.81 &  258.21 & 19 Feb. 2010       & 0.3& HSO& 38.3 &  $<$0.072 &$<$22.48 \\
21--20 & 660.17593148 & 136.28 &  348.75 & 20 Jul. 2010&0.7& HSO& 32.4 &	      $<$0.074 &$<$23.10 \\
22--21 & 691.51905756 & 142.77 &  381.93 & 20 Jul. 2010&0.7& HSO& 31.1 &	      $<$0.128 &$<$39.96 \\
33--32 &1035.29846577 & 214.16 &  846.08 &  3 Feb. 2010&0.2& HSO& 20.6 &	      $<$0.275 &$<$85.93 \\
35--34 &1097.57173239 & 227.13 &  949.93 & 20 Jul. 2010&0.7& HSO& 19.4 &	         0.315 &98.4$\pm$22.9 \\ 
\hline 
\end{tabular}
\tablefoot{
\tablefoottext{a}{Weighted mean frequency of the hyperfine components, i.e. $\langle\nu\rangle=\Sigma\nu_i S_i/\Sigma S_i$, where $S_i$ is the component strength.}
\tablefoottext{b}{For non-circular beams a geometric mean of the minor and major axes is given.}
\tablefoottext{c}{Integrated intensity of the line or 3$\sigma$ upper limit in $T_{\rm mb}$ or $T_b$ units.}
\tablefoottext{d}{Brightness temperature calculated from flux within the Rayleigh-Jeans approximation.}
}
\end{table*}

\begin{figure*} 
\sidecaption
\begin{minipage}{12cm}
\includegraphics[angle=270,width=12cm]{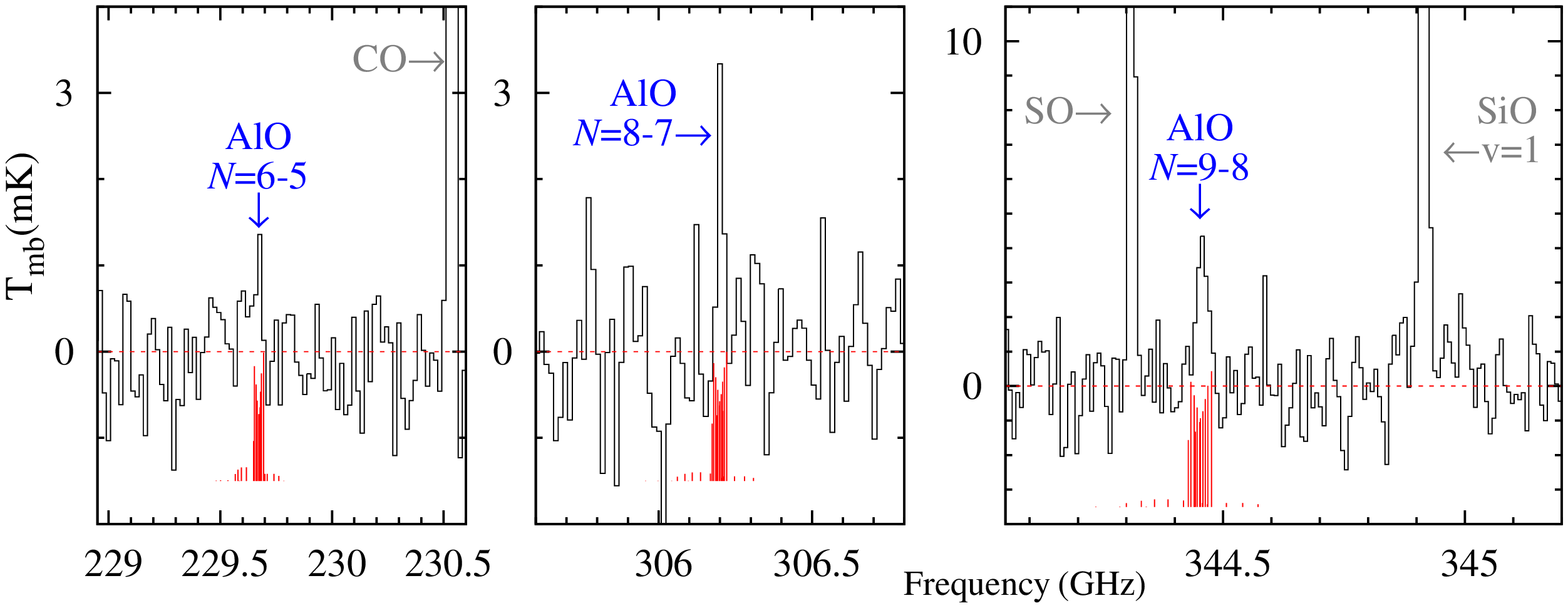}\\[3pt]
\includegraphics[angle=270,width=12cm]{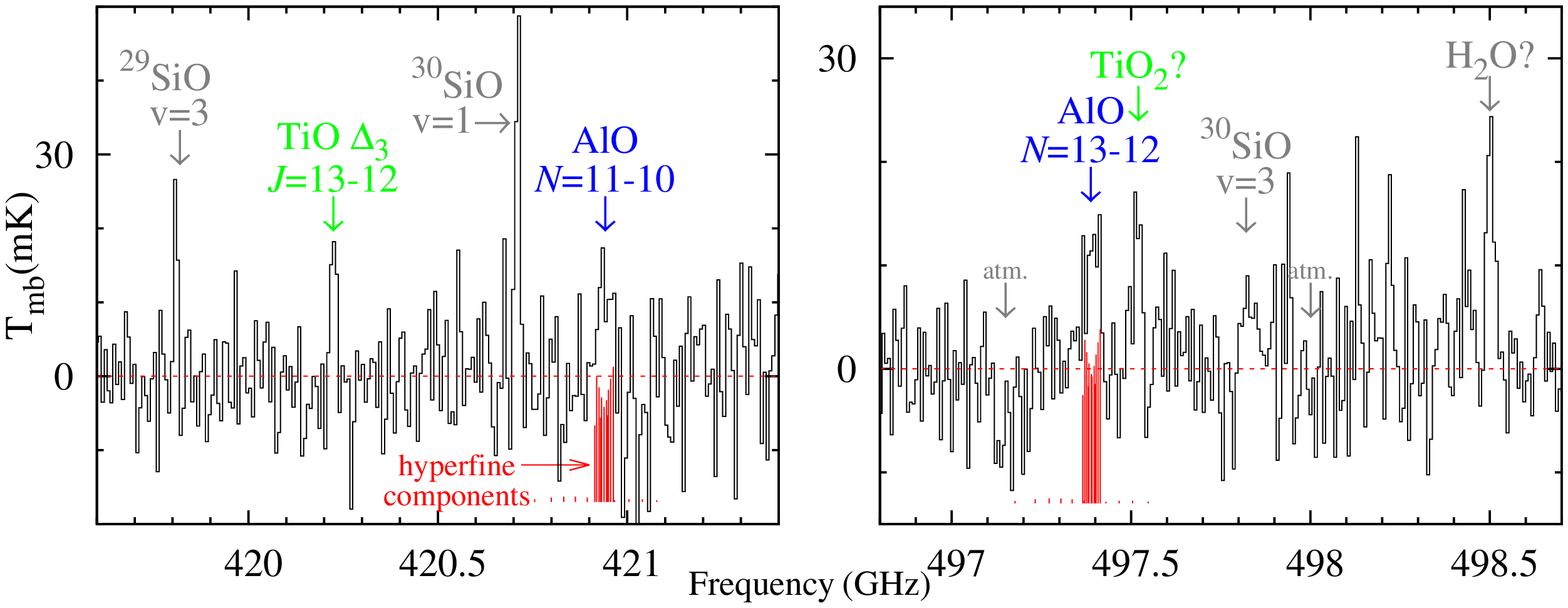}\\[3pt]
\includegraphics[angle=270,width=12cm]{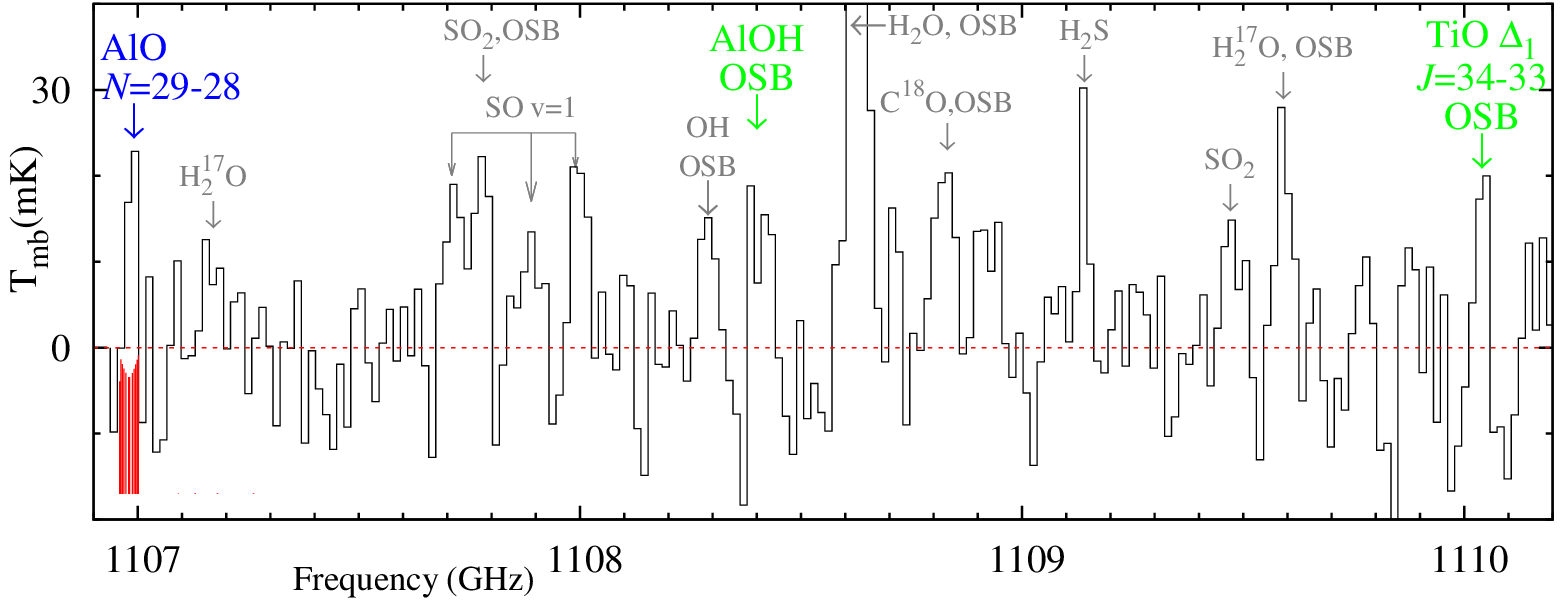}
\end{minipage}
\caption{{\bf Top and middle panels:} Five lines of AlO observed in the submillimeter-wave, single sideband spectra of Mira obtained in 2013--2015 with APEX. The features of AlO are broader (FWHM) than most other lines owing to unresolved hyperfine splitting (the components are shown as red bars). {\bf Bottom:} An archival spectrum of Mira obtained with {\it Herschel} in July 2010, where one other transition of AlO was observed. This double side-band spectrum contains a large number of emission lines from different species in two overlapping frequency ranges. This spectrum is smoothed to a lower resolution than in the APEX spectra shown in the upper panels.}\label{fig-alospec}
\end{figure*}

\subsection{ALMA}
The most sensitive submm observations of Mira to date are those obtained by ALMA at multiple epochs in 2014 and 2015. 
The earliest ALMA data for Mira were obtained within the ALMA 2014 Long Baseline Campaign Science Verification (hereafter, SV data) \citep{alma}. The data are described in detail in \citet{wong}\footnote{see also \url{https://casaguides.nrao.edu/index.php?title=ALMA2014_LBC_SVDATA}}. One spectrum which covers the AlO $N$=6$\to$5 line was observed on 29 October and 1 November 2014 in the ``continuum mode'', i.e. with a poor spectral resolution of 20.5\,\kms. These high angular-resolution observations with a beam of about  34$\times$24\,mas are the most detailed observations of Mira to date. 

The first ALMA observations at submm frequencies were reported in \citet{ramstedt}. They covered two frequency ranges within ALMA Band\,7 at about 330.25--334.00 and 342.35--346.09\,GHz that included the $N$=9$\to$8 line of AlO. From the ALMA archive, we extracted the raw data obtained with the 12-m array only and reduced them with the standard pipeline. The observations were obtained twice, on 24 Feb. 2014 and 3 May 2014, in nearly the same frequency setup but with different array configurations resulting in spatial resolutions (synthesized beams) of 1\farcs59$\times$0\farcs48 and 0\farcs51$\times$0\farcs37, respectively, at the frequency of the AlO line. Unlike in \citet{ramstedt}, we imaged the data from different dates separately. The flux levels of continuum and line emission differ in both datasets indicating variability at a level above the flux calibration uncertainties. The first dataset was calibrated using observations of the primary flux calibrator, Ganymede, with a model flux accurate to $<$5\%, while the second used the secondary calibrator J0334-401 whose flux is monitored at about 15\% accuracy. The calibration of phase and amplitude in the data used for the analysis here were further improved by a self-calibration procedure performed on the cumulative continuum emission. The spectra were recorded at a resolution equivalent to about 0.43\,\kms.

ALMA observed Mira again in Band\,7 on 12, 14, and 15 June 2014 in the same antenna configuration reaching an angular resolution of 0\farcs34$\times$0\farcs32 in the combined data. The observations are described in \citet{planesas}. Four spectral ranges were observed: 330.4--330.7 and 345.6--345.9\,GHz -- at a velocity resolution of 0.1\,\kms\ -- and 331.1--332.8 and 343.7--345.5\,GHz, at a velocity resolution of about 13.6\,\kms. One of the latter ranges -- observed with the low spectral resolution -- covers the $N$=9$\to$8 line of AlO. We imaged the data combining visibilities from the three dates. The data were calibrated in flux using observations of J0334-401, J0238+166, and J2258-279, whose fluxes are known to within 15\%. Further flux equalization between the three datasets was performed by setting the flux of the phase calibrator, J0217+0144, to the same average level.


Mira was observed by ALMA again in Band\,7 on 21 July 2015 reaching the best resolution and sensitivity so far at these wavelengths. Because these data were originally collected for the purpose of our study here, they are described in greater detail than the earlier ALMA observations. The spectra cover four ranges, 342.1--344.0, 344.1--346.0, 354.2--356.1, and 356.1--358.0\,GHz at a resolution of 0.98\,\kms. The spectrum covers the $N$=9$\to$8 line of AlO. The observations with 42 antennas of the 12-m array arranged on baselines between 15 and 1574\,m resulted in an angular resolution of about 0\farcs158$\times$0\farcs127 near the line of AlO. The observations were arranged into two consecutive execution blocks, each $\sim$1.1-h long. The calibrators were J0224+0659 for bandpass, J0217+0144 for phase, and J0238+166 for flux calibration. For the second execution, the flux-calibration scan failed and the bandpass calibrator was used for flux calibration instead. Its fluxes were assumed to be the same as in the first execution calibrated with J0238+166. We checked that the flux calibration in the two blocks was consistent and the data were combined. In order to improve the complex gain calibration of Mira images, we performed extra optimization of phase and amplitude using an iterative self-calibration procedure performed on channels dominated by continuum emission. The procedure increased the signal-to-noise of the data by a factor of a few. 

All the ALMA data were imaged with Briggs weighting with the robust parameter set to 0.5. However, for reference, we also produced images with uniform and natural weighting to increase the nominal angular resolution or sensitivity, respectively. 



\section{Results of submm/FIR observations}\label{sect-mm-results}
\subsection{Identification of AlO}
Most of the APEX spectral ranges were arranged to cover lines of AlO resulting in observations of six different transitions. Three transitions were also covered by ALMA and {\it Herschel}, most of them serendipitously. Two lines were observed in multiple epochs each. All the covered lines of AlO, with their spectroscopic parameters and measured intensities, are listed in Table\,\ref{tab-alo}. The line frequencies were taken from \citet{YamadaAlO}.

The $N$=6$\to$5 line was first covered in 2013 by APEX spectra but was not detected. When the observation was repeated with APEX in 2015, the line was detected and the emission peak was significantly above the corresponding rms noise levels of the earlier APEX attempt. As argued in Sect.\,\ref{sect-var}, the `emergence' of the line must be due to intrinsic variability of the source. The line was also observed in 2014 with ALMA within the SV observations at the high angular resolution but at the poor spectral resolution. Table\,\ref{tab-alo} presents the line intensities corresponding to a region where the \emph{absolute} intensity was above the 3$\times$rms noise level in the profile-integrated map. 

The rotational transitions from $N_{up}$=8, 9, 11, 12, 13 were observed with APEX in several runs between 2013 and 2014. All but 12$\to$11 were firmly detected. The 9$\to$8 line was observed twice in that period, 11 months apart, and showed identical intensity in both APEX spectra (and they were combined, Sect.\,\ref{sect-var-apex}). The same 9$\to$8 transition was observed with ALMA at four different epochs and showed clear changes in the integrated flux (Table\,\ref{tab-alo}). The variability is discussed in later sections.  

%
%
%

One extra transition of AlO, $N$=29$\to$28, was covered close to the edge of a spectrum from {\it Herschel}/HIFI. This line was detected using the same instrumental setup as in VY\,CMa \citep{vy_hifi}. The AlO emission is detected in Mira at a level of about 3$\sigma$ and is clearly apparent after smoothing the spectrum to a resolution of a few \kms. The position of the feature agrees very well with that of the stellar radial velocity. 

\begin{figure}
\centering
\includegraphics[width=0.99\columnwidth]{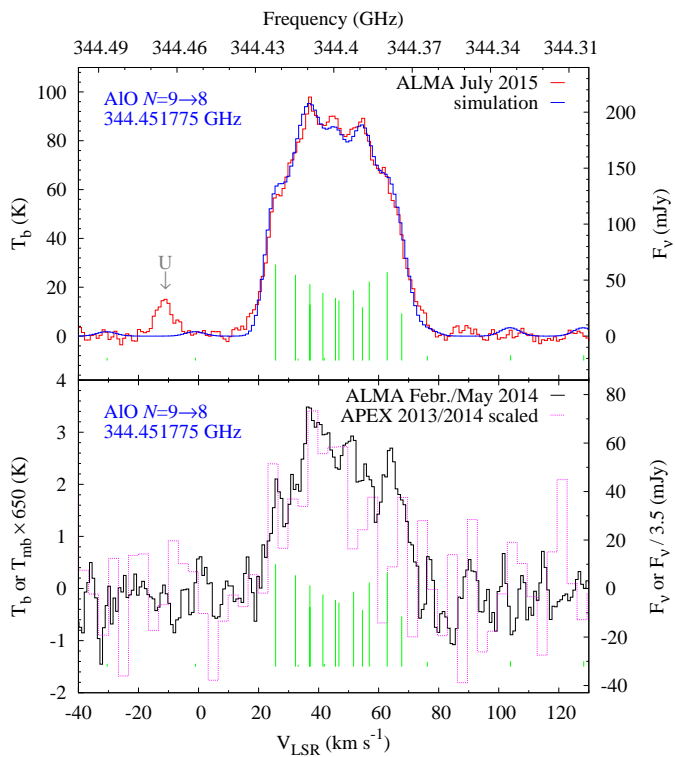}
\caption{
{\bf Top:} The AlO $N$=9--8 spectrum observed with ALMA on 12 July 2015 (red line). This is the best-quality (S/N and spectral resolution) spectrum of this transition to date. It was extracted from a region where profile-integrated flux is above the 3$\times$rms noise level. Our simulation of the profile is drawn in blue (see text for details). The green vertical bars mark individual hyperfine components with the height of the bars corresponding to the relative line strength, $S_i$, at $V_{\rm LSR}$ of 46.9\,\kms.   
{\bf Bottom:} The profiles of the same line in the APEX (magenta) and ALMA data from Feb. and May 2014 (black). Both spectra were smoothed for clarity. The APEX spectrum is scaled in intensity units.}\label{fig-AlO-profile}
\end{figure}

Most of the detected lines are shown in Figs.\,\ref{fig-alospec} and \ref{fig-AlO-profile}. Because the rotational lines of AlO have considerable hyperfine splitting their FWHMs are broader (typically $\Delta\nu$=60\,MHz in the submm region) than those of other thermal lines originating from the same region of the envelope. This is clearly observed in our submm spectra of Mira (Fig.\,\ref{fig-alospec}). The $N$=29$\to$28 line in the FIR  \emph{appears} narrower than those at lower frequencies owing to smaller hyperfine splitting of this high-frequency line. The positions and widths of lines leave little doubt that they belong to AlO. Our APEX observations covered a substantial range of frequencies within the available 345 and 460\,GHz atmospheric windows, which combined with the {\it Herschel} FIR spectra allowed us to perform a comprehensive identification of spectral features. As a result we can confidently state that none of the detected transitions assigned to AlO is significantly contaminated by other species observed in Mira.

With the observations in hand, in the following sections we attempt to characterize the AlO-containing gas that gives rise to the rotational emission.


\subsection{Location and extent of the AlO emission}\label{AlOmap}
\begin{figure}
\centering
\includegraphics[angle=0, trim=50 110 0 90, clip=true, width=0.99\columnwidth]{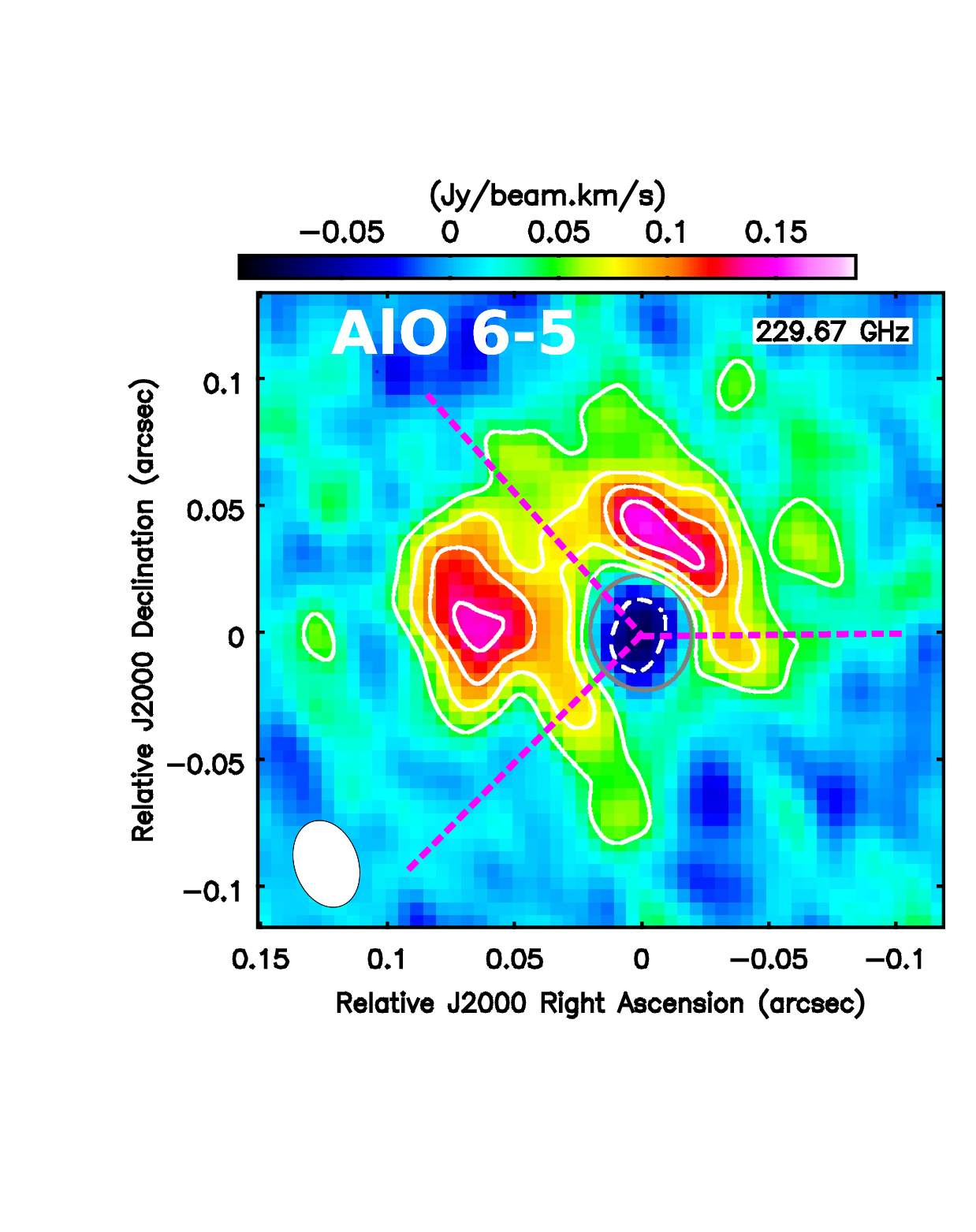}
\caption{The integrated-intensity map of AlO emission in the $N$=6$\to$5 transition as observed with ALMA with a beam of 34$\times$24\,mas, shown as a white ellipse. White contours are drawn at --3 (dashed), 3, 4, 5, 6 times the map rms noise level of 15\,mJy/beam\,\kms. For comparison, the continuum emission is shown with a gray contour at 50\% of its peak emission which represents the extend of the beam-smeared radio photosphere. The dashed lines (magenta) show the directions of spatial cuts which are presented in Fig.\,\ref{fig-projections}. The map was produced by using data processed in CLEAN after continuum subtraction.}
\label{fig-mom0AlO65}
\end{figure}

The high angular resolution of the ALMA SV observations of the $N$=6$\to$5 line provide the most detailed view of the AlO spatial structure. Figure\,\ref{fig-mom0AlO65} presents a map of the continuum-subtracted and profile-integrated flux in the line (the appearance of the region depends on the details of data processing as explained in Appendix\,\ref{appendix-clean}). These observations resolved the stellar radio photosphere \citep{lynn,wong} and we definitely observe AlO absorption toward the resolved stellar disk. It is represented by the negative signal in Fig.\,\ref{fig-mom0AlO65}. The absorption against the stellar disk must be formed in gas with an excitation temperature lower than that of the mm-wave photosphere, i.e. below 2600\,K \citep{wong}. 

The absorption region is surrounded by patchy emission with the strongest discrete components located approximately 63\,mas east and 42\,mas north from the stellar center, i.e. very near the edge of the mm-wave photosphere (2\,R$_{\rm mm,ph}$=51.2$\times$41.0\,mas or 2.9--4.4\,R$_{\star}$). Toward the southwest, we see only weak AlO emission, if at all. The S/N of the map is not good enough to state precisely how far from the stellar disk the AlO emission spreads out but all emission above three times the map rms noise level is encompassed by a circle of a $\sim$100\,mas radius. 

\begin{figure*}
\centering
\includegraphics[angle=0, width=0.75\textwidth]{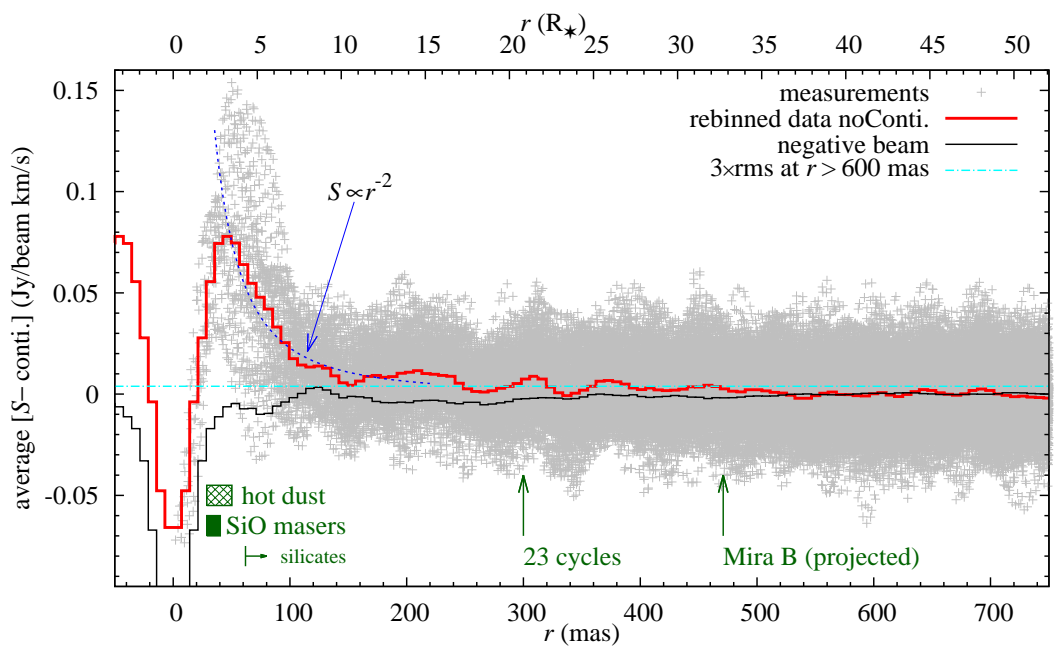}
\caption{The gray points are brightness measurements, $S$=$\int S_{\nu}$d$\nu$, of the AlO(6$\to$5) line with ALMA plotted against the distance from the continuum peak. (They represent values of each pixel of the integrated-intensity map plotted against the radial distance of the pixel with respect to the star.) The red line is a binned average of those measurements mirrored around zero. The cyan dash-dot horizontal line marks the 3$\times$rms level measured beyond 0\farcs6. The AlO emission can be traced above this 3$\times$rms level up to about 400\,mas and the overall profile is well reproduced by a power law, with an example drawn with a dashed curve. Drawn with a black line is the azimuthally averaged dirty beam which was scaled and inverted in intensity  to show the errorbeam structures corresponding to the strongest observed component of the profile. At the bottom of the plot, in green, are shown some characteristic locations within the envelope: the region closest to the star where hot dust has been observed in minimum (inner edge) and maximum light (outer edge) \citep{degiacomi}; the range of SiO maser rings \citep{cotton}; the innermost radius where silicate dust is observed in Mira stars marked as ``silicates'' \citep{karovicova,wong}; the radius at which the wind arrives after 23 cycles (or 20.8 yr) at the terminal wind speed of 5\,\kms; and the projected location of Mira\,B.}
\label{fig-profile}
\end{figure*}

In order to increase the dynamic range of the observations and infer the overall distribution of the AlO gas, we created a radial profile of the AlO region by averaging the map in the full azimuthal angle around the position of the continuum peak. The profile, shown in Fig.\,\ref{fig-profile}, demonstrates that the brightness nearly follows a power-law distribution and rises just next to the edge of the radio photosphere (the effect is smeared by the restoring beam). It can be traced as far as 150\,mas (10\,R$_{\star}$) from the star center. Some bumps seen at larger radii are unlikely to be real.

\begin{figure}
\centering
\includegraphics[angle=0,trim=40 55 0 10, clip=true, width=0.99\columnwidth]{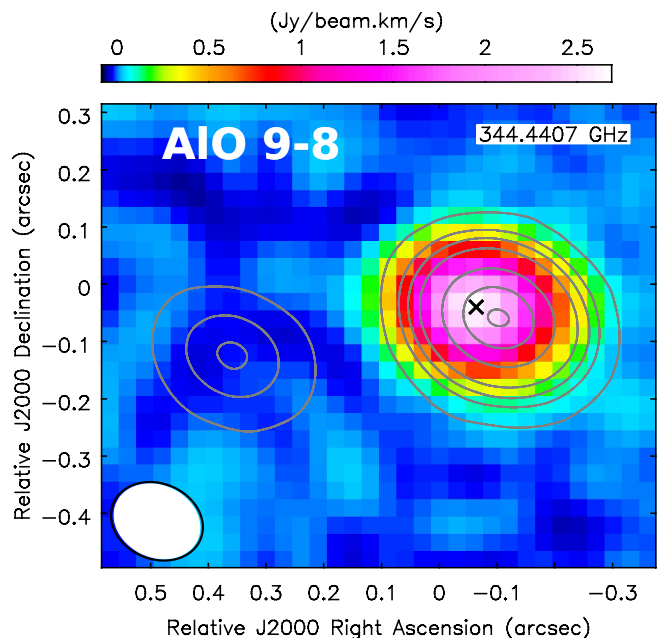}
\caption{The same as Fig.\,\ref{fig-mom0AlO65}, but for the $N$=9$\to$8 transition of AlO observed with a lower angular resolution of 163$\times$132\,mas. Note that the map shows a region of the sky $\sim$4 times larger than in Fig.\,\ref{fig-mom0AlO65} and also encompases Mira\,B (left continnum source). Gray contours show continuum emission of the binary at levels of 2.5, 5, 10, 30, 60, and 90\% of the maximum. The center of the AlO emission, marked with a black cross, does not coincide with that of the continuum of Mira\,A.}\label{fig-mom0AlO98}
\end{figure}

The minimum and maximum fluxes of the AlO(6$\to$5) emission are --73.4 and +153.8\,mJy/beam\,\kms. The emission above the 3$\sigma$ noise level occupies a solid angle about 27 times larger than that of absorption below the --3$\sigma$ level. This dominance of emission over absorption should produce a net pure emission feature if the source was not spatially resolved. Indeed, the $N$=9$\to$8 transition of AlO when observed with ALMA at $\sim$5 times lower angular resolution than the SV data appears as an emission region, i.e. no negative component is apparent (Fig.\,\ref{fig-mom0AlO98}). A Gaussian fit to the profile-integrated map of AlO(9--8) gives a source size of 123$\times$74\,mas ($\pm$6\,mas) at PA=95\fdg1$\pm$5\fdg3. As can be seen in Fig.\,\ref{fig-mom0AlO98}, the emission center is offset from the center of the submm-wave continuum by 40.2$\pm$2.3\,mas (2.8\,R$_{\star}$) toward the northeast (PA=45\degr$\pm$3\degr). What we observe in $N$=9$\to$8 is likely a combination of emission and absorption similar to that seen in $N$=6$\to$5. Indeed, the SV $N$=6$\to$5 data smoothed to the resolution of the $N$=9$\to$8 data from 2015 show almost exactly the same morphology and relative offset of the net AlO emission with respect to continuum. The AlO peak in the smoothed data is at an offset of 33.2$\pm$0.9\,mas (2.3\,R$_{\star}$) along a PA of 42\degr$\pm2$\degr. This comparison suggests that the overall distribution of the emission clumps did not change considerably over the nine months between the observations and at the different phases (0.5 vs. 0.2). We also note that the emission area of the 6$\to$5 line is equivalent to a Gaussian source with a FWHM of 107\,mas which is very close to the geometric mean of the size we determined for the 9$\to$8 line, FWHM=95\,mas. The latter line is observed at a S/N of 140 so we recover essentially all of its emission. That they cover the same effective area implies that we recover most of the emission in $N$=6$\to$5, even though it is observed at a lower S/N of 20 (when smoothed to the same angular resolution). The two transitions have $E_u$ values that differ by only about 44\,K and thus should not show drastically different morphologies. 

In the ALMA data for the $N$=9$\to$8 line from February, May, and June 2014, the angular resolution is not good enough to fully resolve the binary and the location of AlO emission with respect to Mira\,A alone cannot be investigated in detail. Nevertheless, simple single-component fits to AlO emission and to the combined continuum of Mira\,A and B\footnote{The cumulative continuum flux is largely dominated by the A component as their submm flux ratio is 13.5 \citep{planesas}.} imply that the AlO emission in June 2014 was located 30$\pm$4\,mas northeast (PA=44\fdg5$\pm$7\fdg2) off the continuum peak. The size of the AlO emission in this dataset is 149.9($\pm$9.7)$\times$117.5($\pm$12.5)\,mas and the source is elongated along a PA of 129\degr$\pm$7\degr.

In the data from May 2014, at an even poorer angular resolution, the offset is also present and is measured to be 51$\pm$28\,mas at PA=14\degr$\pm$23\degr. The source size is 176($\pm$22)$\times$85($\pm$29)\,mas at PA=94\degr$\pm$12\degr. The angular resolution and S/N of the data from February are too poor to put any reliable constraints on the size and location of AlO.

It can be concluded that the general distribution of the AlO gas around the stellar disk did not change considerably over the entire period of the ALMA observations discussed here (29 Oct. 2014--21 July 2015) with the net emission being strongest at 30--50\,mas (2.0--3.5\,R$_{\star}$) northeast from the star and with its longer axis extending approximately along a PA of 90\degr--130\degr. The typical size of the emission region is (120--175)$\times$(75--120)\,mas. 

Among all the species observed with ALMA longest baselines so far \citep{wong}, only one weak line of SO ($^3\Sigma^-$ $7_8-7_7$ with $E_u$=81\,K near 214.33\,GHz) exhibits a morphology similar to that of AlO. The similarity of AlO emission to the spatial distribution of SO might suggest a similar excitation mechanism.   

\subsection{Analysis of the spectral profiles of AlO}\label{sect-AlOprof}

The spectral resolution of the ALMA SV data of the $N$=6$\to$5 line is rather poor and the entire line is only covered by seven resolution elements. Nonetheless, it still reveals interesting information about AlO around Mira\,A. We extracted spectra from the continuum-subtracted cube within three apertures with the size of the synthesized beam centered at ({\it i}) the absorption minimum, ({\it ii}) maximum emission east from the stellar disk and ({\it iii}) north from it. The absorption trough and eastern emission reach their extrema at exactly the same velocity of 53.4$\pm$0.5\,\kms, while the northern emission peaks at 42.4$\pm$0.4\,\kms. We next find that in a net spectrum averaged over the entire AlO(6--5) region, i.e. including emission and absorption, the profile is centered at the same velocity as the northern emission, 42.0$\pm$0.5\,\kms\ (the errors given here are 1$\sigma$ uncertainties of a Gaussian fit). The shift between the bulk of emission and the northern emission with respect to absorption and eastern emission of 11\,\kms\ is much smaller than one resolution element of 20.4\,\kms. With such a coarse binning and our modest S/N, these results are tentative, but we believe that there is an actual shift. 


Our highest-quality spectrum of the $N$=9$\to$8 line is the one acquired with ALMA on 12 July 2015 and is shown in the top panel of Fig.\,\ref{fig-AlO-profile}. The spectral resolution and S/N ratio are sufficient to observe a multi-peak structure of the AlO profile imposed by the numerous overlapping hyperfine components. The hyperfine structure and intrinsic broadening form a profile with FWHM of 45\,\kms. We performed a simple simulation of the profile where each hyperfine component was represented by a Gaussian of intensity proportional to the component's strength, $S_i$.  The central velocity was derived by cross-correlating the spectrum with the simulation in the {\tt rv.fxor} task of IRAF and additional $\chi^2$ model testing was performed in CASSIS\footnote{\url{http://cassis.irap.omp.eu/}}. This yielded a radial velocity $V_{\rm LSR}$=46.9$\pm$0.1\,\kms\ and intrinsic line width (FWHM) in the range 8--14\,\kms. The simulation reproduces the observed profile satisfactorily as shown in Fig.\,\ref{fig-AlO-profile}. The central velocity of $N$=9$\to$8 is consistent to within the uncertainties with the center-of-mass velocity of Mira of $V_{\rm LSR,sys}$=46.8$\pm$0.5\,\kms.

All other ALMA and APEX spectra of the $N$=9$\to$8 transition have too poor spectral resolution or S/N for a detailed analysis. Nevertheless, their positions and widths are generally consistent with what we have derived from the ALMA 2015 data. Two of the profiles are shown in Fig.\,\ref{fig-AlO-profile}. The shape of the profile from combined ALMA observations from February and May 2014 displays sub-peaks that appear sharper than in 2015. This may indicate that the intrinsic broadening was smaller in 2014. 


\subsection{Interpretation} 
The AlO observations can all be understood in a scenario in which inhomogeneously distributed AlO gas is infalling on the star: the strong northern emission component of $N$=6$\to$5 has nearly the same radial velocity as the star because it is located primarily in the plane of the sky and its motions are dominated by the tangential component. The redshifted absorption must be in front of the stellar disk and provides the strongest support for the infall interpretation; the eastern emission of AlO 6$\to$5, which also seems to be redshifted, must then be located closer to us than the star. However, we cannot entirely exclude here, that the velocity field may show an irregular pattern and that the individual clumps move randomly, with some moving out from the star while others fall on it. Then, the location of the emitting regions along the line of sight cannot be defined by the radial motions alone. (The redshifted absorption would, invariably, indicate infall motions in the part of the envelope seen against the stellar photosphere.)


The large broadening of 8--14\,\kms\ is likely related to the projected velocity dispersion within the AlO envelope. The maximum radial velocity probed by the emission in a spherically symmetric infalling (or outflowing) envelope of a radius $r$ seen partially against a stellar disk of a radius $R_{\star}$ is $V_{\rm r,max}=V_{\max} \sqrt{1 - \sin^2(R_{\star}/r)}$. The expected projected velocity dispersion is twice that figure. We observe the AlO emission at radii of 3--5.5\,$R_{\star}$ (Sect.\,\ref{AlOmap}) for which the velocity dispersion simplifies to $V_{\rm r,max}\approx V_{\rm max}$. This means that the observed maximum radial velocity is already a good measure of the maximum deprojected velocity. From this, we imply that the AlO gas captured by the observations is most likely infalling on Mira with a typical velocity of about 4--7\,\kms. If we are only seeing random clumps which do not reach the maximum possible velocity allowed by the dynamic process responsible for the motion, then the range of 4--7\,\kms\ is only a lower limit on the maximum velocity. We note, however, that the detailed model of Mira's envelope of \citet{wong}, constrained by  high-quality ALMA data, also includes infall motions with an amplitude of 7\,\kms.

\subsection{Excitation analysis and abundance determinations}\label{sect-excitation}
\begin{figure}
\centering
\includegraphics[angle=0,width=0.99\columnwidth]{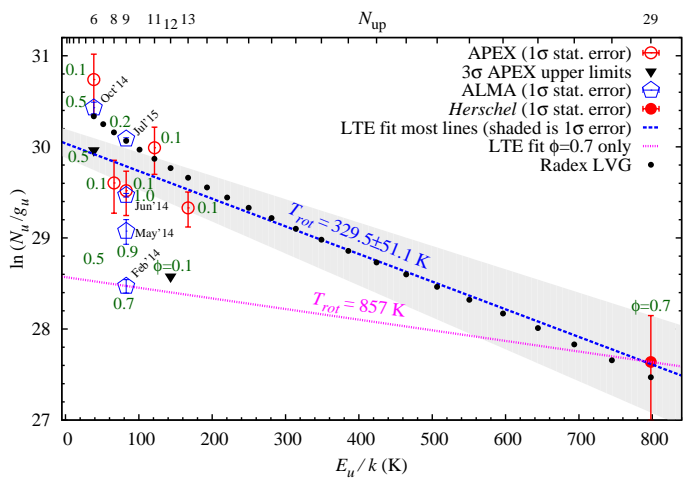}
\caption{Population diagram for AlO. Circles and pentagons respresent detected lines, while triangles mark upper limits. The dashed blue line is the LTE linear fit to most of the detected lines (see text). The dotted magenta line is a fit to the two points with $\varphi$=0.7. Black filled circles correspond to a non-LTE RADEX model with nearly the same parameters as constrained in the LTE fit and with H$_2$ density of 10$^9$\,cm$^{-3}$. Numbers in green show the visual phase corresponding to the datapoints.}\label{fig-rotdiagr}
\end{figure}

The physical conditions of the AlO-bearing gas were constrained by using a population-diagram \citep{popdiagr}. All our measurements, corrected for the beam filling factor, are shown in Fig.\,\ref{fig-rotdiagr}. The scatter in point positions is large even for a single transition which was measured several times. We interpret this scatter as a consequence of true variability of the emission and discuss it in more detail in Sect.\,\ref{sect-var}. First, we treat the upper limits and the two measurements of $N$=9$\to$8 with the lowest fluxes, both from the Ramstedt et al. dataset, as outliers and will ignore them for now. We obtain a linear fit to the remaining points using a weighted least-square regression procedure. This yielded the excitation temperature of $T_{\rm ex}$=329$\pm$51\,K and a column density of $N$(AlO)=(4.0$\pm$0.6)$\times$10$^{15}$\,cm$^{-2}$. The linear fit is shown in Fig.\,\ref{fig-rotdiagr}. 

The derived $T_{\rm ex}$ and $N$(AlO) are subject to many uncertainties \citep[cf.][]{mangum}. In our simple excitation analysis, it was assumed that the lines are optically thin and arise in isothermal gas under local thermodynamic equilibrium (LTE) conditions. To correct the measurements for beam filling factors, we assumed a source size equal to a circular Gaussian of 0\farcs12 FWHM. This is the typical size we found in Sect.\,\ref{AlOmap}. 

An important difficulty in analyzing the AlO data is a possible violation of the Boltzmann distribution in level populations. Even if only the APEX data collected before 2015 are taken into account, it is clear that under thermal equilibrium the $N$=12$\to$11 line ($E_u$=143.3\,K) should have been detected significantly above the noise level. We can reject any possibility that this line was not detected owing to instrumental or data-reduction problems. 
Because the observation of $N$=12$\to$11 was nearly simultaneous with the observation of the 11$\to$10 line, which has a similar value of $E_u$ and which was detected with the same instrument, the only explanation of the observations is that the excitation of the AlO gas violates the assumption of LTE. The population of rotational levels may be strongly influenced by optical/IR excitation. This scenario is supported by the presence of AlO bands in optical spectra of Mira and their irregular variability (Sect.\,\ref{opt}). In the face of these arguments, the excitation temperature we obtain is ill-defined and does not have to be equal to the local kinetic temperature. Indeed, for the bulk of AlO gas located within a radius of $\sim$70\,mas (Sect.\,\ref{AlOmap}), models of Mira's envelope predict a kinetic temperature above about 900\,K \citep[][{\tt o54} models ]{ireland2011} or above 570\,K \citep{wong}.  

We tested some of the assumptions of our population-diagram analysis by performing additional calculations with the radiative-transfer code RADEX \citep{radex} which applies the escape-probability formulation to solve the radiative transfer problem in one dimension under non-LTE conditions (statistical equilibrium is assumed). It still assumes a homogeneous and isothermal medium, but takes into account the line opacity effects. We used it in a mode that is equivalent to the large-velocity-gradient (LVG) method for a spherical envelope. To account for collisions of AlO with molecular hydrogen, we used the scaled collision-rate coefficients derived for SiO and He \citep{collRates} in the form in which they appear in the LAMDA database\footnote{\url{http://home.strw.leidenuniv.nl/~moldata/}}. All the observed AlO lines become thermalized for hydrogen densities exceeding $\log(n/{\rm cm}^3)$=8. At higher H$_2$ densities and at the kinetic temperature and column density derived from the population-diagram analysis, we find that all the observed lines are optically thin with an optical depth of $\tau$=0.1--0.4. 
In Fig.\,\ref{fig-rotdiagr}, we present a RADEX model with a slightly higher column density, $N$(AlO)=6$\cdot$10$^{15}$\,cm$^{-2}$, which fits the fluxes of our best ALMA measurements very well. 

A severe shortcoming of the rotational-diagram analysis is the sparse coverage of $E_u$, especially for higher rotational transitions. Our temperature fit relies strongly on the {\it Herschel} measurement at the high-energy end. Without this single observational point, the excitation temperature would be poorly constrained (194$\pm$143\,K) for the selected measurements. Furthermore, if we assume that the line fluxes change regularly with phase, the {\it Herschel} measurement at $\varphi$=0.7 does not match most of the other considered points collected mainly close to the visual maximum. There is only one other measurement of AlO emission at $\varphi$=0.7 which was obtained with ALMA in February 2014 and was discarded in our earlier analysis because its fluxes were significantly below all other measurements. Combining the two measurements yielded an excitation temperature of 857\,K and a column density of 2$\cdot$10$^{15}$\,cm$^{-2}$. While the column density is practically the same as that obtained in our earlier approach, the excitation temperature is significantly higher and is closer to the kinetic temperature expected in a region near radii of 3\,R$_{\star}$. If we now naively assume this is the actual excitation temperature at phase 0.7 and the temperature does not change significantly with phase, one needs two orders of magnitude higher column density of AlO to explain most of the measurements obtained near maximum light in the mm and submm range. This interpretation of our data would indicate significant changes in the amount of AlO with phase. We, however, disfavor this scenario because the line variability is not correlated with phase (see below) and hydrodynamic models predict drastic changes in gas temperature close to the star \citep[e.g.][]{ireland2011}. 

\paragraph{Abundance} The derived AlO column density should be corrected for the effect from absorption (less than 5\% of the derived $N$(AlO)) and increased by AlO residing behind the radio photosphere (at most 10\%). These corrections are much smaller than the uncertainties in the estimated value of $N$(AlO). We use the value of $N$(AlO)=5$\cdot$10$^{15}$\,cm$^{-2}$, as the most representative, i.e. time- and area\footnote{Meaning the area in which $S_{\nu}$ is above the 3$\sigma$ noise level.}-averaged column density. Its full uncertainty -- including that of the source size, actual temperature, radiative transfer details, data processing -- is at least an order of magnitude. A calculation of a realistic abundance of AlO from this figure requires knowing the distribution and density of hydrogen. This information is not available from observations. It is even unknown whether hydrogen is mainly in the molecular or atomic form \citep[see discussion in][]{wong}. One can attempt to calculate the abundance using the density profiles of theoretical and semi-empirical models of Mira's environment but those usually assume spherical symmetry and differ in the absolute density scales at a given radius by orders of magnitude. For instance, for the hydrodynamic CODEX models of the {\tt o54} series\footnote{This series was constructed to fit the physical parameters of $o$\,Ceti. We converted the mass densities to number densities assuming a composition of 30\% He and 70\% of H$_2$. The H$_2$ column density was calculated for twice the distance from the radio photosphere to 90\,mas.} of \citet{ireland2011}, we get average H$_2$ column densities of the order of 10$^{22}$--10$^{24}$\,cm$^{-2}$ which give an AlO abundance of the order of 10$^{-7}$--10$^{-9}$. Referring to the semi-empirical model of \citet{wong}, we obtain an AlO abundance of 5$\cdot$10$^{-10}$. These values are significantly lower than the cosmic elemental abundance of aluminum with respect to hydrogen of 3$\times$ 10$^{-6}$ (or twice that value with respect to molecular hydrogen). If our calculations are correct, a significant fraction of Al, at least $\sim$90\%, is locked in species other than AlO.  

\subsection{AlO variability at submm wavelengths}\label{sect-var}
It should be noted that our excitation analysis has the complication that we combined data from different epochs but we have provided proof that the emission is likely variable at least in some of the lines. The current data suggest that the emission region does not change much in shape and size on a timescale of years, but the absolute flux in some lines vary considerably. This is best illustrated in the population diagram in Fig.\,\ref{fig-rotdiagr}. The magnitude of the variability indicated by the spread in the measurements of each of the $N$=6$\to$5 and 9$\to$8 transitions is a factor a few or more, which can strongly influence the temperature determination. It is not known whether the variability is due to changes in the excitation temperature or the column density of AlO.

The variability of AlO lines may be a phenomenon related to that observed in SiO masers in Mira variables. The two lowest rotational lines of SiO $\varv$=1 and $\varv$=2 near 43 and 86\,GHz show variability that correlates closely with phase. The maximum maser emission occurs near visual phases 0.05--0.2 (0.13 for $o$\,Ceti) and thus coincides with the maximum of infrared light curves \citep{Pardo}. This correlation is likely to be related to radiative pumping through vibrational bands of SiO in the near-infrared \citep{masers1,masers2,Pardo}. The SiO amplitude  changes from cycle to cycle, in particular for $o$\,Ceti. Higher rotational lines of SiO at mm and submm wavelengths vary strongly with phase too.  In the first systematic study of this phenomenon, \citet{SiOvar} found some lines to disappear completely for phases in a range 0.4--0.7. Interestingly and contrary to SiO masers, radio maser emission of OH seems to show variability that anti-correlates with optical light variations \citep{GerardBeougois1993} so there may be different phenomena driving variable line emission. 

Our submm observations of AlO provide poor coverage of the stellar variability phase (see Table\,\ref{tab-alo}) because most of the observations were conducted near optical maxima, i.e. near visual phase 0.1. All but one spectrum obtained near this phase resulted in an AlO detection, but the two single-dish observations obtained at a later phase near 0.5 did not show AlO emission. At still later phases of 0.7 and 0.9, the very sensitive ALMA observations showed lower intensity than measurements at phases 0.1--0.2. One can claim that there is a general trend of AlO intensity decreasing with phase which would correspond to what is observed for the mm/submm masers of SiO. However, the pattern of AlO variability in Mira seems very erratic. There are multiple exceptions that differ from the above trend: ({\it i}) the $N$=29$\to$28 line observed with {\it Herschel} was detected at a very late phase of 0.7 (visual light minimum); ({\it ii}) the $N$=12$\to$11 line was not detected although it was observed near infrared maximum (phase 0.1) simultaneously with other AlO lines which were detected; and ({\it iii}) the $N$=6$\to$5 line observed with ALMA and APEX shows very consistent intensities although probed at phases of 0.1 and 0.5. Also, the 9$\to$8 line observed in August 2014 ($\varphi$=0.1), June 2014 ($\varphi$=0.0), and July 2014  ($\varphi$=0.1) shows excellent agreement although it was measured in different cycles. For comparison, the SiO maser lines hardly ever have the same flux in different cycles. 

Although we took  every precaution to use the best possible data, our picture of the variability in AlO transitions may be slightly distorted by instrumental and calibration issues. Better phase coverage and observations in multiple cycles are needed to obtain more conclusive variability studies of submm AlO lines in Mira. However, erratic variability is also observed in the optical bands of AlO discussed in Sect.\,\ref{sect-optAlO-var}.

\subsection{Identification and analysis of AlOH}\label{sect-aloh}
Rotational lines of AlOH were measured accurately in the laboratory \citep{AlOHlab} but it has only been identified in the circumstellar envelope of the red supergiant VY\,CMa \citep{ref-AlOH,kami_surv}. Our APEX observations covered three transitions of AlOH ($J$=11$\to$10, 13$\to$12, 15$\to$14) and the collected {\it Herschel}/HIFI spectra covered five more transitions (18$\to$17, 21$\to$20, 22$\to$21, 33$\to$32, and 35$\to$34). Only three were found, $J$=13$\to$12, 15$\to$14, and 35$\to$34, at an integrated intensity 3$\sigma$--5$\sigma$ above the corresponding noise levels. The hyperfine splitting in the rotational lines of AlOH is smaller than that of AlO, and the width of the lines does not provide additional support for the identification as it was the case for AlO. Nevertheless, the identification is likely correct as there are no other good candidate carriers that would explain these features in a consistent manner. All AlOH observations are summarized in Table\,\ref{tab-alo}. Unfortunately, none of the sensitive ALMA observations covered AlOH.

Combined into a rotational diagram, the emission features imply a temperature of 1960$\pm$170\,K (assuming a source size of 0\farcs2). Upper limits on the non-detected lines are consistent with this fit. The derived excitation temperature is high but is plausible in this source if the gas is located close to the pulsating photosphere of Mira. The drastically different temperatures for AlO and AlOH gas (330\,K vs. 1960\,K) are rather surprising. In the models of IK\,Tau of \citet{gobrecht}, AlO is converted in AlOH at late pulsation phases close to the star, making then AlOH the prevalent Al-bearing species at large radii. The bulk of gas containing AlOH should therefore appear cooler than the gas abundant in AlO. Our results suggest this is not the case in $o$\,Ceti but non-thermal level population and line variability may influence our analysis of AlOH, just as discussed for AlO in Sect.\,\ref{sect-excitation}. 


Because of the many uncertainties in the AlOH excitation and the unknown spatial origin of its emission, we did not derive the molecular abundances. It is important for Al chemistry in Mira, but a quantitative analysis will require more observations.  

\subsection{Other Al-bearing species in submm spectra}
We searched for other Al-bearing species in the submm spectra of Mira using the standard spectral line catalogs \citep{jpl,cdms2}. The only other Al-bearing molecules observed so far in circumstellar environments are AlCl and AlF. Multiple transitions with $E_u\!>$162\,K of AlCl and $E_u\!>$44\,K of AlF were covered by the most sensitive APEX, ALMA, and {\it Herschel} observations, but none was firmly detected. We also covered but did not detect lines of AlCN, AlNC, and AlS. In Sect.\,\ref{sect-AlH}, we report that AlH was observed in Mira at optical wavelengths, but no pure rotational transitions were covered \citep{AlH}. An observation of this molecule at submillimeter wavelengths is challenging, because the lowest transition near 387\,GHz is within a deep telluric water band. In the current data, AlO and AlOH are the most prominent mm/submm tracers of Al in $o$\,Ceti. 




%
%

\section{Optical spectroscopy of Al-bearing species}\label{opt}
The optical spectra of material surrounding cool evolved stars contain useful information about gas-phase species thought to be important for dust condensation, as has been shown in \citet{kami_alo} and \citet{kami_tio} for VY\,CMa. In the following, we present results of our search of those species in multiepoch spectra of Mira.

Of primary interest is AlO because it is the direct gas-phase precursor of alumina dust, Al$_2$O$_3$ \citep{sarangi,gobrecht}. The electronic $^2\Sigma^+ B-^2\Sigma^+ X$ system of AlO has been reported to show anomalous behavior in optical spectra of Mira variables. Not only do the strengths of the absorption bands change with phase and from cycle to cycle, but the absorption bands were observed to turn into emission features in several sources, including Mira \citep[][and references therein]{keenan,garrison}. 
Here, we investigate the behavior of the optical bands of AlO in Mira using long-term spectroscopic data with the aim of identifying patterns which would allow us to establish the origin of AlO. 
We then used the same observational material to investigate the presence of spectral signatures of other interesting species, including AlOH, AlH, and \ion{Al}{I--II}.

\subsection{Optical observations}
In order to investigate the time variation of the optical bands of AlO in the spectrum of Mira, we performed an extensive search for archival spectra and literature data. We searched through all major public archives for optical spectra covering any sequence of the $B$--$X$ system. Only spectra at high and intermediate resolutions turned out to be useful in our analysis. 
A resolution better than $R\!\approx$5000 was needed to disentangle the bandheads from nearby spectral features of other species and to investigate the shape of the rotational profile of the bands. Consequently, most of the collected data are from echelle spectrographs. The high-resolution spectra we have collected are listed in Table\,\ref{tab-obslog-opt}. A major portion was acquired in spectropolarimetric mode and in these cases the total-power (Stokes $I$) data were analyzed. Whenever possible, we used reduced (`pipelined') data products offered by some of the archives, while in the remaining cases the data reduction was performed manually with IRAF using standard reduction techniques \citep[e.g.][]{echellereduction}. Particular care was taken to obtain good wavelength calibration. Exposures from the same night were averaged. Most of the spectra in our sample are not calibrated in flux. Although most of the echelle spectra were corrected for a blaze response function, the slope of the pseudo-continuum in most of the spectra is unknown at wavelength ranges covering multiple orders. Visual magnitudes for each observed spectrum are available from the densely-spaced observations of the American Association of Variable Star Observers (AAVSO). The nearest measurement for the given date is included in Table\,\ref{tab-obslog-opt}, as well as information on the spectral coverage (ignoring gaps in some of the echelle spectra) and resolution, references to papers describing the observing run or the name of the principal investigator of the observing project if no literature reference could be identified.

%
\begin{sidewaystable*}
\caption{High-resolution optical spectra of Mira collected from archives.}\label{tab-obslog-opt}
\centering
\begin{tabular}{cc cc cc cc cc cc}
\hline\hline
Instrument                &Telescope/observatory   & Date	   & JD        &Nominal& \multicolumn{2}{c}{Spectral coverage}&Vis. & Vis. & Reference	  &	Notes\\
                          &			               &  	       &	       & $R$   & start (\AA)& end (\AA)& phase & mag  & or project PI&		\\
\hline\hline
McKellar Spectrograph     &Dominion Obs. 1.2m     & 1965-12-14 & 2439108.7 &     21000&	  3900 &      5100 &     0.94 & 3.5 &  PI: E. Griffin  & \tablefootmark{a}\\
ELODIE                    &OHP 		       	      & 1998-11-07 & 2451124.0 &     42000&	  4000 &      6800 &     0.89 & 7.0 &  1	  & \tablefootmark{b}\\
Echelle Hi-Res Spectrogr. &Mt. Ekar 182 cm, Asiago& 1999-02-11 & 2451221.2 &     20000&	  4600 &      9470 &     0.18 & 6.0 &  2	  & \tablefootmark{c}\\
ELODIE                    &OHP 		       	      & 1999-09-29 & 2451450.0 &     42000&	  4000 &      6800 &     0.86 & 7.0 &  1	  & \tablefootmark{b}\\
ELODIE                    &OHP 		       	      & 1999-12-16 & 2451528.9 &     42000&	  4000 &      6800 &     0.10 & 3.4 &  1	  & \tablefootmark{b}\\
HIDES                     &Okayama-NAOJ 188cm     & 2001-08-29 & 2452151.3 &     95000&	  4620 &      5820 &     0.97 & 3.2 &  3      &				\\
FEROS                     &ESO MPI 2.2 m	      & 2004-10-02 & 2453280.3 &     48000&	  3600 &      9200 &     0.36 & 7.2 &  4	  & \tablefootmark{d}\\
SARG                      &TNG 		       	      & 2005-12-11 & 2453716.4 &     11800&	  5000 &      8000 &     0.67 & 9.0 &  PI: E. Verdugo	  & \tablefootmark{e}\\
ESPaDOnS                  &CFHT 3.6m		      & 2006-08-14 & 2453962.1 &     68000&	  3690 &     10480 &     0.41 & 7.2 &  PI: Dinh-V.-Trung   & \tablefootmark{f}\\
SOPHIE                    &OHP 1.93m		      & 2007-01-12 & 2454113.3 &     75000&	  3870 &      6940 &     0.86 & 5.6 &  PI: D. Gillet	  & \tablefootmark{b}\\
SOPHIE                    &OHP 1.93m		      & 2007-01-20 & 2454121.2 &     75000&	  3870 &      6940 &     0.89 & 4.3 &  PI: D. Gillet	  & \tablefootmark{b}\\
SOPHIE                    &OHP 1.93m		      & 2007-01-30 & 2454131.2 &     75000&	  3870 &      6940 &     0.92 & 3.3 &  PI: D. Gillet	  & \tablefootmark{b}\\
ISIS                      &WHT, INT la Palma	  & 2007-07-31 & 2454312.7 &      4500& 3650-4560 & 6630-7300 &  0.46 & 8.5 &  PI: F. Leone	  &\tablefootmark{f}\\
MMCS                      &ZEISS-2000, Terskol Astr. Obs.& 2007-08-15 & 2454327.5 &     13000&	  3970 &      7485 &     0.51 & 8.3 &  PI: O. Andriyenko   &				\\[5pt]
NARVAL                    &2m TBL, Pic du Midi           & 2007-09-04 & 2454348.7 &     65000&	  3700 &     10480 &     0.57 & 8.8 &  5	  & \tablefootmark{f}\\
                          &			      & 2008-01-20 & 2454486.3 &	  &	       & 	   &     0.98 & 3.6 &				\\
                          &			      & 2008-02-10 & 2454507.3 &	  &	       & 	   &     0.05 & 4.2 &				\\
                          &			      & 2008-08-29 & 2454708.7 &	  &	       & 	   &     0.65 & 9.2 &				\\
                          &			      & 2009-02-26 & 2454889.3 &	  &	       & 	   &     0.19 & 5.1 &				\\
                          &			      & 2009-07-24 & 2455037.6 &	  &	       & 	   &     0.64 & 9.2 &				\\
                          &			      & 2009-09-23 & 2455098.5 &	  &	       & 	   &     0.82 & 8.8 &				\\
                          &			      & 2009-10-03 & 2455107.4 &	  &	       & 	   &     0.85 & 8.0 &				\\
                          &			      & 2009-10-25 & 2455130.5 &	  &	       & 	   &     0.92 & 4.7 &				\\
                          &			      & 2009-11-24 & 2455160.3 &	  &	       & 	   &     0.01 & 3.5 &				\\
                          &			      & 2009-12-10 & 2455176.3 &	  &	       & 	   &     0.05 & 4.0 &				\\
                          &			      & 2009-12-20 & 2455186.3 &	  &	       & 	   &     0.08 & 4.2 &				\\
                          &			      & 2010-01-06 & 2455203.4 &	  &	       & 	   &     0.14 & 4.6 &				\\
                          &			      & 2010-01-18 & 2455215.3 &	  &	       & 	   &     0.17 & 4.7 &				\\
                          &			      & 2010-02-10 & 2455238.3 &	  &	       & 	   &     0.24 & 5.4 &				\\[5pt]
HARPS                     &ESO 3.6m		  & 2012-08-09 & 2456148.4 &    115000& 3780-5300  &5340-6910  &     0.97 & 3.7 &  PI: F. Leone	  & \tablefootmark{f}\\
\hline
\end{tabular}
\tablebib{
(1)~\citet{elodieobs};
(2)~\citet{Asiago};
(3)~\citet{hides};
(4)~\citet{ferosobs};
(5)~\citet{narval}.
}
\tablefoot{
\tablefoottext{a}{Scanned plate.}
\tablefoottext{b}{Archive product.}
\tablefoottext{c}{Interorder gaps above 6880\AA.}
\tablefoottext{d}{Interorder gaps above 8530\AA.}
\tablefoottext{e}{Underexposed in blue part.}
\tablefoottext{f}{Spectropolarimetric observation.}
}
\end{sidewaystable*}

The oldest spectra we were able to access in digitized form are scanned spectral plates from the Dominion Astronomical Observatory. In addition to these, we found reproductions (scanned GIF files of paper copies) of Mira spectra covering the AlO band in \citet{joy} (later reproduced in \citet{bax} and \citet{merrill62}) and \citet{keenan}, which show two-dimensional plates. Additionally, \citet{garrison} showed an extracted one-dimensional spectrum. Further useful information on the appearance of AlO bands in Mira is found in \citet{keenan}, \citet{kipper}, and \citet{garrisonAbstract}. 

\begin{figure}
\centering
\includegraphics[angle=270,width=0.99\columnwidth]{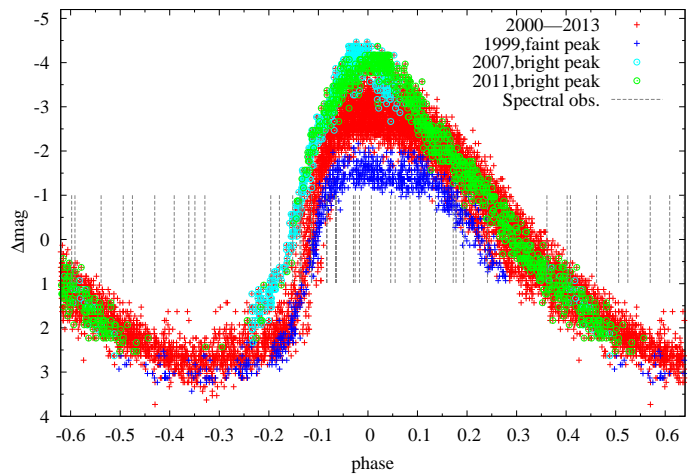}
\caption{Phased AAVSO light curve of Mira in 1999--2013. The faint-maximum cycle of 1999 and the two recent cycles with bright maxima (2007 and 2011) are shown with blue, green, and cyan symbols, respectively. The dates of spectroscopic observations (Table\,\ref{tab-obslog-opt}) are marked with verical bars. For some dates, the corresponding bars appear twice in the graph, i.e. at $\varphi$ and $\varphi \pm 1$.}\label{fig-phase}
\end{figure}

Table\,\ref{tab-obslog-opt} also lists the phase corresponding to the given observing date. Although the shape of the Mira's visual light curve deviates strongly from a sinusoidal pattern and its period is known to change by a few days on time scales of years \citep{period}, we use a simple sine function to define the phase for data since 1999, with a period of 333 days \citep[cf.][]{period} and date of maximum on JD=2452161.0. The date of maximum was measured for the 2001 cycle which showed a very well defined maximum. Such a parametrization of the stellar periodicity is also consistent with our sinusoidal fit to the AAVSO visual data for 1999--2013. The shape of the phased light curve within that time span together with the markings representing the coverage of the collected visual spectroscopy (Table\,\ref{tab-obslog-opt}), are shown in Fig.\,\ref{fig-phase}. The maxima occur close to phase $\varphi$=0, while -- due to an asymmetry of the actual light curve -- the minima occur at $\varphi\!\approx\!-0.33$=0.67. Using $JHKL$ photometry from \citet{shenavrin} and \citet{whitelock}, we found that the near-infrared fluxes ($JHKL$) peak at $\varphi=0.17\pm0.05$ on our periodicity scale.  

We also searched through the archives for near-infrared spectra of Mira that covered the $A$--$X$ band of AlO \citep[cf.][]{kami_v1309,baner}. We found only one spectrum with a sufficient spectral resolution, $R$=37\,600, that was obtained on 26 August 2001 ($\varphi$=0.96) with the NIRSPEC instrument on Keck by A. Nelson. A simulation of the $A$--$X$ system by a similar method as in \citet{kami_v1309} revealed that the (4,0) band with heads at 1225 and 1243\,nm is the strongest near-infrared band for a broad range of excitation temperatures. We reduced and inspected only the part of the Keck spectrum that covers the (4,0) band. We found no signatures of the electronic system of AlO. However, it should be noted that the spectrum was acquired only three days before optical observations with HIDES in which the AlO bands of the $B$--$X$ system were very strong. This leads us to a conclusion that the NIR system is not particularly useful for studies of AlO in $o$\,Cet. Indeed, the NIR band is expected to be much weaker than the optical bands \citep{exomolAlO}. 

\subsubsection{Observed variations in the B-X band of AlO}\label{sect-optAlO-var}
\begin{figure*}
\centering
\includegraphics[angle=270,width=0.79\textwidth]{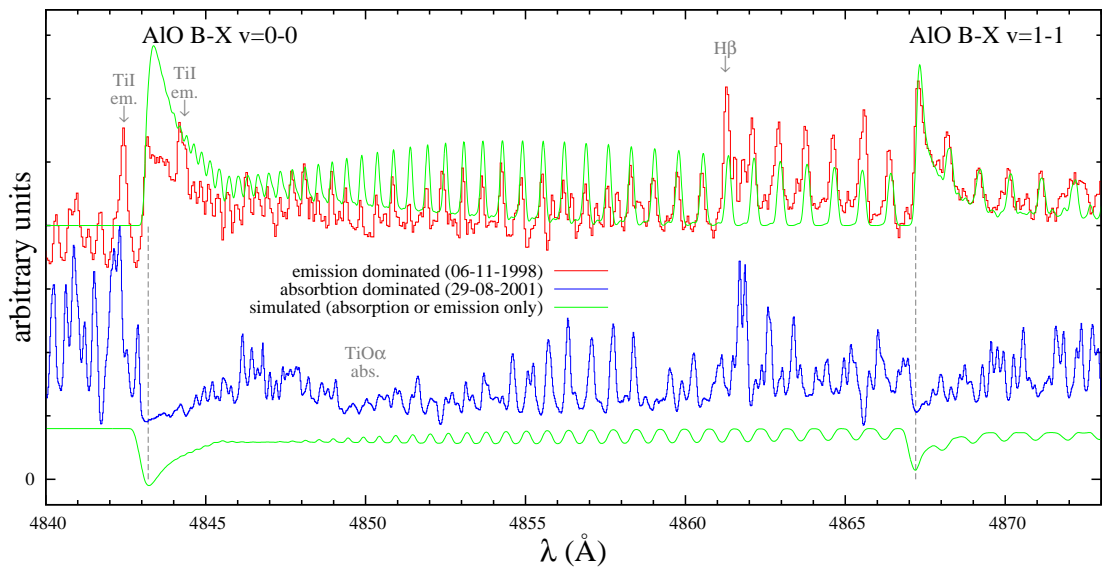}
\caption{Sample optical spectra of AlO in two epochs (red and blue), covering the heads of the (0,0) and (1,1) bands and their rotational combs. Simple simulations (green) illustrate the location of the emission and absorption components. Many spectral features are of atomic origin. The reason for the poor correspondance of the observation and simulation at the (0,0) bandhead of AlO is the lack of an underlying absorption feature in the similation and a limited correction for optical thickness, which is highest near the head. All spectra covering these and other bands of AlO at different dates are shown in Figs.\,\ref{fig-AlOp2}--\ref{fig-AlOm1}.}\label{fig-AlOband}
\end{figure*}

We were able to identify features in the AlO $B$--$X$ system involving the vibrational levels up to at least $\varv_{\rm up}$=3 in the upper electronic state in all the collected high- and mid-resolution spectra of Mira. The main bandheads appear in absorption, occasionally in emission, and as a combination of absorption and emission features. Examples of emission- and absorption-dominated spectra are shown in Fig.\,\ref{fig-AlOband}. Independent of whether they appear in emission or in absorption, the bandheads are rather weak, but the characteristic shape of the band is readily recognizable in the spectra. 


The collected spectra are sufficient to study the long-term variability within the bands which is not feasible with the current mm and submm data. Results of this analysis are presented separately for spectra dominated by absorption and emission in Sects.\,\ref{sect-opt-abs} and \ref{sect-opt-abs}. All spectra covering the AlO sequences $\Delta\varv$=2, 1, 0, --1 are shown in the spectral atlas in Figs.\,\ref{fig-AlOp2}--\ref{fig-AlOm1}.

\subsubsection{Absorption in AlO}\label{sect-opt-abs}

The $\Delta\varv$=$-2$ sequence, located near 5340\,\AA, and all negative sequences at longer wavelengths -- although covered by many of our spectra -- are too weak for variability studies. The $\Delta\varv$=$-1$ sequence near 5080\,\AA\ is dominated by absorption features which show definite variability. Episodes when AlO absorption becomes weaker than normal were identified. The weakest absorption band was observed on 2012-08-09. The weakening of those bands is not correlated with phase or time. The apparent weakening of those absorption features may be due to an overlying AlO emission component of variable intensity. A presence of such a component is evident in the higher progressions, $\Delta\varv$=0 and 1. Bands within the $\Delta\varv$=0 sequence are weak but still definitely present.

We used the spectra of the $\Delta\varv$=--1 sequence, which is longward of 5080\,\AA\ and is not contaminated or just weakly so by emission, to measure the representative velocities of the absorbing AlO gas. First, we measured the velocity of the spectrum from 2010-02-10 by cross-correlating it with a simulated spectrum of the absorption. The cross-correlation algorithm {\tt rv.fxcor} implemented in IRAF was used \citep[e.g.][]{fxcor} and the simulated spectrum was calculated in an analogous way as in \citet{kami_alo}. 
The typical velocity for our sample of spectra is 57.5\,\kms\ 
with the highest measured  deviations from this value of +2.0 and --1.4\,\kms, and a standard deviation of 1.0\,\kms. The phases corresponding to the spectra densely cover the full pulsation cycle and hence $V_h$=57.5$\pm$2.0\,\kms\ characterizes the full dispersion well. The central velocity is consistent with the center-of-mass velocity of Mira ($V_{h,\rm sys}$=57.0$\pm$0.5\,\kms) and the central velocity of the submm emission of AlO in pure rotational lines ($V_{h,\rm sys}$=57.1\,\kms, Sect.\,\ref{sect-AlOprof}). The variations in velocity that we measured are small but greater than our uncertainties. 

A rare event recorded in the AlO spectra is the appearance of extra absorption components that are strongly redshifted. This is observed in the (1,0) and (2,1) heads of $\Delta\varv$=+1 in 24 July, 23 September, and 3 October 2009 (but not 23 days later) -- see Fig.\,\ref{fig-AlOp1}. In the same spectra, the (0,0) band is unusually broad, the (0,1) bands are slightly broader than usual, and there are no unusual changes in the other sequences. The extra absorption components therefore involve only $\varv_{\rm low}$=0 and 1 levels, suggesting that they arise in material cooler than that producing usual absorption bands of AlO and observed in progressions involving higher-$\varv_{\rm low}$. Similar apparent broadening or splitting of spectral features is simultaneously observed for other species, including \ion{Al}{I} (Sect.\,\ref{sect-al}). In the profile of the AlO\,(2,1) band, we measured that, in addition to the usual component very close to stellar center-of-mass velocity, there are extra components at heliocentric velocities of about +95\,\kms\ and possibly at +140\,\kms. Such high velocities reaching 80\,\kms\ with respect to the star are surprising, but not impossible in this source. The motions are likely not associated directly with the usual pulsation-driven shock wave, because they are absent in most of the cycles covered by our spectra. 

The usual absorption profiles of AlO also show considerable variability in their widths. We infer the intrinsic Doppler broadening as the kernel width that we used to convolve our simulated spectrum to fit the observed features. When the bands are strong, their half-width is nearly 30\,\kms, as in 2001-08-29 ($\varphi$=0.97). Such a high velocity dispersion in Mira can only be explained as arising in the shock wake. Macroturbulence is expected to be greatly enhanced behind a shock \citep[cf.][]{macroturbulance}, however, the broad profile may also be an opacity effect which we were not able to simulate with our simple tools. In a great majority of our spectra the absorption is narrower, often comparable to the instrumental profile of a few \kms. (It cannot be measured accurately owing to a much lower S/N of these narrow and weak features.) This may indicate that the AlO absorbing gas is present both in the shocked gas and in material that had time to dissipate the shock energy.

We could not constrain precisely the rotational temperature of the gas from the shapes of the bands. The major difficulty is how to define the level of the local pseudocontinuum underlying the bands. Another obstacle is the presence of numerous atomic absorption features that distort the band shape. Our simulation of the bandheads are nevertheless consistent with the observations within a broad range of rotational temperatures from a few hundred to $\sim$1500\,K. The vibrational temperature, although variable, is likely equal to a few hundred K.

\paragraph{Spatial origin of AlO absorption}
Because of the complex behavior of the AlO features, we shall consider several locations of the absorbing AlO gas. If the absorption was produced in the accelerated wind seen in front of the stellar disk, the features would be blueshifted by the magnitude of the wind terminal velocity, i.e. by --5\,\kms\ from the stellar velocity (Sect.\,\ref{sect-intro-structure}). Such a large offset is not observed. In addition, features arising in a homogeneous, stationary wind would not show temporal velocity variations contrary to what we have measured. Therefore, the AlO absorption features do not come from the fully-accelerated wind, and must arise closer to the star at radii $\lesssim$5\,\rstar\ \citep[cf.][]{Nowotny2010}. 


We also exclude the possibility that the AlO absorption is formed in the photosphere on the basis of the low amplitude of the velocity variations in the AlO features. Rovibrational lines $\Delta\varv$=3 of CO which are observed in absorption in Mira are believed to be formed in the stellar photosphere. They show velocity variations with an amplitude of 24\,\kms\ \citep{IRveloCO} which are much higher than what we have measured for the AlO absorption bands. Similarly, high-excitation atomic lines show considerable shifts with phase. By cross-correlating the spectra in the region between 4035 and 4095\,\AA, that is dominated by numerous sharp absorption lines, we determined that the relative displacements between different dates reach 9.4\,\kms. (This value is higher than the maximal displacements found by \citet{joy2}, but is smaller than those in \citet{joy}). The amplitude of velocity changes for atomic lines is known to be correlated with the brightness of the cycle's maximum \citep{joy2} and since we covered cycles with very bright maxima (e.g. 2007) as well as faint minima (1998/99), the low amplitude of AlO displacements indicates that the AlO gas is nearly stationary, independent of this inter-cycle  effect. The AlO-bearing gas therefore must be located high enough above the pulsating atmosphere to not be influenced by the regular high-amplitude motions, say at $>$2.5\,\rstar. 


We conclude that the visual absorption features must arise between the photosphere and the fully accelerated wind, or roughly between 2.5 and 5\,\rstar. The kinetic temperature in this region is $\sim$1000\,K (as in the CODEX models of \citealp{ireland2011}) which is consistent with our rough constraints on the rotational temperature of AlO as traced by the $B$--$X$ band. It is very likely that the electronic bands are formed in the same AlO gas as that observed in pure rotational transitions at mm and submm wavelengths. The variable nature of the electronic absorption is consistent with the observation that the AlO gas is clumpy, as seen in the $N$=6$\to$5 line with ALMA. The observed variability can arise as clumps of different opacity pass in front of the stellar disk.

\subsubsection{Emission in AlO}\label{sect-opt-em}
Our spectroscopic records contain two episodes of emission-dominated profiles of AlO on: ({\it i}) 1998-11-06 with strong emission in the $\Delta\varv$=0 and +1 sequences, and ({\it ii}) 2007-09-04 with the strongest emission signatures in the $\Delta\varv$=1 band. Much weaker, but still easily recognizable emission was seen in a sequence of spectra registered immediately after the 1998-11-06 ($\varphi$=0.9) flare, i.e. on 1999-02-11 ($\varphi$=0.2) and 1999-09-28 ($\varphi$=0.9). Chronologically the next spectrum, from 1999-12-16, shows very weak absorption bands, but the absorption profile is definitive only in the spectrum from 2001-08-29 ($\varphi$=1.0). For the AlO flaring event which was most evident on 2007-09-04 ($\varphi$=0.6), the emission was observed mainly in the $\Delta\varv$=+1 sequence, and was preceded and followed by spectra with emission nearly filling the absorption bands. From these two events, we conclude that bright, easily observable emission in AlO bands may appear at any phase, may last for at least one year, and is not necessarily limited to a single cycle. This behavior is different than that observed in most atomic/ionic lines which show regular changes in intensity with peaks near $\varphi$=0.9--0.1 and minima near $\varphi$=0.6 \citep{richter2001}. The two strong emission episodes took place at phases 0.9 and 0.6, i.e. near maximum and minimum visual light. This erratic variability resembles that of pure rotational lines of AlO observed at mm and submm wavelengths (Sect.\,\ref{sect-var}).

The central velocities of the AlO emission component could only be measured for the 1998-11-06 and 2007-09-04 spectra. On the earlier date, not only the bandheads are useful for velocity measurements but also the individual features of the rotational comb far from the head are very well reproduced by our simulation and constrain the radial velocity (Fig.\,\ref{fig-AlOband}). The velocity of the band is $V_h$=53.8$\pm$0.5\,\kms. The central velocity of the emission seen on 2007-09-04 is less certain but our best estimate is 56$\pm$2\,\kms. The AlO emission therefore appears stationary or blueshifted by at least 3\,\kms\ with respect to the stellar velocity (57.0$\pm$0.5\,\kms). We believe that the underlying absorption is too weak to influence the position measurements we have performed. Although the AlO-emission episodes are too rare to form definitive statements, the kinematical characteristics resemble that observed in most atomic/ionic optical emission lines which also always appear blueshifted \citep{richter2001}. The atomic features are thought to be associated with a shock propagating outward in the stellar atmosphere. Low-excitation atomic lines show low velocity displacements of a few \kms, while emission features requiring high excitation can be shifted by as much as 20\,\kms. We conclude that the visual emission of AlO arises when the shock front reaches the higher parts of the stellar atmosphere where the shock-imposed gas motions are small, thereby explaining the small shift of only 3\,\kms. 



The intrinsic broadening of the emission features is small. To reproduce the rotational comb of the AlO emission observed on 06-11-1998, we convolved the simulated spectrum with a Gaussian of FWHM of 8\,\kms, a value that is comparable to the resolution of the spectrum. The measured width is also consistent with that derived for the submm lines of AlO (Sect.\,\ref{sect-AlOprof}).

\subsubsection{Excitation mechanism of electronic AlO bands in Mira}

The upper electronic state of the $^2\Sigma^+ B-^2\!\Sigma^+ X$ system of AlO is 28\,800\,K above the ground and therefore is most likely excited by shocks. Excitation temperatures of the order of 10$^4$\,K are definitely possible in Miras since strong emission in recombination lines of hydrogen are observed on every cycle. \citet{richter2003} were able to model weak and episodic emission of \ion{Fe}{II} ($E_u$=32\,570\,K) and [\ion{Fe}{II}] (63\,950\,K) on the assumption that the upper levels are excited in the shock at a radius $\sim$1.5\,R$_{\star}$. The gas giving rise to the AlO emission might be excited in a similar way, but probably at slightly higher distances from the star. Also, if AlO emission is associated with the shock, then its presence for more than one cycle, as in 1998/1999, may indicate that the emission arises in gas of very low density that is not able to efficiently cool down only through radiation in a single period \citep[cf.][]{richter2003}. 

Analogous to the shock-excited lines of \ion{Fe}{II}, AlO emission is sporadic. One may speculate that the AlO bands show an emission component only when the shock reaches the AlO material that is usually seen in Mira only in the absorption bands. It is unclear why this occurs only occasionally. The episodic emission in \ion{Fe}{II} and [\ion{Fe}{II}] lines was linked to a shock wave that is stronger than typical and is associated with cycles that show bright maxima \citep{richter2003}. However, as we discuss in Sect.\,\ref{sect-AlO-brightMax}, it is doubtful whether there is a link between the occurrence of AlO emission and bright-maxima cycles. Hydrodynamic pulsation models calculated for long time series show, however, that the positions of shock fronts and the extend of shocked gas change chaotically in Mira stars \citep{irelandErratic}. 

The absorption by AlO is most likely a result of resonant scattering of stellar light off the circumstellar gas. The gas is clumpy (Sect.\,\ref{AlOmap}) and the observed variability may be explained by different clumps passing in front of the stellar disk. Most of the time, however, the gas is stationary with respect to the center-of-mass velocity what suggests it may be located in a region sometimes referred to as a MOLsphere \citep{molsphere} or may correspond to the "warm molecular layer" the existence of which was proposed before e.g. by \citet{WC} and \citet{woitke}. Interferometric observations in the infrared support the existence of such a molecular layer \citep[][and references therein]{wittkowski2015}. The vibrational temperature derived from the AlO bands in \emph{absorption} of a few hundred K is in agreement with such a location. The \emph{emission} episodes would then be associated with events when, for some reason, the volume of AlO gas gets larger than usual so its apparent size is bigger than the stellar disk and scattered emission dominates the band appearance. This scenario is more speculative than direct shock excitation, but if it is correct it would still indicate that AlO is located in the upper atmosphere of Mira.


There must be a difference in the vibrational temperature that characterizes the absorbing and emitting gas because we observe discrepancies in the relative strengths of emission and absorption in different sequences and bands. It is also evident that the excitation conditions of the emitting gas change with time. For instance, on 2007-09-04 when the $\Delta\varv$=+1 bandheads appeared as strong emission features, the corresponding $\Delta\varv$=0, +2 bands showed cancel-out bands. On 1998-11-06, in turn, AlO flared in the $\Delta\varv$=0 and +1 sequences but this time the $\Delta\varv$=0 sequence was by far the stronger one. In other words, the $\varv_{\rm up}$=0--$\varv_{\rm low}$ progression of the emission system peaked for $\varv_{\rm low}$=0 on 1998-11-06, and for $\varv_{\rm low}$=1 on 2007-09-04, suggesting a higher vibrational temperature of the emitting gas in 2007. 



We conclude that the most likely excitation mechanism of the gas that leads to the emission episodes is shocks which occasionally reach the AlO gas which we usually see only in absorption and in the mm/submm rotational lines, and which resides at a few stellar radii from the star. We interpret the appearance of AlO emission as a result of changes in excitation, but changes in abundance (either due to formation or destruction of AlO) and relocation of gas within the envelope cannot be excluded. 



\subsection{Electronic systems of other Al-bearing molecules}\label{sect-AlH}
\paragraph{AlOH} The submm detection of AlOH in Mira (Sect.\,\ref{sect-aloh}) encouraged us to look for electronic transitions of this molecule. We could not find in the literature the spectroscopic constants nor even approximate positions of electronic bands of AlOH in the optical. However, a few electronic systems of AlOH are known in the ultraviolet, i.e. close to 2395\,\AA\ and 2495\,\AA\ \citep{Pilgrim,Li}. Spectra of Mira\,B acquired by the International Ultraviolet Explorer (IUE) between 1979 and 1983 and by the Hubble Space Telescope (HST) in 1995 \citep{Reimers1985,HSTspec} covered this wavelength range and show broad absorption features at the expected location of the AlOH bands. The absorption structures are not identified in the original studies and are too broad to belong to atomic transitions. Moreover, they are unlikely to be blends of known atomic lines as most strong multiplets were identified in the spectra. Furthermore, we find the same wide features in some of the archival UV spectra obtained by HST/STIS (1999 and 2004) and  Galaxy Evolution Explorer (GALEX; 2006), available through the MAST archive\footnote{\url{http://archive.stsci.edu}}. Although both Mira A and B were located in the instruments apertures, it is the B component that was thought to be the major source of the continuum at these wavelengths. If AlOH is the carrier of the broad features, the gas would have to be located between us and Mira\,B so at least 0\farcs48 or 8$\times$10$^{14}$\,cm away from Mira\,A. This would indicate that some aluminum is locked into AlOH at quite large distances from the pulsating atmosphere. To verify this finding will require a dedicated study of the electronic bands of AlOH in Mira. 

\paragraph{AlH} We were surprised to see an identification of an AlH band near 4242\,\AA\ in the historical spectral plates of Mira presented in \citet[][their Figs.\,9 and 10]{merrill62}, because hydrides are generally thought to be rare in spectra of giants. However, AlH has been reported in emission in another Mira variable, $\chi$\,Cyg, near minimum light \citep{AlH-Herbig,AlH-Herbig2}. The emission in the spectral region between 3066\,\AA\ and 4412\,\AA\ in $\chi$\,Cyg was limited to a specific range of rotational lines belonging to the (0,0), (1,0) and (1,1) bands of the $A^1\Pi-X^1\Sigma$ system. 
Many of our spectra of $o$\,Ceti cover these three bands and there is a great number of absorption lines in this spectral region which coincide with the positions of AlH lines in \citet{AlHzawiejaAX}. A detailed simulation of the $A$--$X$ band was obtained in {\tt pgopher} \citep{pgopher} using the spectroscopic constants from \citet{AlHzawiejaAX}. Each band consists of well separated rotational lines and the shapes of individual lines are nearly indistinguishable from atomic features in the observed spectra. The density of lines in this spectral region is very high and, in principle, chance coincidences may occur for many of the lines. We found, however, that for each strong transition of AlH there is a corresponding absorption feature in the spectra of Mira with a perfectly matching radial velocity, leaving little doubt that AlH is present in $o$\,Ceti. The AlH bands appear in absorption in all our spectra and the features seem to be slightly stronger with respect to the local continua near minimum visual light. The AlH bands were not found to turn into emission features in the 19 spectra we examined, indicating that Mira is unlikely to undergo AlH-flaring episodes such as those reported for $\chi$\,Cyg. 

The line positions of AlH vary slightly with time. We investigated the temporal variations quantitatively using cross-correlation techniques and band simulations at different excitation temperatures. The most blueshifted spectrum of the (1,0) band was observed on 2009-07-24 ($\varphi$=0.6) with $V_h$=55.2$\pm$0.7\,\kms\ and the highest velocity of 59.2\,\kms\ was measured in the spectrum from 2009-10-03 ($\varphi$=0.9). Such low variations of the radial motions of $\pm$2\,\kms\ with respect to the center-of-mass velocity, signify that the absorbing gas must be located in the parts of Mira's atmosphere that do not participate in the high-amplitude pulsation motions such as those observed in CO $\Delta \varv$=3 lines \citep{IRveloCO}. The AlH gas may be partially placed in the same layer where we observe AlO in absorption.


We were not able to constrain precisely the rotational temperature of AlH, because ({\it i}) the line ratios are greatly affected by the shape of the local pseudo-continua and ({\it ii}) of the presence of other unidentified lines in these spectral regions. However, our rough estimates suggest a few hundred K, consistent with AlH being located in the outer atmosphere of Mira, i.e. 2.5--4\,R$_{\star}$. 

Another known electronic system of AlH, $C^1\Pi-X^1\Sigma$, is located in the UV range near 2240\,\AA\ \citep{zachwiejaCX}, but we found no evidence for it in the UV spectra of Mira from IUE, HST, and GALEX. 



\subsection{Atomic aluminum}\label{sect-al}
\paragraph{\ion{Al}{I} absorption:} 
We examined the behavior of the resonance doublet $3p^2P$--$4s^2S$ of \ion{Al}{I} at 3944.01 and 3961.5\,\AA\ (i.e. in between the two strong H\&K lines of \ion{Ca}{II}) which is covered by many of our spectra. These are shown in a time sequence in Fig.\,\ref{fig-AlI}. Unlike most lines of neutral atoms in Mira's spectrum, lines of \ion{Al}{I} show strong and very broad absorption profiles. The large width of these lines in $o$\,Ceti and other Mira variables was noticed previously by \citet{joy2} and \citet{merrill62}. The profiles change dramatically in shape and depth over time. Typically, they are triangular in shape and considerably sharper than a Lorentzian profile; the full base-widths are $<$2.5\,\AA, but the actual width is likely higher as the observed pseudocontinuum is below the true continuum level. Often, close to minimum  light ($\varphi\!\sim$0.6), the lines turn into parabolic features which can be as broad as 10\,\AA\ (full width in the stronger $\lambda$3961 line). On occasion (e.g. on 2009-07-24 \& 2009-09-23), they appear as multicomponent irregular absorption features. The blue and red wings show strong asymmetries independent of the overall shape of the \ion{Al}{I} profile whether parabolic or sharp.


In stars of spectral types G--K, the blue resonance lines of \ion{Al}{I} are formed just above the photosphere \citep{Mauas}. No detailed models exist for stars of later spectral types but one might expect that in M-type stars the lines form very close to the photosphere as well. The large apparent width of the lines support this expectation. The \ion{Al}{I} lines are broadened mainly by the radiative and van der Waals effects \citep[cf.][]{Mauas}, where the latter is dependent on the local density. The large width of the observed \ion{Al}{I} lines in Mira requires high local densities such as those expected close to the photosphere. The observed absorption is therefore a signature of photospheric \ion{Al}{I}. The occasional broadening and reshaping of these spectral features may be associated with a  shock passing the photosphere. We were able to reproduce the observed shape of the \ion{Al}{I} features with stationary model atmospheres of MARCS \citep{MARCS} after increasing the continuum opacity in the spectral range investigated\footnote{The extra opacity was introduced to explain the profiles of the H\&K lines of \ion{Ca}{II}. The problem of missing opacity around the lines in G and K giants was identified by \citet{short1,short2}.}, thus assuring us that the \ion{Al}{I} features form close to the photosphere. Although we attempted to derive the \ion{Al}{I} abundances with model atmospheres, we encountered problems in the definition of the true continuum level and all opacity sources in this spectral region. As a result, we were unable to determine a reliable quantitative estimate of the Al abundance. 

\paragraph{\ion{Al}{I} emission:} In at least six out of the 23 spectra that covered the \ion{Al}{I} lines, weak emission components in the cores of the absorption profiles were identified. The emission is blueshifted with respect to the center of the absorption profile. On dates when the absorption profiles are very narrow, a separate emission line is not seen, but an emission component that partially fills the core of the absorption profile is implied from the asymmetry of the feature. We measured the location of the emission peaks for three dates when the emission component was well separated from the absorption wings, but did not correct for the underlying and highly asymmetric absorption profile: $V_h$=+37\,\kms\ on 2007-01-30 ($\varphi=0.92$); +57\,\kms\ on 2007-09-04 ($\varphi=0.57$); and +47\,\kms\ on 2008-08-29 ($\varphi=0.65$), where the formal uncertainties in the centroid positions are less than 1\,\kms. The measurements may be affected by a shift caused by the underlying absorption core. To show displacements of up to 20\,\kms\ from the center-of-mass velocity, the blueshifted emission must arise next to the shock when it is still deep in the stellar atmosphere. The displacements are comparable to or even higher than the blueshift observed in CO $\Delta\varv$=3 lines ($\Delta V_{\rm blue}\!\approx$13\,\kms) which is usually associated with the deepest observable layers of the star \citep{IRveloCO,Nowotny2010}. Because the details of the excitation mechanism that gives rise to the emission is not known, we are unable to assess the amount of material responsible for the emission.

We did not identify either absorption or emission features that might suggest there is a substantial amount of atomic Al in the outer atmosphere and wind of Mira, implying that Al is locked in molecules (and dust) above the photosphere. 

\paragraph{\ion{Al}{II} and \ion{Al}{III}} We also searched in the optical spectra for ionic forms of Al. The lowest-excitation line of \ion{Al}{II} covered by our spectra has $E_{\rm low}$=83\,800\,K. A very weak absorption line is found at the expected wavelength $\lambda_{\rm lab}$=4663.05\,\AA\ but shifted to the stellar radial velocity. This line is expected to be weak owing to a small Einstein coefficient, $A_{\rm up,low}\!\approx$10$^5$\,s$^{-1}$, and very large excitation potential. With the local spectral line density, it is very likely that the observed feature belongs to another atomic ion. Indeed, in the atlas of Arcturus \citep{atlasArcturus}, the line is identified as belonging to \ion{Fe}{II}. The feature changes its velocity systematically with a highest excursion toward the blue of $\Delta V$=--7\,\kms\ near $\varphi$=0.6. The next lowest-excitation line of \ion{Al}{II} covered in our spectra has $E_{\rm low}$=120\,000\,K and $A_{\rm up,low}$ that is two orders of magnitude larger. Despite the much larger $A_{\rm up,low}$, there is nothing conspicuous near its location ($\lambda_{\rm lab}$=4663.05\,\AA). The line may require too high an excitation to be effectively populated. The best test for the presence of \ion{Al}{II} would be an observation of the resonance line with the rest wavelength of 2669.1\,\AA. Although it was covered in several UV spectra in the literature or available in the archives, the line does not seem to be present. \citet{Reimers1985} identified one weak emission feature of the resonance doublet of \ion{Al}{III}, at 1854.7/1862.6\,\AA,  in their IUE spectra. Whether this feature arises near Mira A or B is questionable and so is the identification. We conclude that our observational material and the literature data are not suitable to trace the ionic forms of Al in Mira, and their contribution to the total aluminum content are likely insignificant.

%
%

\section{Summary and discussion}\label{discussion}
\subsection{Currently accessible Al-bearing species}
In the previous sections, we have attempted to identify the main gas-phase carriers of Al in the circumstellar environment of Mira. We found neutral atomic aluminum, \ion{Al}{I}, in the photosphere and occasionally being excited above the photosphere. The molecules AlH, AlO, and AlOH are present mostly above the photosphere, in a region where the formation of alumina dust is thought to take place ($\gtrsim$2\,R$_{\star}$). There is a tentative indication that AlOH may extend farther from the star and is present in the wind seen against the UV continuum of the companion. Of all these species, we were able to derive quantitative information on the amount of AlO. The observed features are variable in intensity and show modest changes in their Doppler shifts. We excluded the possibility that other simple Al-bearing species are easily accessible for mm-wave observations, such as AlF, AlCl, AlCN, AlNC, and AlS. The contribution of gas-phase Al$_2$, AlO$_2$, and Al$_2$O to the total content of Al in the envelope of Mira remains unknown.  


The observability of a given species alone does not directly imply that it has a high abundance, especially with data obtained at different sensitivities. Our observations show, however, that AlH, AlO, AlOH, and \ion{Al}{I} are viable species for tracing gas-phase aluminum in circumstellar envelopes. Their relative emission distributions and time variability provide important observational constraints that can be directly compared to chemical models. For example, the role of AlH was omitted in some earlier chemical studies \citep[e.g.][]{GS98}, although it is an important observable species. The most recent models of \citet{gobrecht} include AlH in the reaction network and our observations confirm that it is an important tracer of Al. 

\subsection{Circumstellar chemistry: observations vs. models}
We derive an average AlO abundance of 10$^{-9}$--10$^{-7}$ with a high uncertainty owing to the unknown content and form of hydrogen. \citet{kami_alo} obtained a similar abundance of $\sim\!10^{-8}$ for AlO in the envelope of the red supergiant VY\,CMa that was derived from optical emission bands. The agreement between the two sources suggests that the processes governing aluminum chemistry in the envelopes of evolved late-type stars of different masses and different mass-loss rates may be similar. 


The non-equilibrium model with condensation of IK\,Tau  \citep{gobrecht} show that the abundances of Al-bearing species vary with the stellar pulsation phase. Our observations of $o$\,Ceti show that the lines are variable but we are unable to distinguish between the variable excitation conditions and the change in the column of gas. Also, the variability we observe is not well correlated with phase, contrary to the key assumption in the model. The model gives an AlO abundance of $10^{-6}$ in the shocked photosphere which decreases to $10^{-10}$--$10^{-8}$ between 2 and 5\,R$_{\star}$ for IK\,Tau (Cherchneff, priv. comm.). The latter abundance is close to $10^{-9}$--$10^{-7}$ we derived for the extended atmosphere of $o$\,Ceti (Sect.\,\ref{sect-excitation}). A physico-chemical model of $o$\,Ceti is required to better reproduce the observed characteristics of Al-bearing species in this object.


The chemical models predict that most Al-bearing molecules are destroyed above the photosphere by the pulsation shock and then reformed higher up in the atmosphere \citep{gobrecht}. Therefore, one would expect to see in our best-resolution data a gap (depletion zone) in the AlO distribution just above the pulsation shock if the discontinuity is thick enough. The region from 1\,R$_{\star}$ to $\sim$3\,R$_{\star}$, where one would expect such a discontinuity, is very difficult to interpret in our data. The presence of absorption toward the stellar photosphere complicates the picture at angular scales comparable to the ALMA restoring beam, because emission and absorption are indistinctly convolved and mutually affect the image reconstruction of each other (Appendix\,\,\ref{appendix-clean}). Another complication is the patchy structure of the emission at angular scales comparable to the stellar size. Our data do not allow us to verify the presence of such a gap and it may well be that this region is within the radio photosphere, at $r\!<$1.6\,R$_{\star}$, where we had no access with the current ALMA observations. Since the ALMA observations were obtained at a visual phase of about 0.5, near minimum light when the star is expected to be largest, corresponding observations closer to maximum light would peer deeper into the material surrounding the star.   

Much more astronomical spectroscopic observations (preferably taken with a high angular resolution), more physically realistic models, and supporting laboratory spectroscopic measurements are needed to advance our understanding of the complex chemistry of AGB stars. 

\subsection{AlO and alumina dust formation in Mira}
AlO is the a crucial gas-phase Al-bearing species for alumina dust formation, because the dimerisation of AlO followed by the oxygenation of the AlO dimers produce Al$_2$O$_3$, which in turn forms dimers and small alumina clusters \citep{sarangi}. Therefore, the chemical production of AlO determines the efficiency of alumina-dust formation and, consequently, is also linked to the gas-to-dust mass ratio. The formation of alumina dust is expected mainly at very high \emph{dust} temperatures of about 1700\,K. Interferometric observations in the MIR show that the alumina dust indeed forms very close to the photospheres of AGB stars, i.e. at $r\!\approx$2\,R$_{\star}$ \citep[e.g.][]{karovicova}. In $o$\,Ceti, the hottest dust is observed as close as 30\,mas=2\,R$_{\star}$ \citep{lopez,degiacomi}, but it has not been definitely established whether it is alumina or silicate dust. Here we assume that at such close proximity to the star the hot dust is Al$_2$O$_3$. Silicates are unlikely to condense there because \citet{wong} have recently shown that considerable depletion of SiO takes place in Mira at radii above 4\,$R_{\star}$. It is important to note here that 2\,R$_{\star}$ is only the inner radius at which the alumina condensation and AlO depletion starts and that the process may continue farther from the star. 

Our high-resolution ALMA data provide a view of the AlO distribution in unprecedented detail with the beam corresponding to approximately 2\,R$_{\star}$. Here we discuss how these observations can be understood within the context of alumina dust formation.


\paragraph{Average AlO abundance as a function of radius for $r\!\gtrsim$3\,R$_{\star}$} %
As mentioned in Sect.\,\ref{AlOmap}, the azimuthally-averaged brightness of AlO emission follows approximately a power law at radii $\gtrsim$2\,R$_{\rm ph,mm}$ or $\gtrsim$3\,R$_{\star}$ (Fig.\,\ref{fig-AlO-profile}). The observed brightness profile reflects the total density changes, excitation conditions, and possible abundance variations of AlO. The emission brightness changes only slowly with temperature because the brightness is proportional to $\exp(-1/T_{\rm ex})$ and the temperature changes approximately as $r^{-0.3}$ \citep{wong}. Although the envelope is inhomogeneous and the velocity field is complicated, to a first approximation one might expect that the total density decreases as $r^{-2}$. The models of Mira's envelope suggest even steeper power laws of $r^{-3}$ \citep{wong,ireland2011} while NIR observations of dust in $o$\,Ceti require the dust density to decrease as $r^{-1.5}$ \citep{lopez}. The data can be reproduced by all these power laws equally well owing to the large scatter caused by local inhomogenities (Fig.\,\ref{fig-profile}). As a result, the average profile suggests that no significant AlO depletion takes place at radii $\gtrsim$3\,R$_{\star}$. 


\begin{figure}
\centering
\includegraphics[angle=0,width=0.99\columnwidth]{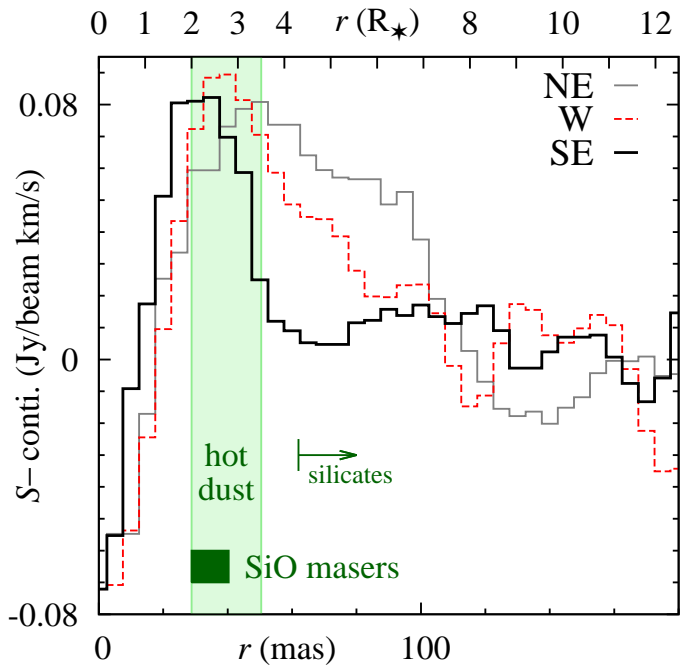}
\caption{Spatial cuts of the AlO 6$\to$5 emission imaged by ALMA. The radial profiles extend from the center of the star and represent three position angles: 54\degr\ (nearly northeast), 133\degr\ (south-east), and $-90$\degr\ (west), shown in Fig.\,\ref{fig-mom0AlO65}. Marked in green are  characteristic locations within the envelope: the region closest to the star where hot dust has been observed in minimum/maximum light (inner/outer edge) (light green); the range of SiO maser rings (dark green); and the innermost radius where silicate dust is thought to condense in Mira stars (marked as "silicates").}\label{fig-projections}
\end{figure}

\paragraph{AlO radial profiles in specific directions--a proof for depletion?} If we ignore the brightest emission clumps east and north/north-west from the star in Fig.\,\ref{fig-mom0AlO65}, the AlO emission is limited to a horseshoe-shaped ring of radius 2\,R$_{\star}$. The emitting material is likely located in the plane of the sky. In Fig.\,\ref{fig-projections}, we show three spatial cuts through this structure along three position angles of 54\degr\ (nearly northeast), 133\degr\ (south-east), and --90\degr\ (west) and extending from the absorption center (Fig.\,\ref{fig-mom0AlO65}). While two of those spatial cuts are in qualitative agreement with the average radial profile, the southeast cut shows distinctively a narrow feature with the line flux that practically disappears at a radius of 3.5\,R$_{\star}$, i.e. exactly in the region where hot dust has been observed in Mira. If we interpret the emission intensity as corresponding directly to the column density of AlO (i.e. the excitation conditions do not change much with radius), then this profile is exactly what one would expect if substantial depletion of AlO began at about 2\,R$_{\star}$. It is tempting to interpret this AlO deficit as depletion onto dust. It would be interesting to verify this depletion scenario by checking whether there is a dust-rich clump at the radial extension of this spatial cut beyond 3\,R$_{\star}$, but it is not possible to perform this test with the currently available data. Although there is no strong proof in our data for global AlO depletion, it may be that the gas-phase AlO is consumed by other species only in particular directions.

\paragraph{AlO variability and clumpiness vs. episodic dust production} %
The ALMA maps of the $N$=6$\to$5 transition (Fig.\,\ref{fig-mom0AlO65}) provide a detailed glimpse of the clumpy distribution of AlO gas very close to the photosphere. Together with data presented in \citet{wong}, they provide the first direct view of the inhomogenities in the cool molecular layer at about 2\,R$_{\star}$. The existence of such inhomogenities has been implied in $o$\,Ceti before  \citep[e.g.][and references therein]{wittkowski2015} and in other oxygen-rich AGB stars \citep[e.g.][]{xavier,ohnaka}. The clumps are close to the star and are likely to be transient, which is strongly supported by the variability of AlO observed in the rotational and electronic lines reported here.  


The question which naturally arises is whether the complex spatial and temporal characteristics of AlO can influence the dust production. The dust produced around Mira has a very inhomogeneous distribution and its spectral features change with time \citep{lopez,lobel}. Mira's flux variability in the range 8--11\,$\mu$m cannot be explained by the light changes of the star alone, rather some variations of the dust density must also occur \citep{lopez}. It has been suggested in the literature that the dust production in AGB stars is in fact episodic and may occur during some phases or only in some cycles \citep[cf.][and references therein]{lopez}. Because the production of alumina dust is chemically directly linked to the production of AlO, the clumpy distribution and time variability of AlO emission presented here may be the direct cause of the observed spatial and temporal characteristics of dust. Identifying the process that makes the molecular emission clumpy is beyond the scope of this paper but it is very likely linked to phenomena in the stellar photosphere. Localized shocks related to convection \citep{schwarzschild} are good candidates to explain the observational features of some AGB stars  \citep[e.g.][]{ohnaka,xavier} and red supergiants \citep[e.g.][]{betelgeuse}. The episodic appearance of visual emission bands of AlO requiring shock excitation supports such an interpretation. 

\paragraph{Link of the AlO emission to faint-maxima cycles?}\label{sect-AlO-brightMax}
The occurrence of emission in electronic bands of AlO in Mira is very rare. It was reported for the first time in 1924. The first identification of AlO bands was made by \citet{bax}, who used spectra of \citet{joy} (these authors assigned the features to Al$_2$O$_3$). Emission was also reported for the 1964 maximum of Mira by \citet{keenan}. The search for AlO emission in Mira was continued by \citet{kipper}, but their observation of the $\Delta\varv$=0 sequence in 1978 did not reveal any emission\footnote{Their spectral resolutions (dispersions of 24 and 12\,\AA/mm) might have been too low  to detect weak emission, especially when mixed with an absorption profile.}. Our observational data add two more episodes of AlO emission to the record.

The occurrences of AlO bands in emission were reported to proceed cycles with fainter-than-average visual maxima (cf. Fig.\,\ref{fig-phase}) which are rare. In the last century, maxima with Vis.$\gtrsim$4\,mag occurred only 5--8 times: in 1924, 1976, 1978, 1984, 1998/99, and possibly also in 1912, 1913, and 1931 (the photometric coverage is incomplete near these maxima). Emission was observed in 1924  over a few weeks \citep{keenan}. \citet{garrison} observed Mira during most maxima in 1916--1997 and noticed that AlO emission appears in weak cycles. Our 1998/1999 observation of the AlO emission also coincided with a faint maximum. However, the 2007 observation of the AlO emission event  proceeded a normal cycle. This raises the question as to whether there is a direct correspondence between AlO emission and faint maxima.

The existence of such a correlation would be very intriguing in the context of metal-oxide formation and variable dust production of Mira. One possibility is that faint maxima occur owing to enhanced dust production which would increase the extinction and reduce the optical fluxes of the star. Alternatively, some cycles may be dimmer due to increased production of TiO and other metal oxides which constitute the main opacity source in Miras and are directly responsible for their high-amplitude light variations in the visual \citep{reid02}. However the literature and our observations of AlO (and TiO) emission show that there is no established direct link between the occurrences of molecular emission and faint maxima. 

Here, we prefer an interpretation in which the emission episodes in the electronic features of AlO and \ion{Al}{I} are associated with shocks which are stronger than average and were also proposed to explain other inter-cycle phenomena in Mira variables \citep[e.g.][]{richter2003,lopez}. 



\subsection{Final words and prospects}
The challenges in tracing gas-phase species important for dust formation are identified in the current study. We have demonstrated that the complex excitation of gas very near the stellar photosphere -- influenced strongly by repetitive, irregular, and merging shocks -- is the main obstacle in deriving  abundances and more sophisticated analysis tools are necessary to interpret the data. A conclusive test of dust nucleation in stellar envelopes is possible but it will require a substantial observing effort. In particular, the most fruitful would be observations of a multitude of spectral lines with the highest angular resolution (as afforded primarily with the most extended ALMA configurations) conducted nearly simultaneously with IR/MIR interferometric observations of dust features.

\begin{acknowledgements}
We thank T. Kipper (Tartu Observatory) for sending us a copy of his article. We are also grateful to O. Anriyenko (Terskol Observatory) for providing us with his spectra and E. Griffin (NRC-Herzberg) for her efforts to digitalize the DAO spectroscopical plates.
The Submillimeter Array is a joint project between the Smithsonian Astrophysical Observatory and the Academia Sinica Institute of Astronomy and Astrophysics and is funded by the Smithsonian Institution and the Academia Sinica.
We acknowledge with thanks the variable star observations from the AAVSO International Database contributed by observers worldwide and used in this research.
Based on data obtained from the ESO Science Archive Facility and made with ESO Telescopes at the La Silla Paranal Observatory under programme IDs 074.D-0114(A) and 089.D-0383(A).
Based on observations obtained at the Canada-France-Hawaii Telescope (CFHT) which is operated by the National Research Council of Canada, the Institut National des Sciences de l'Univers of the Centre National de la Recherche Scientique of France, and the University of Hawaii. 
Based on observations obtained at the Dominion Astrophysical Observatory, NRC Herzberg, Programs in Astronomy and Astrophysics, National Research Council of Canada.
Herschel is an ESA space observatory with science instruments provided by European-led Principal Investigator consortia and with important participation from NASA.
This research has made use of the Keck Observatory Archive (KOA), which is operated by the W. M. Keck Observatory and the NASA Exoplanet Science Institute (NExScI), under contract with the National Aeronautics and Space Administration. 
Based on analysis carried out with the CASSIS software and JPL spectroscopic database. CASSIS has been developed by IRAP-UPS/CNRS (\url{http://cassis.irap.omp.eu}). 
This paper makes use of the following ALMA data: ADS/JAO.ALMA\#2011.0.00014.SV, 2013.1.00047.S, 2012.1.00524.S, 2013.1.00156.S. ALMA is a partnership of ESO (representing its member states), NSF (USA) and NINS (Japan), together with NRC (Canada), NSC and ASIAA (Taiwan), and KASI (Republic of Korea), in cooperation with the Republic of Chile. The Joint ALMA Observatory is operated by ESO, AUI/NRAO and NAOJ.
%
\end{acknowledgements}

\begin{appendix}
\section{Details of the APEX observations}
\subsection{The spectral setups}\label{appendix-log}
The spectral setups used in the APEX observations are listed in Table\,\ref{tab-log}.

\begin{table*}
\centering
\caption{Log of APEX observations of Mira.}\label{tab-log}
\begin{tabular}{lcccc}
\hline\hline
Frequency & Central  &  Noise rms    & Integr. & Observation\\
setup     & frequency&  per 3.8\,MHz & time ON & dates\\
          & (MHz)    &  ($T_A^*$ mK) & (min)   & \\
\hline\hline
\multicolumn{5}{c}{HET230}\\
\hline
TiO-222L  & 222700.000 & 1.3 &203.4 & 03,17,18,19,20-12-2013\\
vCO21-AlO & 229800.000 & 1.3 &186.9 & 06,15,16,18,23-10-2013\\
vCO21-AlO & 229800.000 & 1.2 &213.9 & 09,10,11,12-06-2015\\
\hline
\multicolumn{5}{c}{FLASH+}\\
\hline
AlO306\_OSB   & 294698.830 & 1.4 & ~99.2 & 07,08-07-2014\\    
AlO306        & 306700.000 & 1.4 & ~99.2 & 07,08-07-2014\\    
AlO344\_OSB   & 332452.819 & 1.9 & ~78.9 & 13,14-08-2013\\
AlO344        & 344454.000 & 1.8 & ~80.4 & 13,14-08-2013\\
AlO344TiO     & 344454.000 & 1.7 & ~84.8 & 30-06-2014; 01,02,06-07-2014\\
AlO344TiO\_OSB& 356455.219 & 2.0 & ~84.8 & 30-06-2014; 01,02,06-07-2014\\[5pt]
AlO420\_OSB   & 408898.804 & 4.2 & ~51.7 & 13,14-08-2013\\
AlO420        & 420900.000 &10.5 & ~51.7 & 13,14-08-2013\\
AlOvCO458     & 458020.000 & 5.1 & ~27.3 & 14-08-2013\\
AlOvCO458\_OSB& 470021.181 & 6.2 & ~27.3 & 14-08-2013\\
AlO497\_OSB   & 485398.793 &24.0 & 168.7 & 01,02,06,07,08-07-2014\\
AlO497        & 497400.000 & 4.3 & 168.7 & 01,02,06,07,08-07-2014\\ 
\hline
\end{tabular}
\end{table*}

\subsection{Spectral variability in the APEX data}\label{sect-var-apex}

\emph{Spectral} variability of stellar sources at submillimeter wavelengths has been rarely reported in the literature. It has been convincingly shown for O-rich Mira variables, including $o$\,Ceti, but only for maser species such as SiO \citep[e.g.][]{SiOvar} and OH \citep{GerardBeougois1993} and no convincing evidence has been presented so far for other molecules. This is partially so, because investigation of variability, especially at submm wavelengths, is challenging for technical reasons. In particular, apparent intensity changes can often be explained by imperfect focus, pointing, and flux  calibrations, uncertainties of which are usually difficult to assess quantitatively. Our APEX observations do not have those usual shortcomings. All the observations presented here were obtained after applying focus calibration which was performed on $o$\,Ceti using its bright CO emission. Similarly, pointing was calibrated on the Mira's CO lines and it did not show deviations from the source position greater than 2\arcsec (usually much less), which is insignificant compared to the telescope beam of 18\arcsec or 27\arcsec at our frequencies. Overall, the pointing model and tracking performance at APEX was excellent in the years covered by our observations. We are confident these basic calibrations do not affect the derived intensity of the spectral features. 

The flux calibration can be tested directly in our observations. It can be safely assumed that the rotational transitions of CO at $\varv$=0, $J$=2--1 and 3--2, covered in two different epochs do not vary on the time scales of years \citep{pepe}. Taking the large size of the envelope producing these lines of $r\!\sim\!10^{16}$\,cm and the corresponding wind crossing time of the order of 10$^5$\,yr, any short-scale change in Mira's wind, e.g. due to a potentially variable mass-loss rate, cannot effectively influence the total flux of those lines. Additionally, the lines arise in gas that is collisionally excited and none of the main gas heating terms can be significantly affected by stellar light variability for these low-$J$ lines. We can therefore use them as a benchmark for relative-intensity calibration of the instrument at different dates.

The band centered at 229.8\,GHz was observed in 2013 and 2015 using exactly the same configuration. The same sensitivity levels were reached. This band covers the CO(2--1) line which was observed at a signal-to-noise ratio (S/N) of 4700 (line peak to rms noise level). The line profiles are identical in both spectra and the line flux is consistent to within 1.12\% on both dates. Other lines covered in the spectrum were the CO(2--1) line at $\varv$=1 and AlO $N$=6$\to$5. The two lines were not detected in 2013 but are definitely present in 2015. On the basis the excellent consistency in the observations of the CO(2--1) line, the change in the appearance of the CO $\varv$=1 and AlO features in 2015 must be due to the intrinsic source variability in those lines, where both are expected to arise in the innermost envelope.

The band centered at 344.454\,GHz was observed in the USB in August 2013 and in the LSB in July 2014. The strongest line covered by these spectra, CO(3--2), has a measured flux 1.07 times higher on the earlier date. Two other strong lines covered, $^{29}$SiO(8--7) and SO(8$_8$--7$_7$), show very similar  intensity ratios, i.e. consistent to within 1\% with what was measured for the CO(3--2) line. The effect is therefore systematic and we corrected the later spectrum by the calculated factor. (The most likely source of this slight inconsistency is an error in the sideband rejection factors of FLASH+, as the sideband was the only different aspect in the instrumental configuration. A few \% uncertainty in the sideband ratio is well within the performance specifications of FLASH+.) In the corrected spectra, the flux of the maser emission of the SiO(8--7) $\varv$=1 line decreases by a factor of 2.1 from 2013 to 2014. This decrease is physical \citep[cf.][]{SiOvar}. For other weaker lines covered, including the species of our interest, CO(3--2) at $\varv$=1 and AlO $N$=9$\to$8, the fluxes are consistent with the rms noise levels on the two dates. We therefore combined the data to increase the S/N for these weak lines. 

\section{Processing of the interferometric data for the AlO 6$\to$5 line}\label{appendix-clean}
While processing the ALMA data for the AlO 6$\to$5 transition, we encountered a similar problem with the CLEAN procedure as that described in detail in \citet{wong} (their Appendix\,B). That is, we found significant differences in the depth and shape of absorption features when processed in CLEAN with continuum and after continuum subtraction. This data-processing problem is closely related to the relative distribution of the emission and absorption regions and their relative strengths. In our work, we present only the data processed in CLEAN after continuum subtraction which have a better S/N and all absorption features are \emph{deeper} than in the data processed with the alternative method. The emission regions one beam size away and farther from the strongest absorption feature appear exactly the same after applying each of the processing schemes. Unlike in \citet{wong}, where data of much higher S/N are analyzed and the processing procedure does not influence the data interpretation, we do not analyze in detail the region within 2\,R$_{\star}$ because the two processing schemes may lead to different interpretations. 

\section{Optical spectra of Mira}
\clearpage
\begin{figure*} [tbh]
\centering
\includegraphics[angle=270,width=0.85\textwidth]{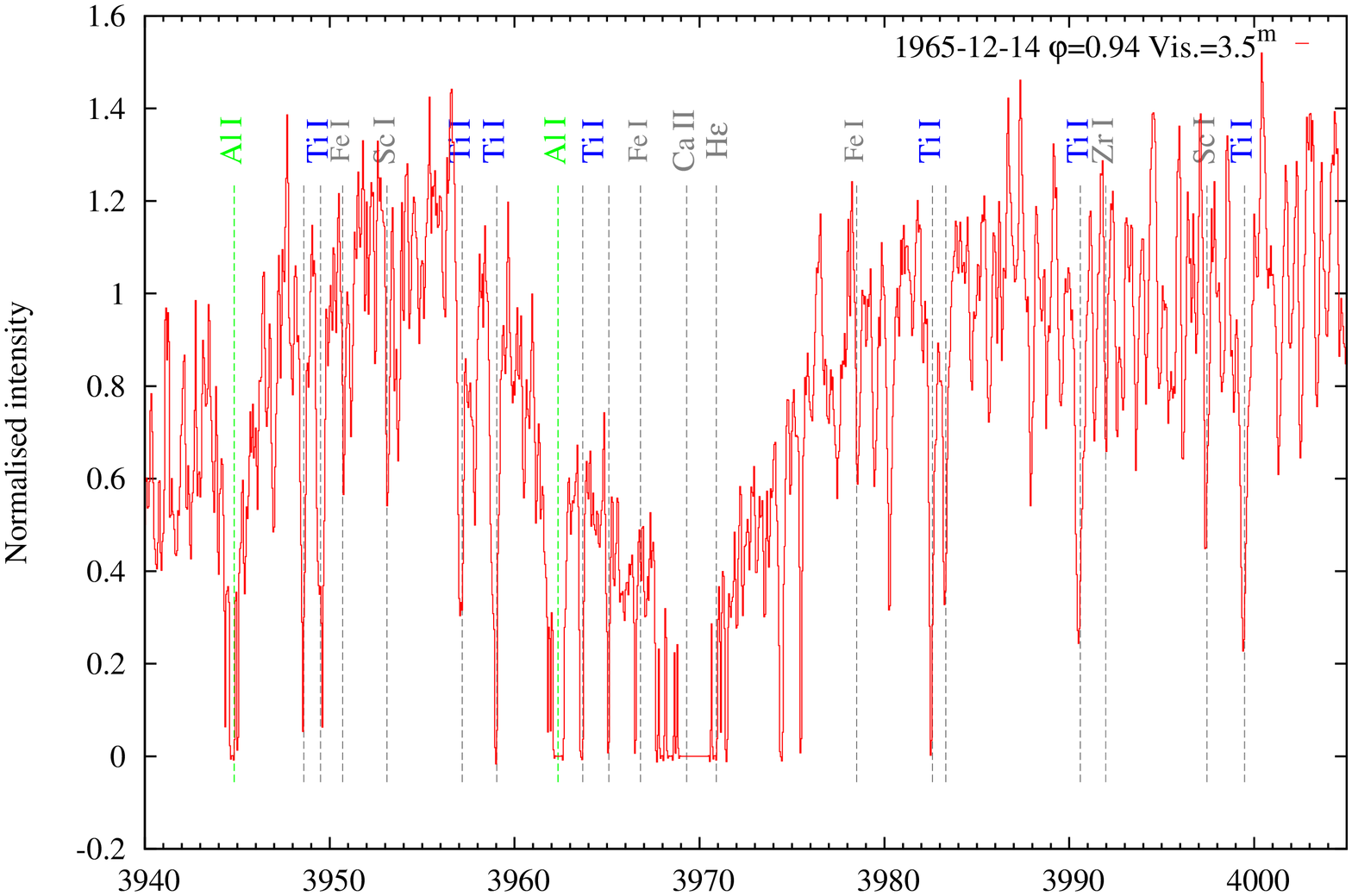}
\includegraphics[angle=270,width=0.85\textwidth]{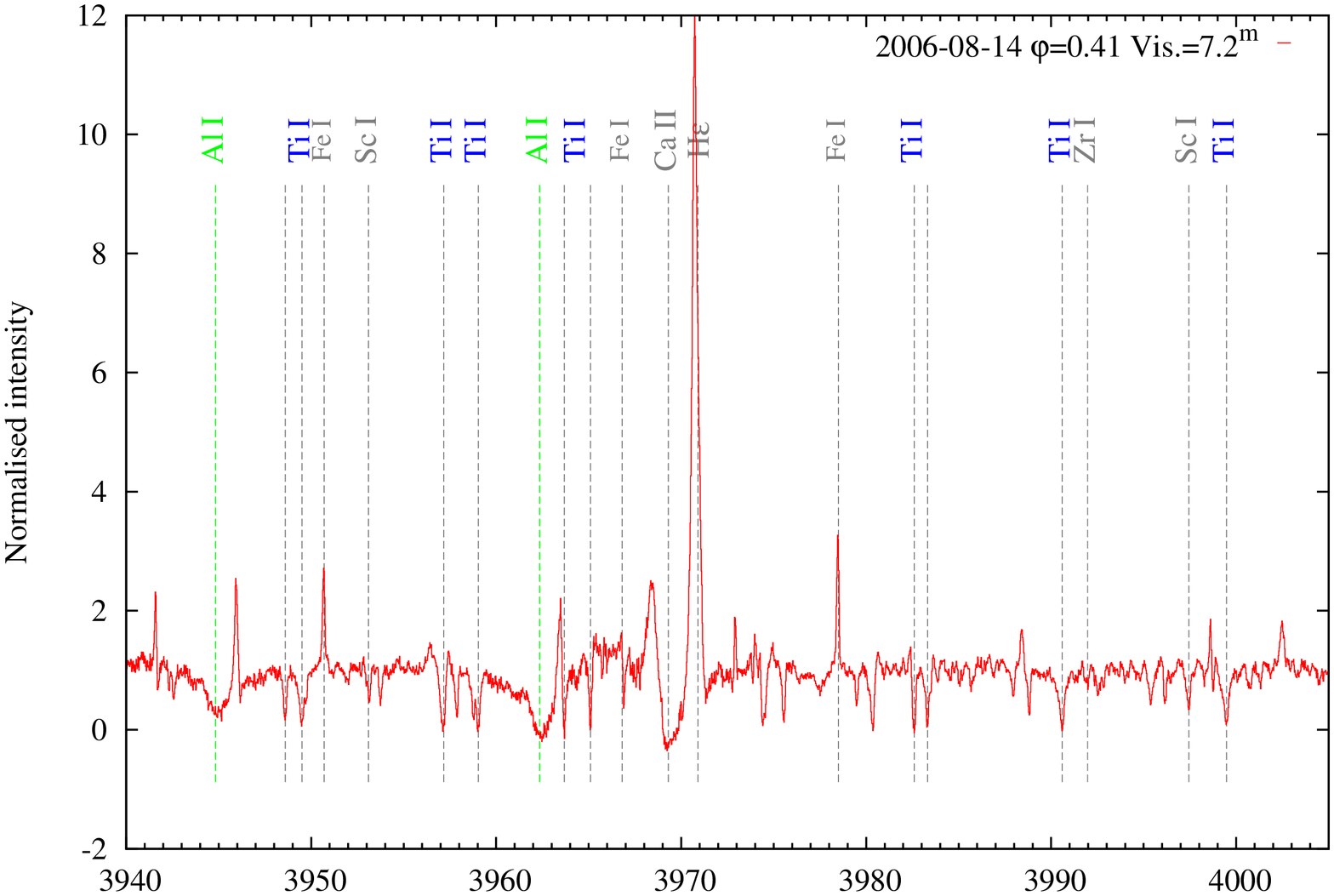}
\caption{High-resolution optical spectra of Mira covering the resonance lines of \ion{Al}{I} and \ion{Ti}{I}. The date of observations and the corresponding phase and visual magnitude are specified in the upper right corner of each panel. All spectra were normalized with high-order polynomials. The pseudo-continuum in this spectral region is dominated by broad wings of the \ion{Ca}{II} doublet and the normalization partially removed its absorption. The normalization was optimised to better illustrate the shape of the \ion{Al}{I} lines and to limit the scale of the figure. The imperfect normalization affects the apparent intensity ratio of the two lines of \ion{Al}{I}. Some spectra were smoothed. Identification labels are shown for selected features at their approximate locations.}\label{fig-AlI}
\end{figure*}

  \setcounter{figure}{0}%

\begin{figure*} [tbh]
\centering
\includegraphics[angle=270,width=0.85\textwidth]{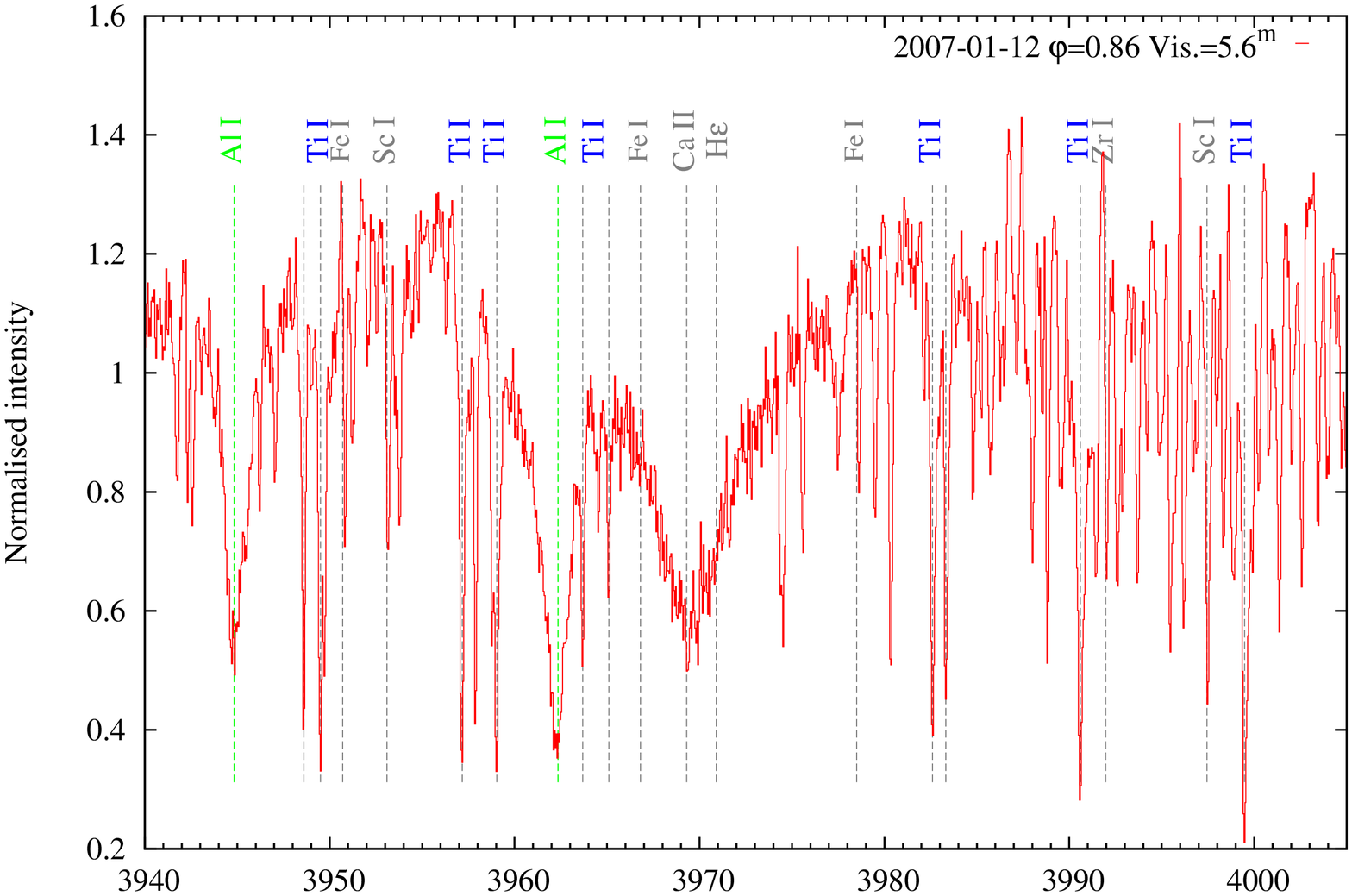}
\includegraphics[angle=270,width=0.85\textwidth]{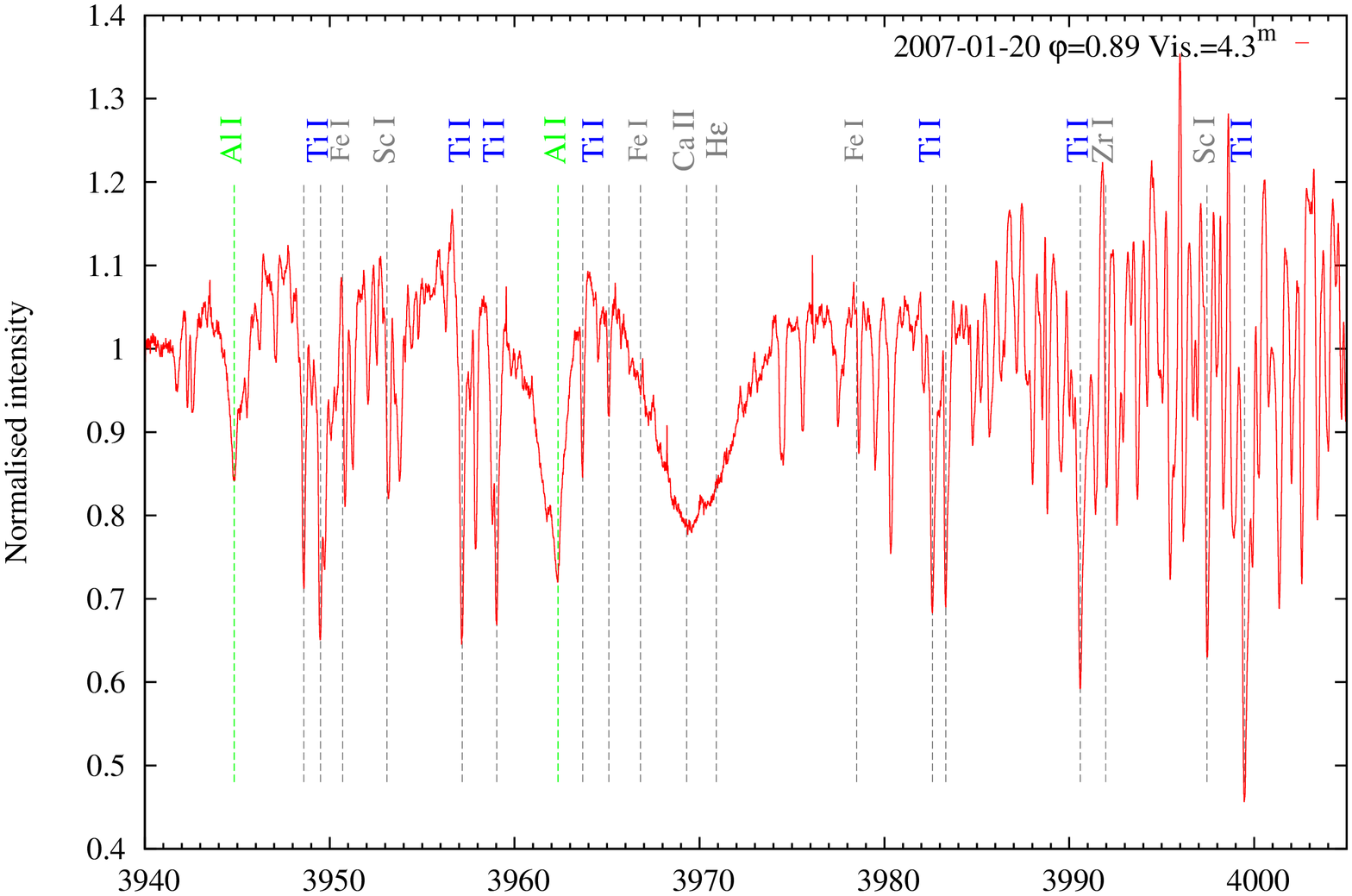}
\caption{Continued.}
\end{figure*}

  \setcounter{figure}{0}%

\begin{figure*} [tbh]
\centering
\includegraphics[angle=270,width=0.85\textwidth]{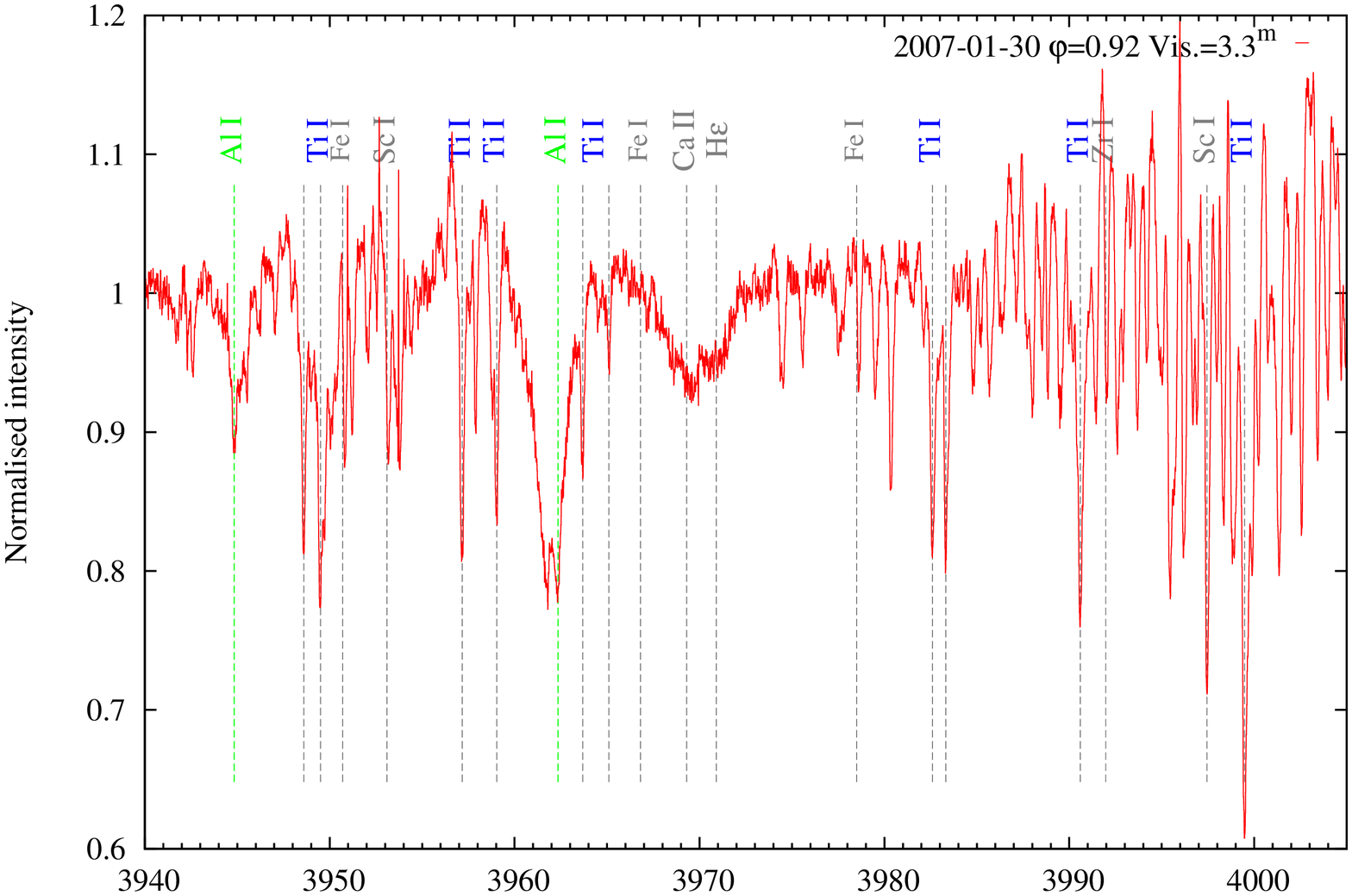}
\includegraphics[angle=270,width=0.85\textwidth]{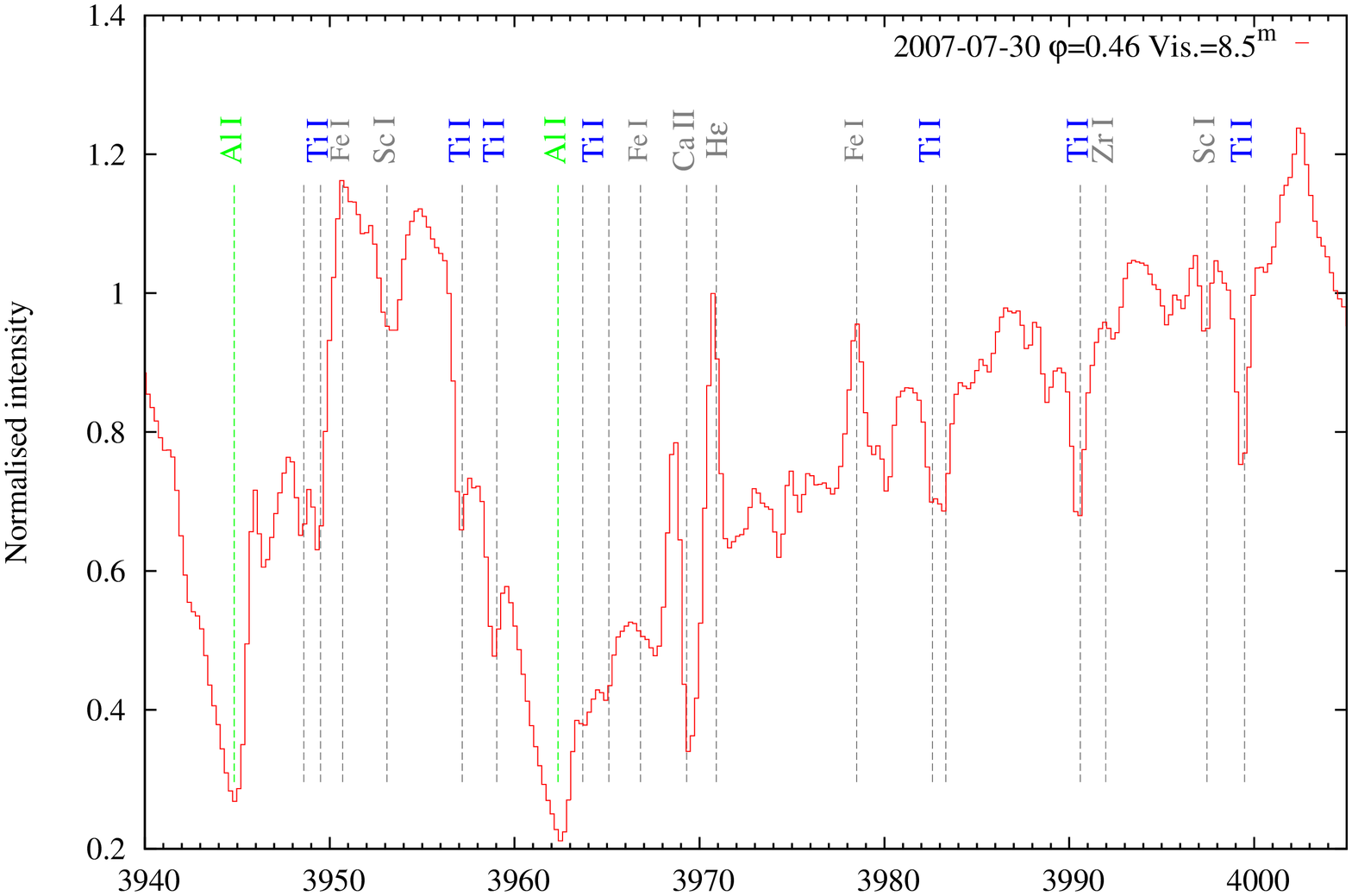}
\caption{Continued.}
\end{figure*}

  \setcounter{figure}{0}%

\begin{figure*} [tbh]
\centering
\includegraphics[angle=270,width=0.85\textwidth]{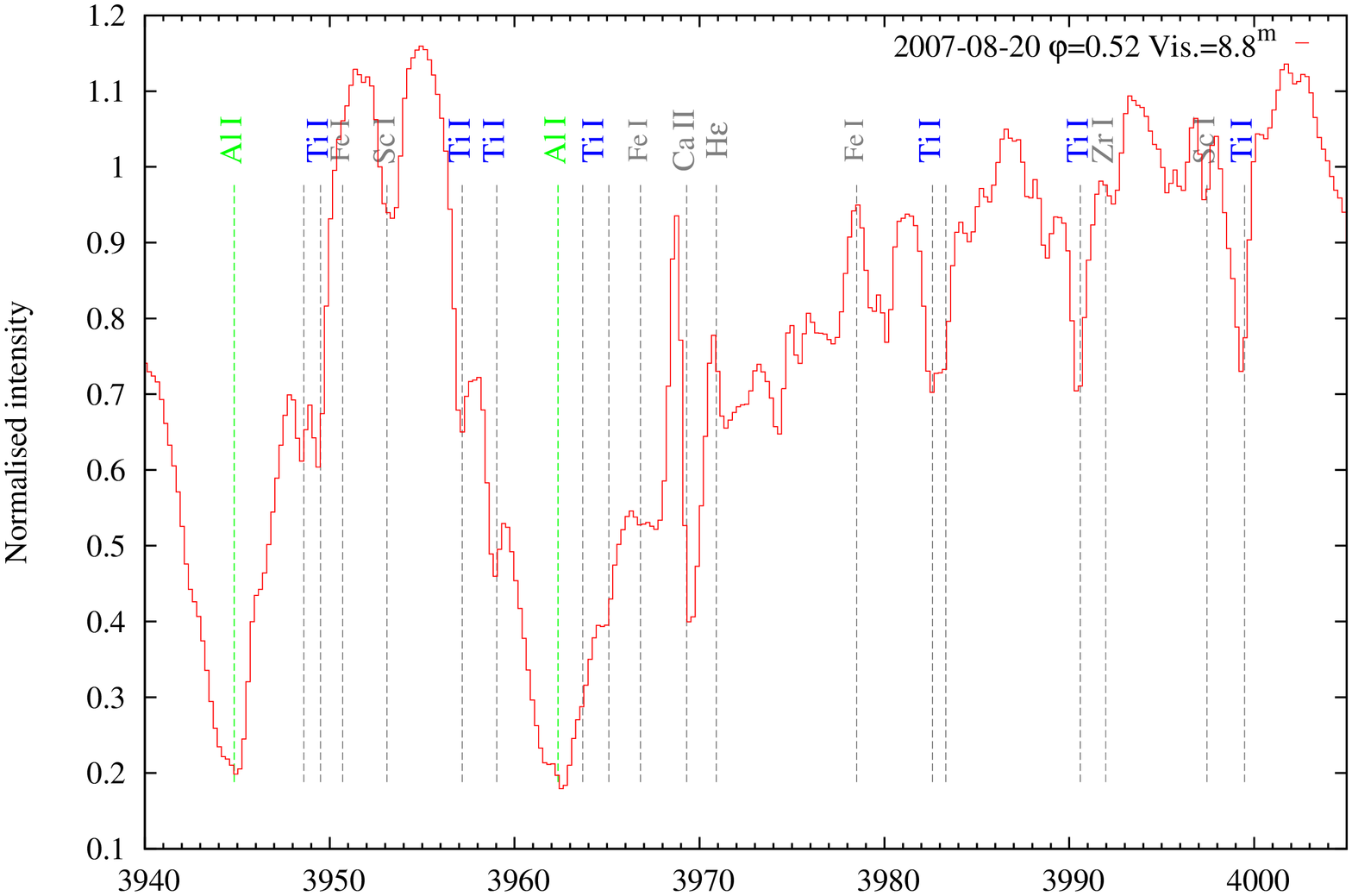}
\includegraphics[angle=270,width=0.85\textwidth]{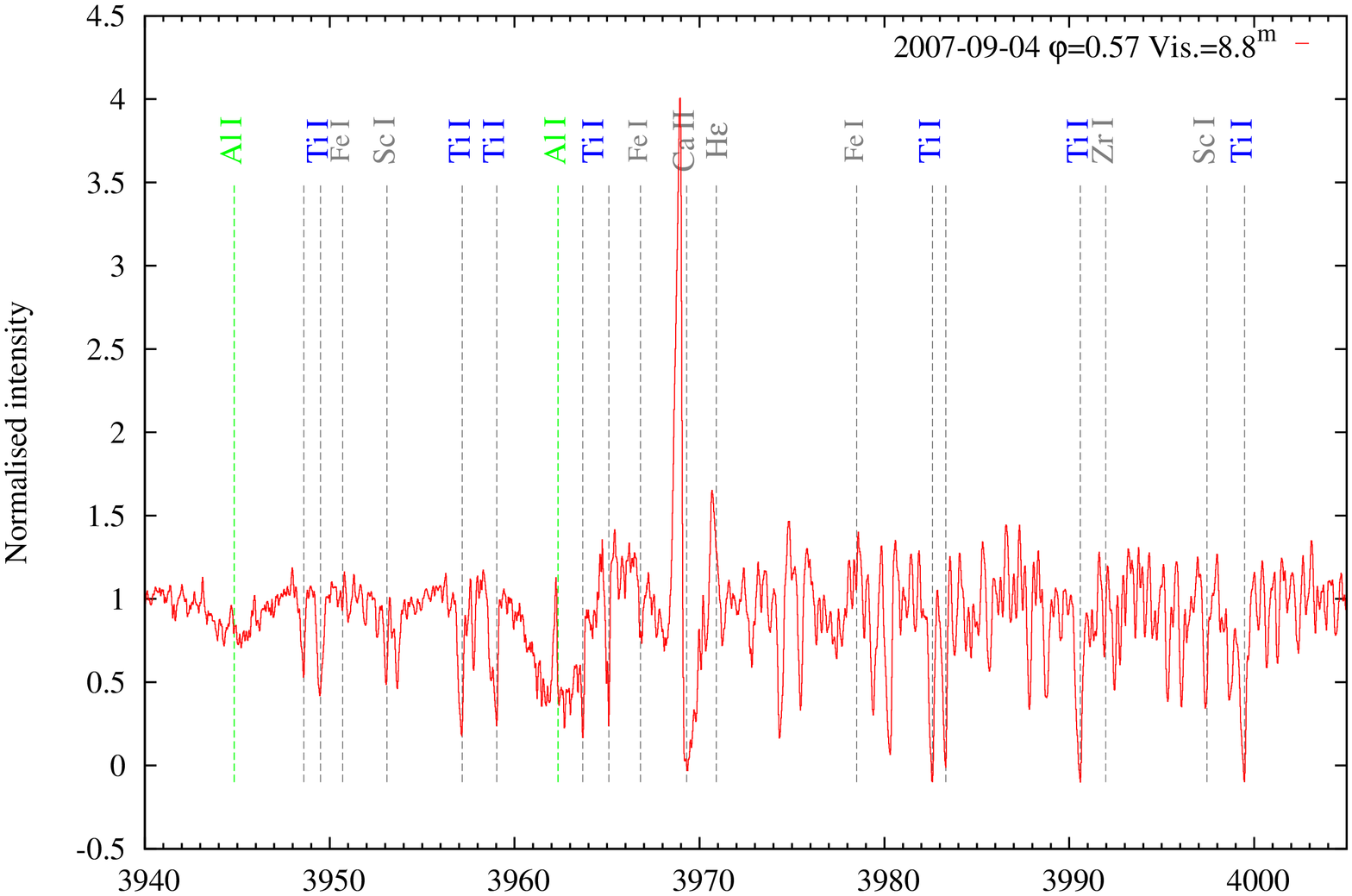}
\caption{Continued.}
\end{figure*}

  \setcounter{figure}{0}%

\begin{figure*} [tbh]
\centering
\includegraphics[angle=270,width=0.85\textwidth]{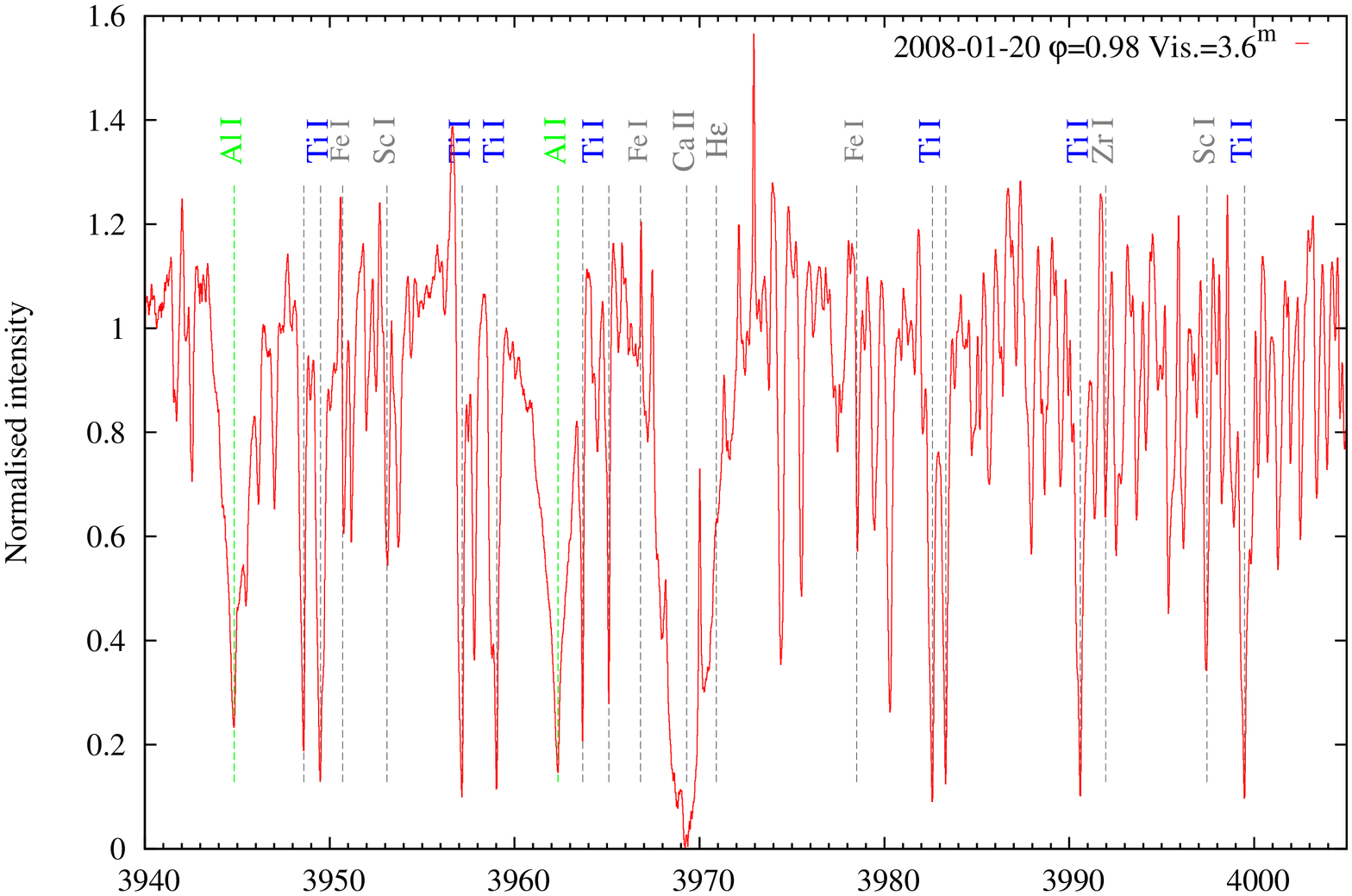}
\includegraphics[angle=270,width=0.85\textwidth]{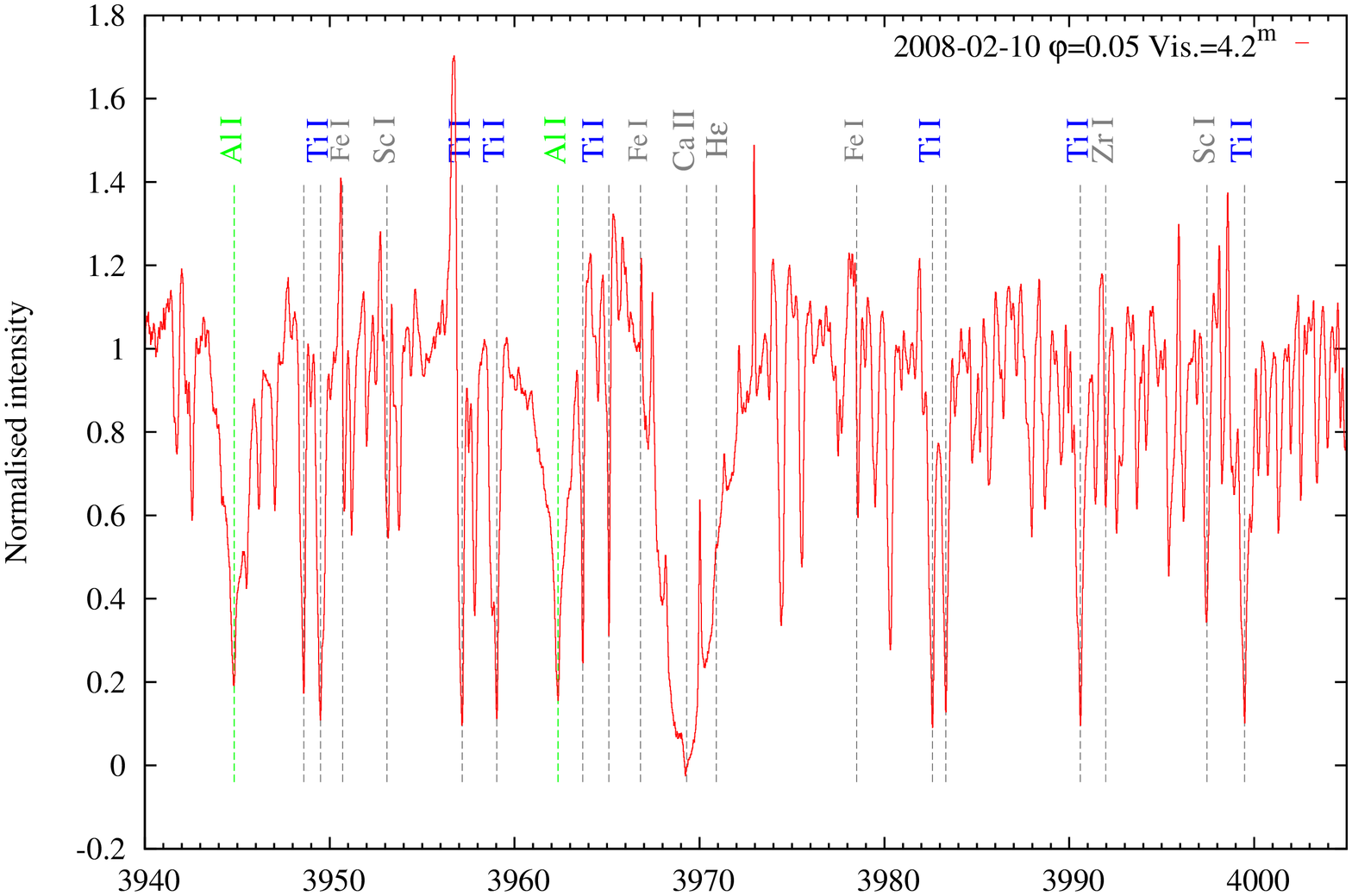}
\caption{Continued.}
\end{figure*}

  \setcounter{figure}{0}%

\begin{figure*} [tbh]
\centering
\includegraphics[angle=270,width=0.85\textwidth]{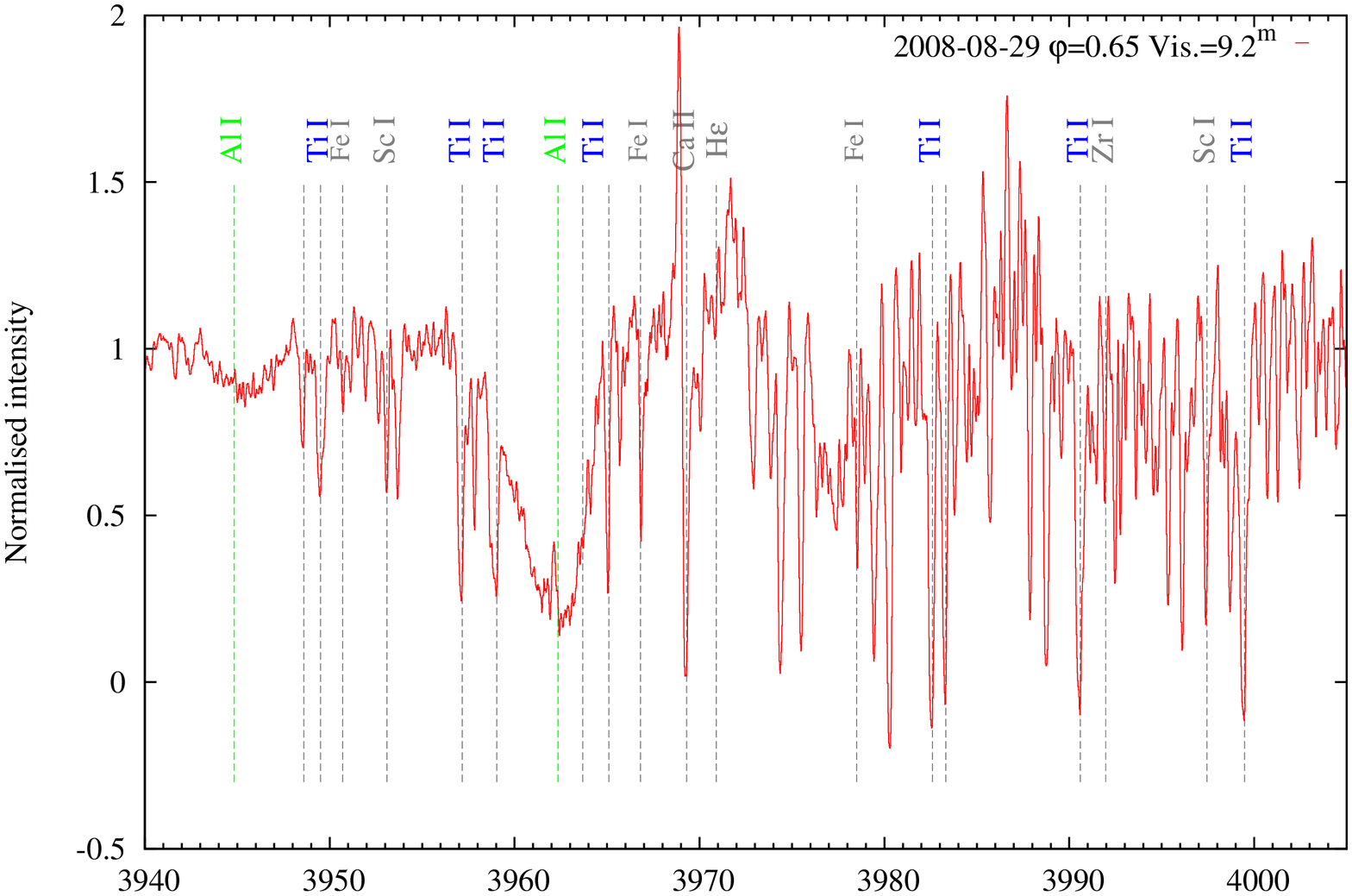}
\includegraphics[angle=270,width=0.85\textwidth]{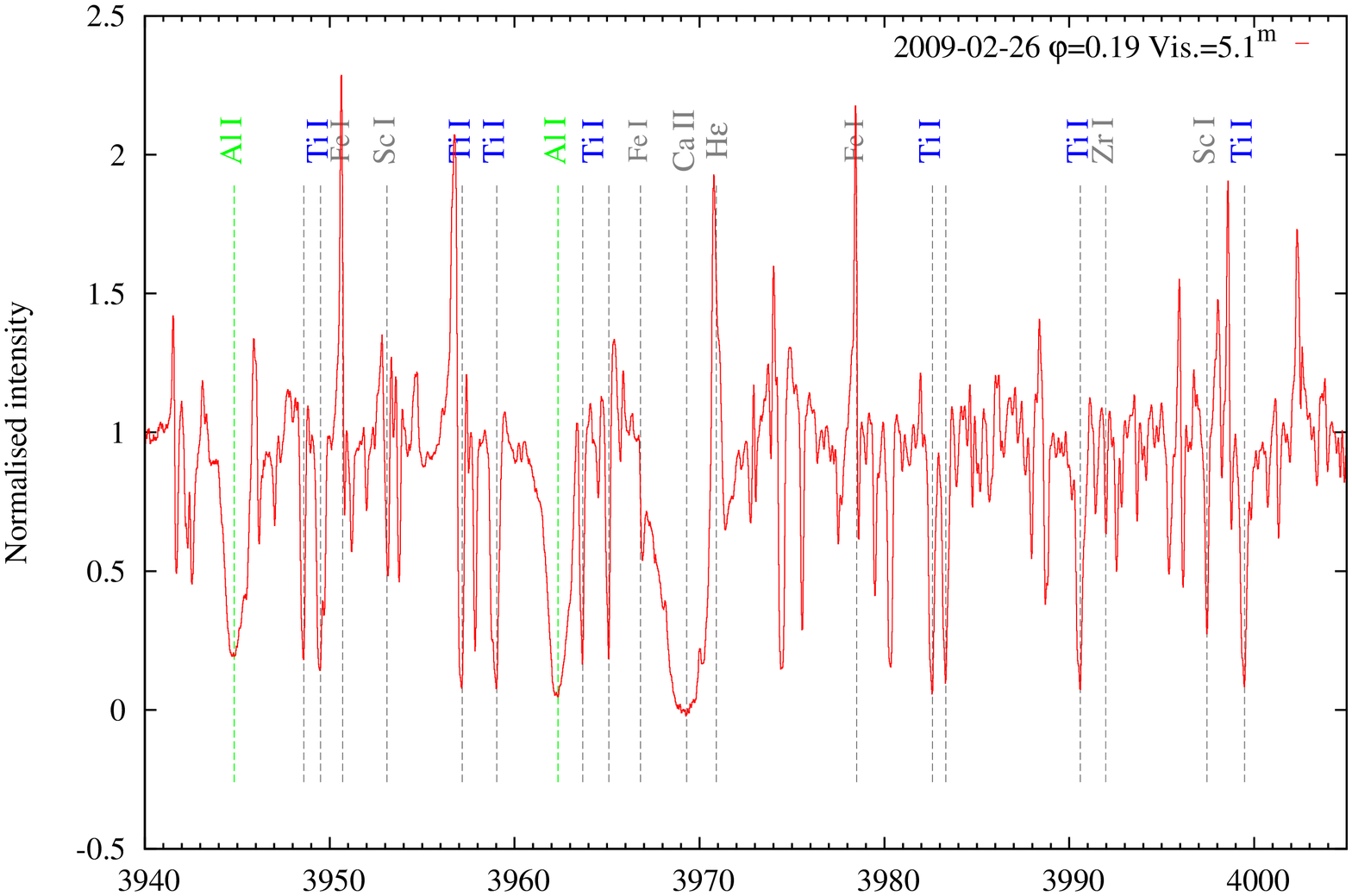}
\caption{Continued.}
\end{figure*}

  \setcounter{figure}{0}%

\begin{figure*} [tbh]
\centering
\includegraphics[angle=270,width=0.85\textwidth]{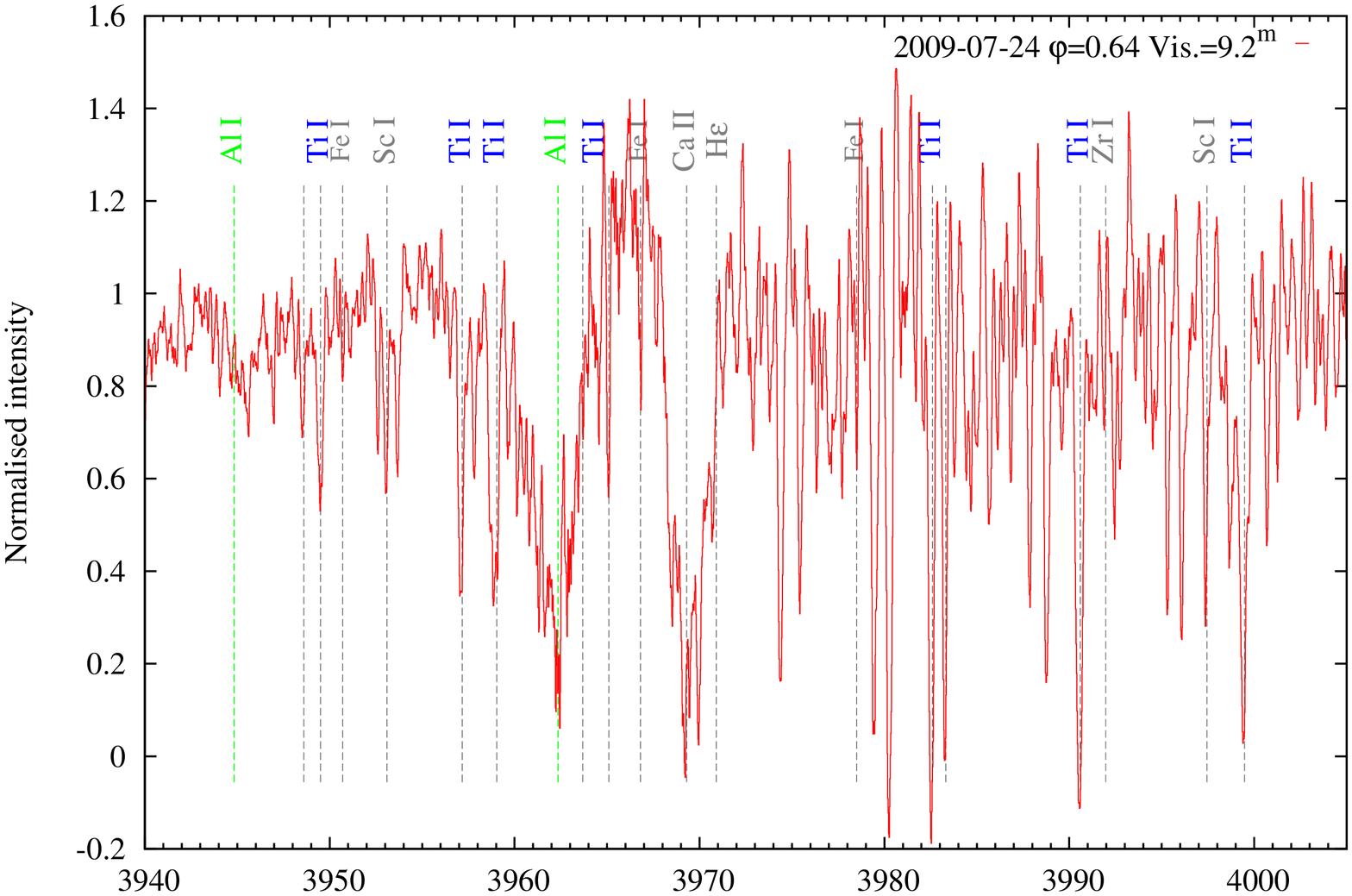}
\includegraphics[angle=270,width=0.85\textwidth]{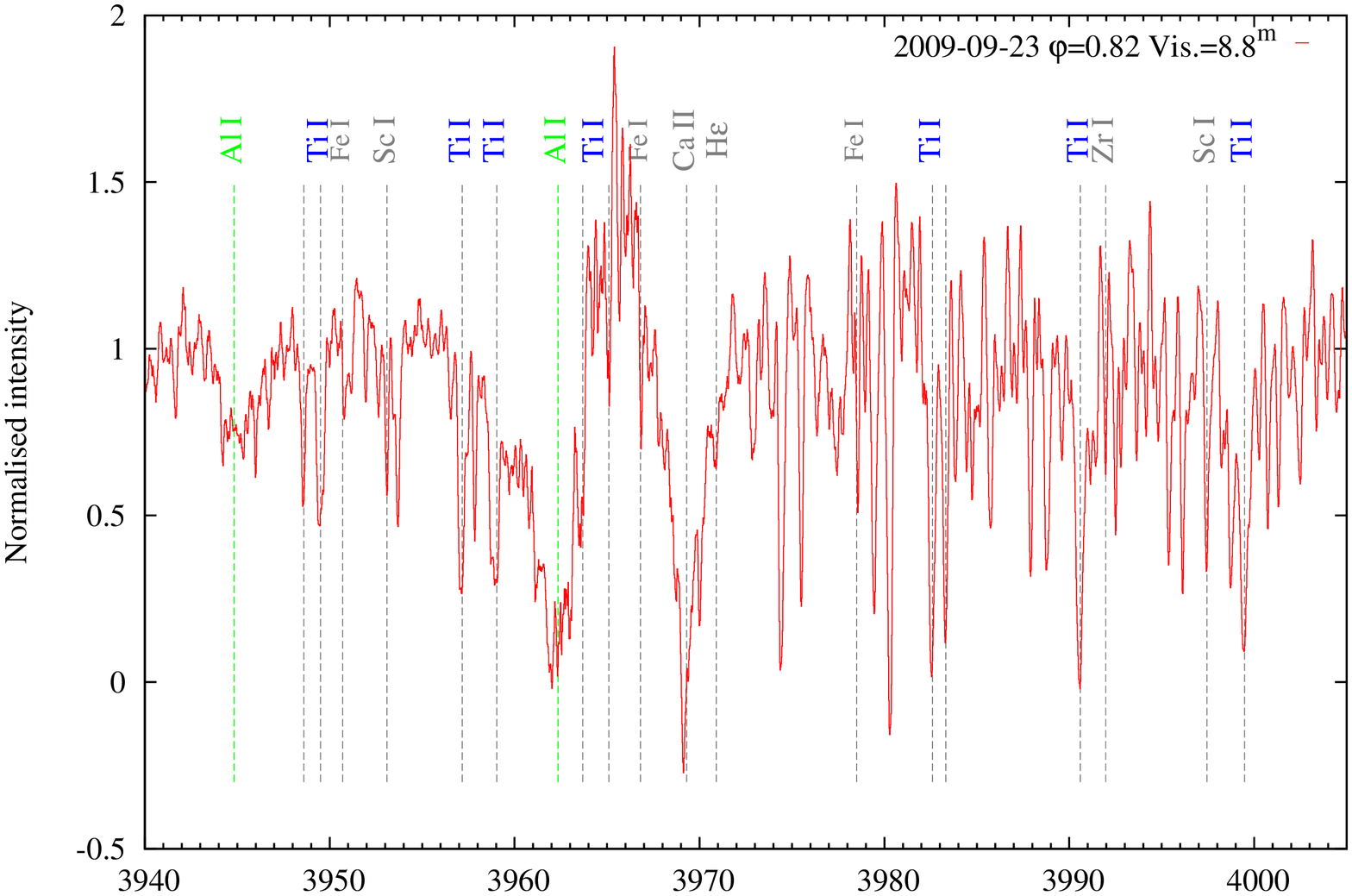}
\caption{Continued.}
\end{figure*}

  \setcounter{figure}{0}%

\begin{figure*} [tbh]
\centering
\includegraphics[angle=270,width=0.85\textwidth]{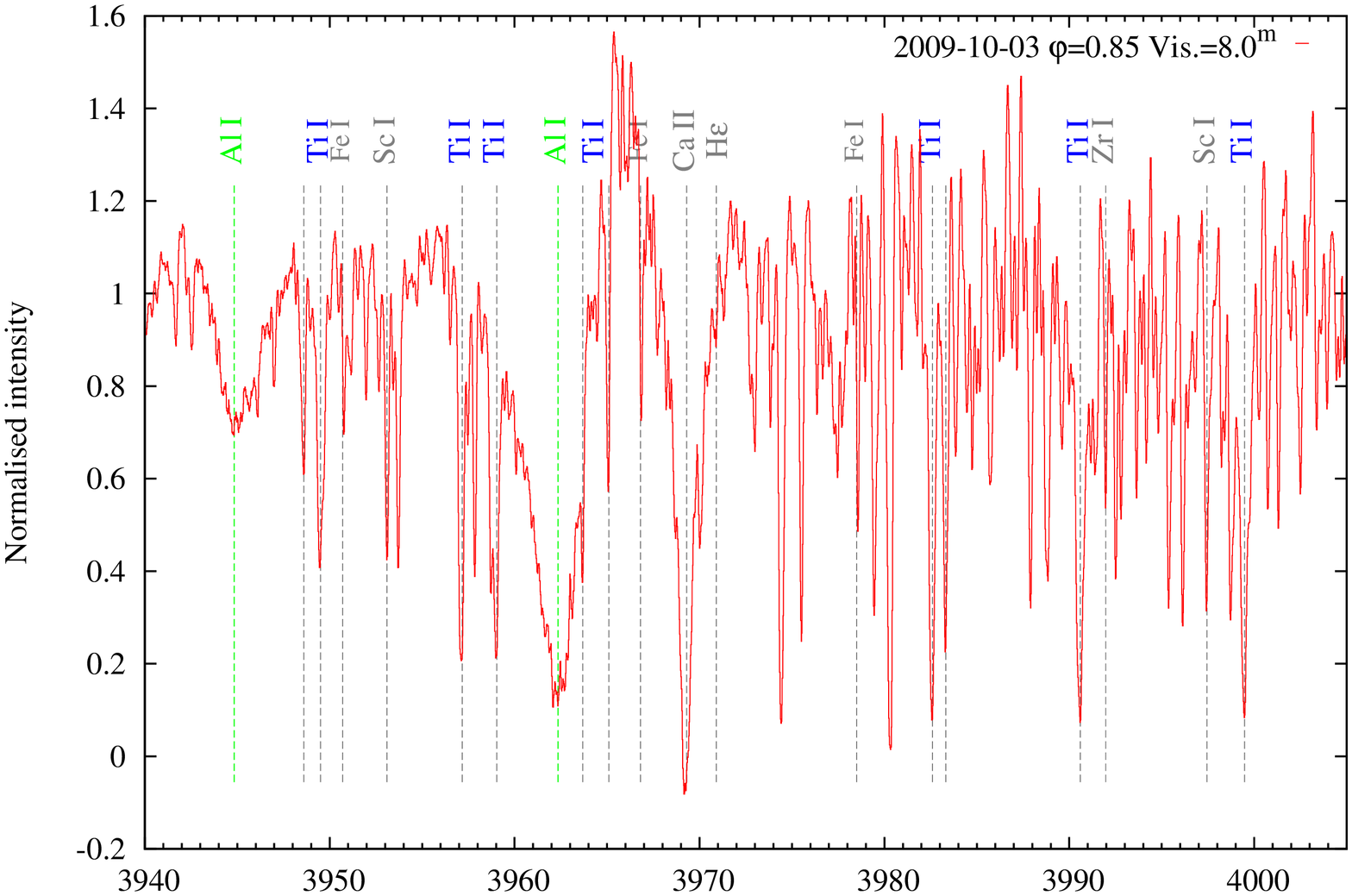}
\includegraphics[angle=270,width=0.85\textwidth]{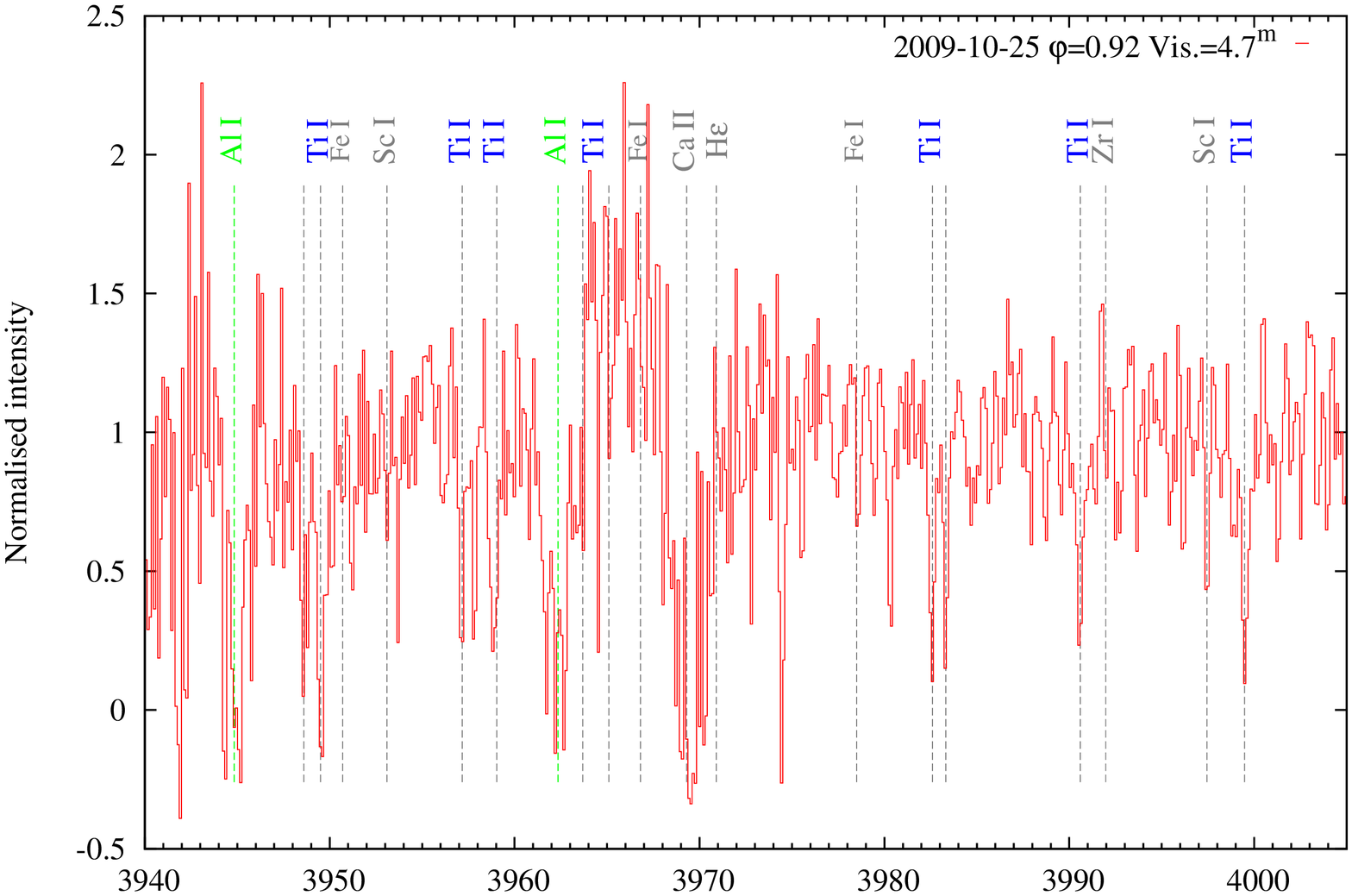}
\caption{Continued.}
\end{figure*}

  \setcounter{figure}{0}%

\begin{figure*} [tbh]
\centering
\includegraphics[angle=270,width=0.85\textwidth]{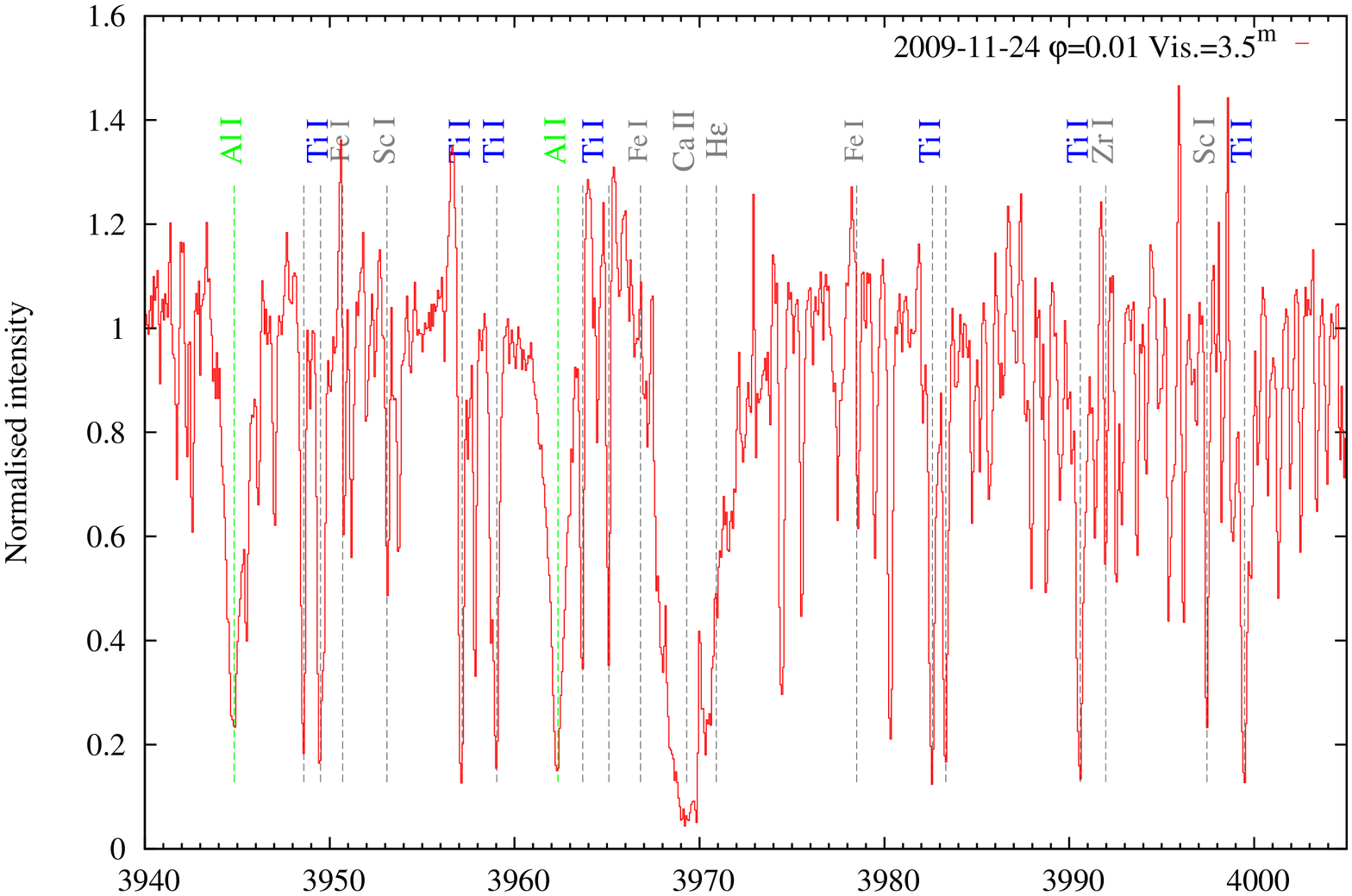}
\includegraphics[angle=270,width=0.85\textwidth]{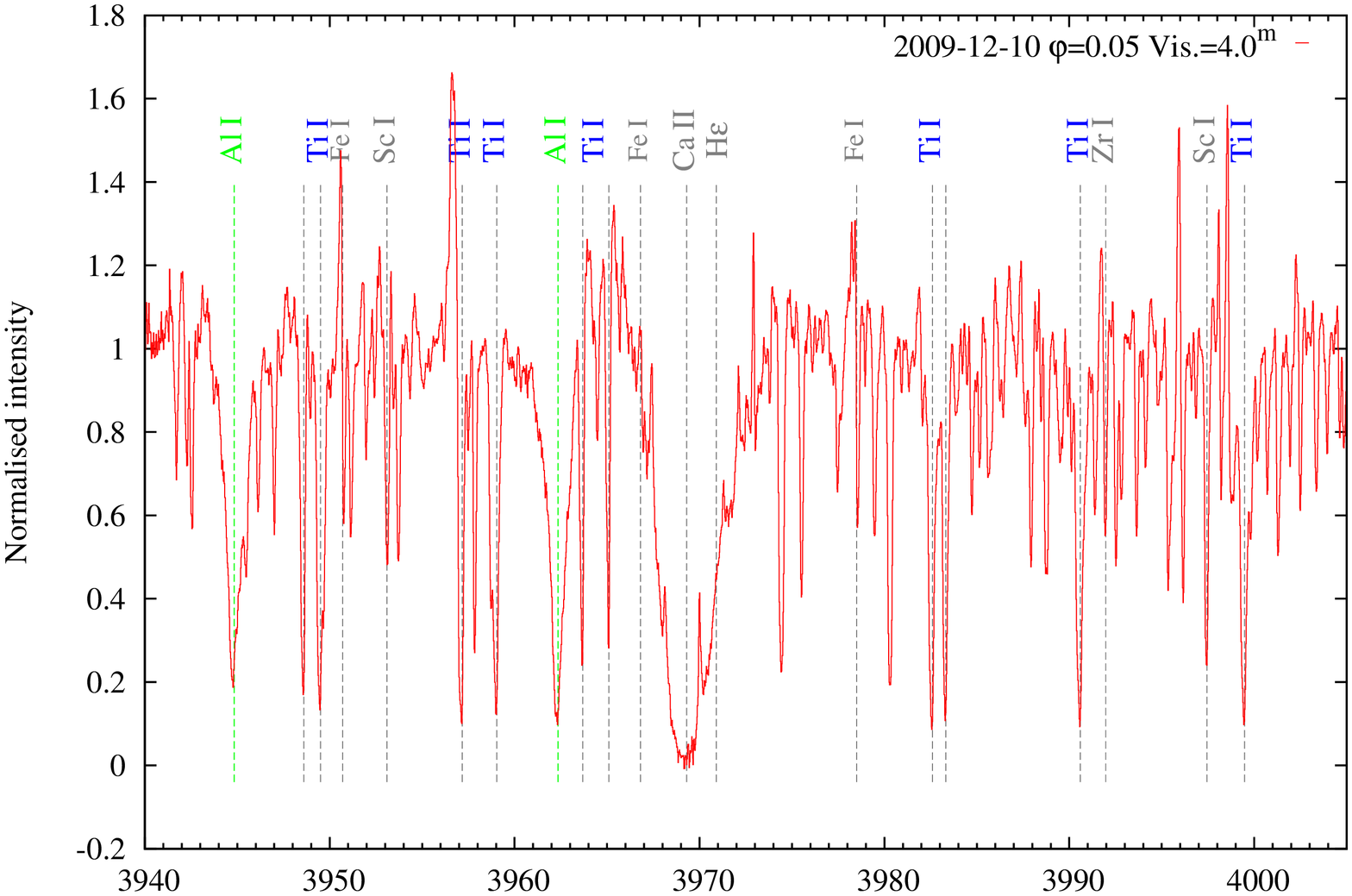}
\caption{Continued.}
\end{figure*}

  \setcounter{figure}{0}%

\begin{figure*} [tbh]
\centering
\includegraphics[angle=270,width=0.85\textwidth]{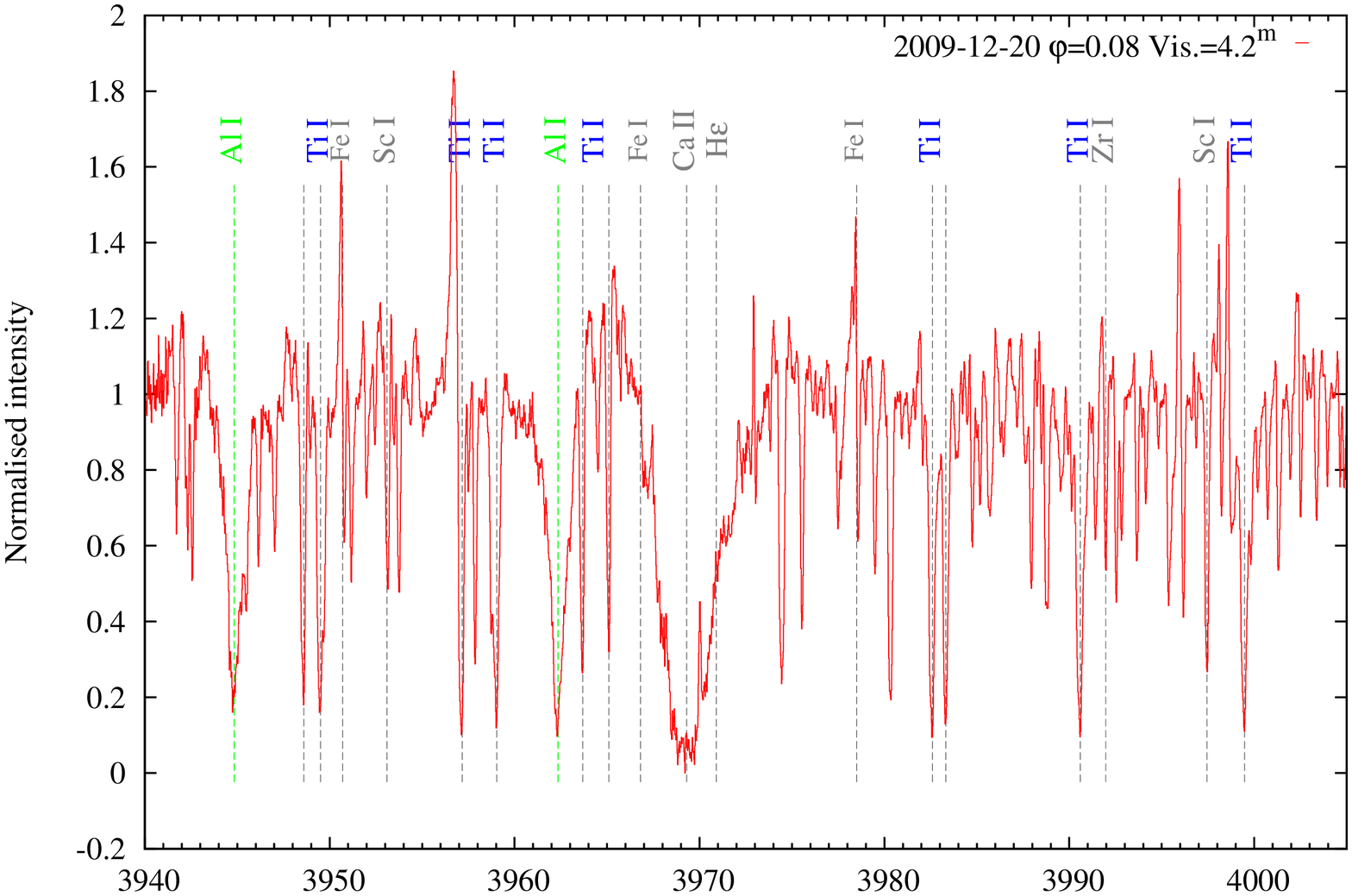}
\includegraphics[angle=270,width=0.85\textwidth]{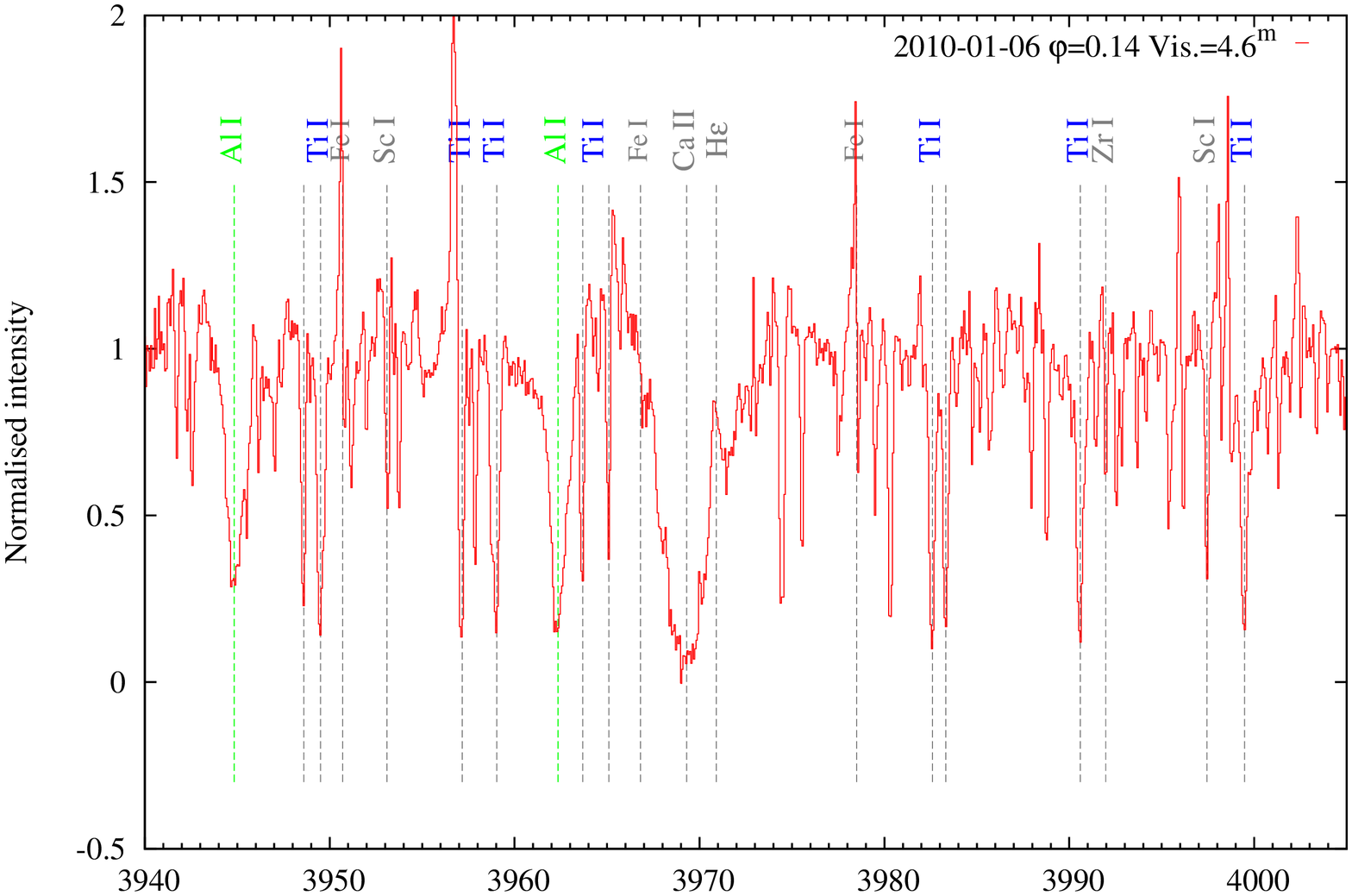}
\caption{Continued.}
\end{figure*}

  \setcounter{figure}{0}%

\begin{figure*} [tbh]
\centering
\includegraphics[angle=270,width=0.85\textwidth]{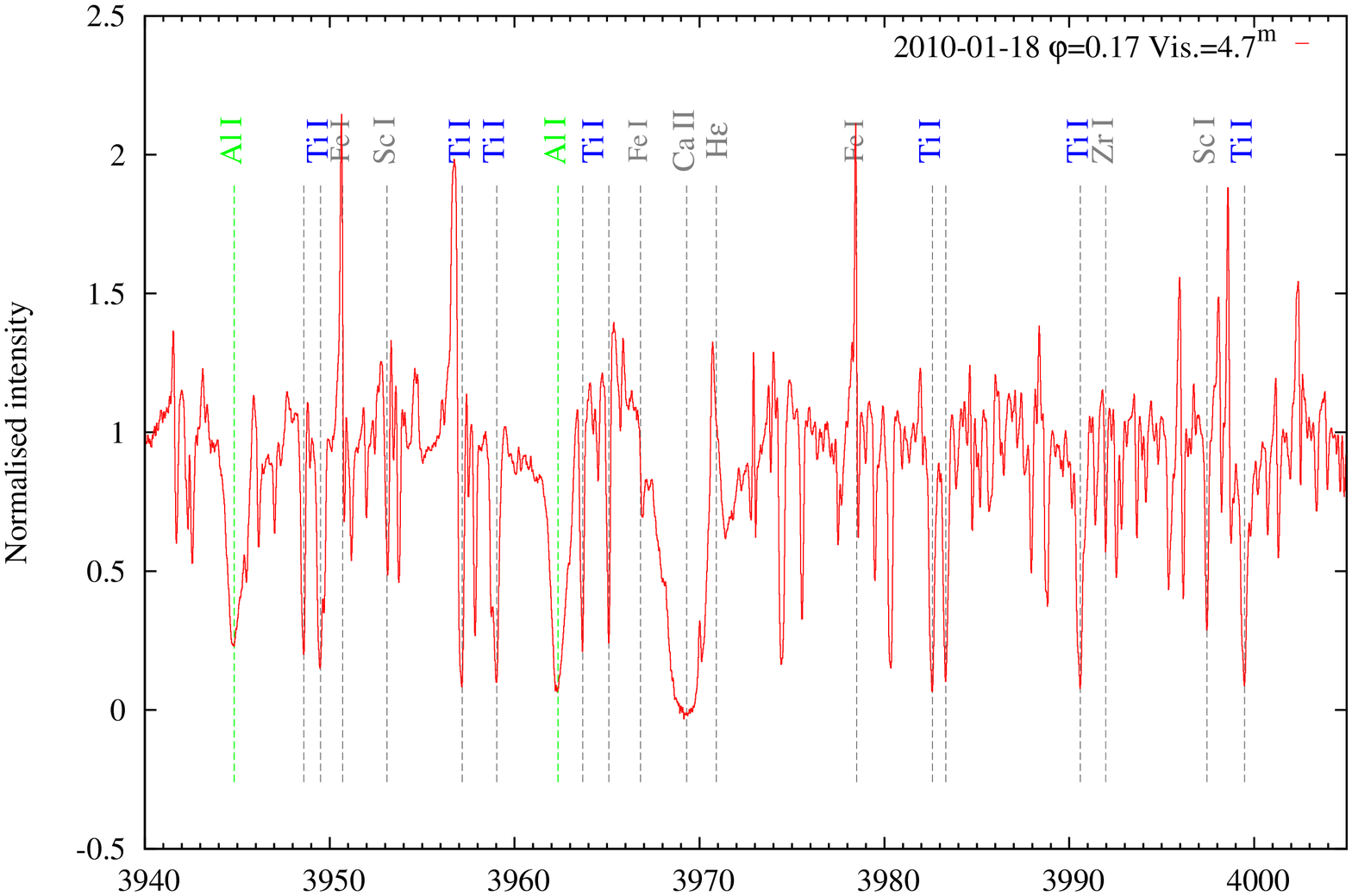}
\includegraphics[angle=270,width=0.85\textwidth]{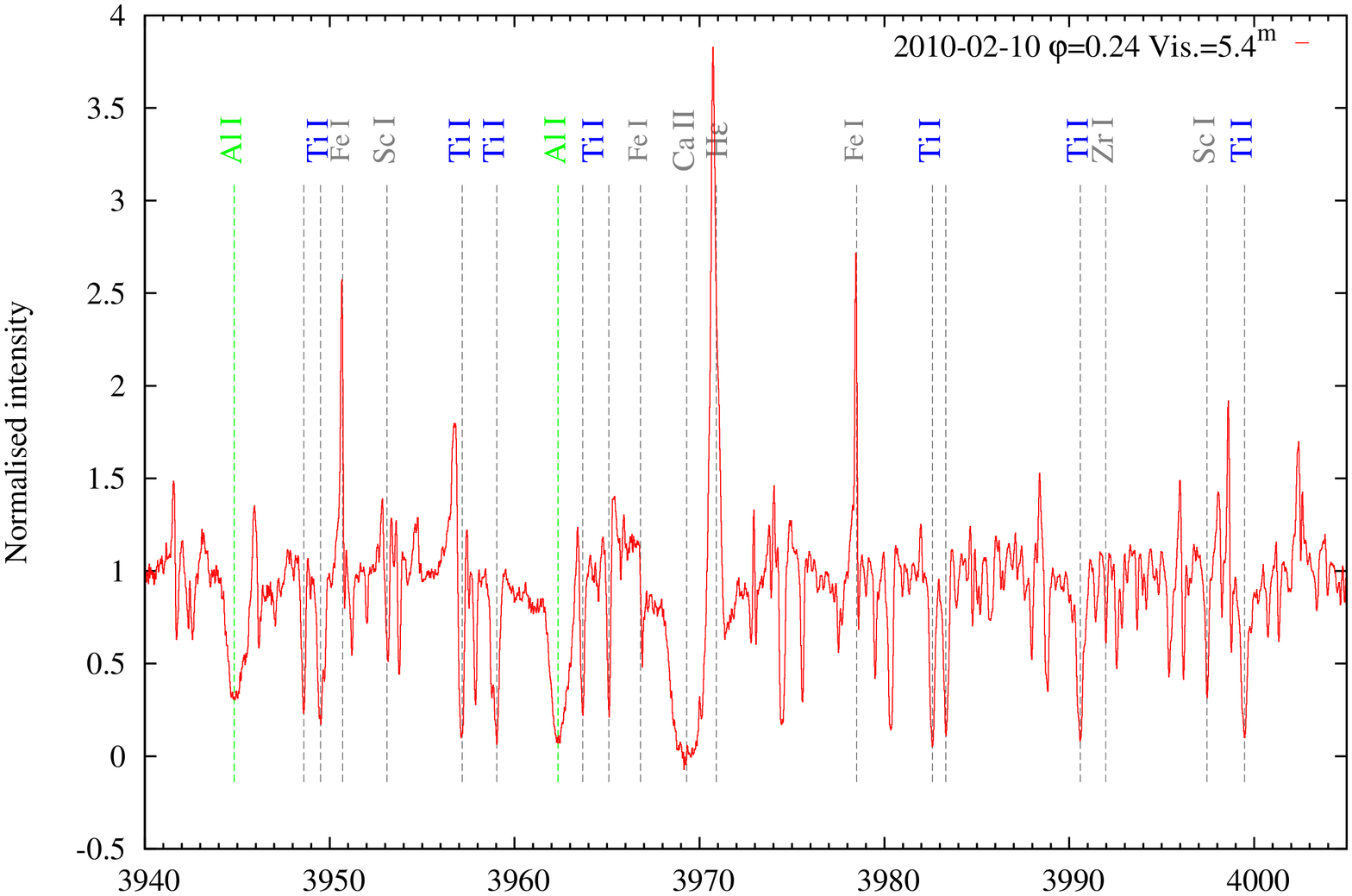}
\caption{Continued.}
\end{figure*}

  \setcounter{figure}{0}%

\begin{figure*} [tbh]
\centering
\includegraphics[angle=270,width=0.85\textwidth]{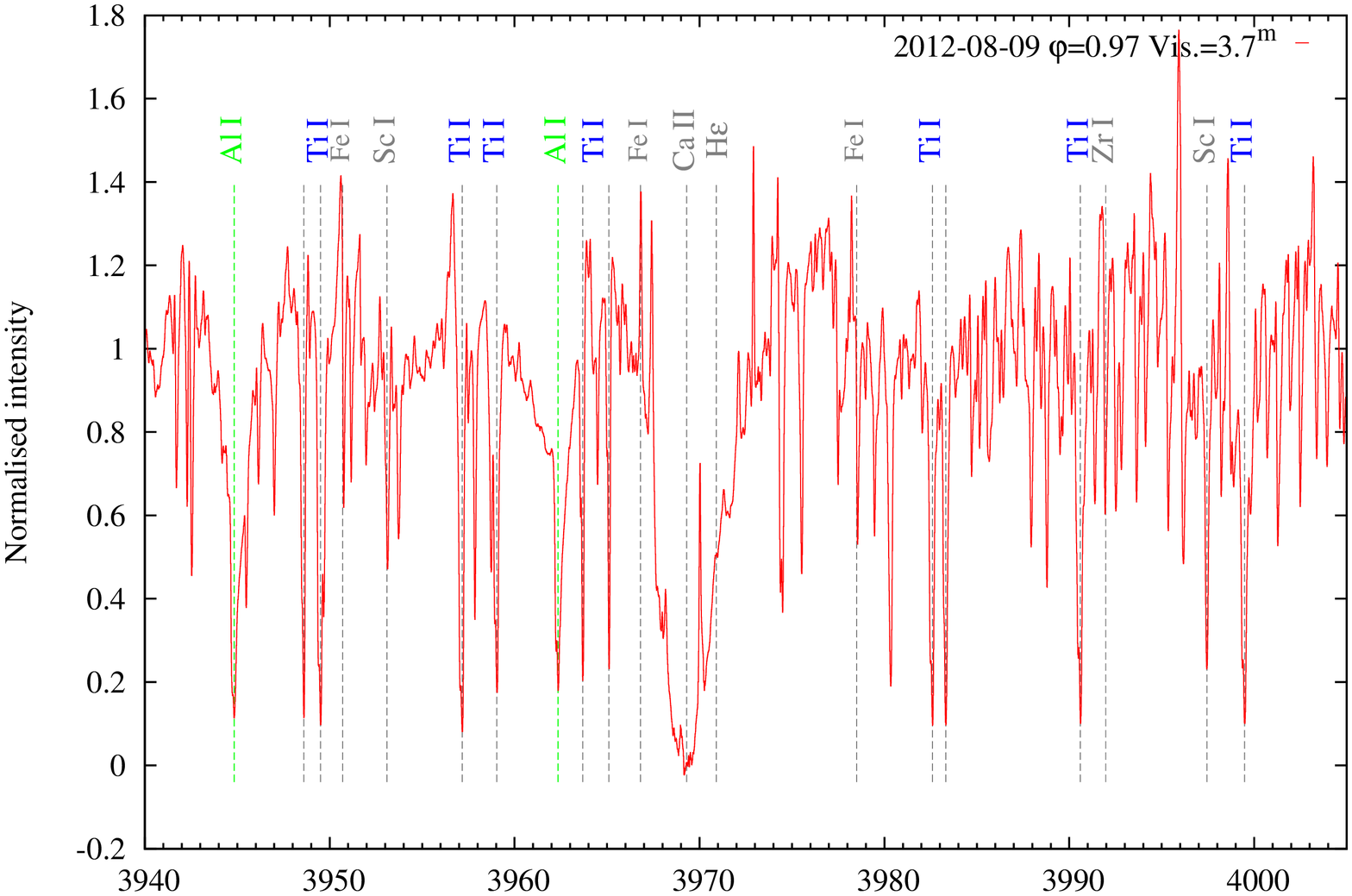}
\caption{Continued.}
\end{figure*}

\clearpage
\begin{figure*} [tbh]
\centering
\includegraphics[angle=270,width=0.85\textwidth]{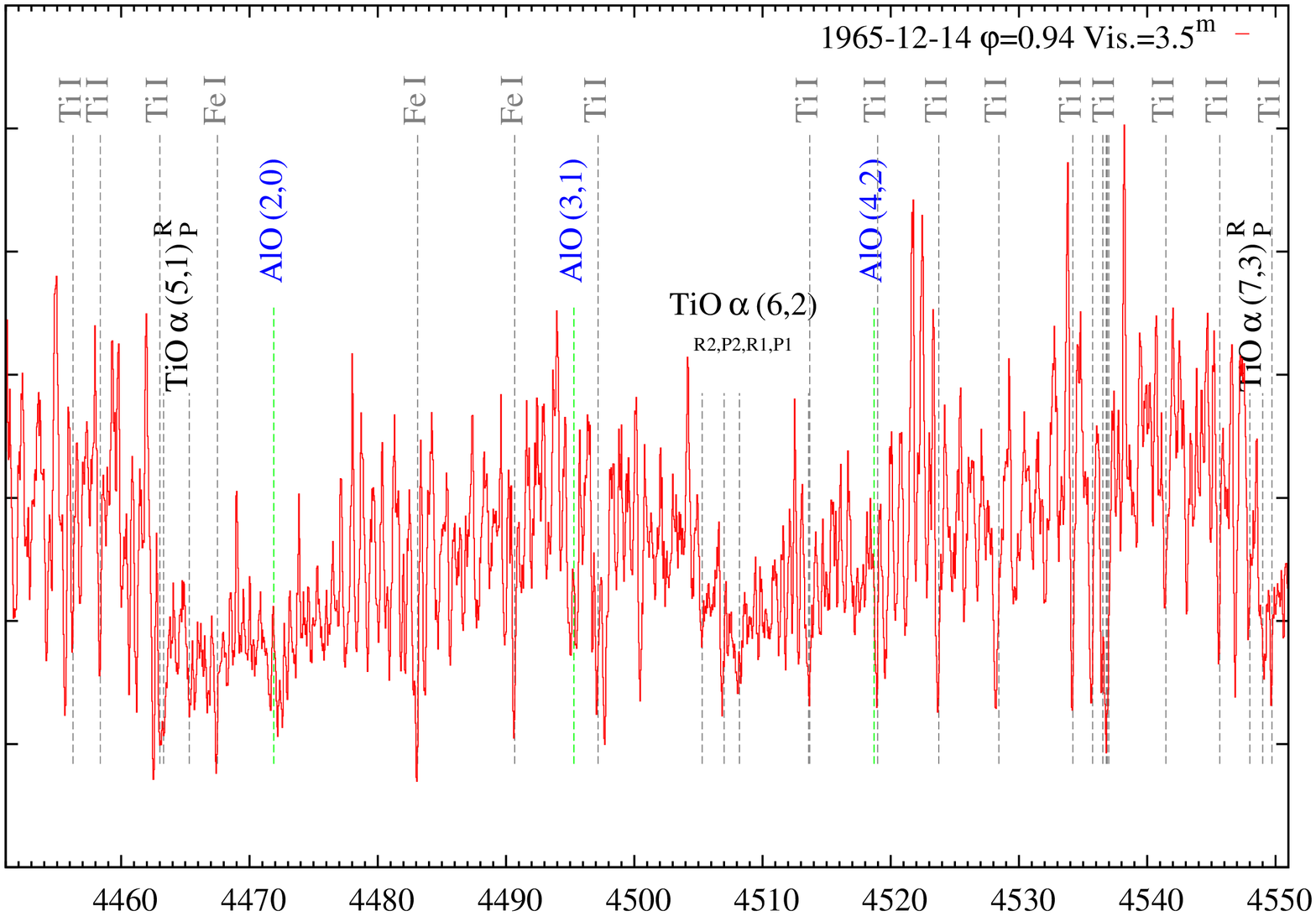}
\includegraphics[angle=270,width=0.85\textwidth]{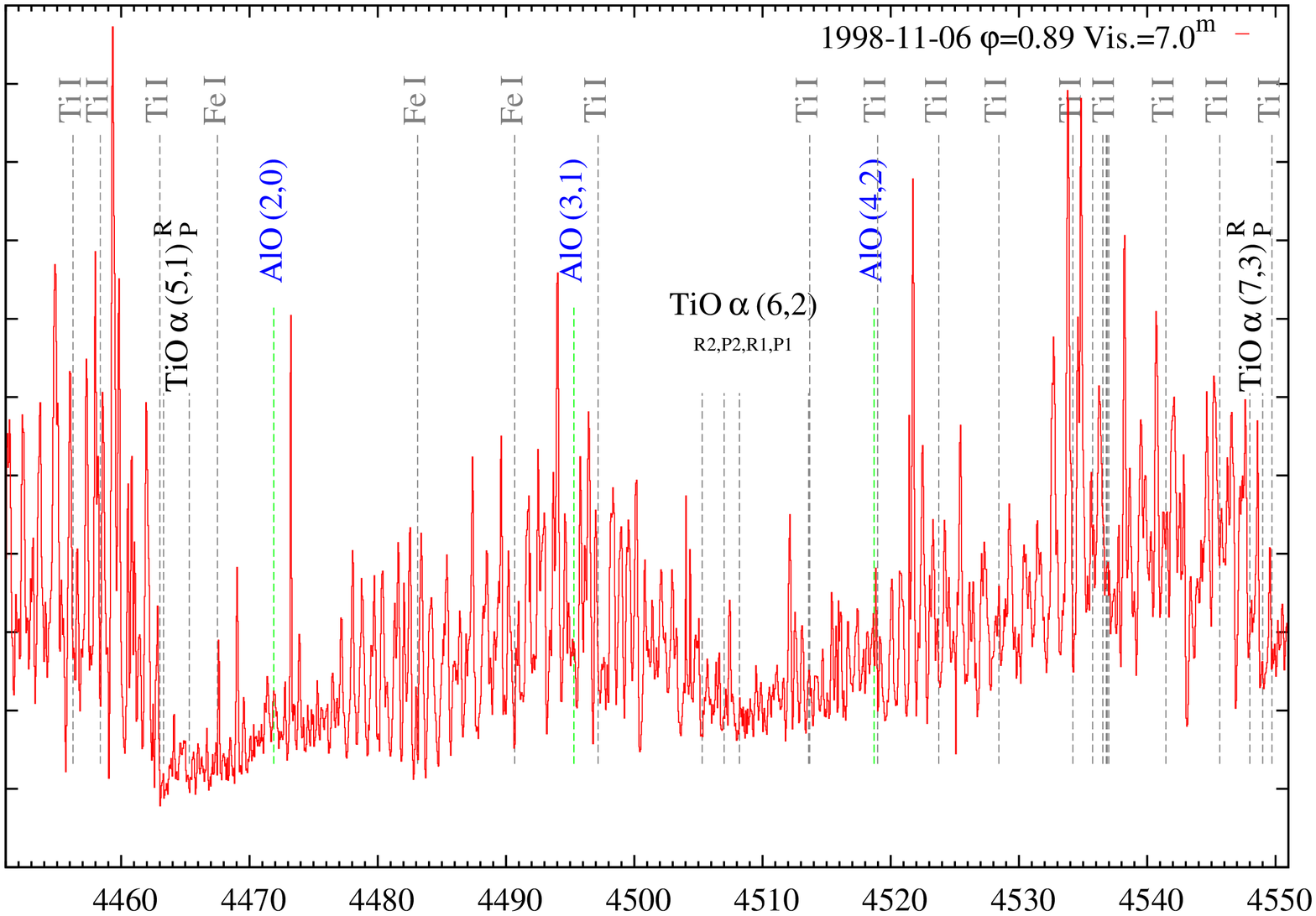}
\caption{High-resolution optical spectra of Mira covering the $\Delta\varv$=+2 sequence of the AlO $B-X$ system. The AlO bands are marked with blue labels. Locations of the identification markings are only approximate. The date of observation and the corresponding phase and visual magnitude are specified in the upper reight corner of each panel. Some spectra were smoothed. Spectra from Narval are affected by an imperfect combination of different echelle orders in the 4485--4495\,\AA\ range.}\label{fig-AlOp2}
\end{figure*}

  \setcounter{figure}{1}%

\begin{figure*} [tbh]
\centering
\includegraphics[angle=270,width=0.85\textwidth]{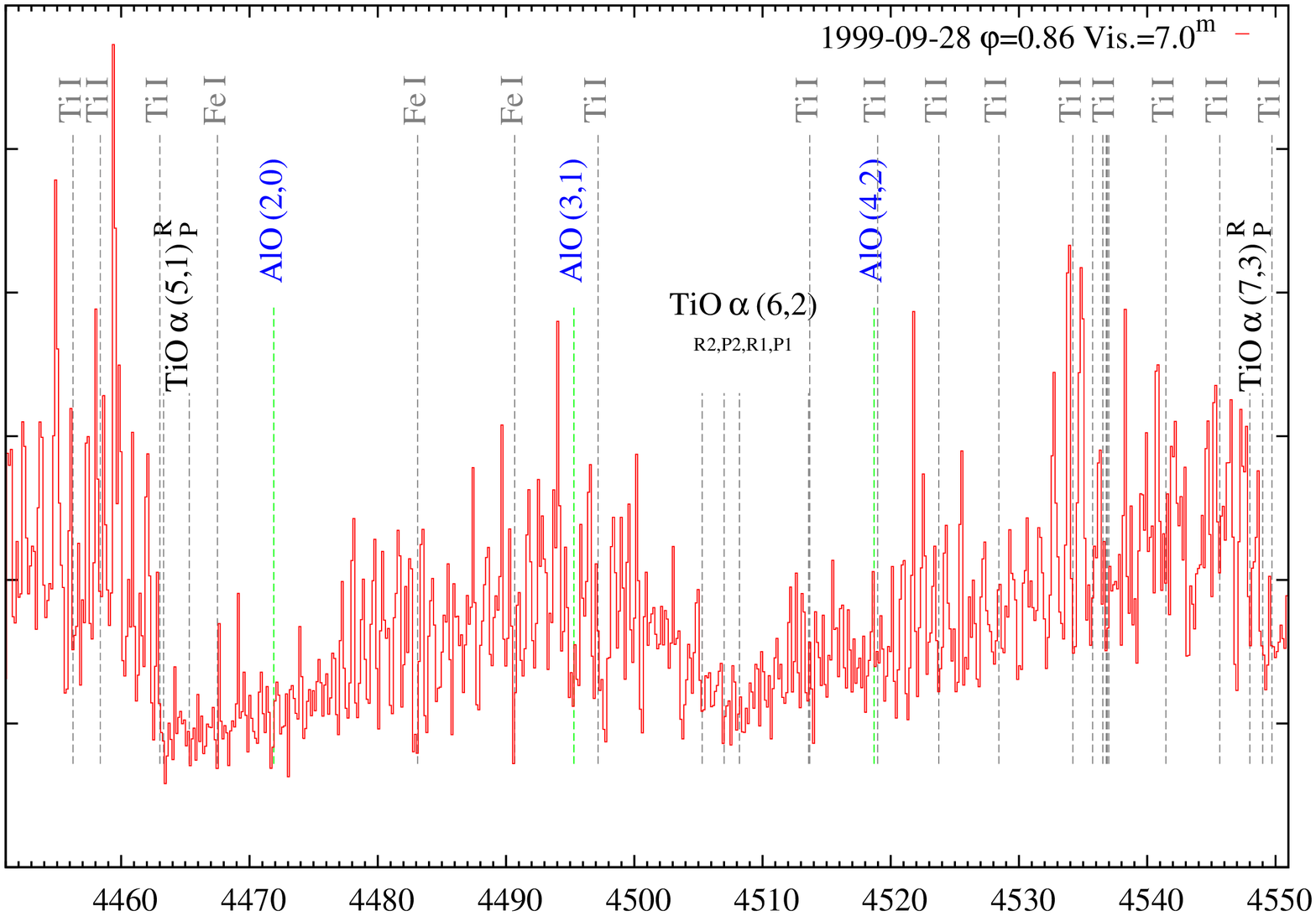}
\includegraphics[angle=270,width=0.85\textwidth]{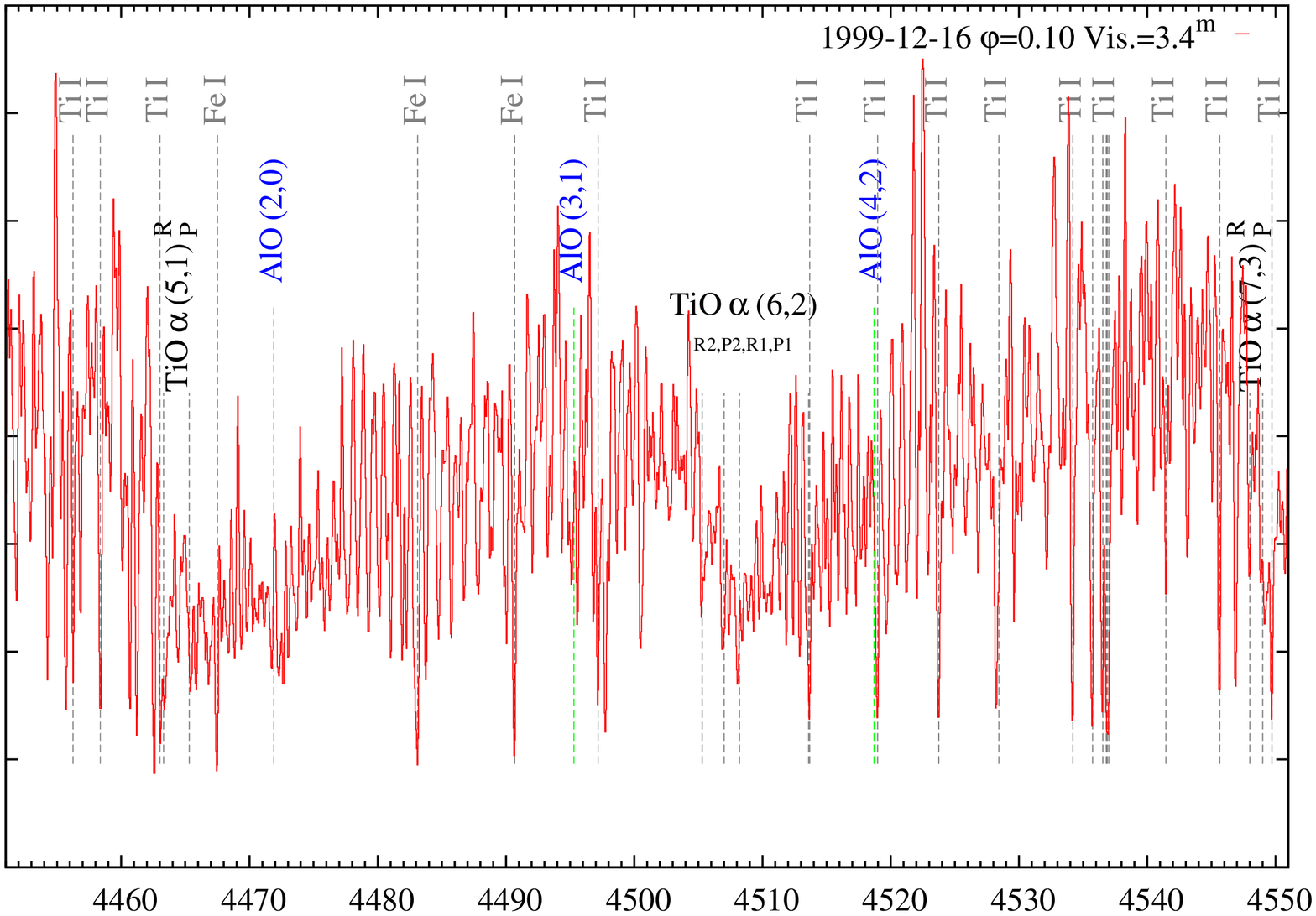}
\caption{Continued.}
\end{figure*}

  \setcounter{figure}{1}%

\begin{figure*} [tbh]
\centering
\includegraphics[angle=270,width=0.85\textwidth]{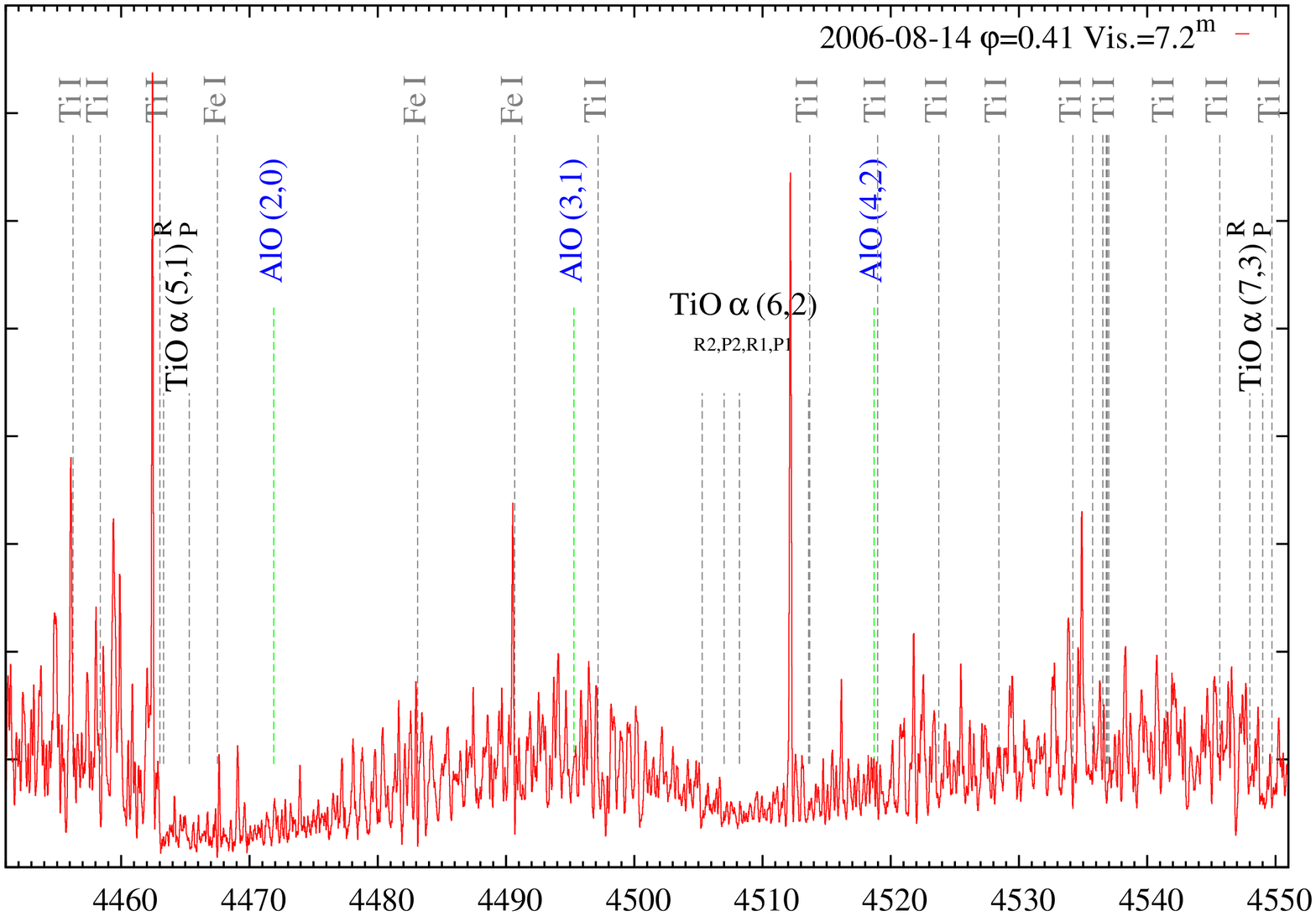}
\includegraphics[angle=270,width=0.85\textwidth]{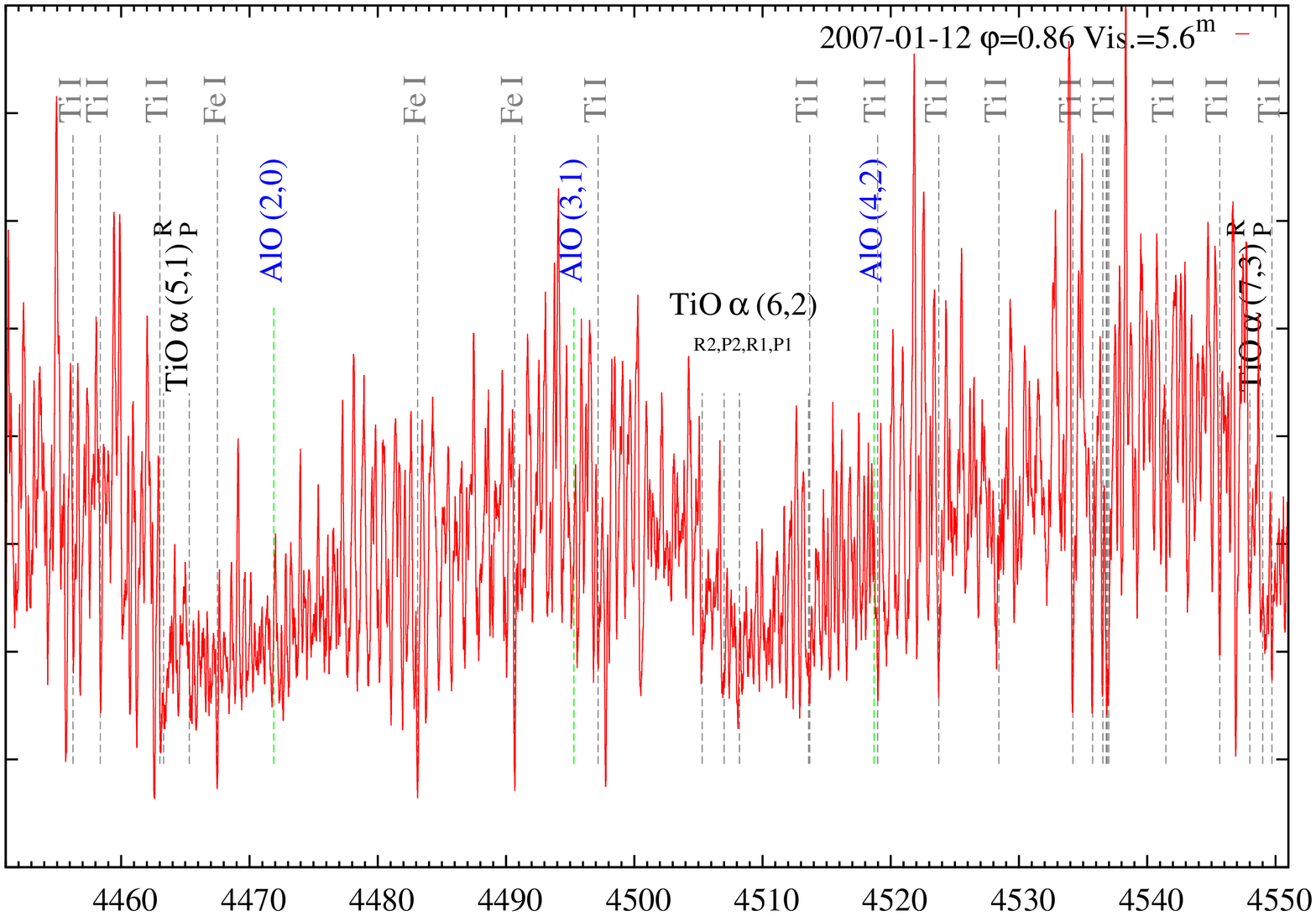}
\caption{Continued.}
\end{figure*}

  \setcounter{figure}{1}%

\begin{figure*} [tbh]
\centering
\includegraphics[angle=270,width=0.85\textwidth]{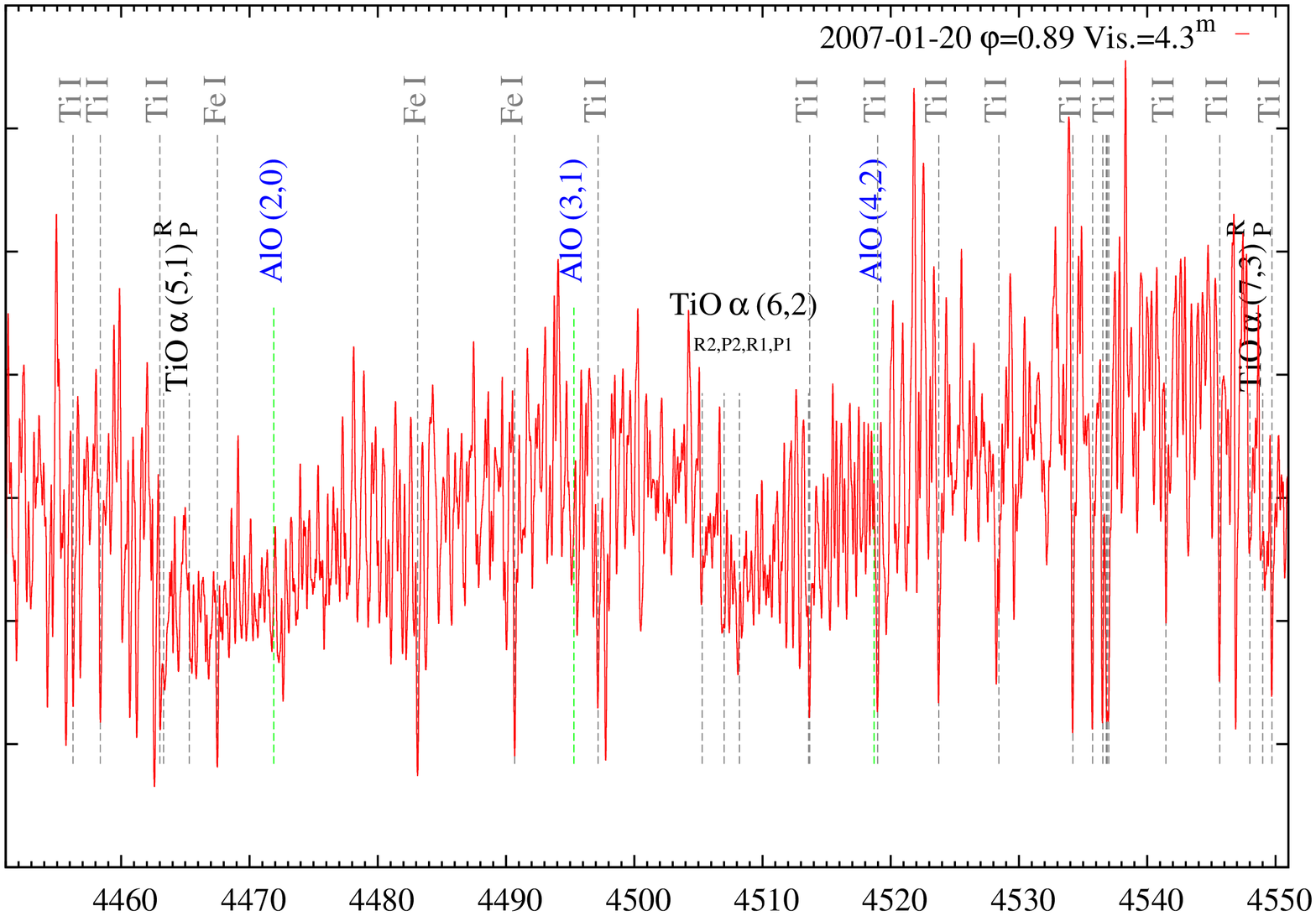}
\includegraphics[angle=270,width=0.85\textwidth]{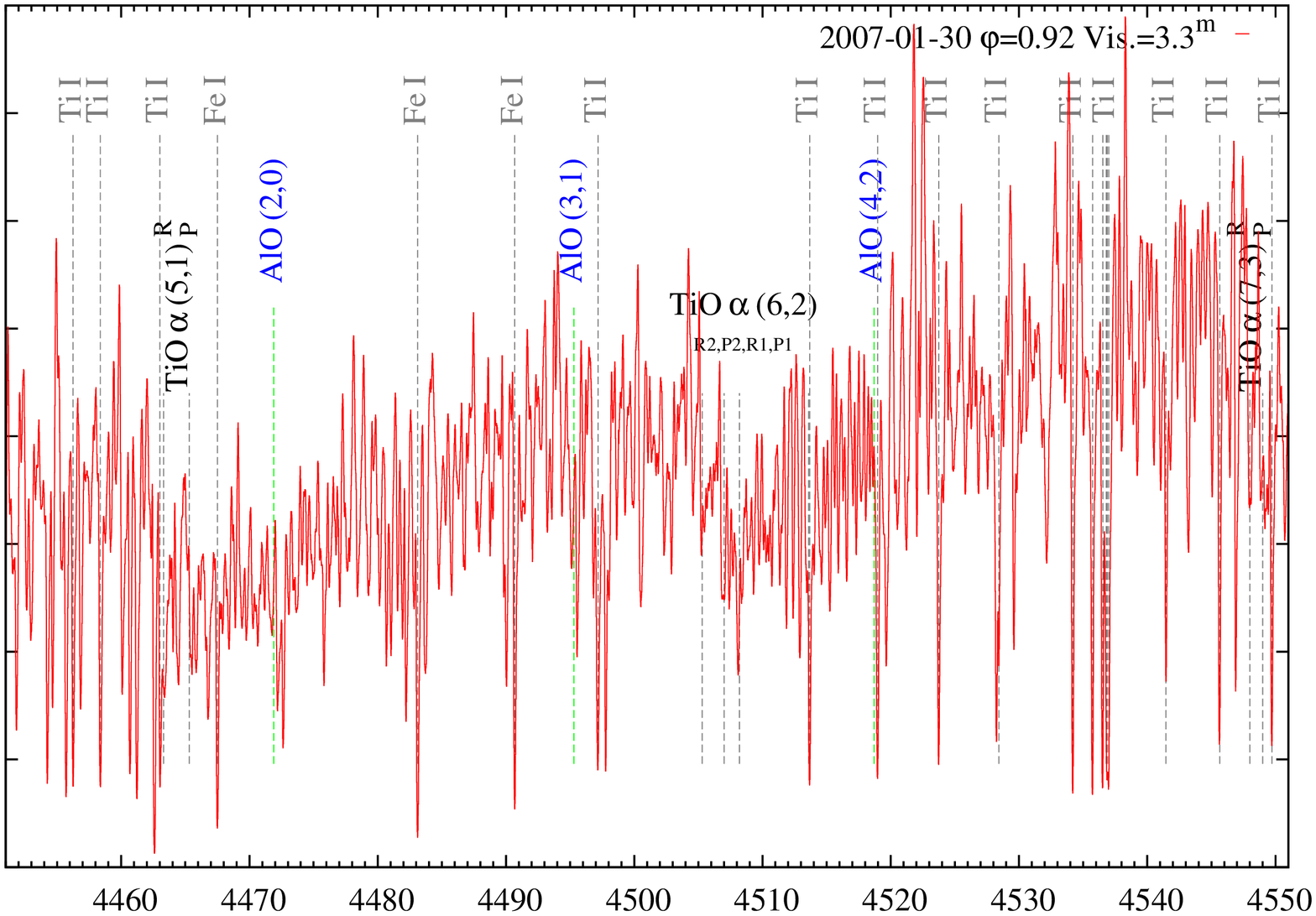}
\caption{Continued.}
\end{figure*}

  \setcounter{figure}{1}%

\begin{figure*} [tbh]
\centering
\includegraphics[angle=270,width=0.85\textwidth]{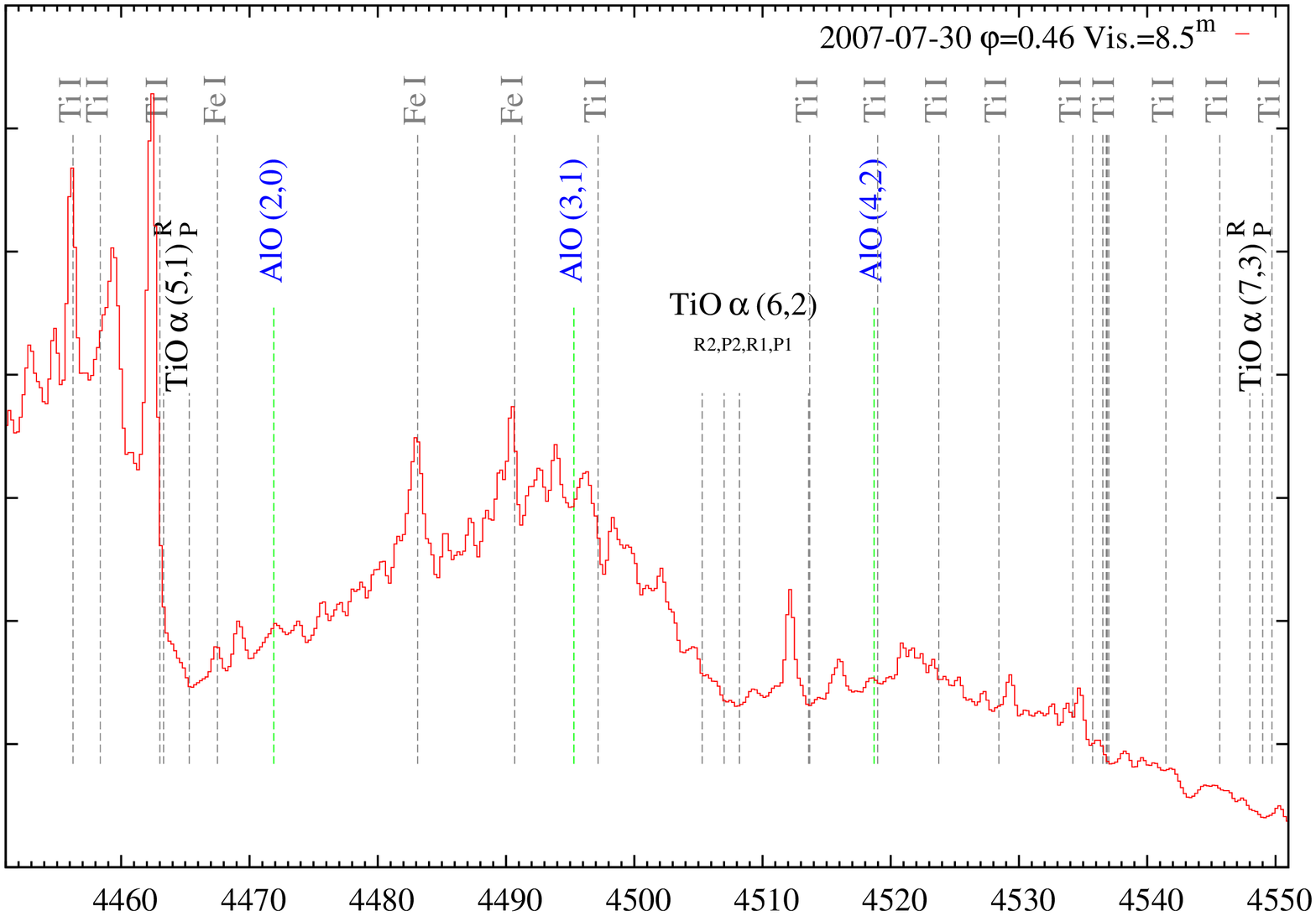}
\includegraphics[angle=270,width=0.85\textwidth]{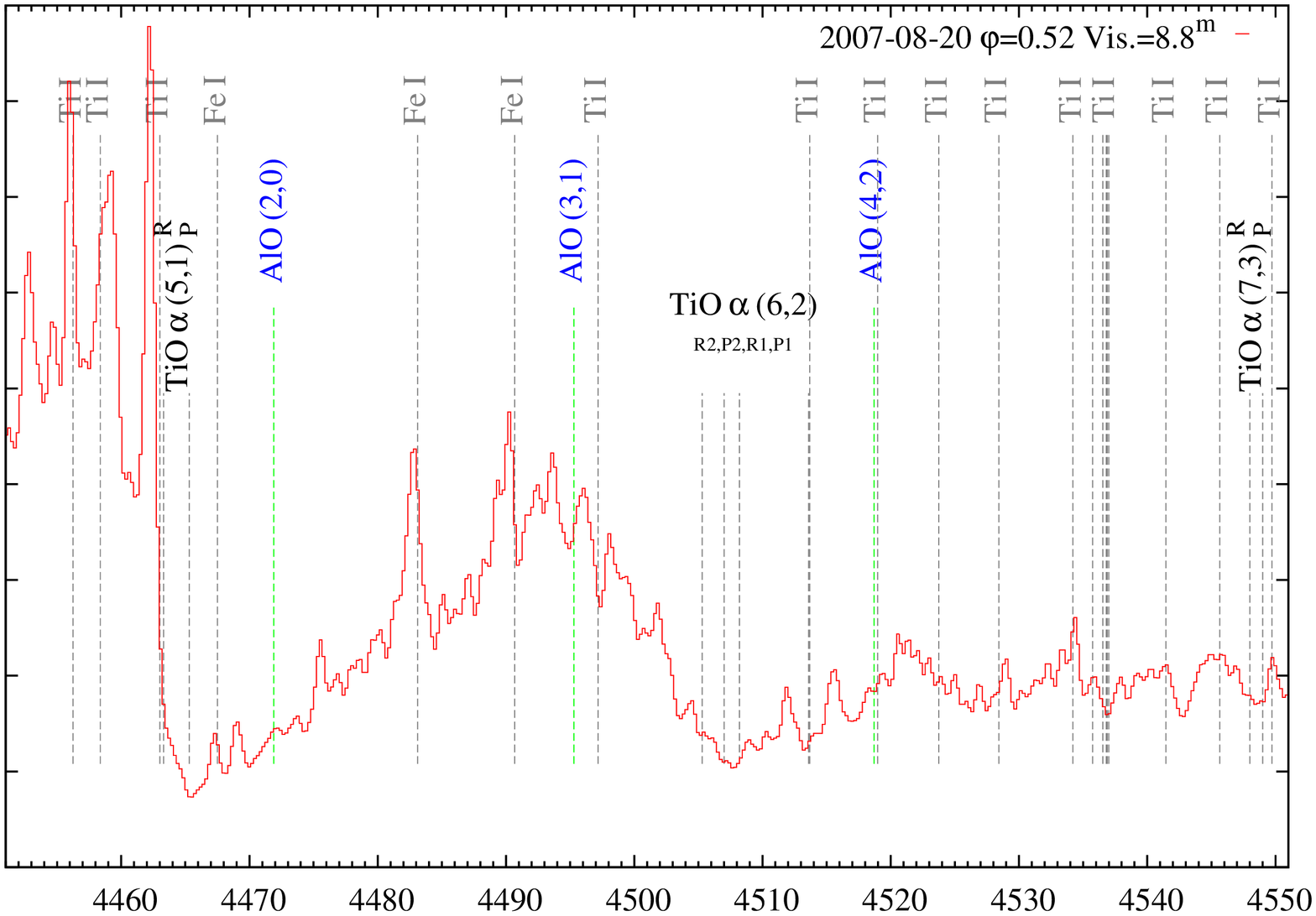}
\caption{Continued.}
\end{figure*}

  \setcounter{figure}{1}%

\begin{figure*} [tbh]
\centering
\includegraphics[angle=270,width=0.85\textwidth]{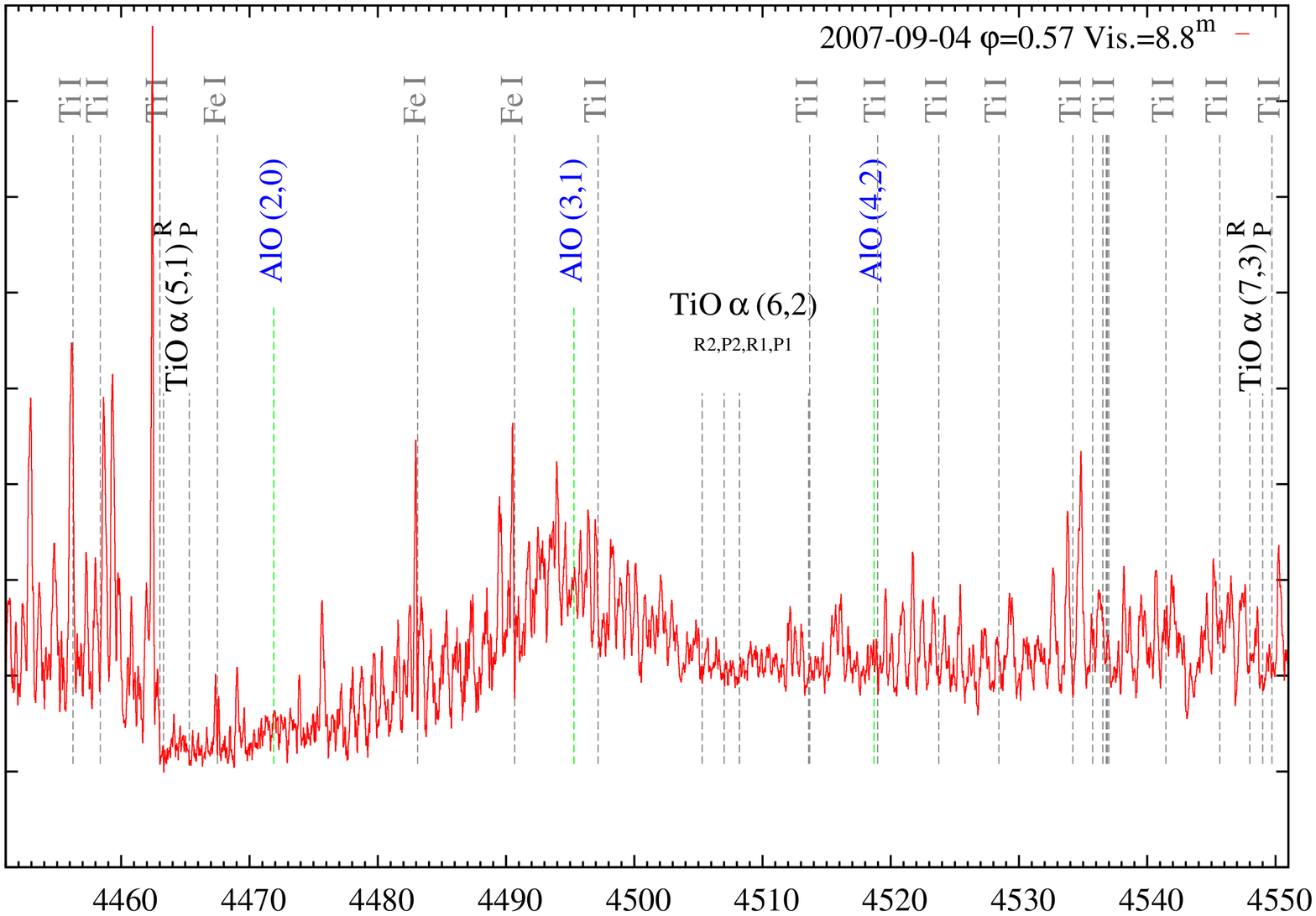}
\includegraphics[angle=270,width=0.85\textwidth]{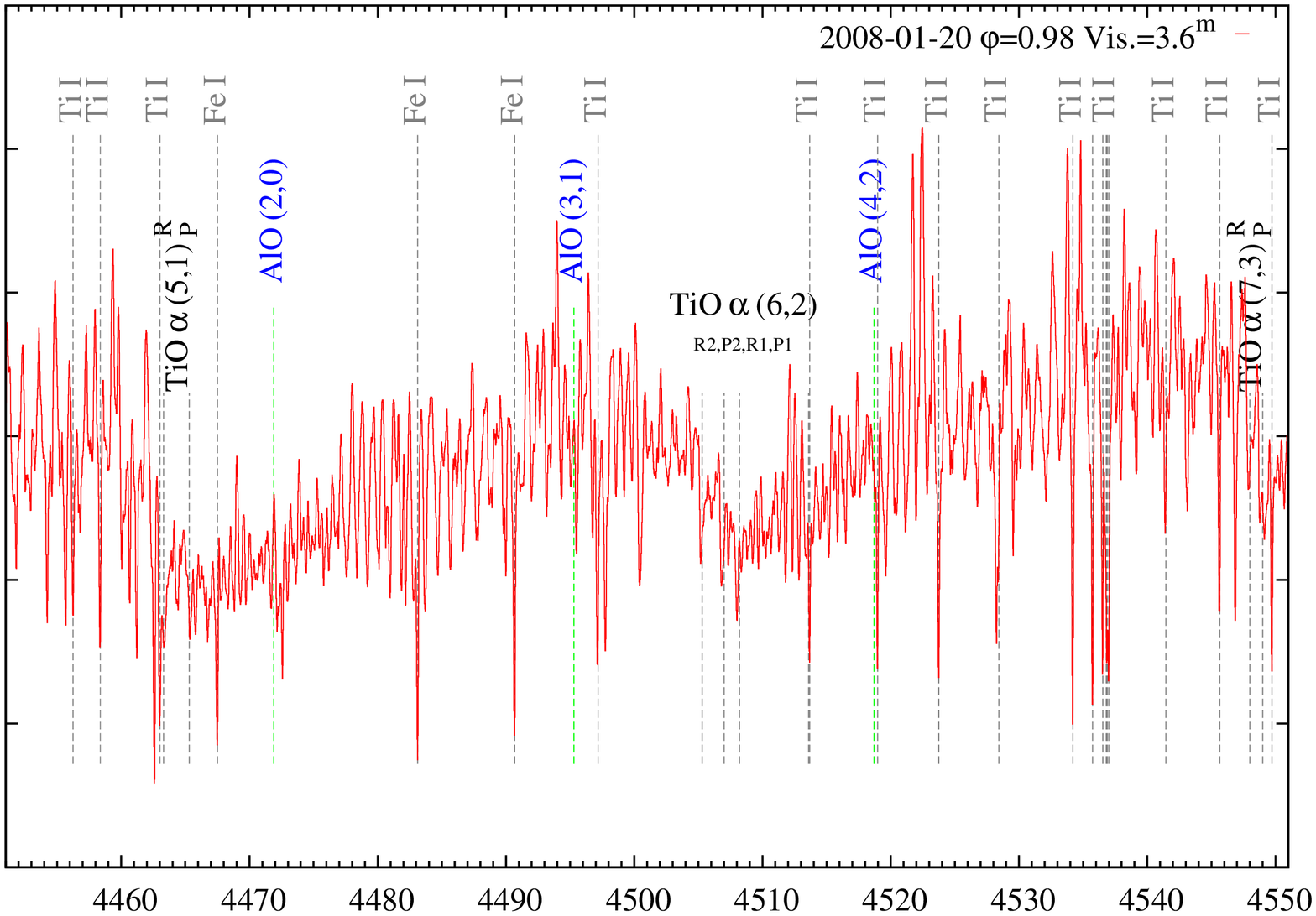}
\caption{Continued.}
\end{figure*}

  \setcounter{figure}{1}%

\begin{figure*} [tbh]
\centering
\includegraphics[angle=270,width=0.85\textwidth]{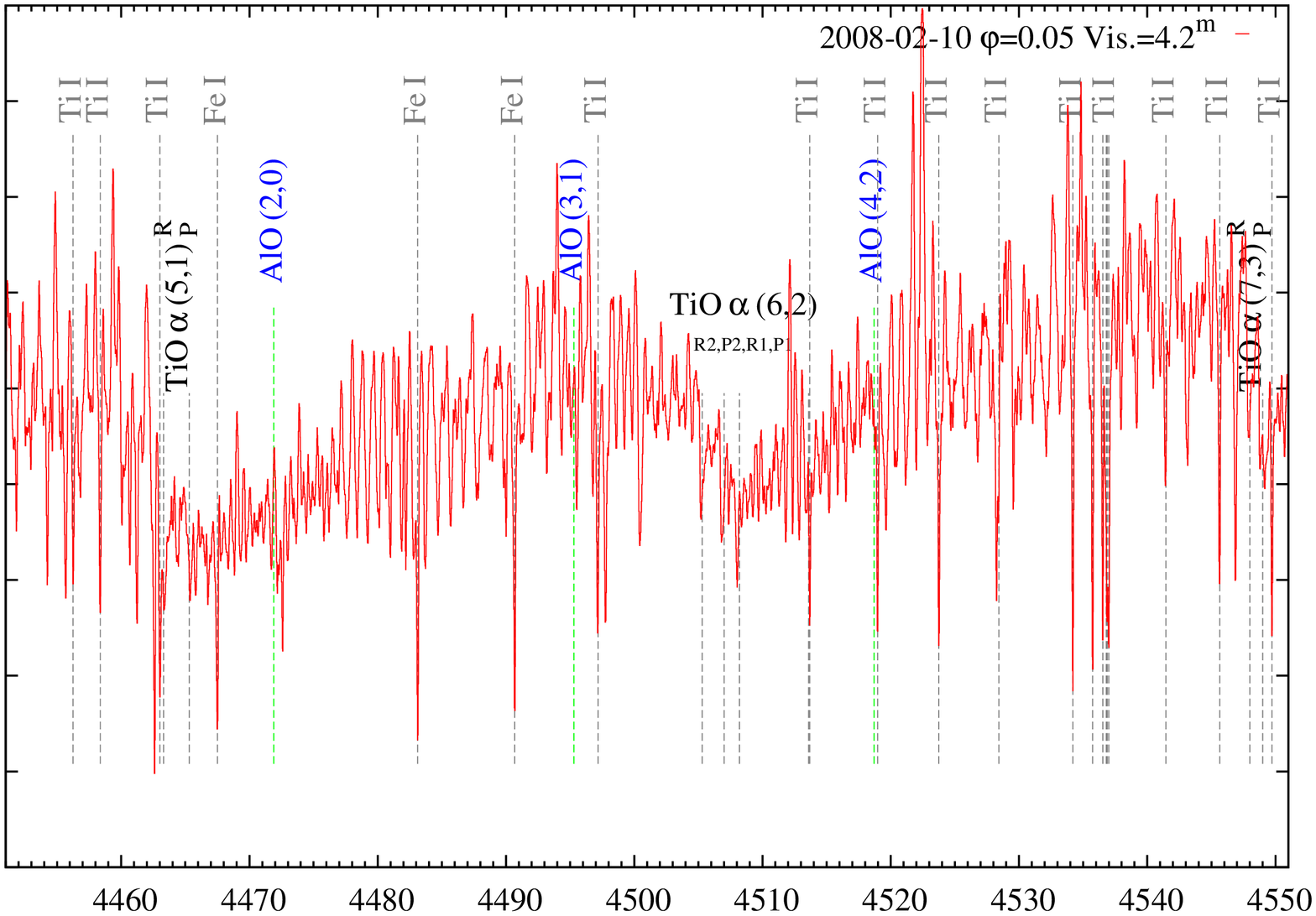}
\includegraphics[angle=270,width=0.85\textwidth]{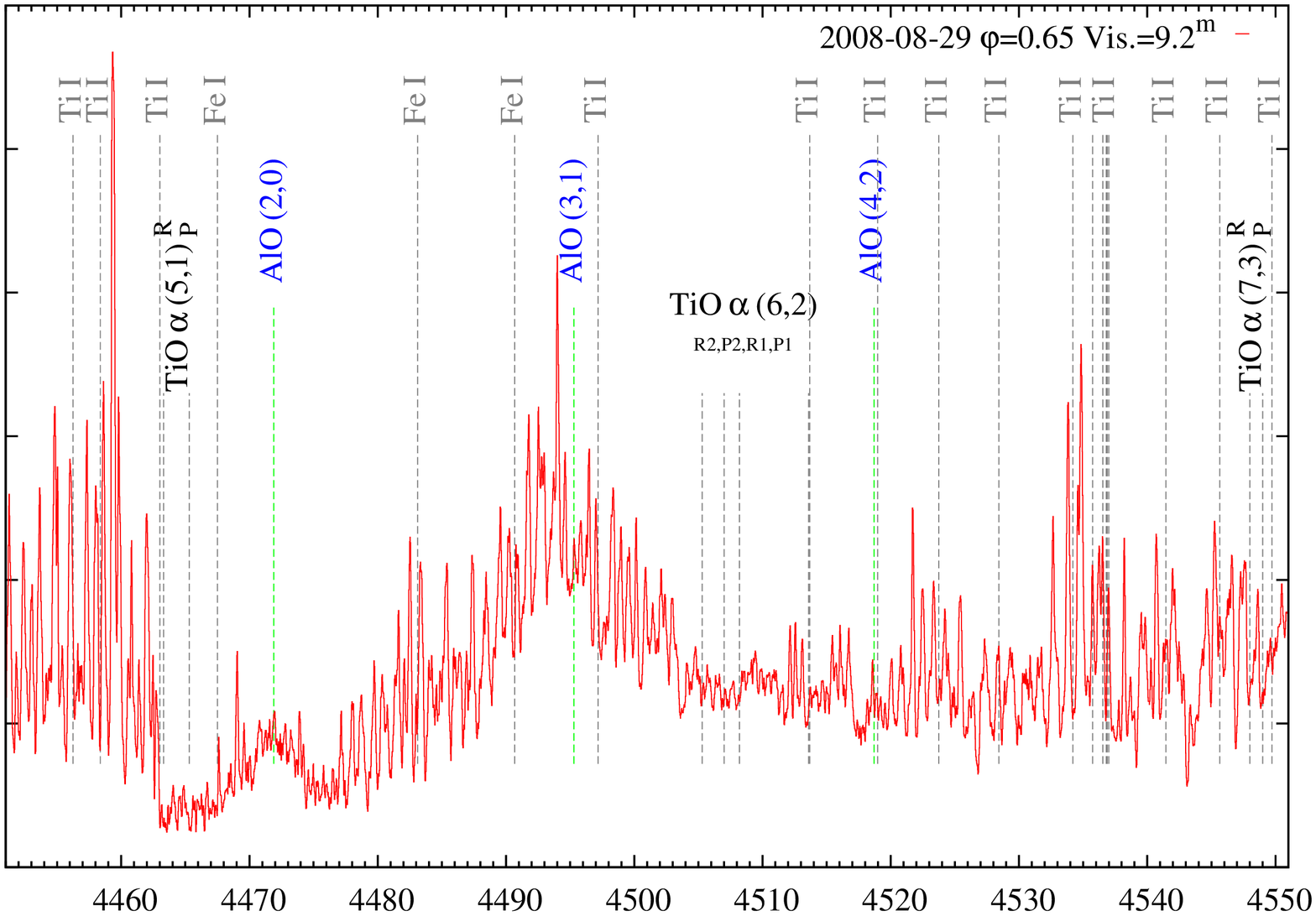}
\caption{Continued.}
\end{figure*}

  \setcounter{figure}{1}%

\begin{figure*} [tbh]
\centering
\includegraphics[angle=270,width=0.85\textwidth]{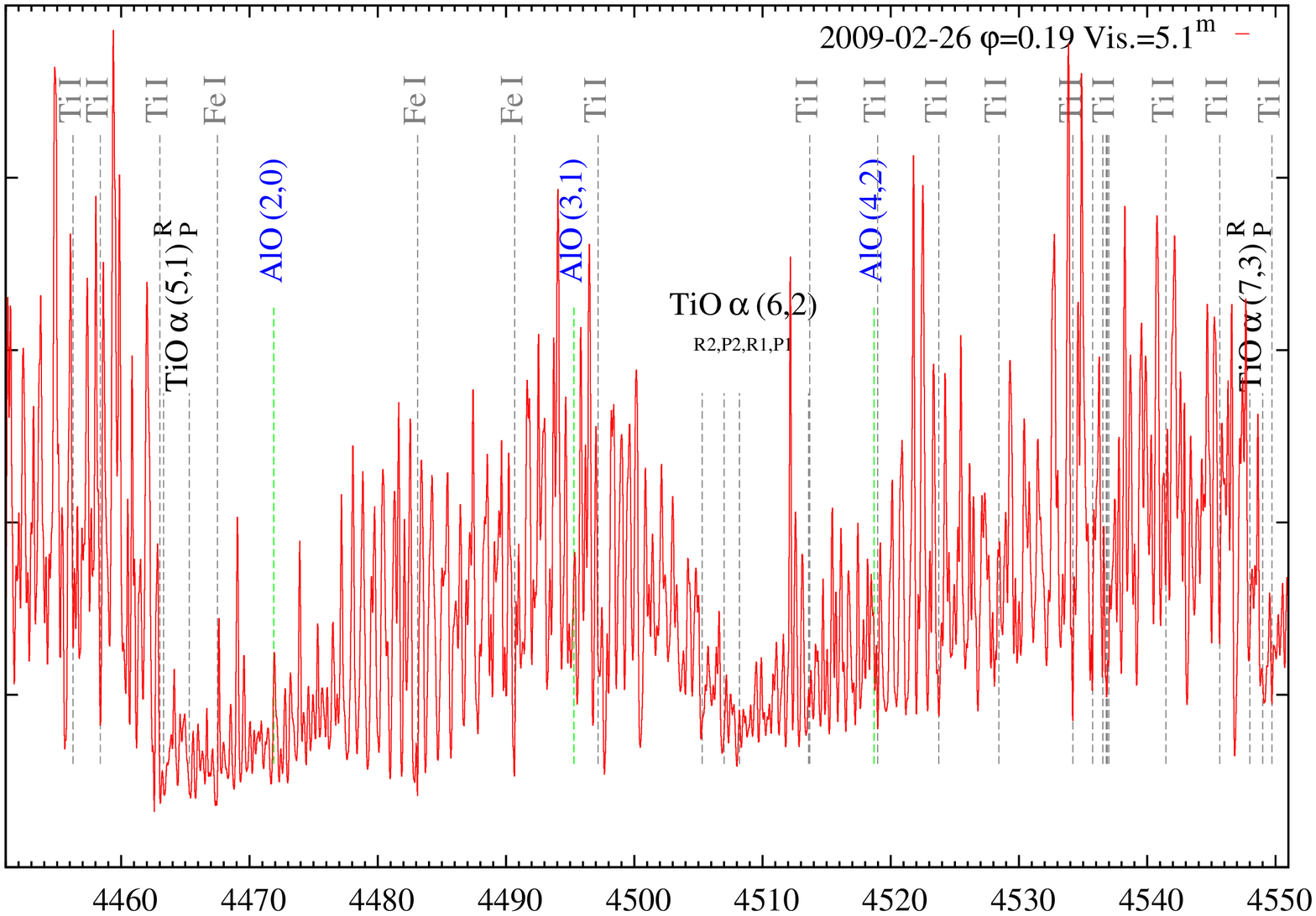}
\includegraphics[angle=270,width=0.85\textwidth]{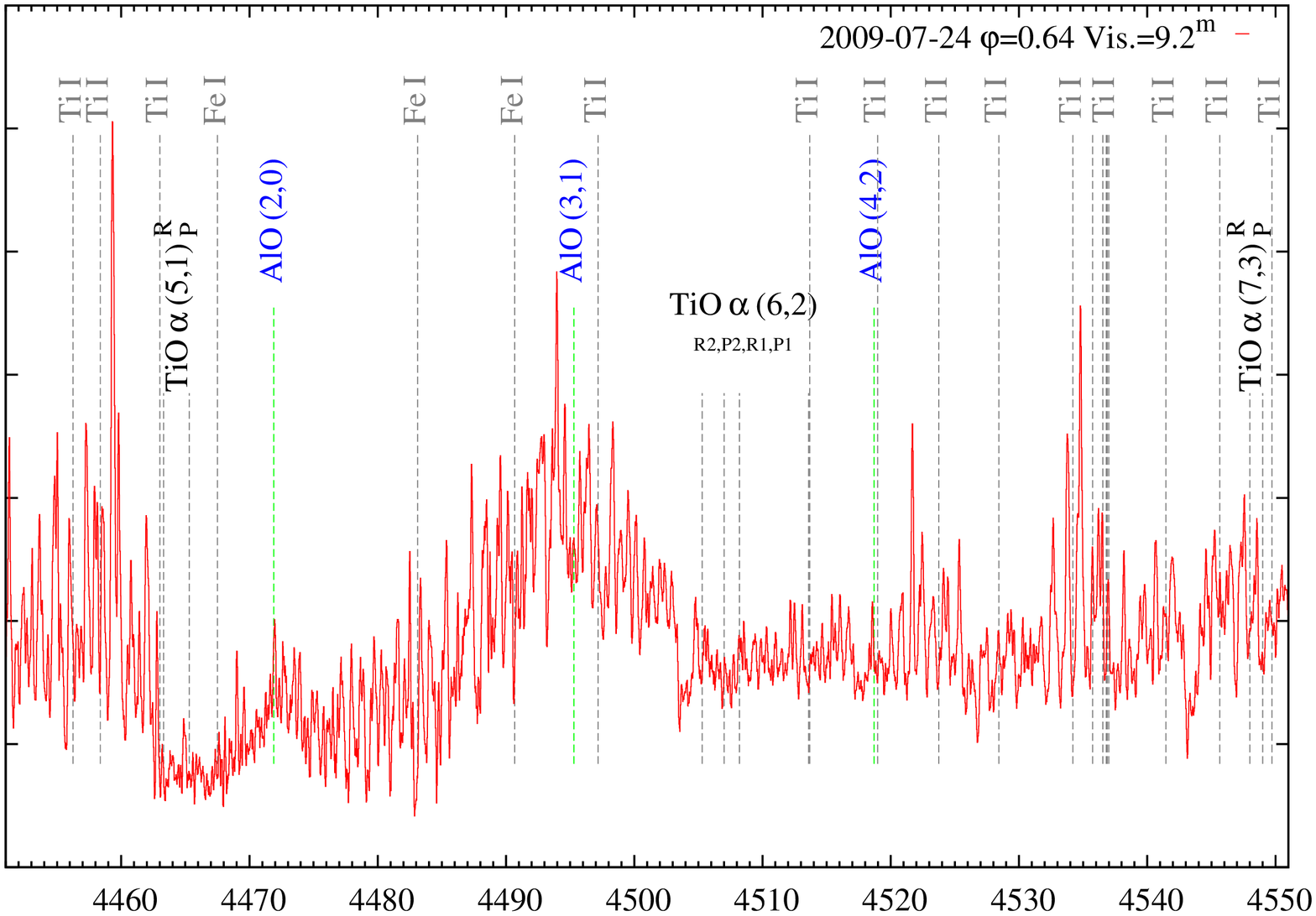}
\caption{Continued.}
\end{figure*}

  \setcounter{figure}{1}%

\begin{figure*} [tbh]
\centering
\includegraphics[angle=270,width=0.85\textwidth]{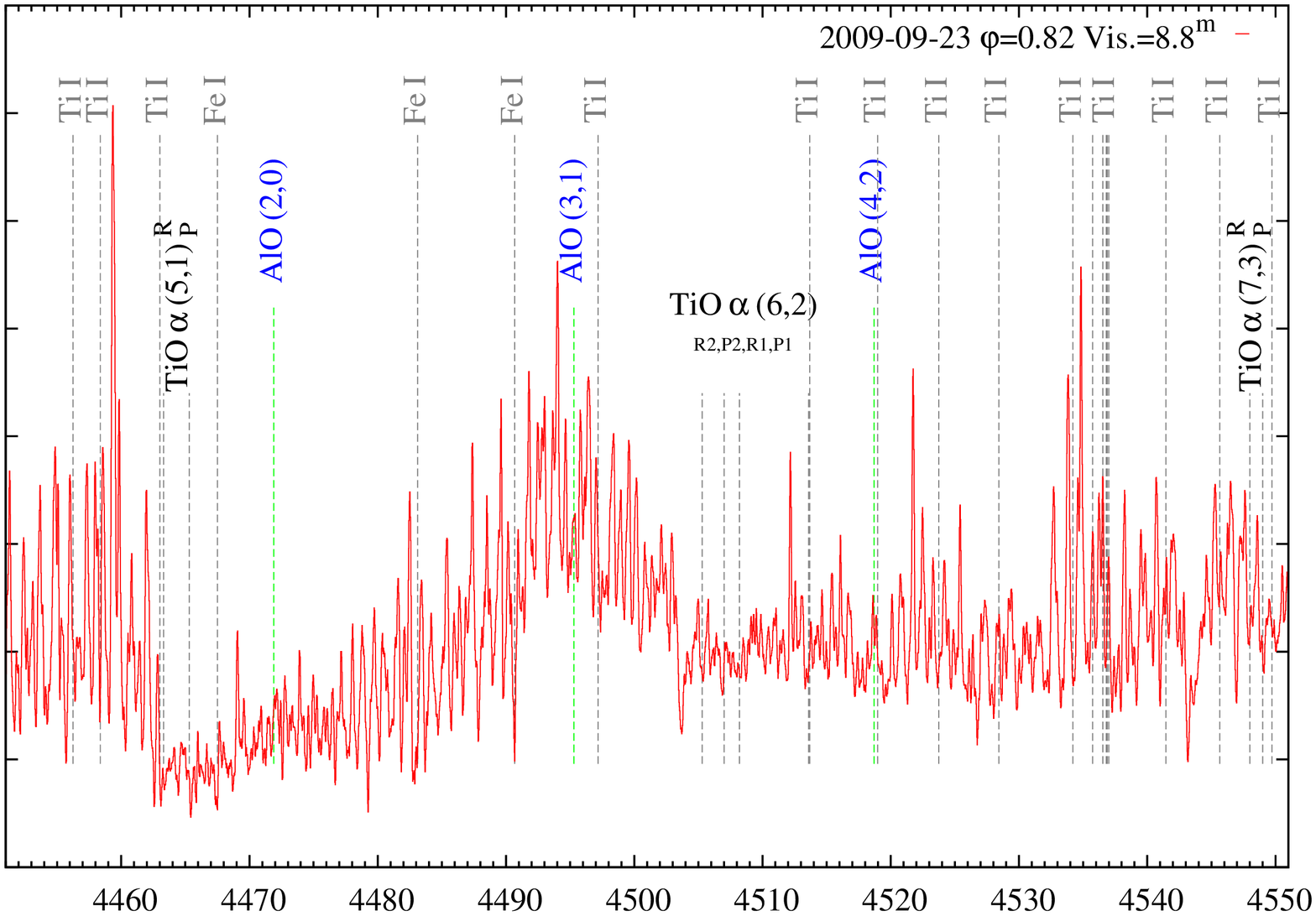}
\includegraphics[angle=270,width=0.85\textwidth]{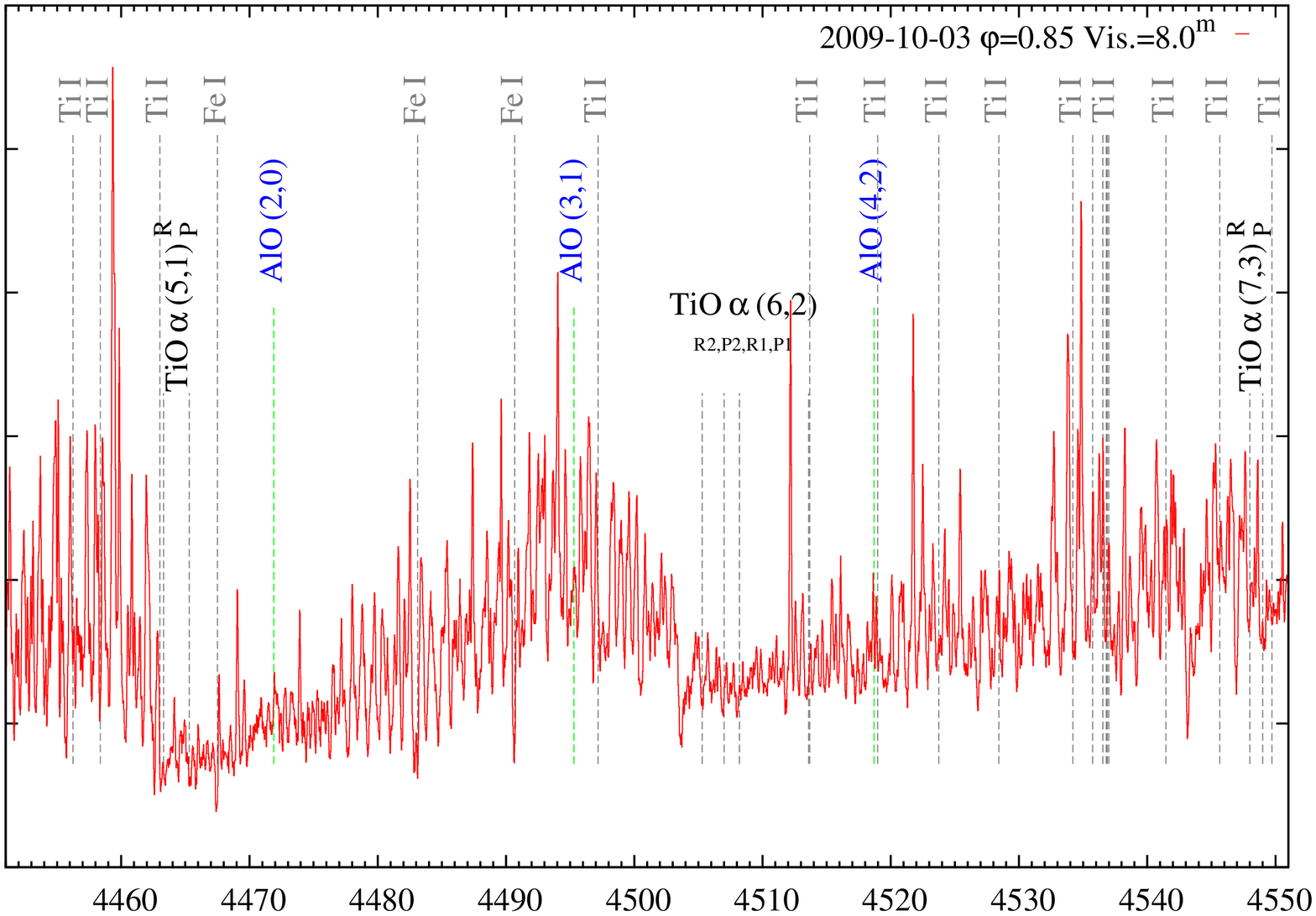}
\caption{Continued.}
\end{figure*}

  \setcounter{figure}{1}%

\begin{figure*} [tbh]
\centering
\includegraphics[angle=270,width=0.85\textwidth]{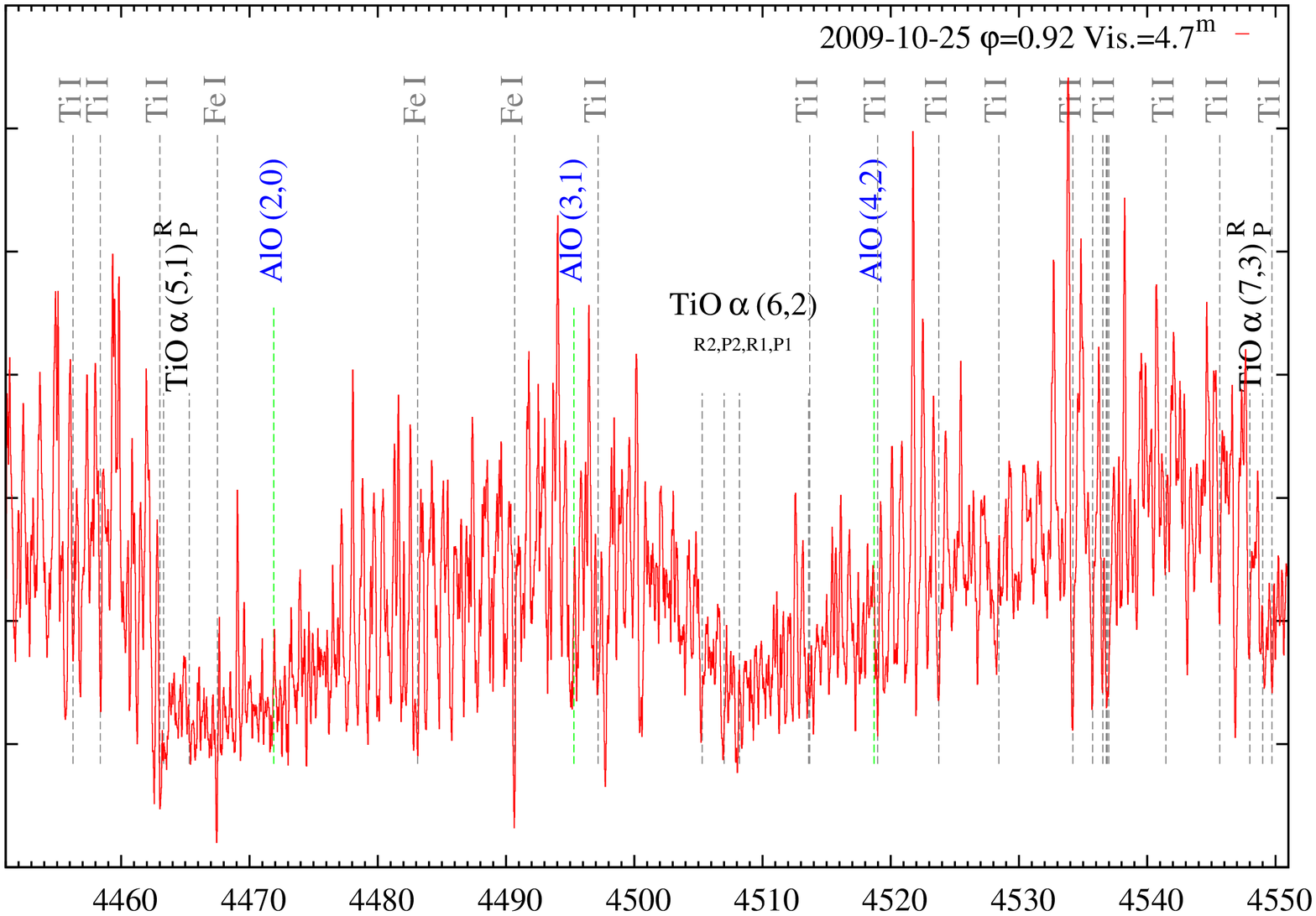}
\includegraphics[angle=270,width=0.85\textwidth]{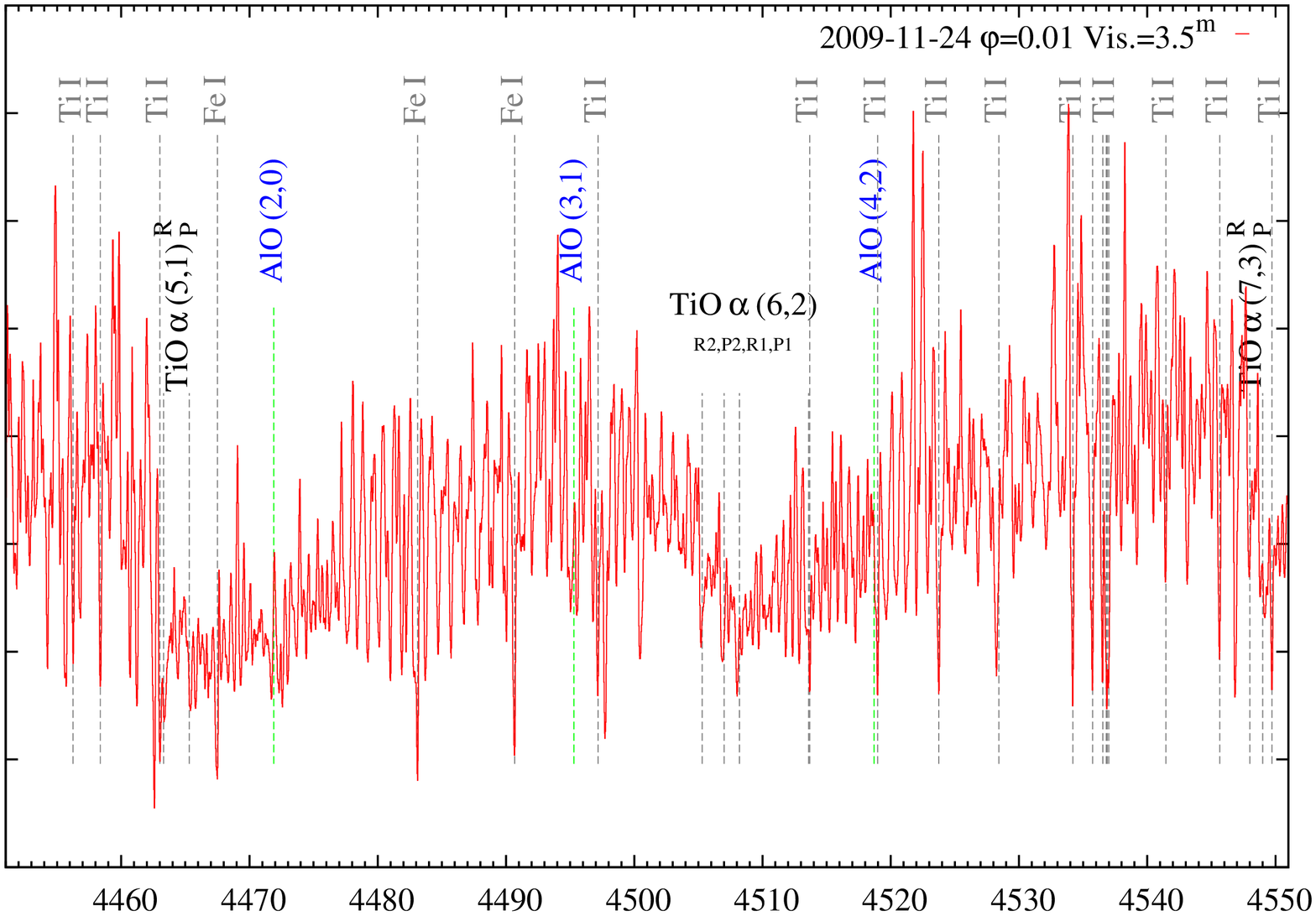}
\caption{Continued.}
\end{figure*}

  \setcounter{figure}{1}%

\begin{figure*} [tbh]
\centering
\includegraphics[angle=270,width=0.85\textwidth]{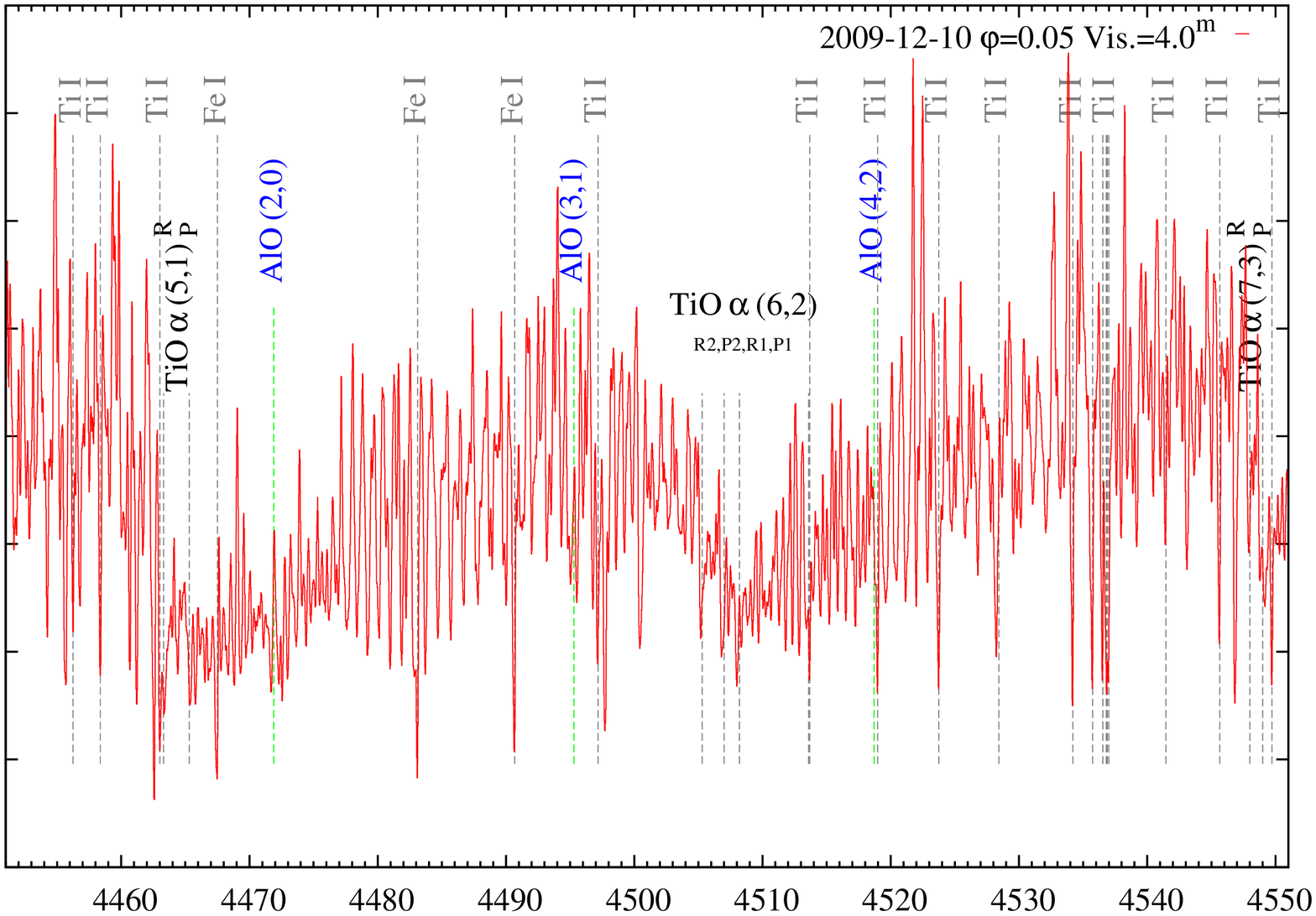}
\includegraphics[angle=270,width=0.85\textwidth]{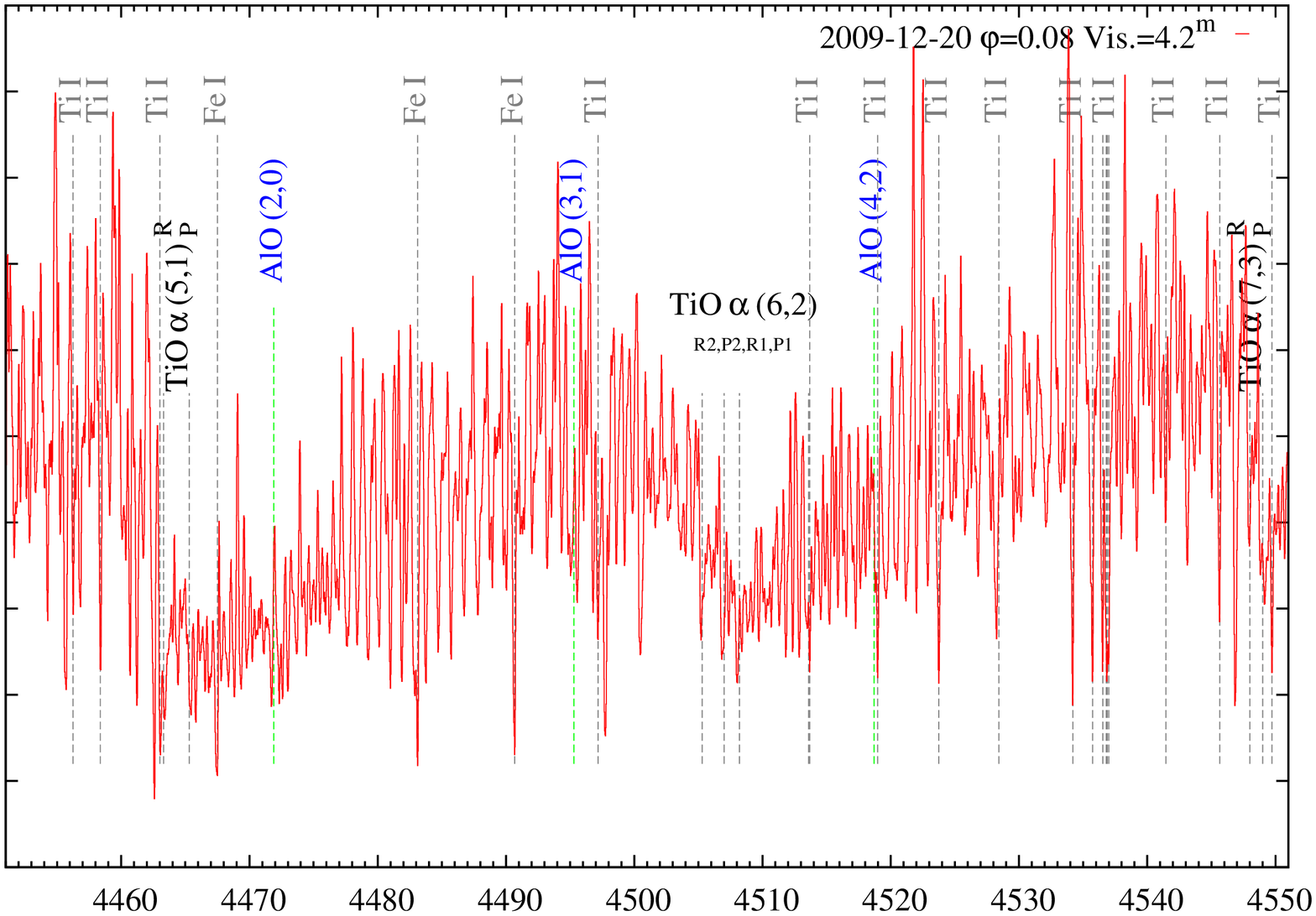}
\caption{Continued.}
\end{figure*}

  \setcounter{figure}{1}%

\begin{figure*} [tbh]
\centering
\includegraphics[angle=270,width=0.85\textwidth]{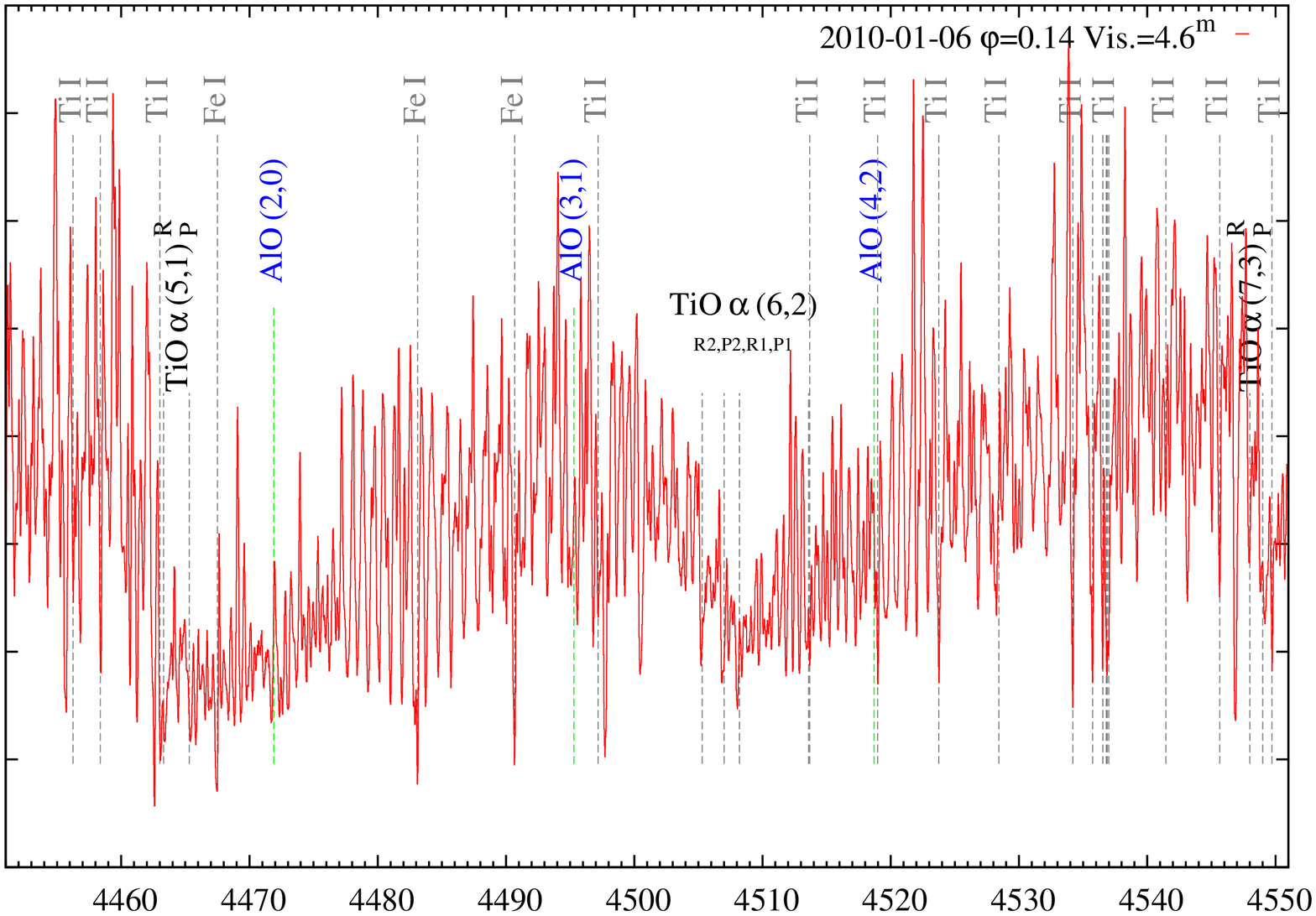}
\includegraphics[angle=270,width=0.85\textwidth]{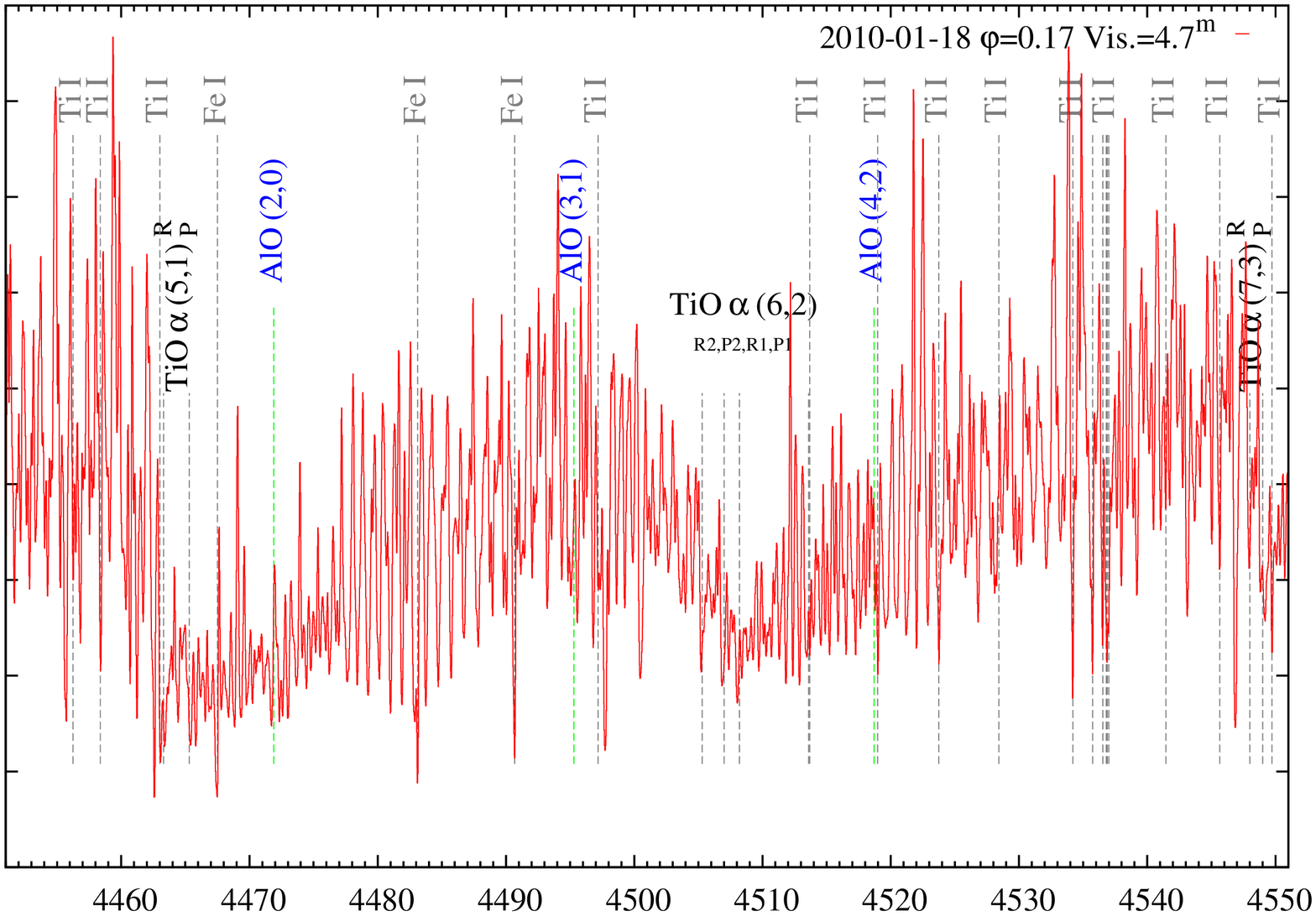}
\caption{Continued.}
\end{figure*}

  \setcounter{figure}{1}%

\begin{figure*} [tbh]
\centering
\includegraphics[angle=270,width=0.85\textwidth]{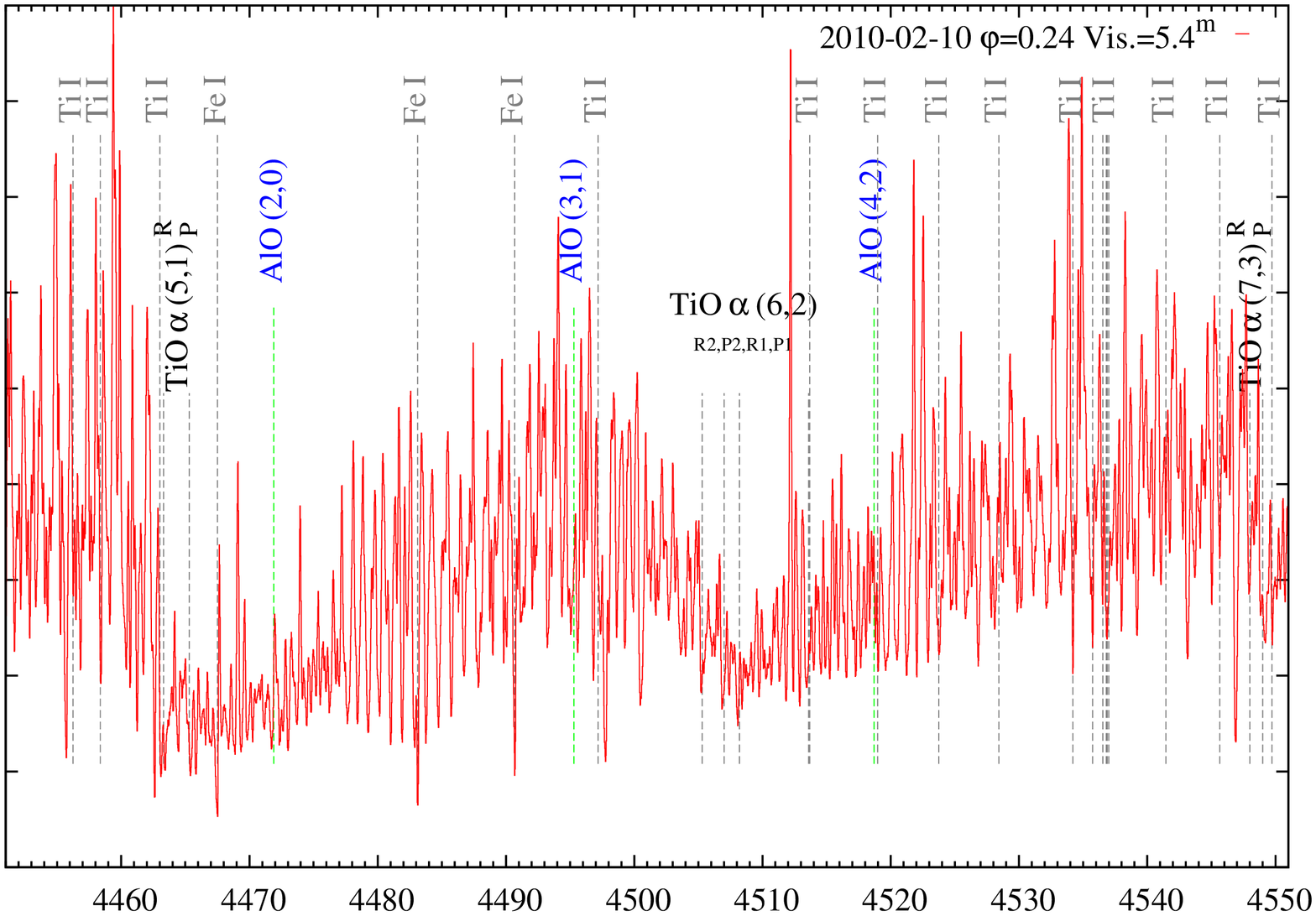}
\includegraphics[angle=270,width=0.85\textwidth]{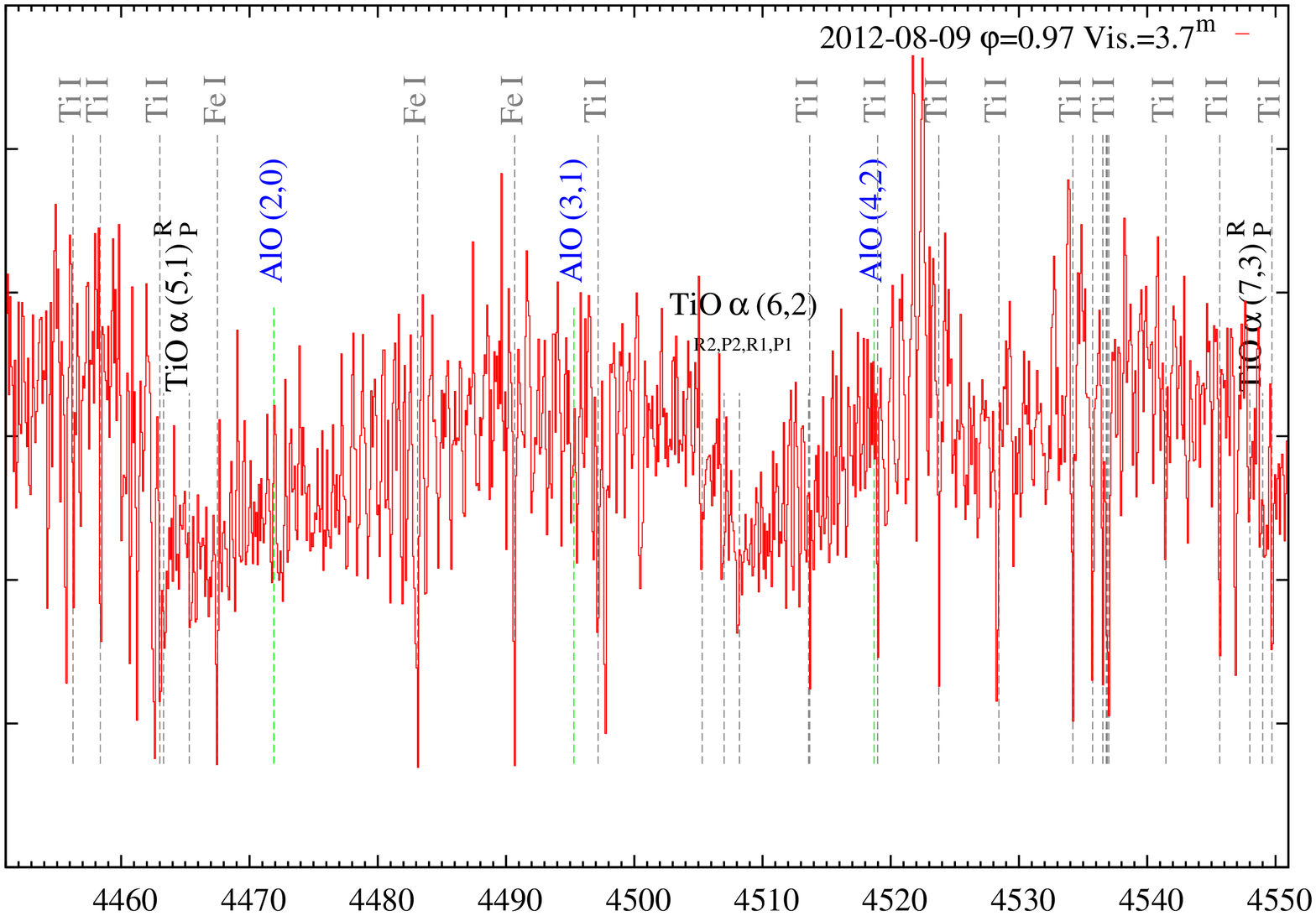}
\caption{Continued.}
\end{figure*}
\clearpage
\begin{figure*} [tbh]
\centering
\includegraphics[angle=270,width=0.85\textwidth]{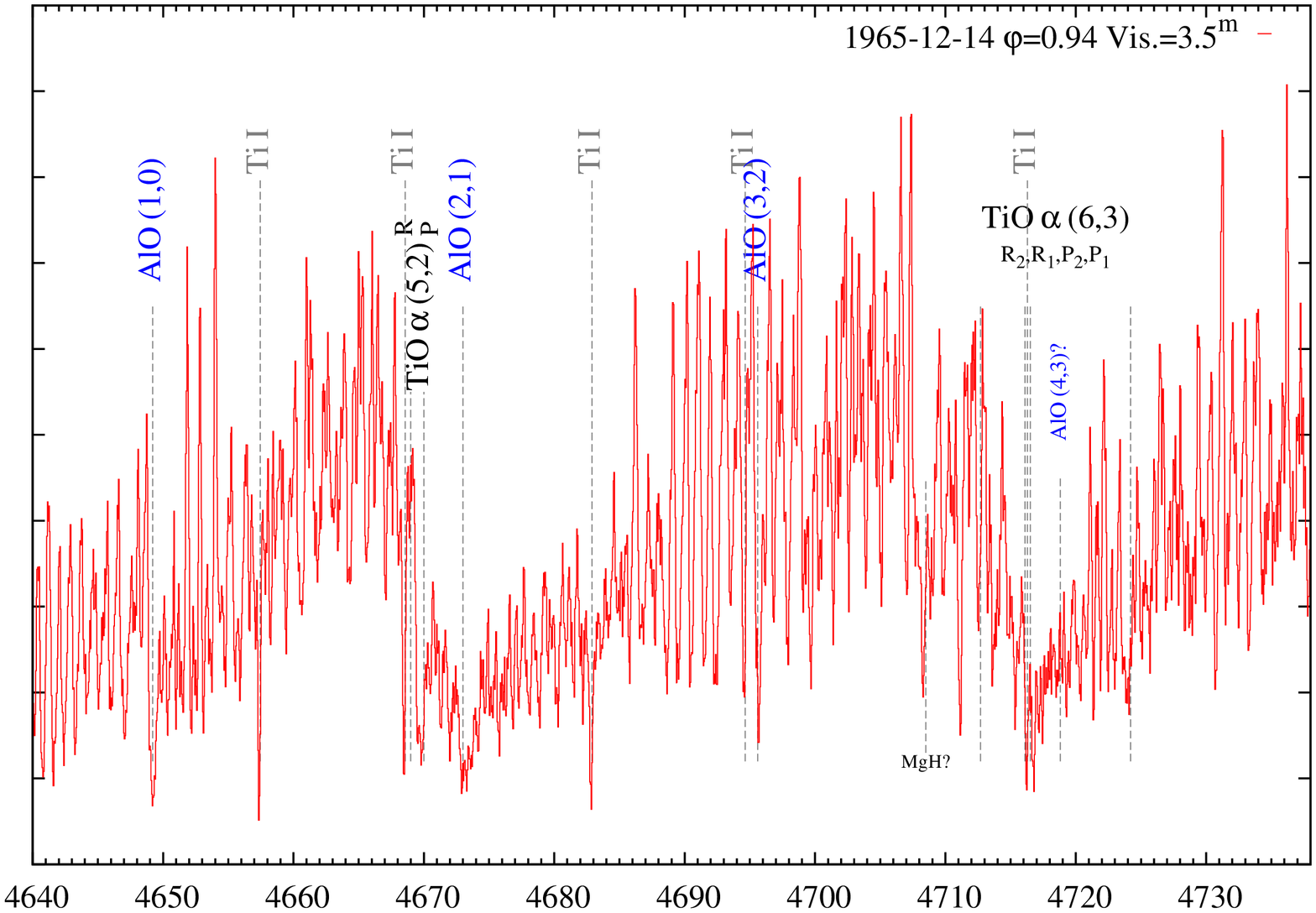}
\includegraphics[angle=270,width=0.85\textwidth]{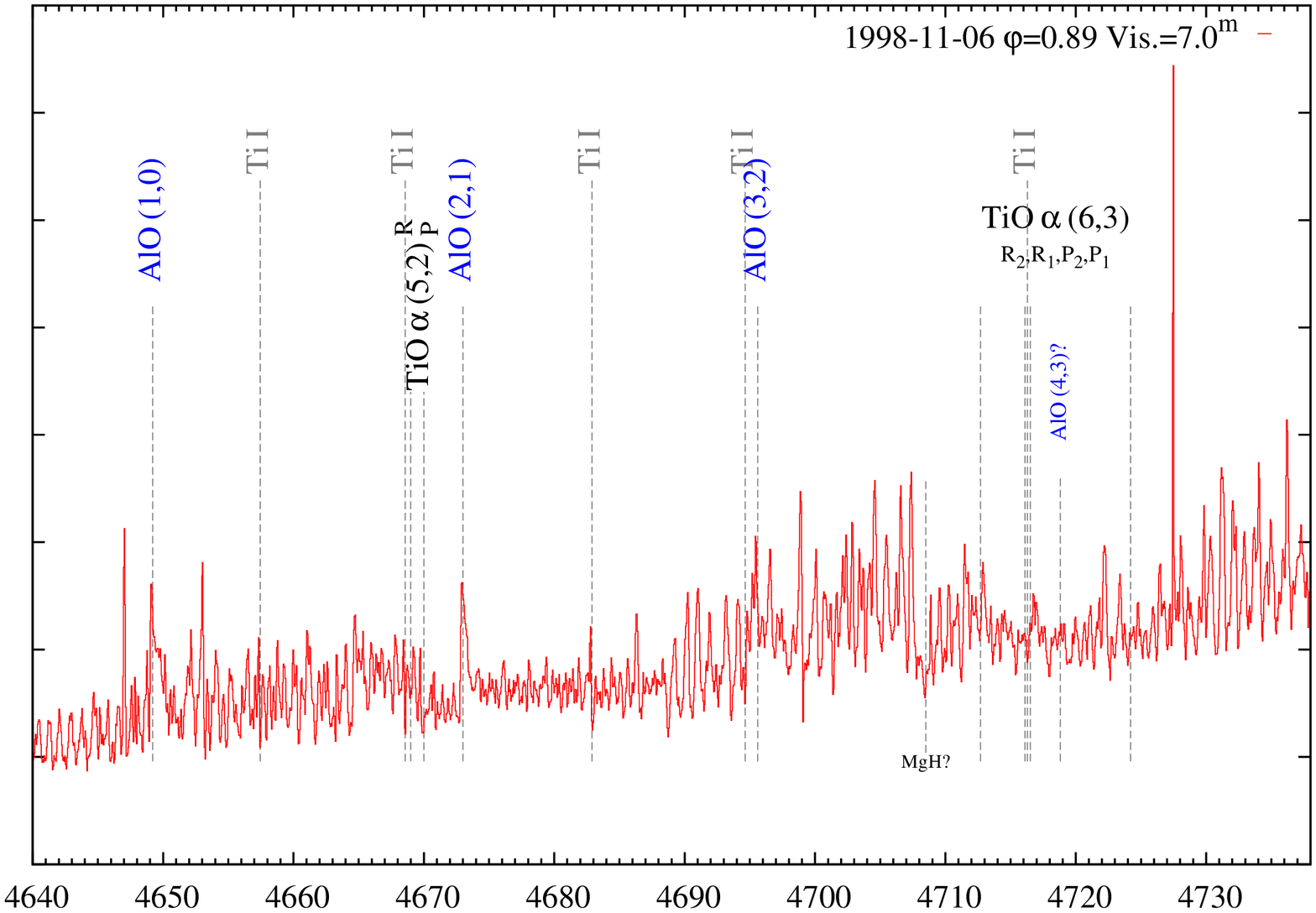}
\caption{The same as Fig.\,\ref{fig-AlOp2} but for the $\Delta\varv$=+1 sequence of AlO $B-X$. Spectra from Narval are affected by an imperfect combination of different echelle orders in the 4665--46755\,\AA\ range.}\label{fig-AlOp1}
\end{figure*}

  \setcounter{figure}{2}%

\begin{figure*} [tbh]
\centering
\includegraphics[angle=270,width=0.85\textwidth]{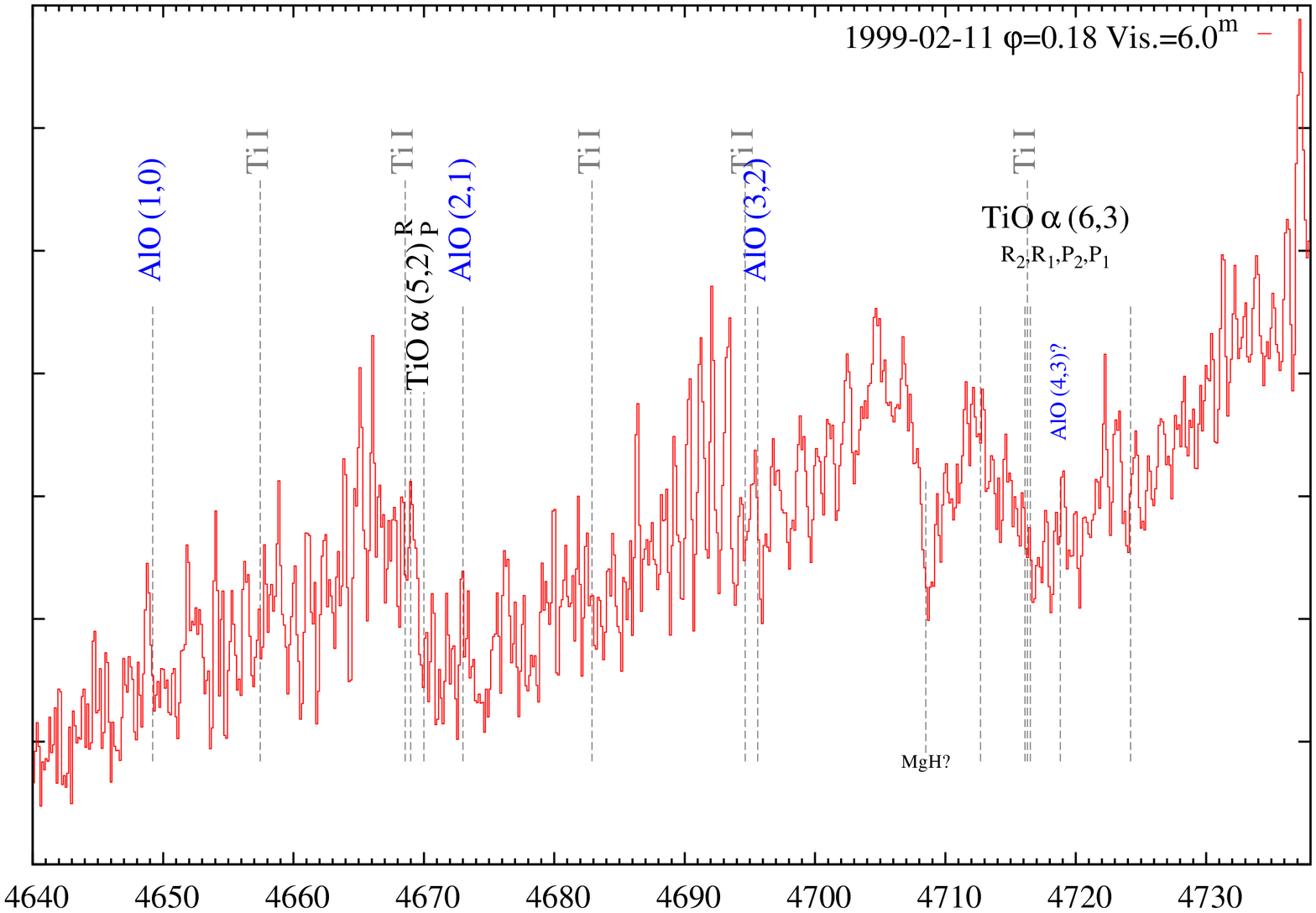}
\includegraphics[angle=270,width=0.85\textwidth]{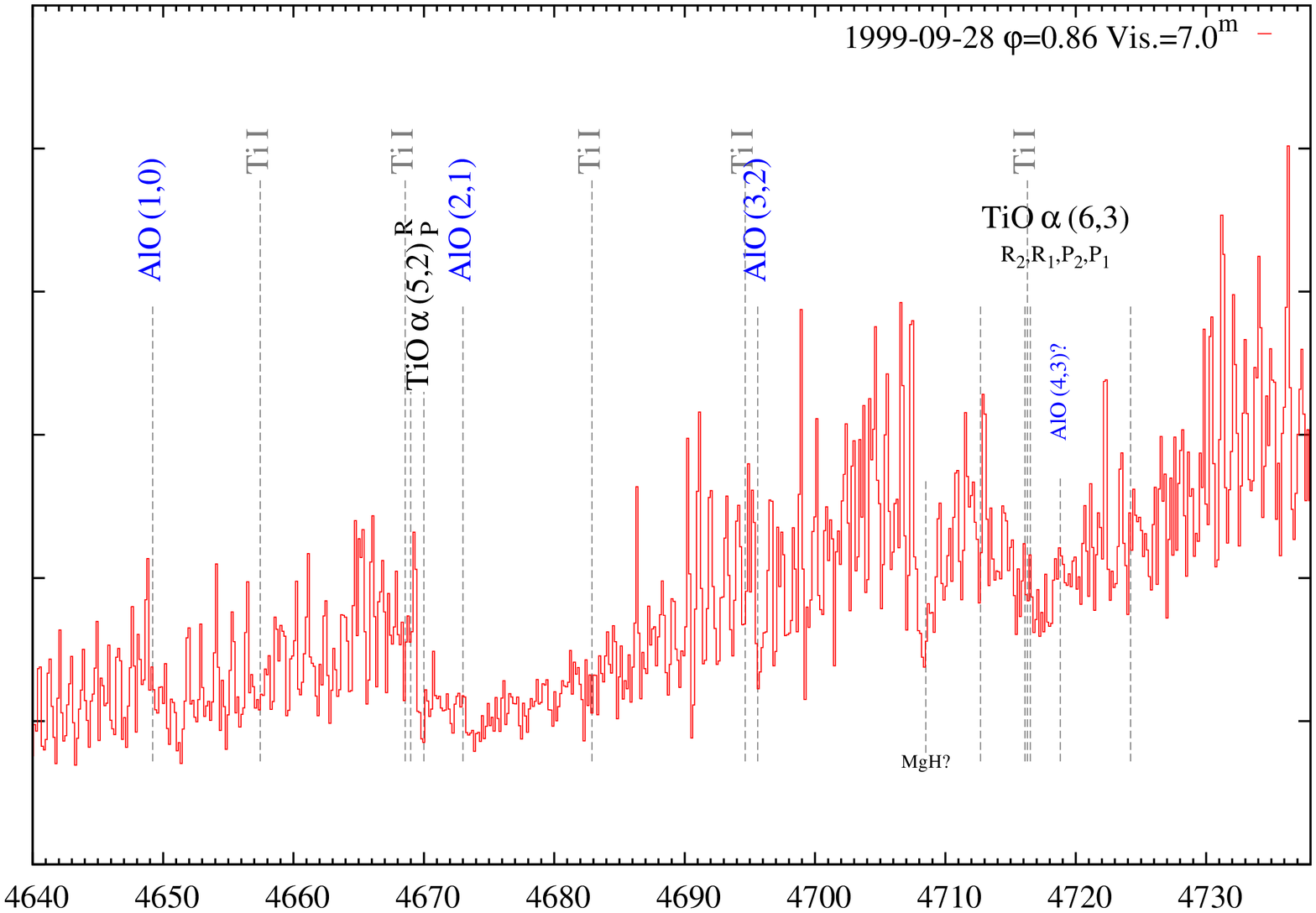}
\caption{Continued.}
\end{figure*}

  \setcounter{figure}{2}%

\begin{figure*} [tbh]
\centering
\includegraphics[angle=270,width=0.85\textwidth]{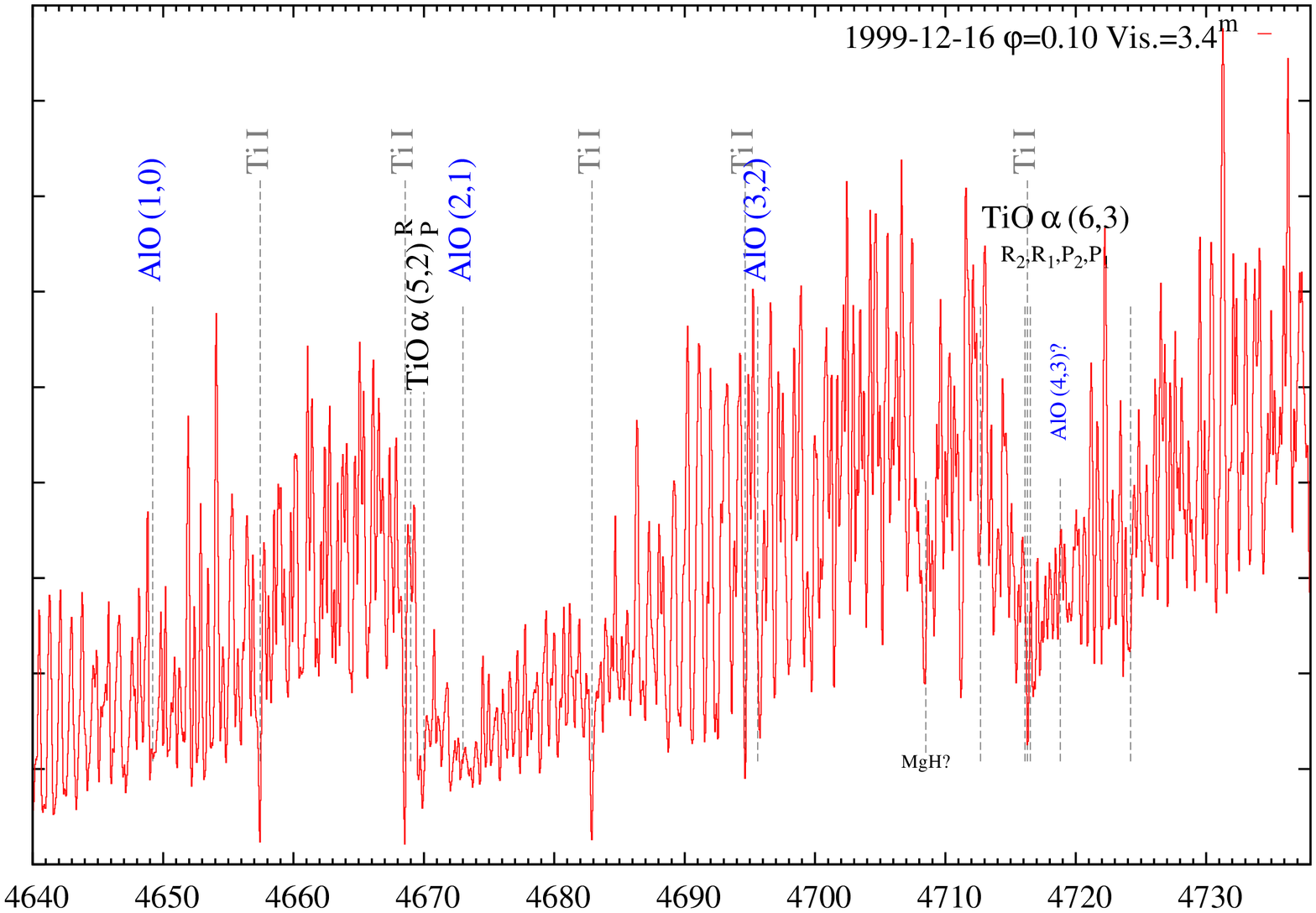}
\includegraphics[angle=270,width=0.85\textwidth]{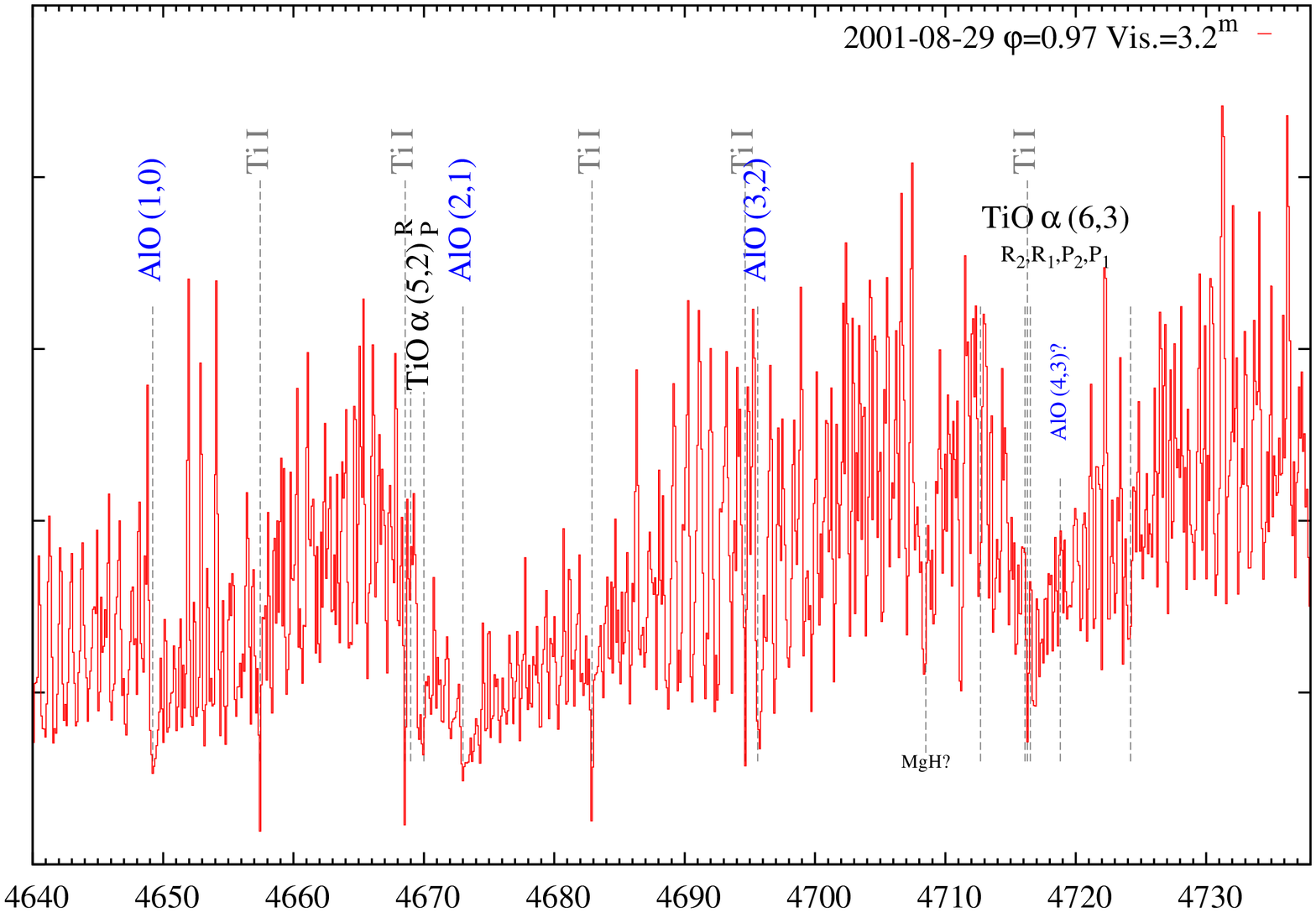}
\caption{Continued.}
\end{figure*}

  \setcounter{figure}{2}%

\begin{figure*} [tbh]
\centering
\includegraphics[angle=270,width=0.85\textwidth]{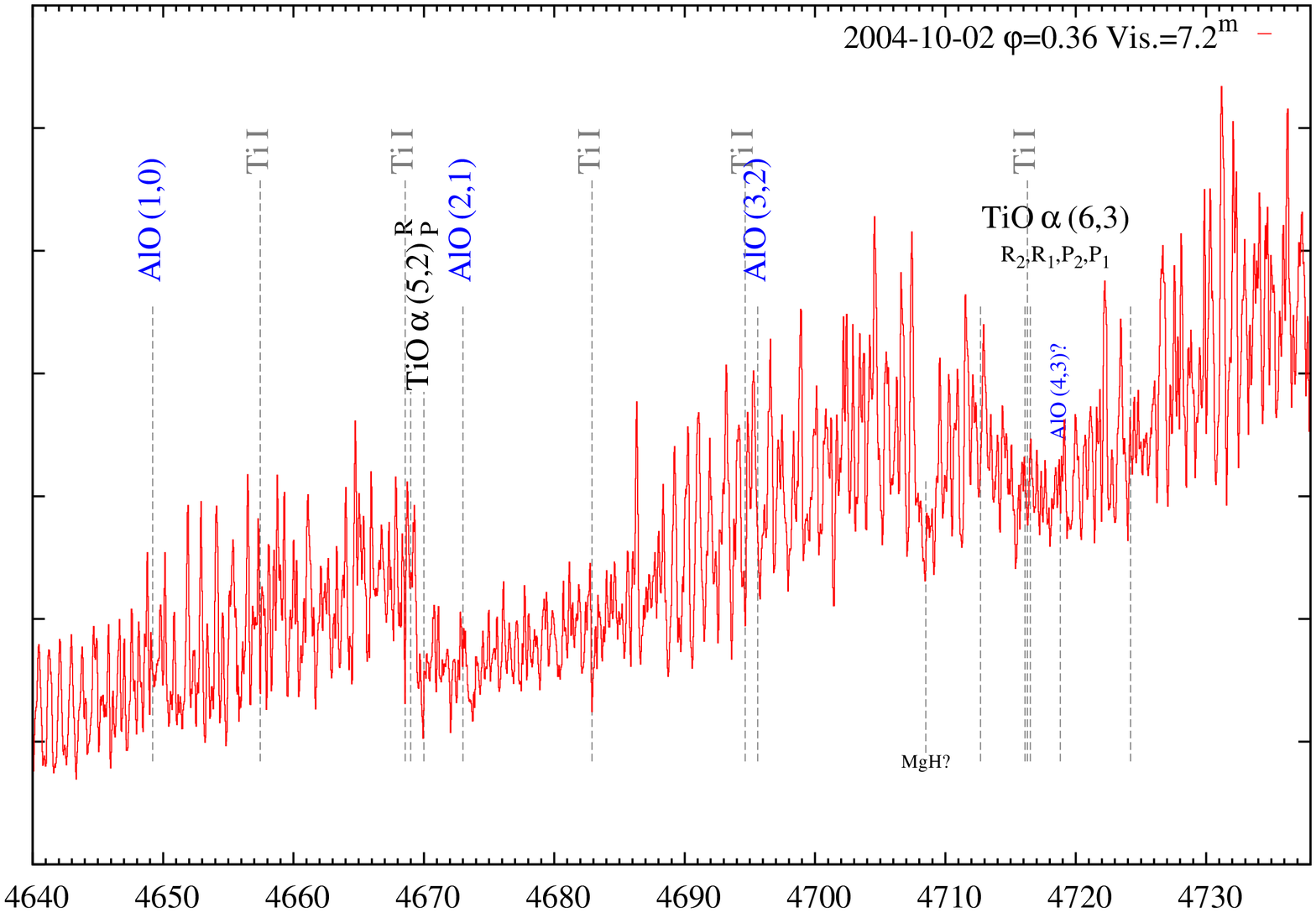}
\includegraphics[angle=270,width=0.85\textwidth]{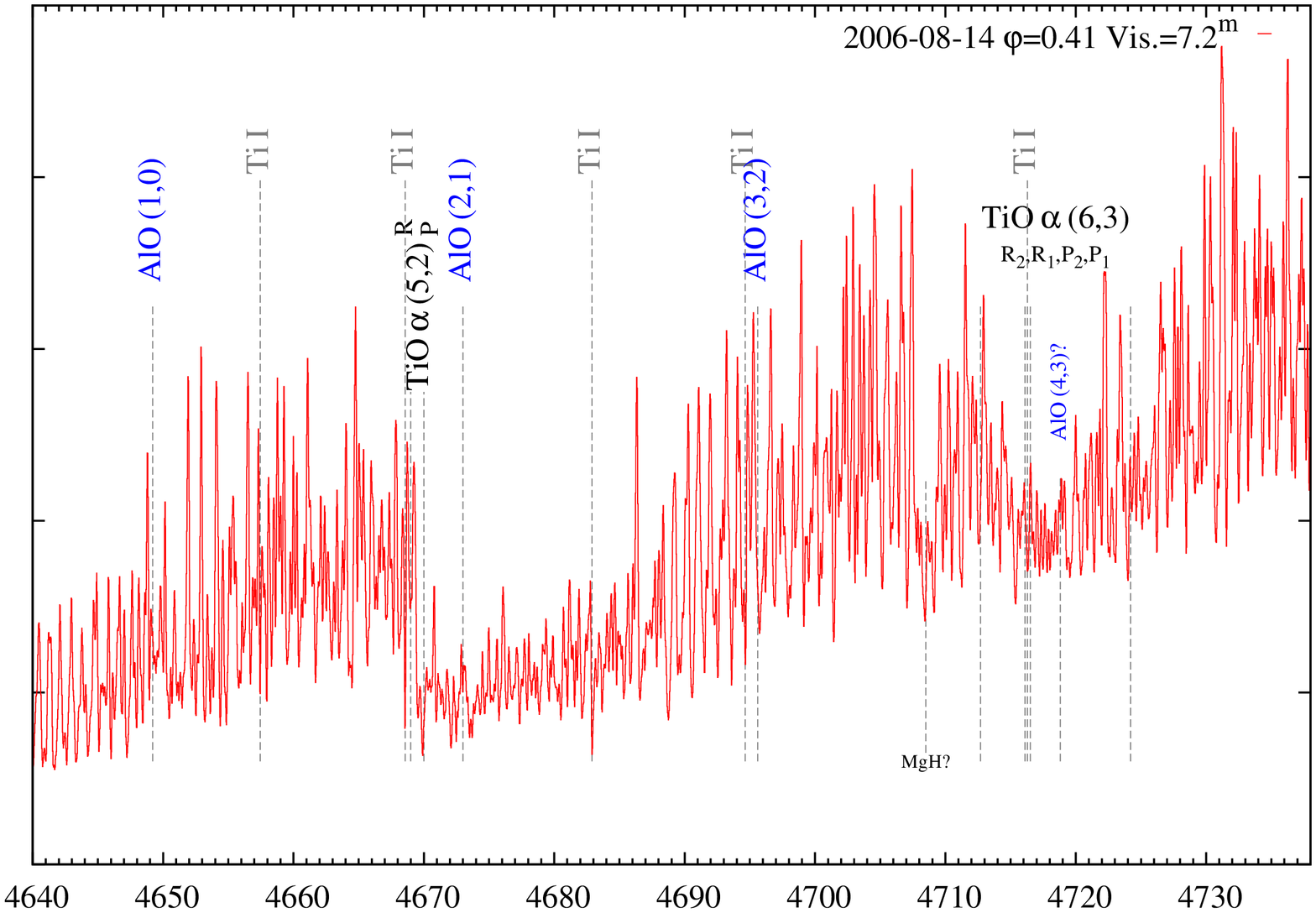}
\caption{Continued.}
\end{figure*}

  \setcounter{figure}{2}%

\begin{figure*} [tbh]
\centering
\includegraphics[angle=270,width=0.85\textwidth]{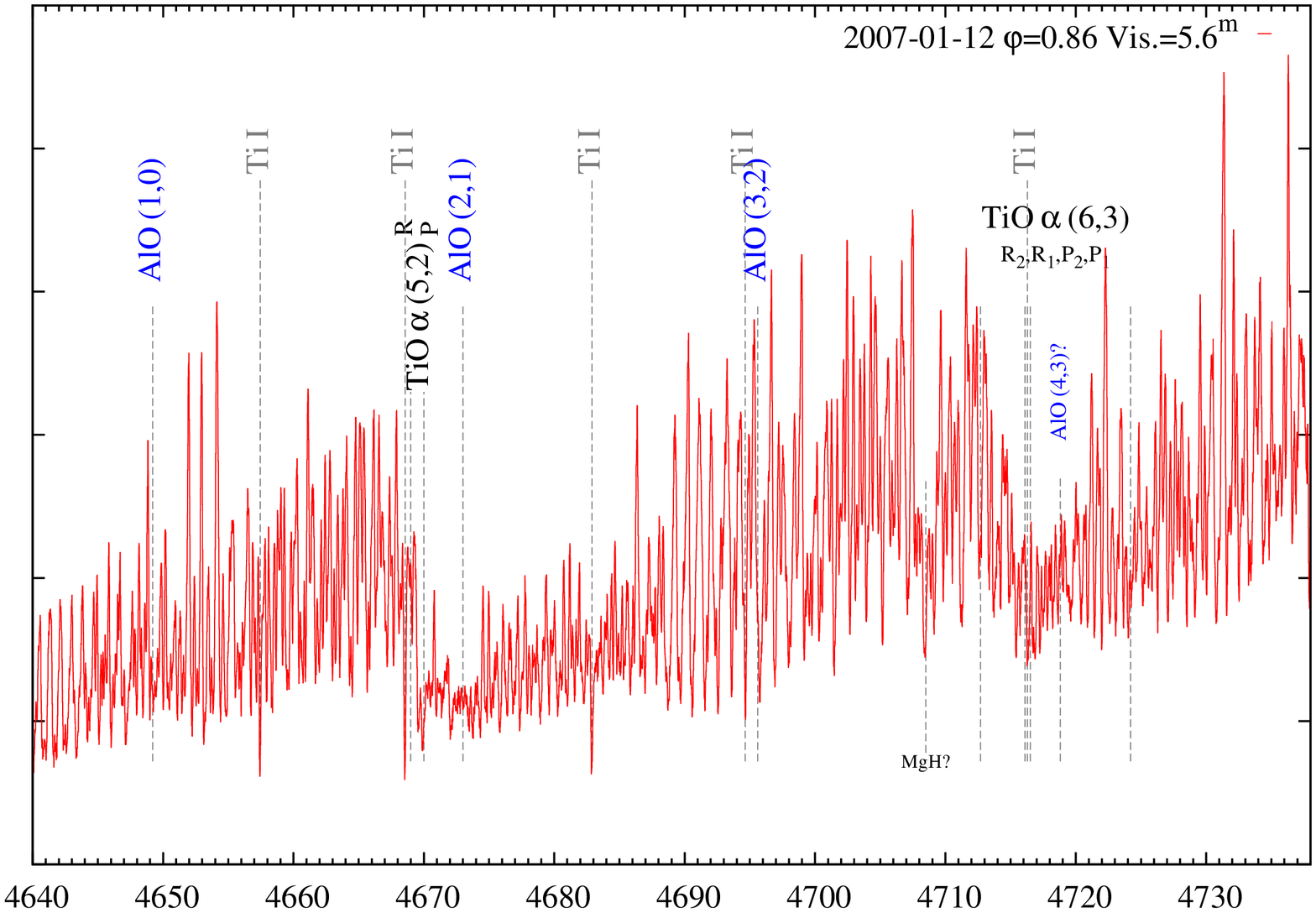}
\includegraphics[angle=270,width=0.85\textwidth]{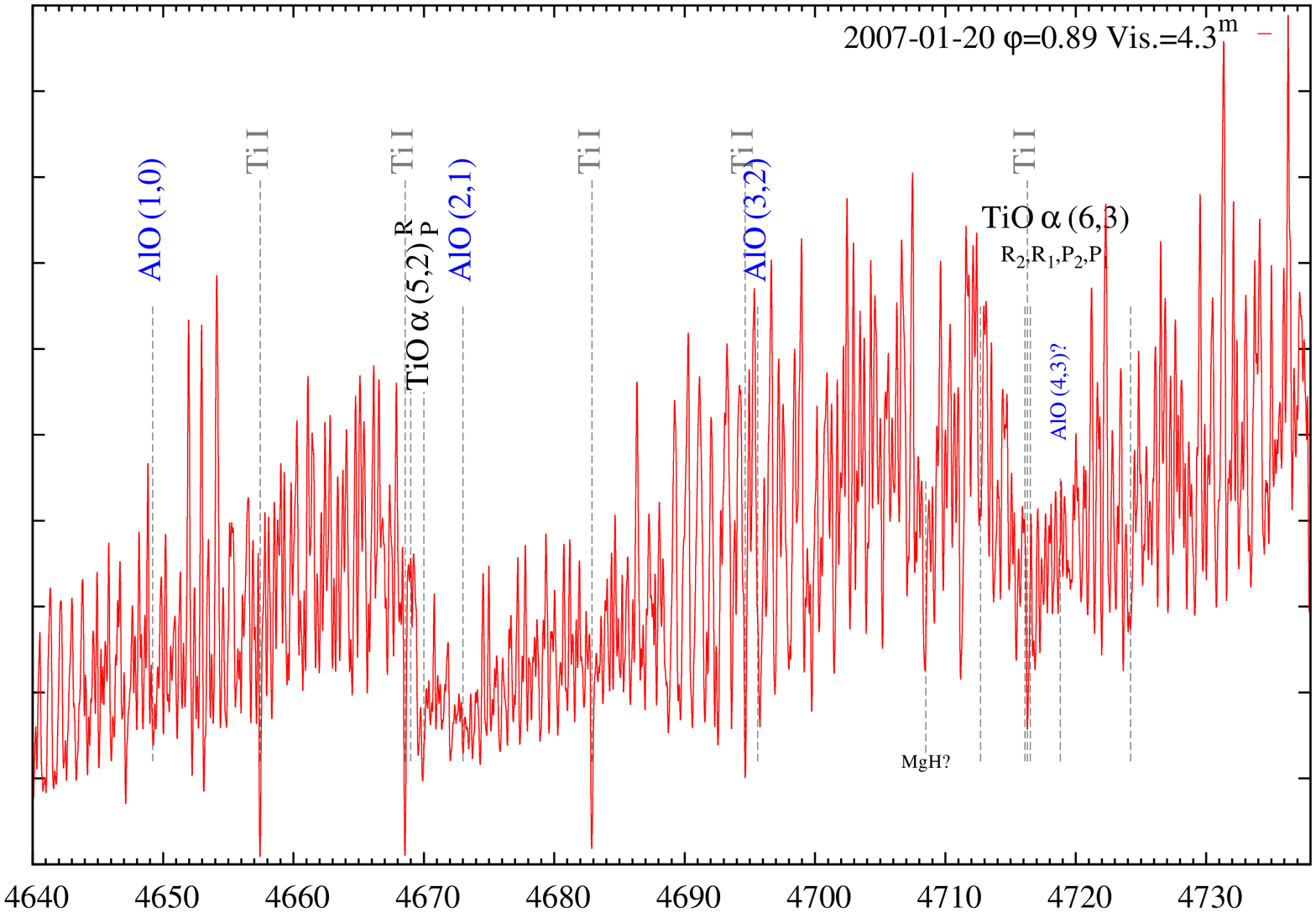}
\caption{Continued.}
\end{figure*}

  \setcounter{figure}{2}%

\begin{figure*} [tbh]
\centering
\includegraphics[angle=270,width=0.85\textwidth]{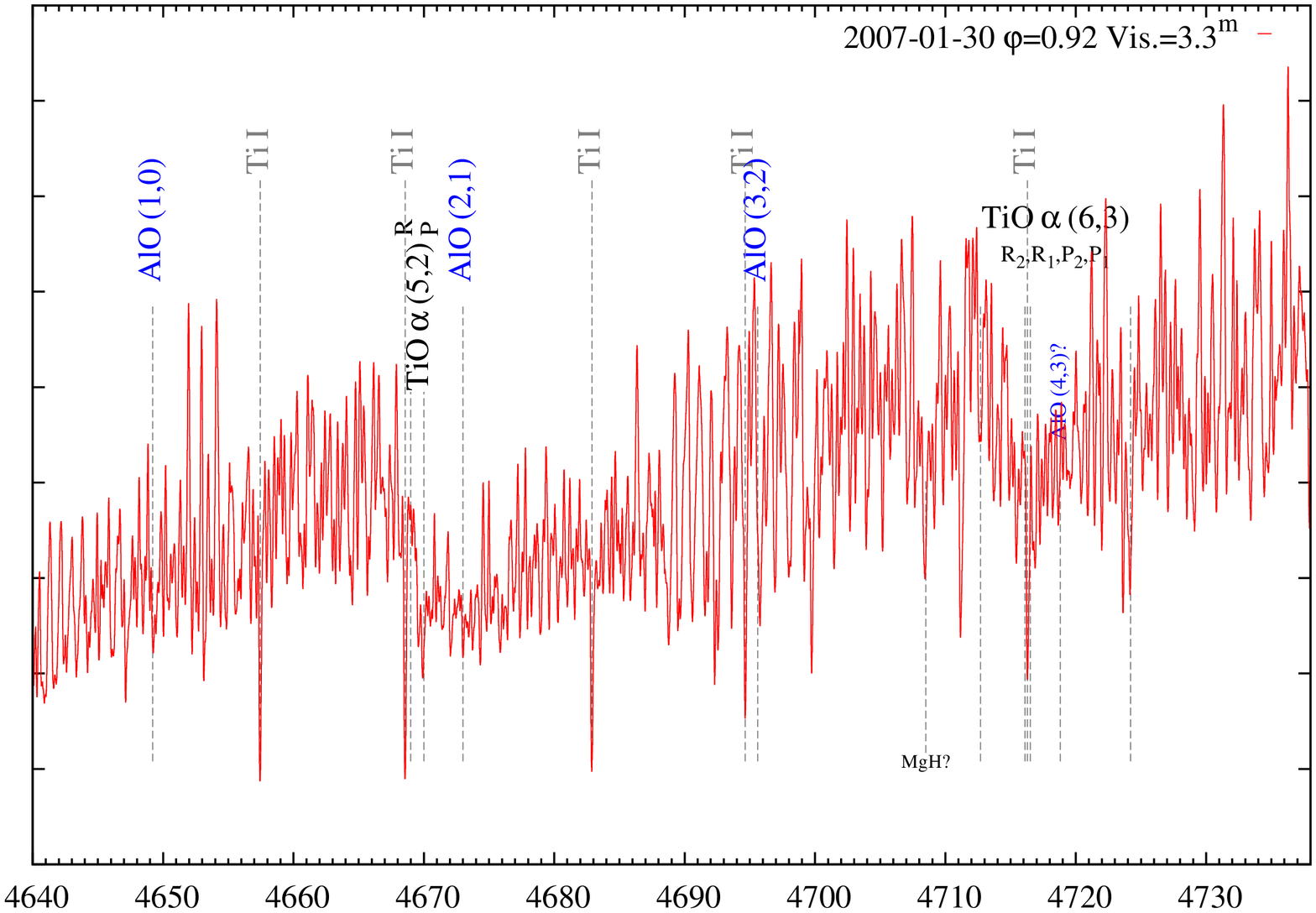}
\includegraphics[angle=270,width=0.85\textwidth]{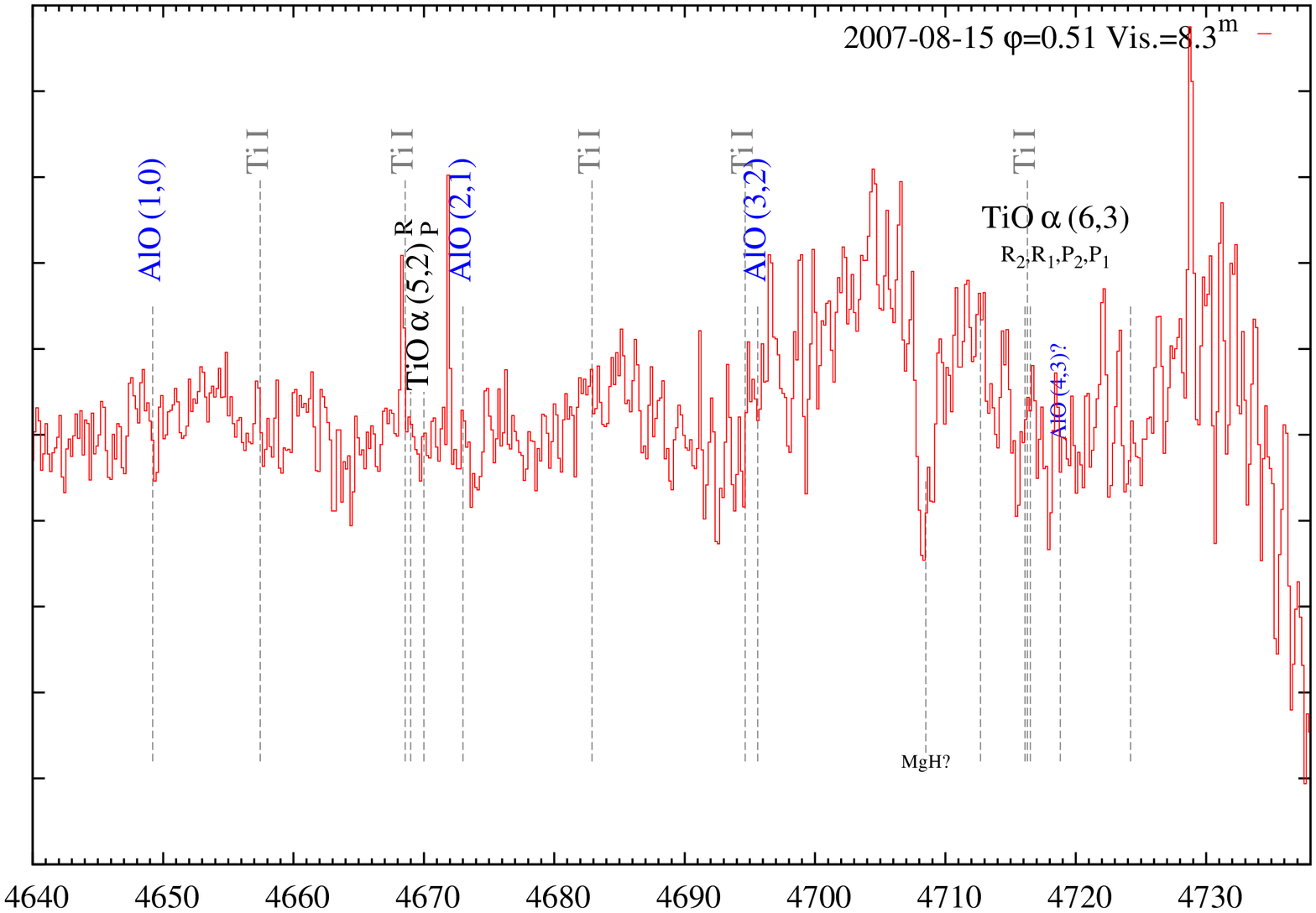}
\caption{Continued.}
\end{figure*}

  \setcounter{figure}{2}%

\begin{figure*} [tbh]
\centering
\includegraphics[angle=270,width=0.85\textwidth]{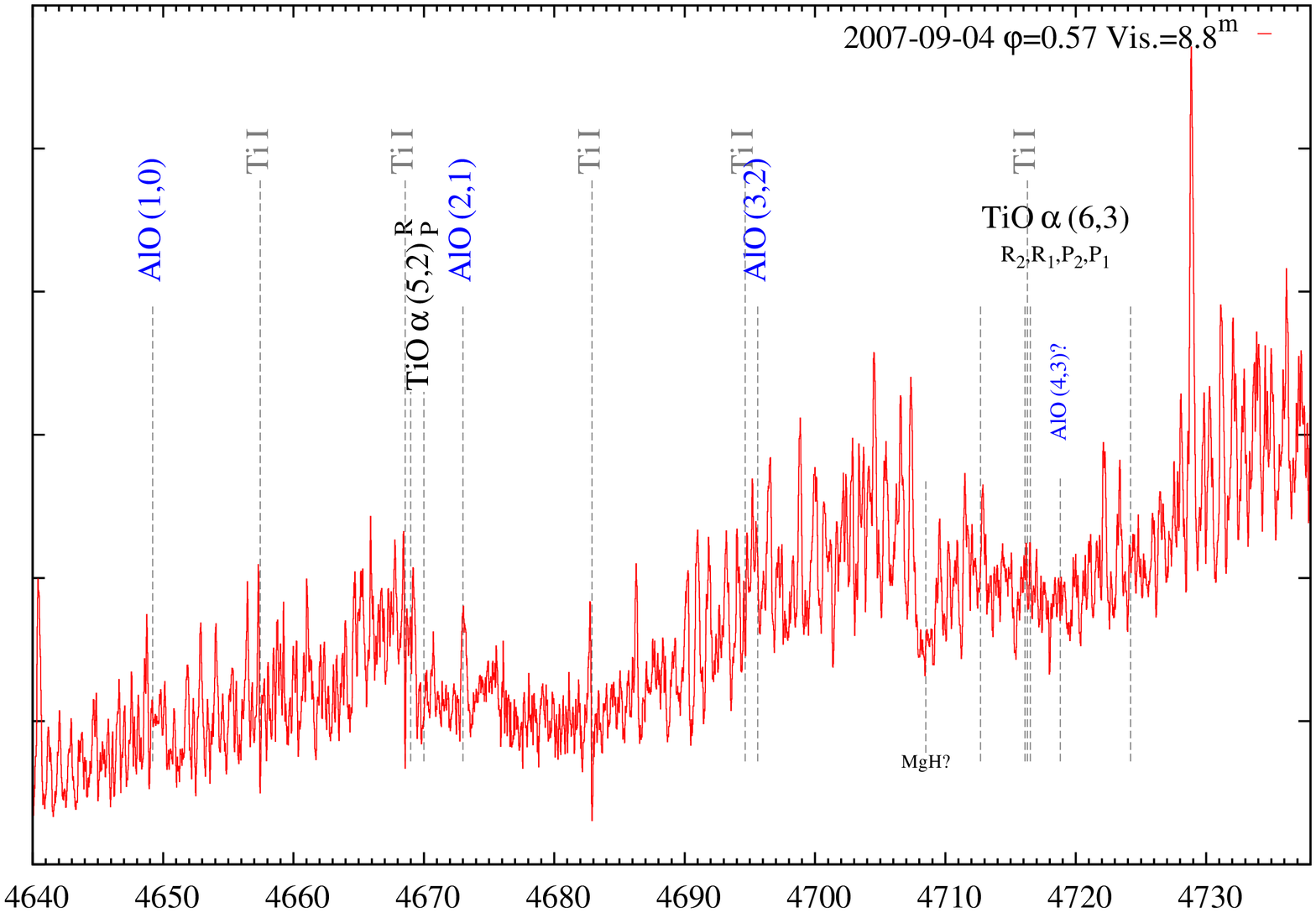}
\includegraphics[angle=270,width=0.85\textwidth]{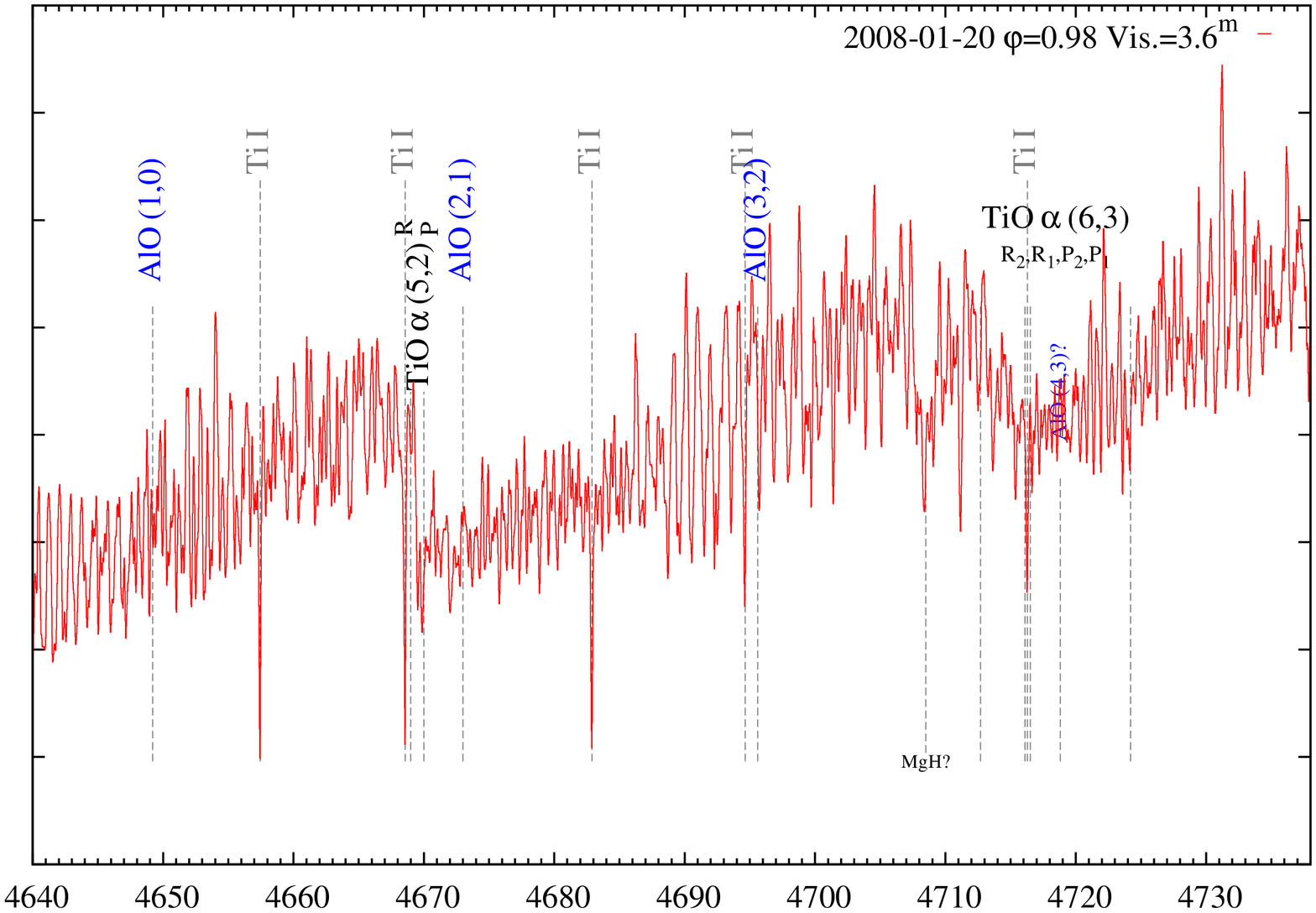}
\caption{Continued.}
\end{figure*}

  \setcounter{figure}{2}%

\begin{figure*} [tbh]
\centering
\includegraphics[angle=270,width=0.85\textwidth]{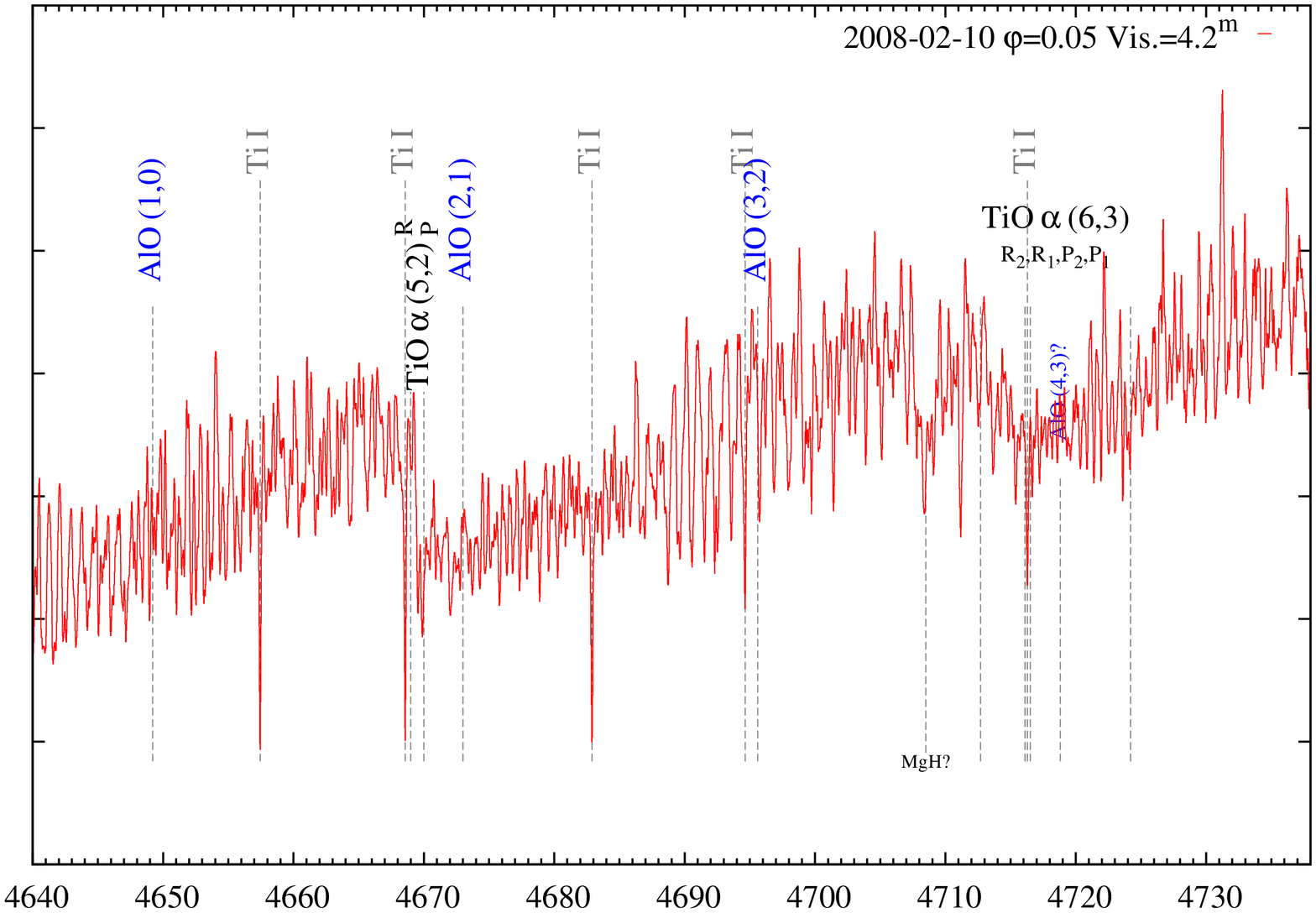}
\includegraphics[angle=270,width=0.85\textwidth]{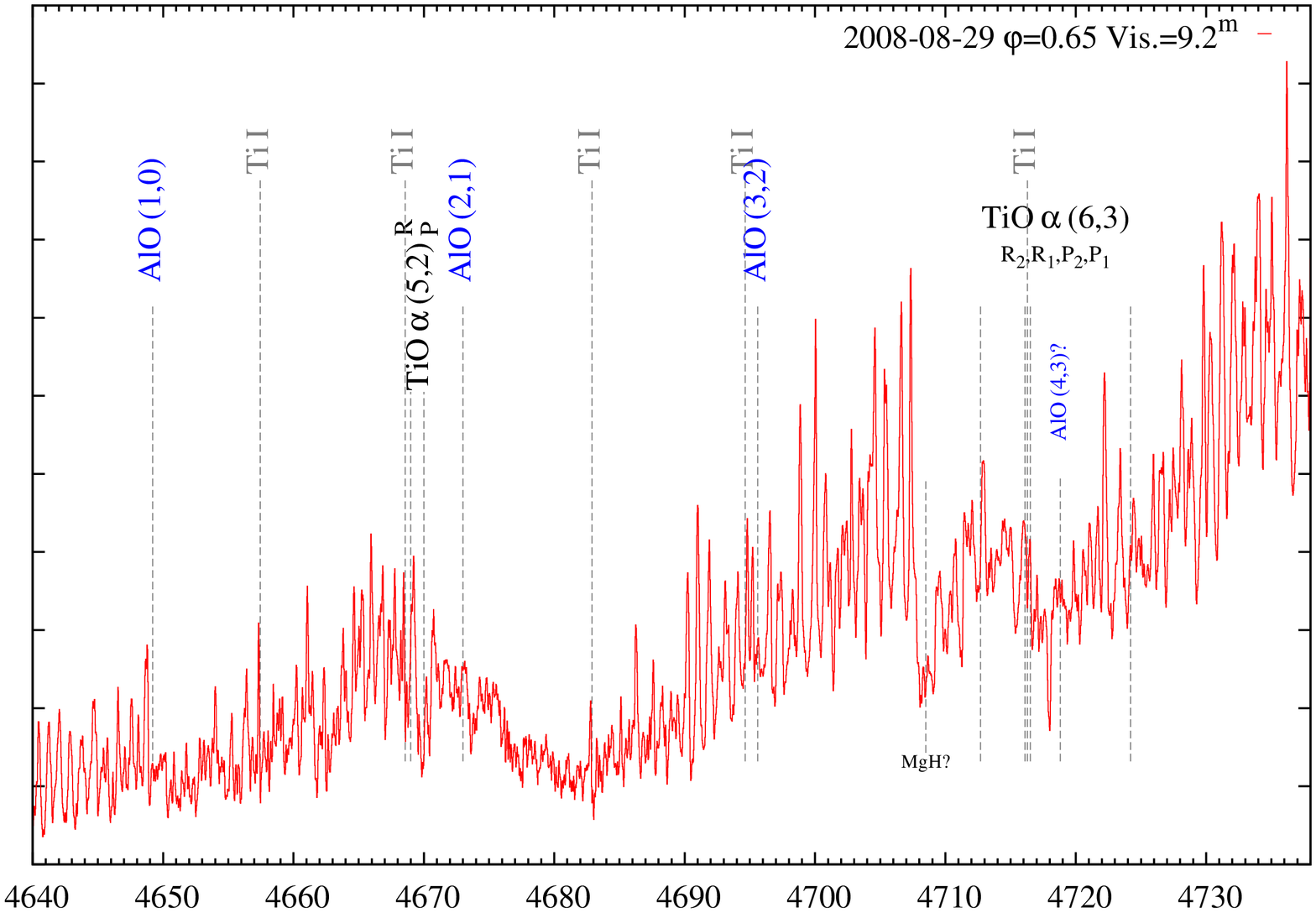}
\caption{Continued.}
\end{figure*}

  \setcounter{figure}{2}%

\begin{figure*} [tbh]
\centering
\includegraphics[angle=270,width=0.85\textwidth]{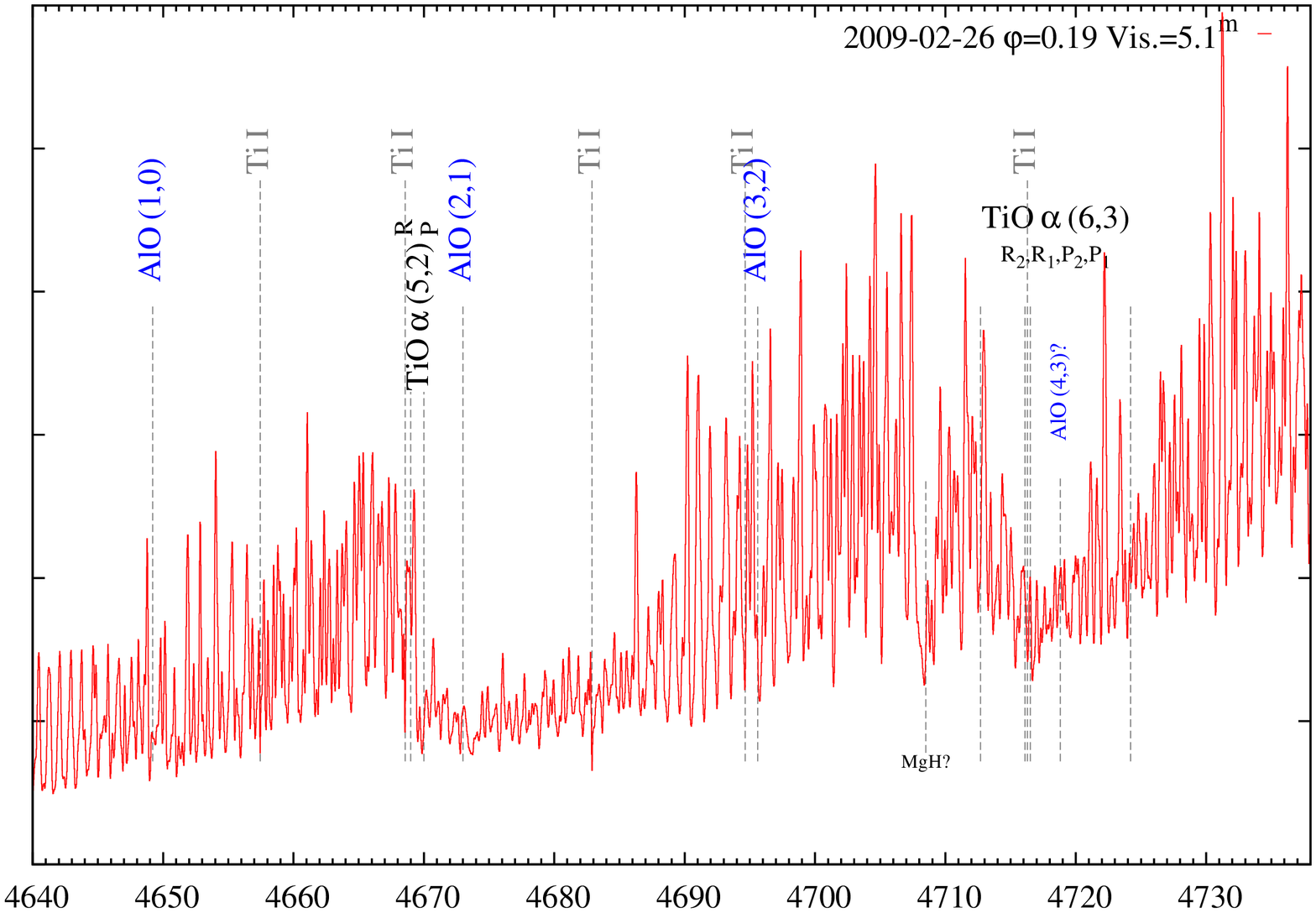}
\includegraphics[angle=270,width=0.85\textwidth]{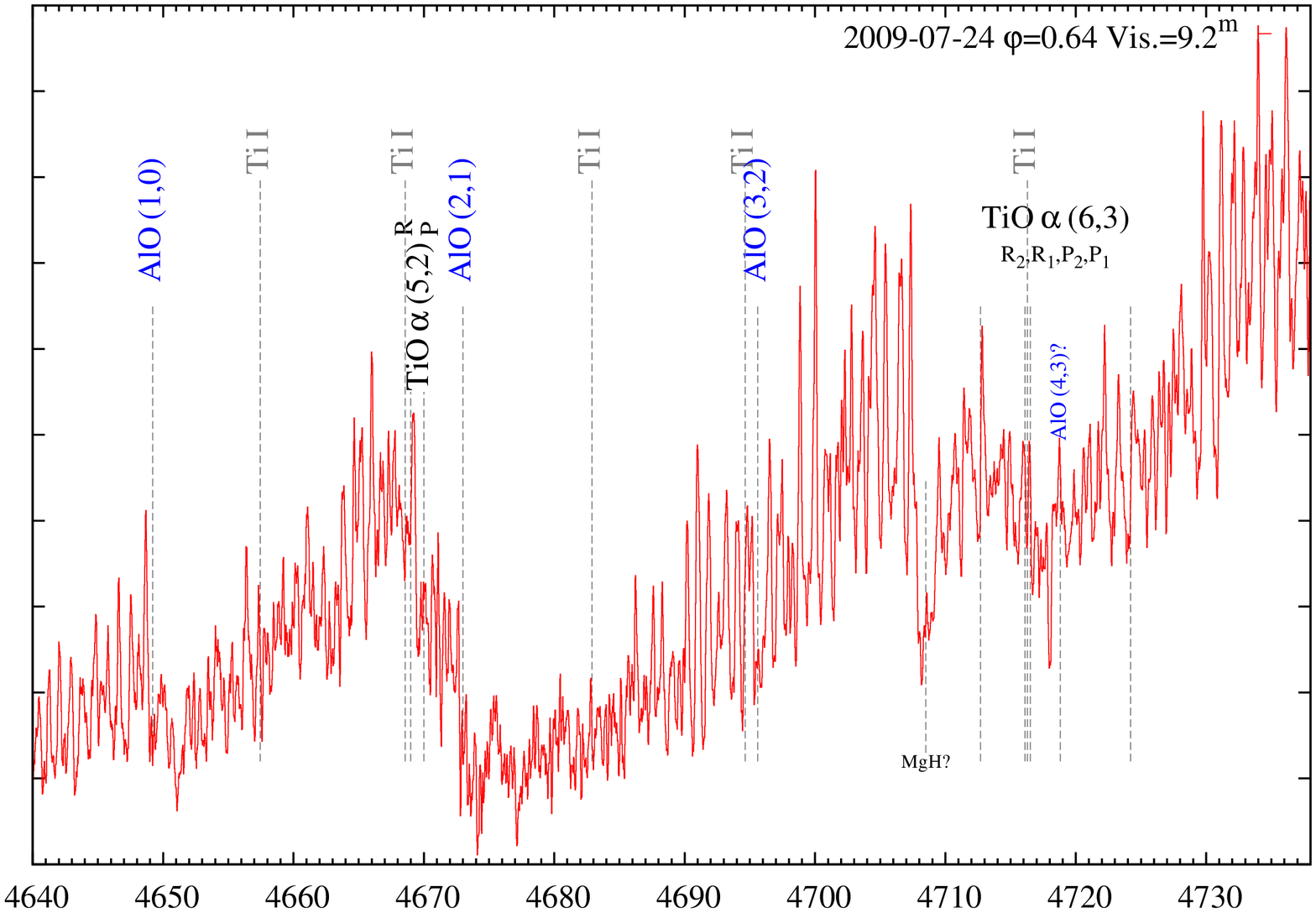}
\caption{Continued.}
\end{figure*}

  \setcounter{figure}{2}%

\begin{figure*} [tbh]
\centering
\includegraphics[angle=270,width=0.85\textwidth]{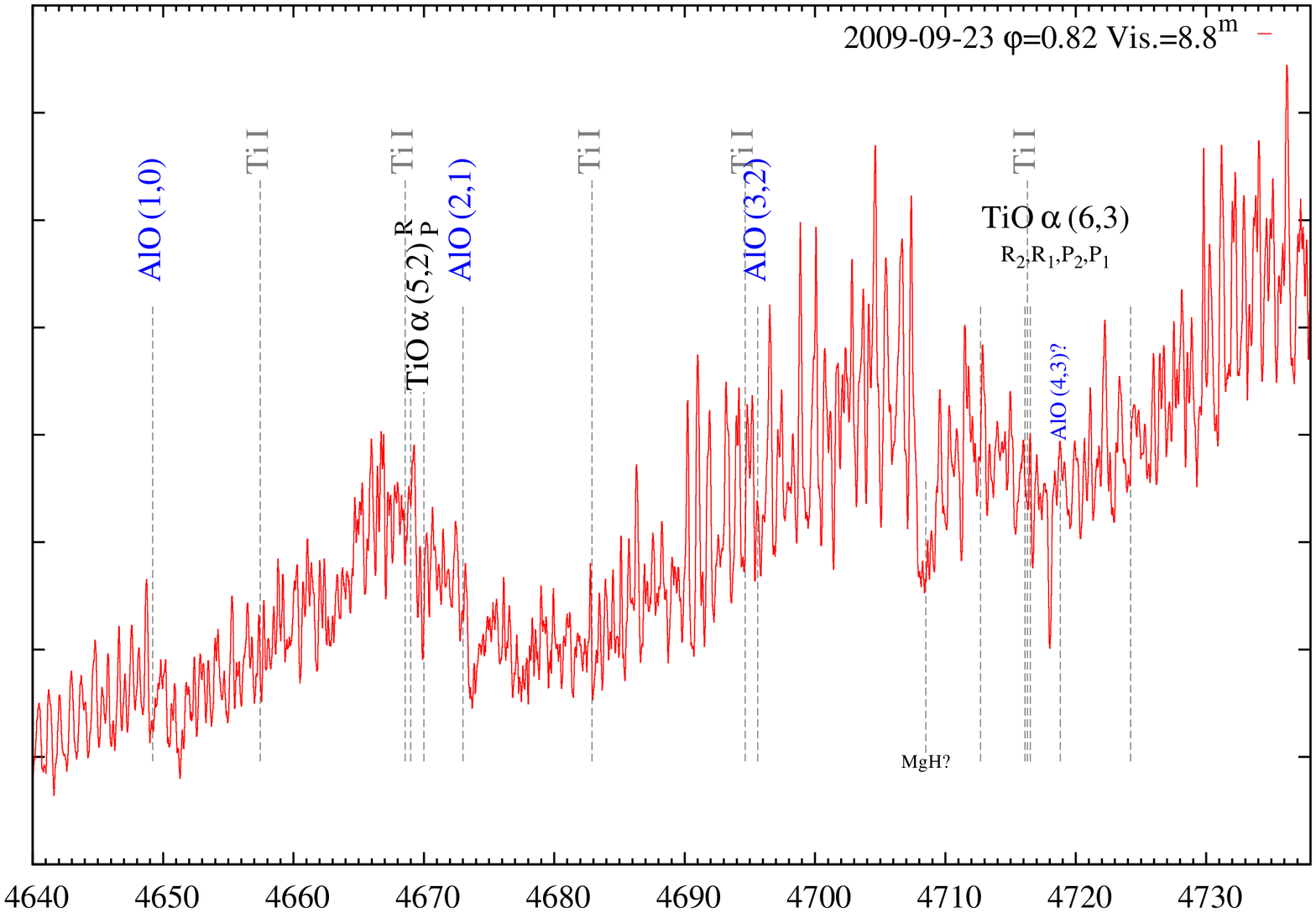}
\includegraphics[angle=270,width=0.85\textwidth]{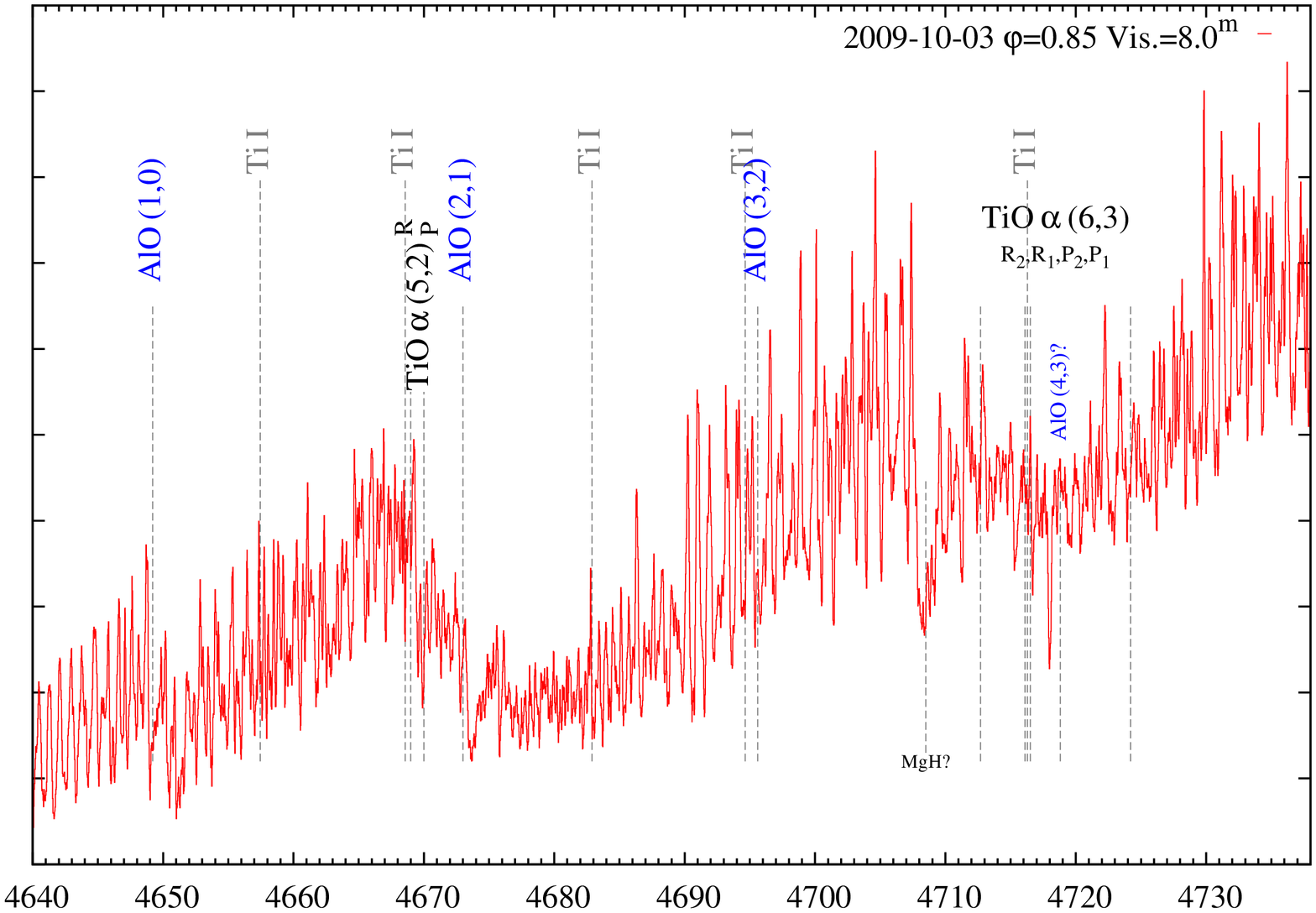}
\caption{Continued.}
\end{figure*}

  \setcounter{figure}{2}%

\begin{figure*} [tbh]
\centering
\includegraphics[angle=270,width=0.85\textwidth]{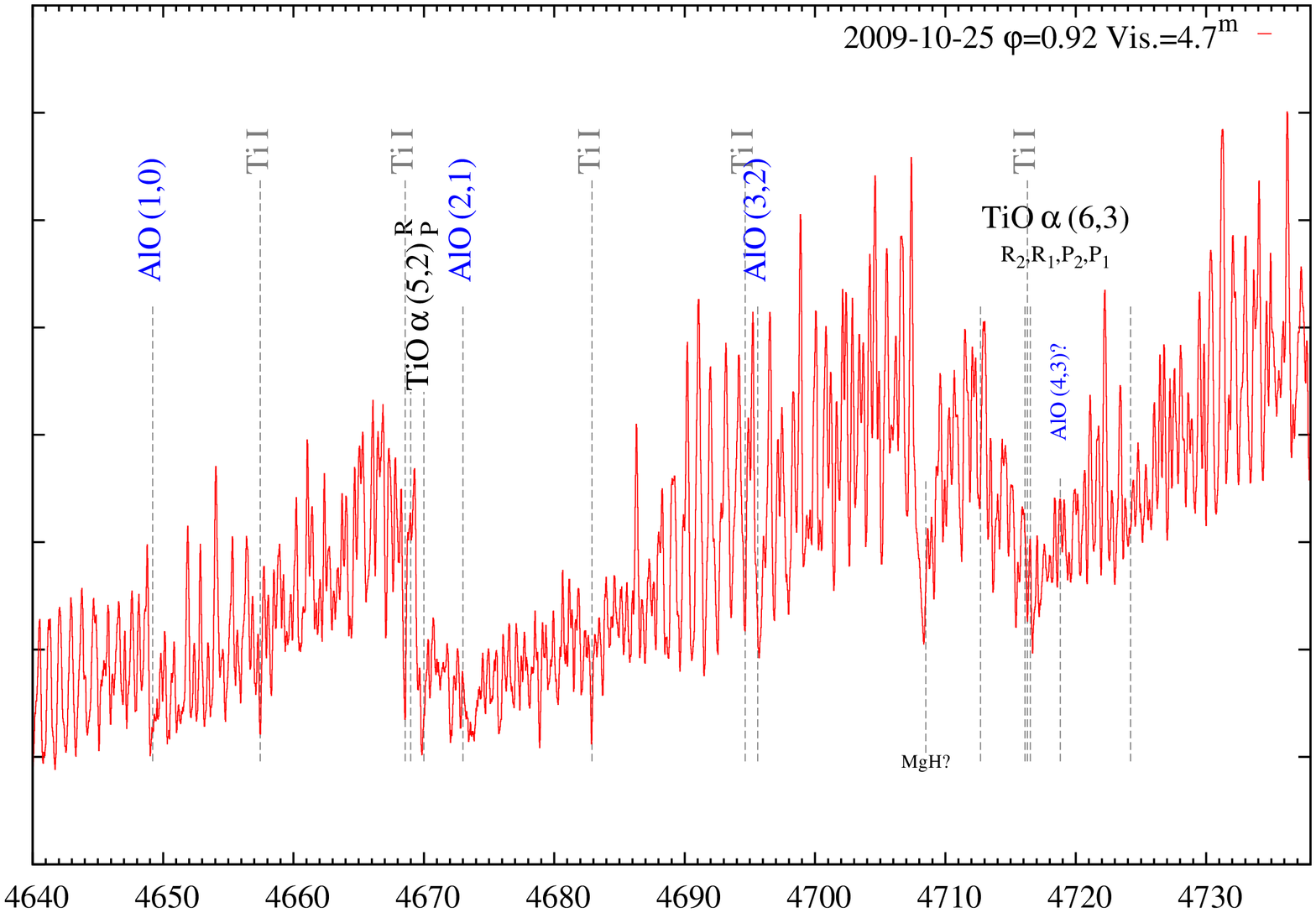}
\includegraphics[angle=270,width=0.85\textwidth]{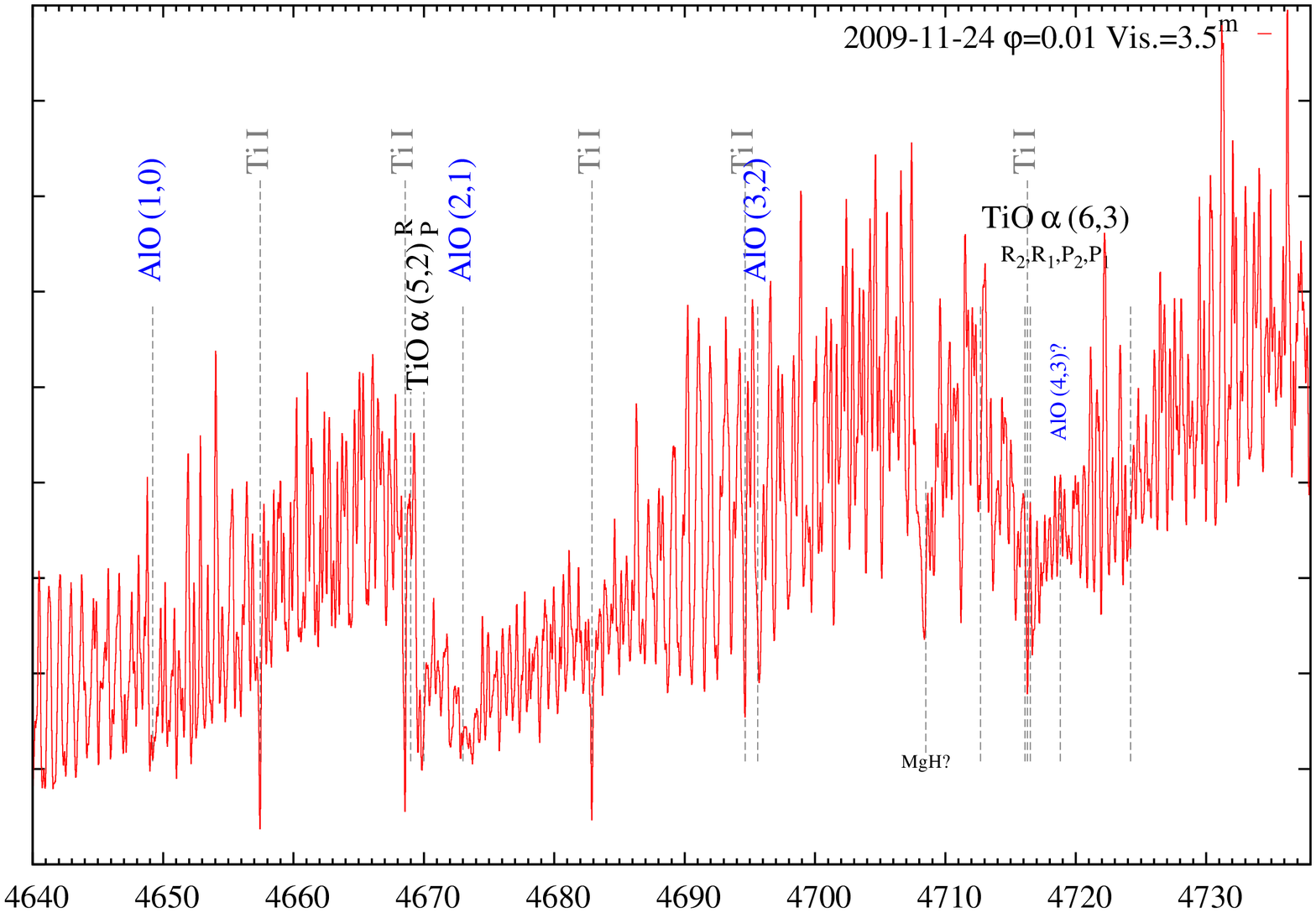}
\caption{Continued.}
\end{figure*}

  \setcounter{figure}{2}%

\begin{figure*} [tbh]
\centering
\includegraphics[angle=270,width=0.85\textwidth]{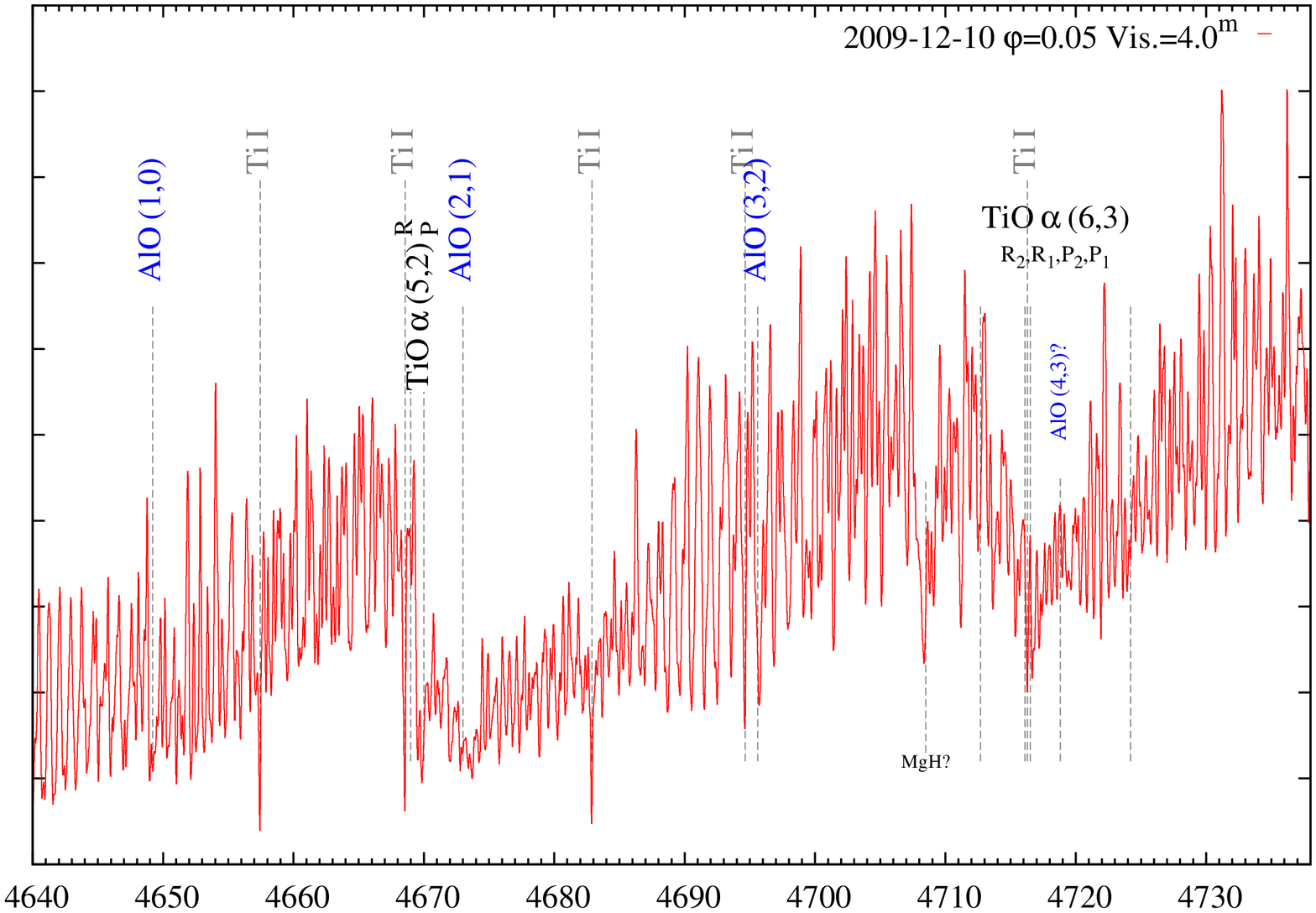}
\includegraphics[angle=270,width=0.85\textwidth]{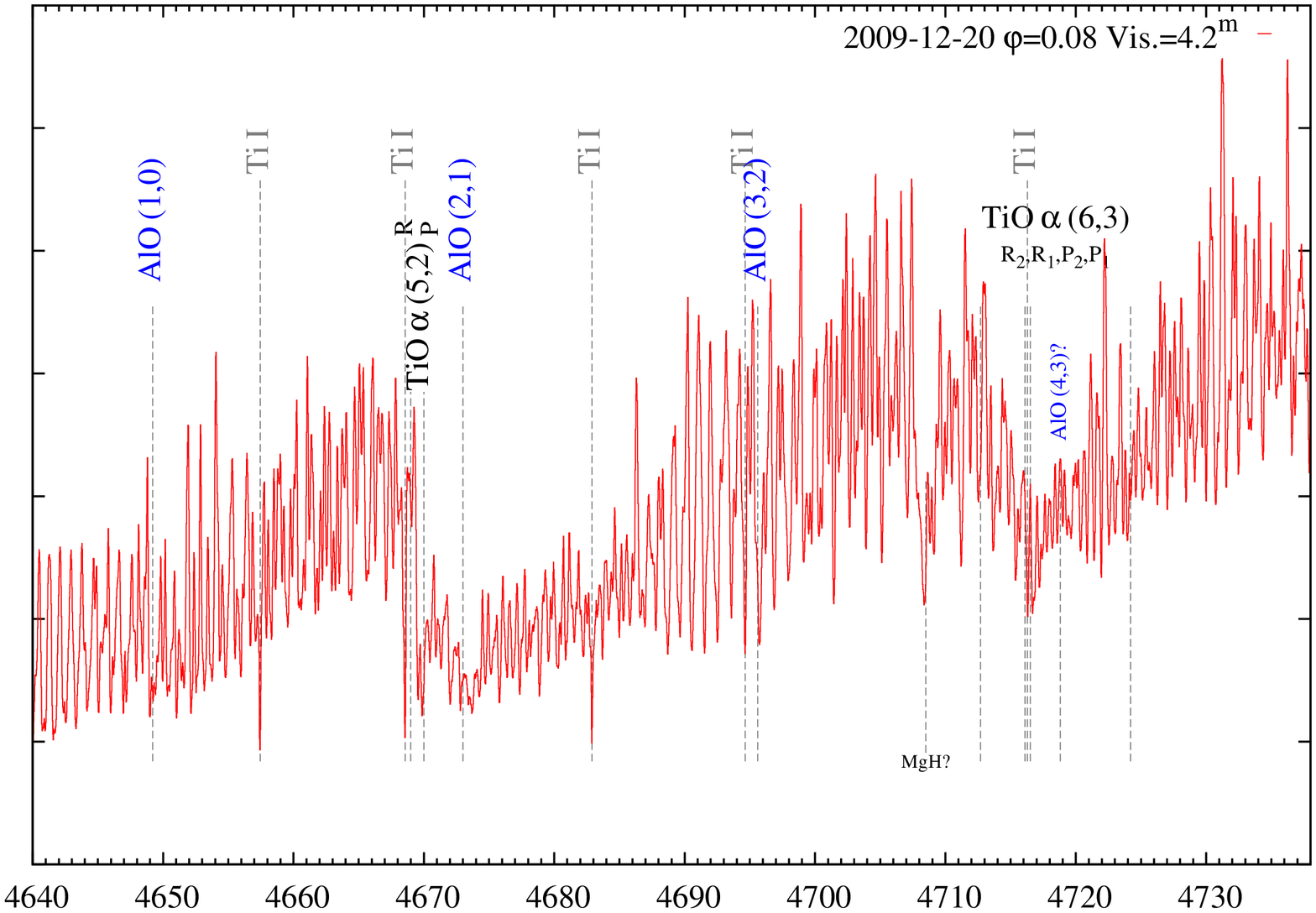}
\caption{Continued.}
\end{figure*}

  \setcounter{figure}{2}%

\begin{figure*} [tbh]
\centering
\includegraphics[angle=270,width=0.85\textwidth]{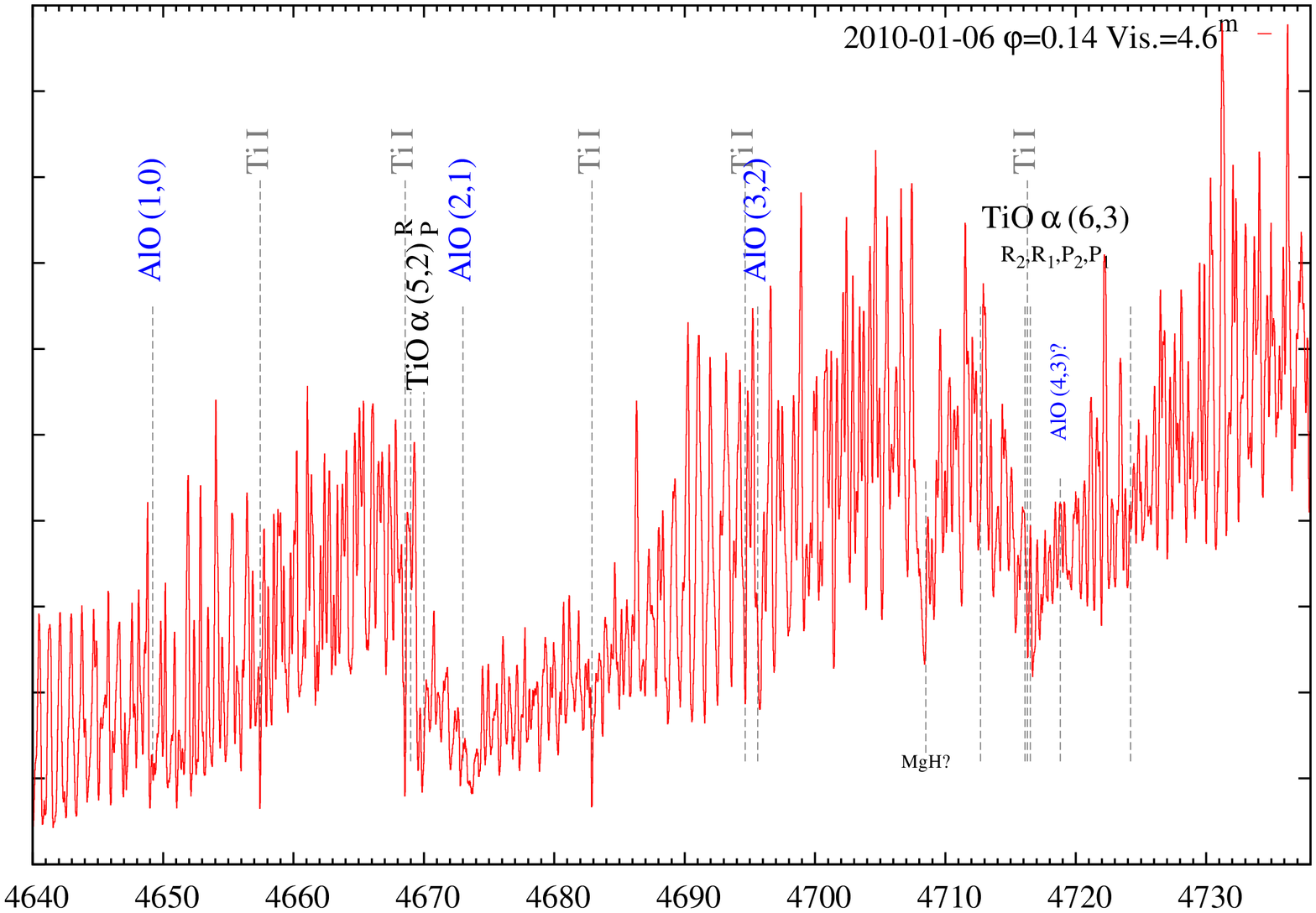}
\includegraphics[angle=270,width=0.85\textwidth]{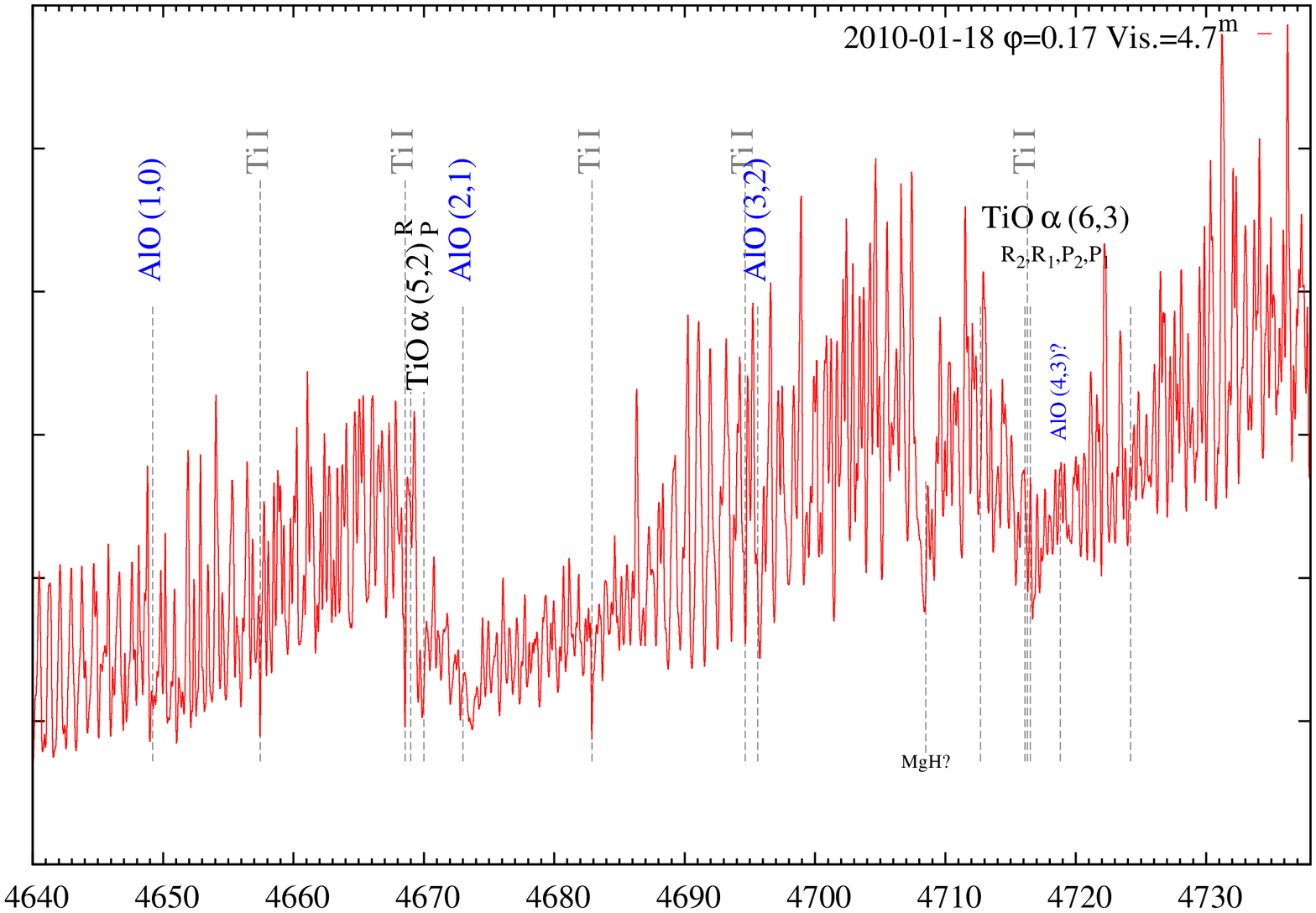}
\caption{Continued.}
\end{figure*}

  \setcounter{figure}{2}%

\begin{figure*} [tbh]
\centering
\includegraphics[angle=270,width=0.85\textwidth]{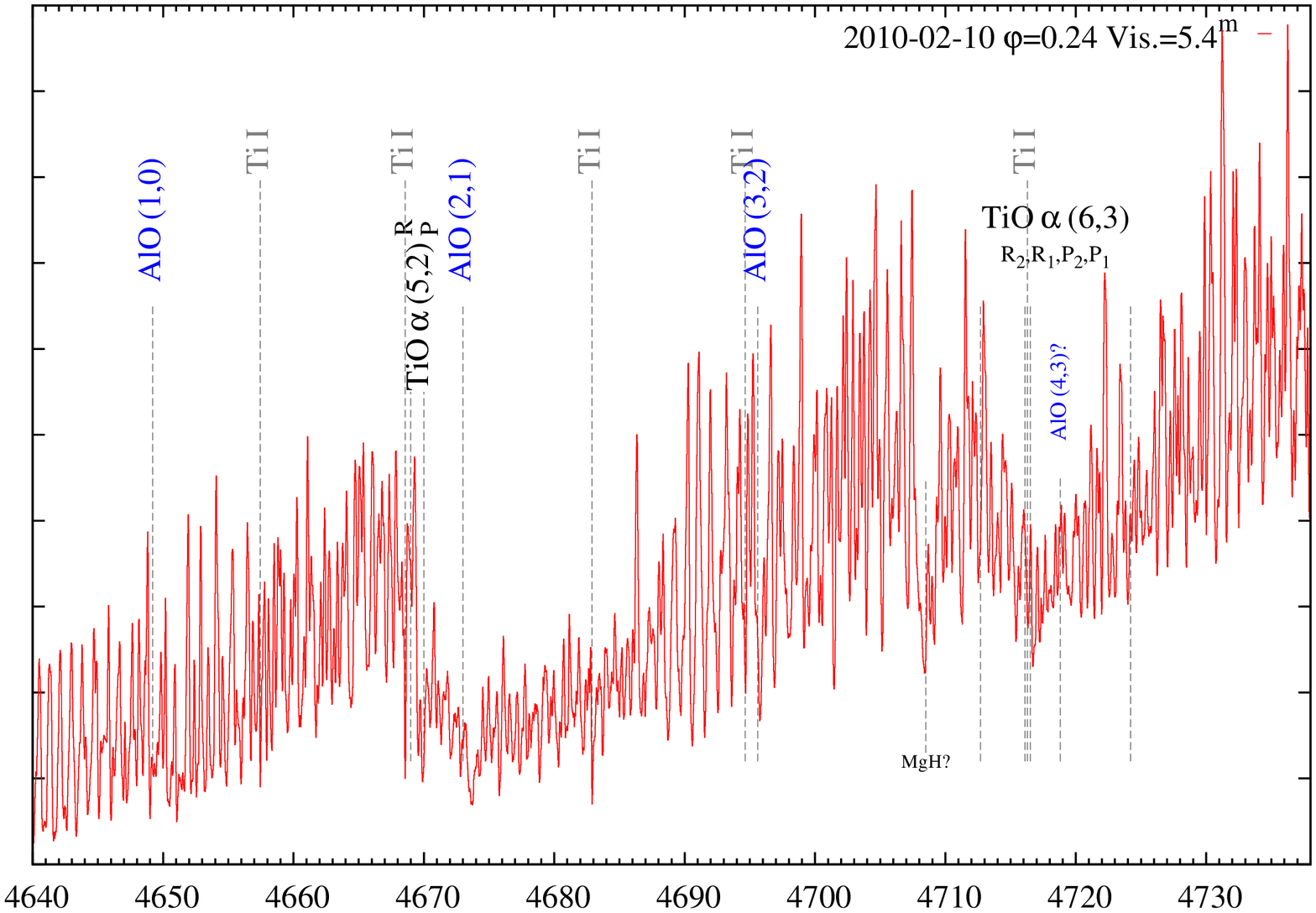}
\includegraphics[angle=270,width=0.85\textwidth]{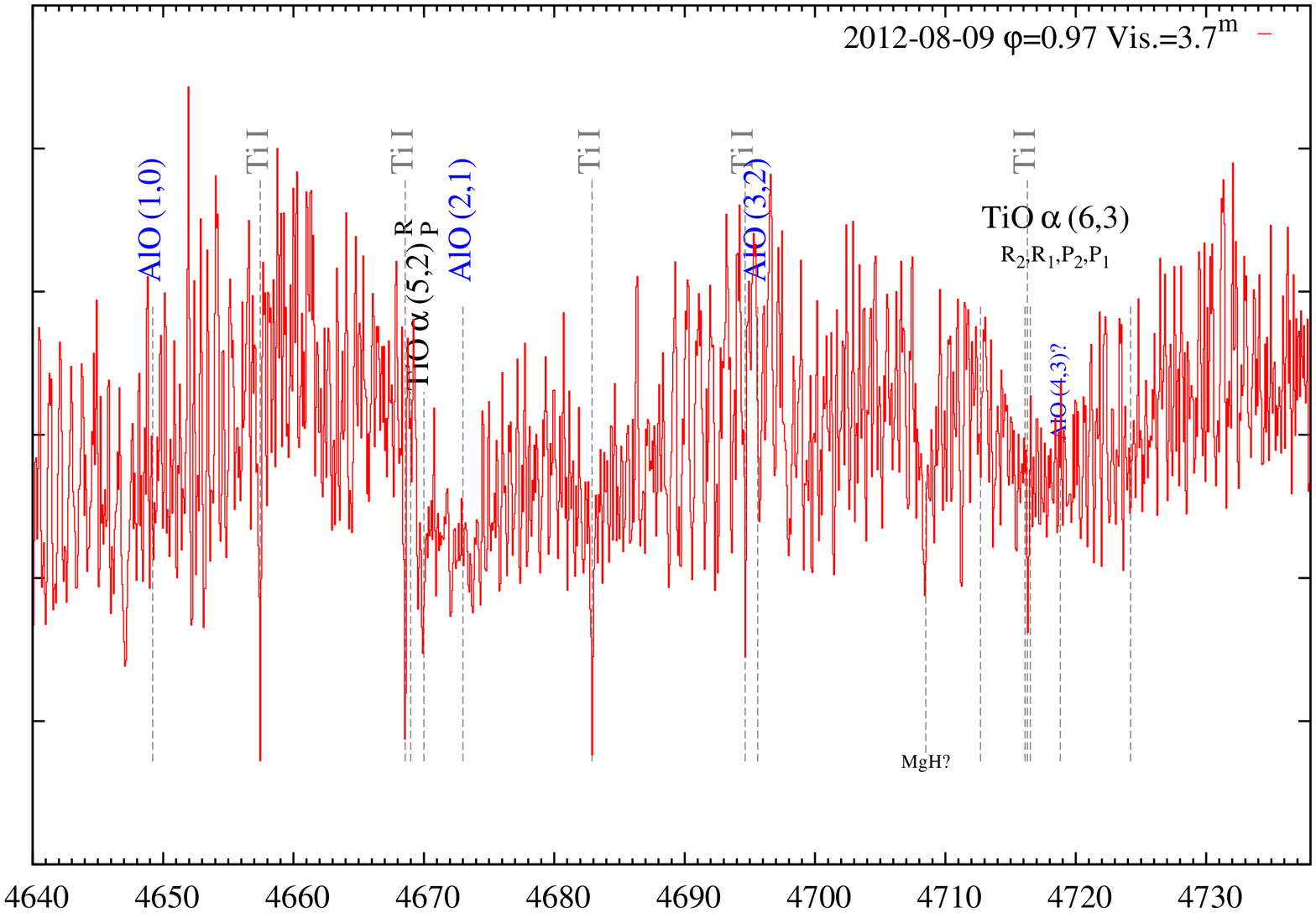}
\caption{Continued.}
\end{figure*}
\clearpage
\begin{figure*} [tbh]
\centering
\includegraphics[angle=270,width=0.85\textwidth]{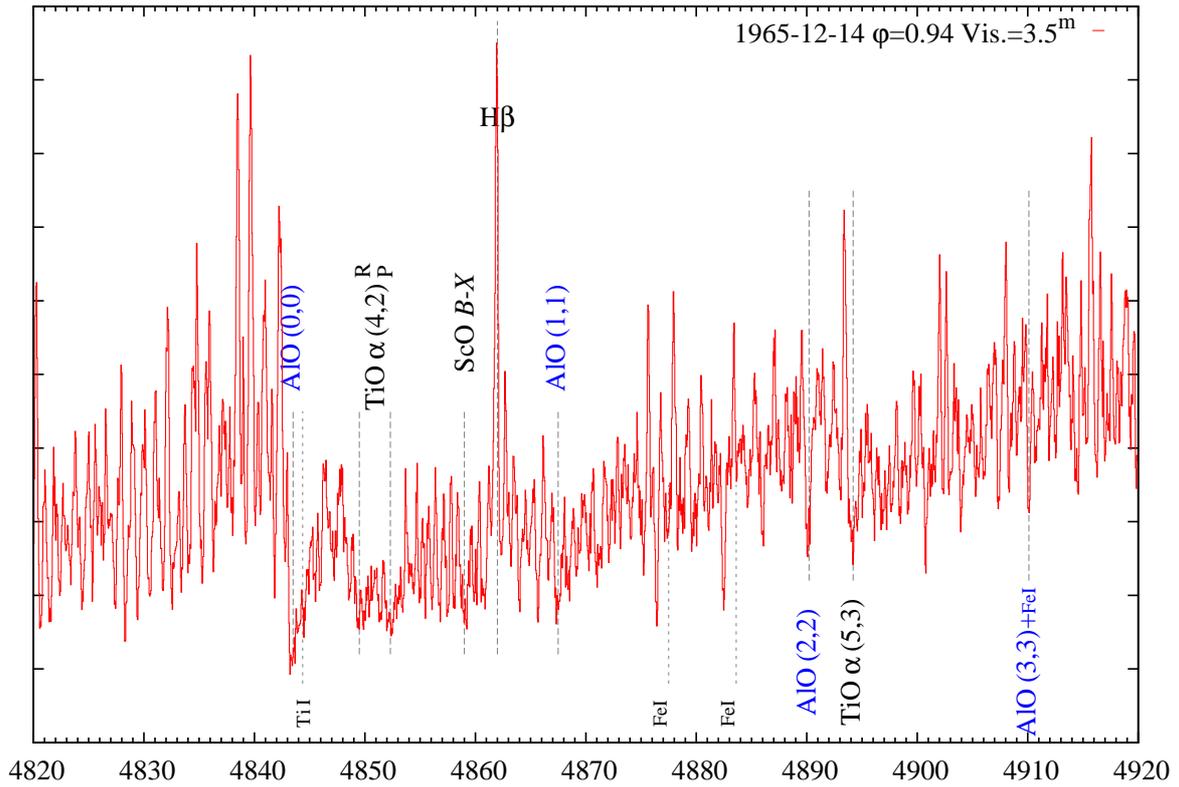}
\includegraphics[angle=270,width=0.85\textwidth]{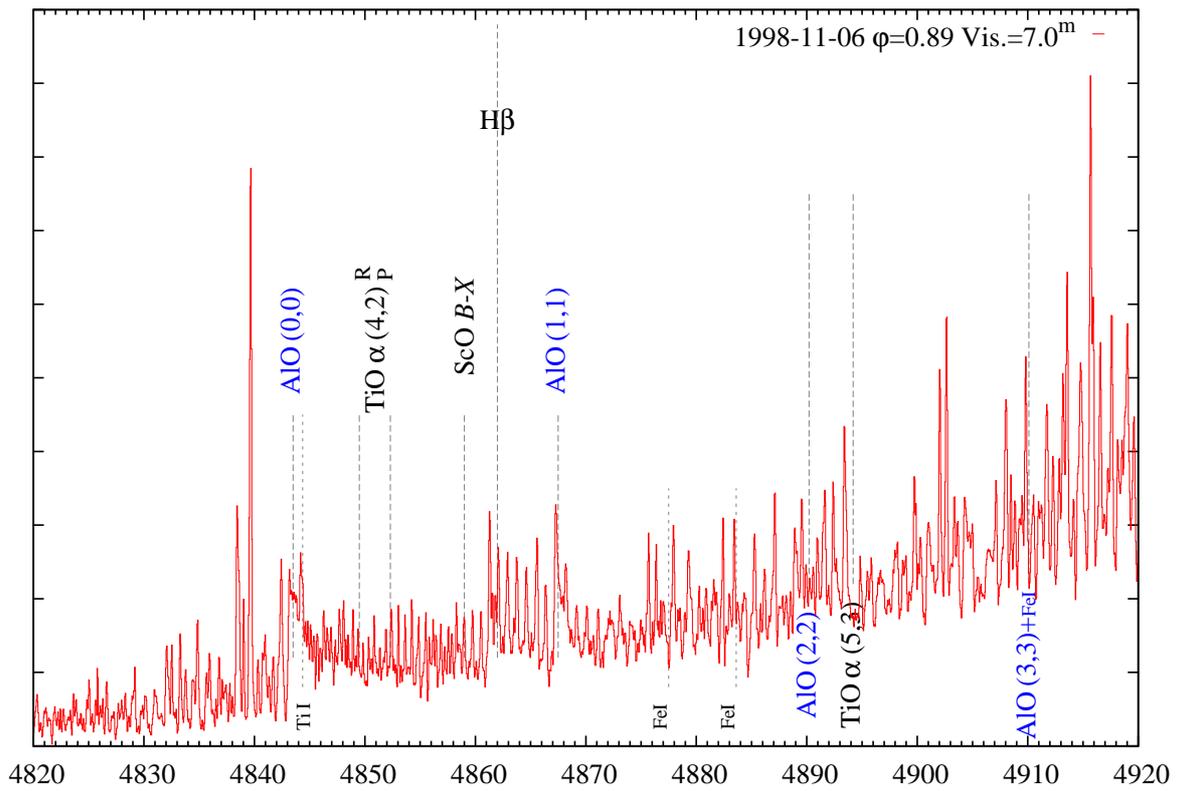}
\caption{The same as Fig.\,\ref{fig-AlOp2} but for the $\Delta\varv$=0 sequence of AlO $B-X$. Spectra from Narval are affected by an imperfect combination of different echelle orders in the 4877--4881\,\AA\ range. The spectrum from 2007-08-15 was diveded by a high-order polynomial.}\label{fig-AlOdv0}
\end{figure*}

  \setcounter{figure}{3}%

\begin{figure*} [tbh]
\centering
\includegraphics[angle=270,width=0.85\textwidth]{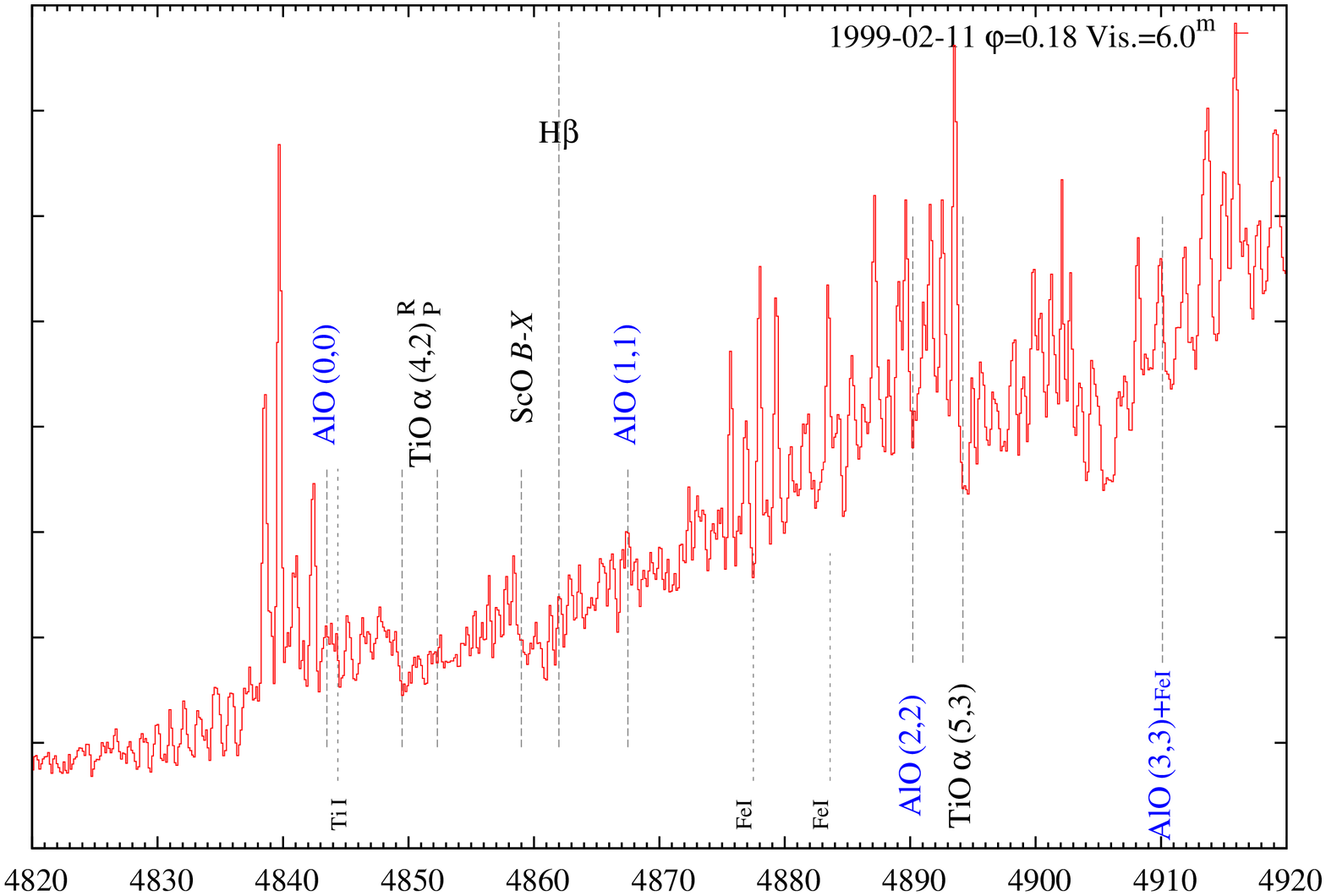}
\includegraphics[angle=270,width=0.85\textwidth]{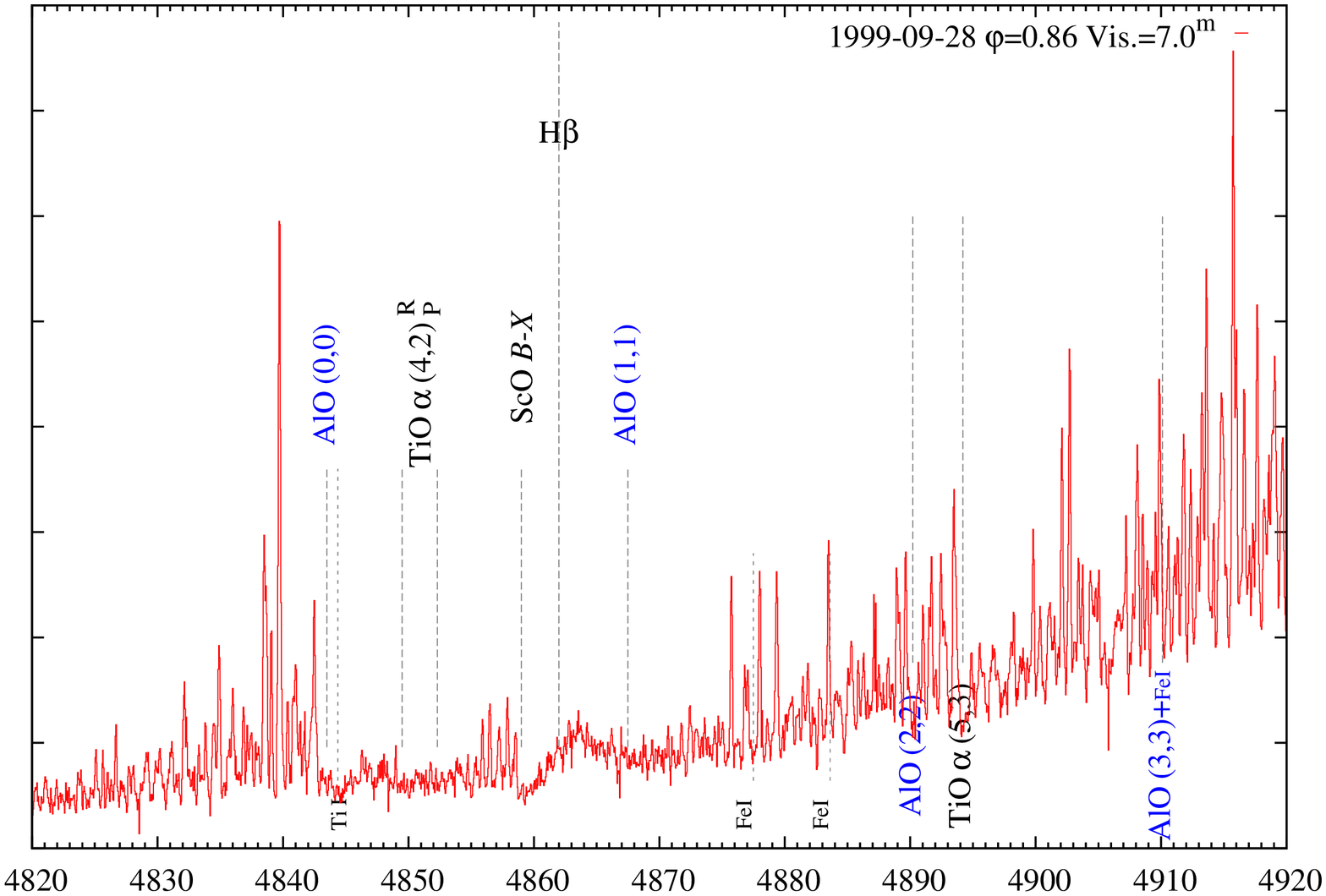}
\caption{Continued.}
\end{figure*}

  \setcounter{figure}{3}%

\begin{figure*} [tbh]
\centering
\includegraphics[angle=270,width=0.85\textwidth]{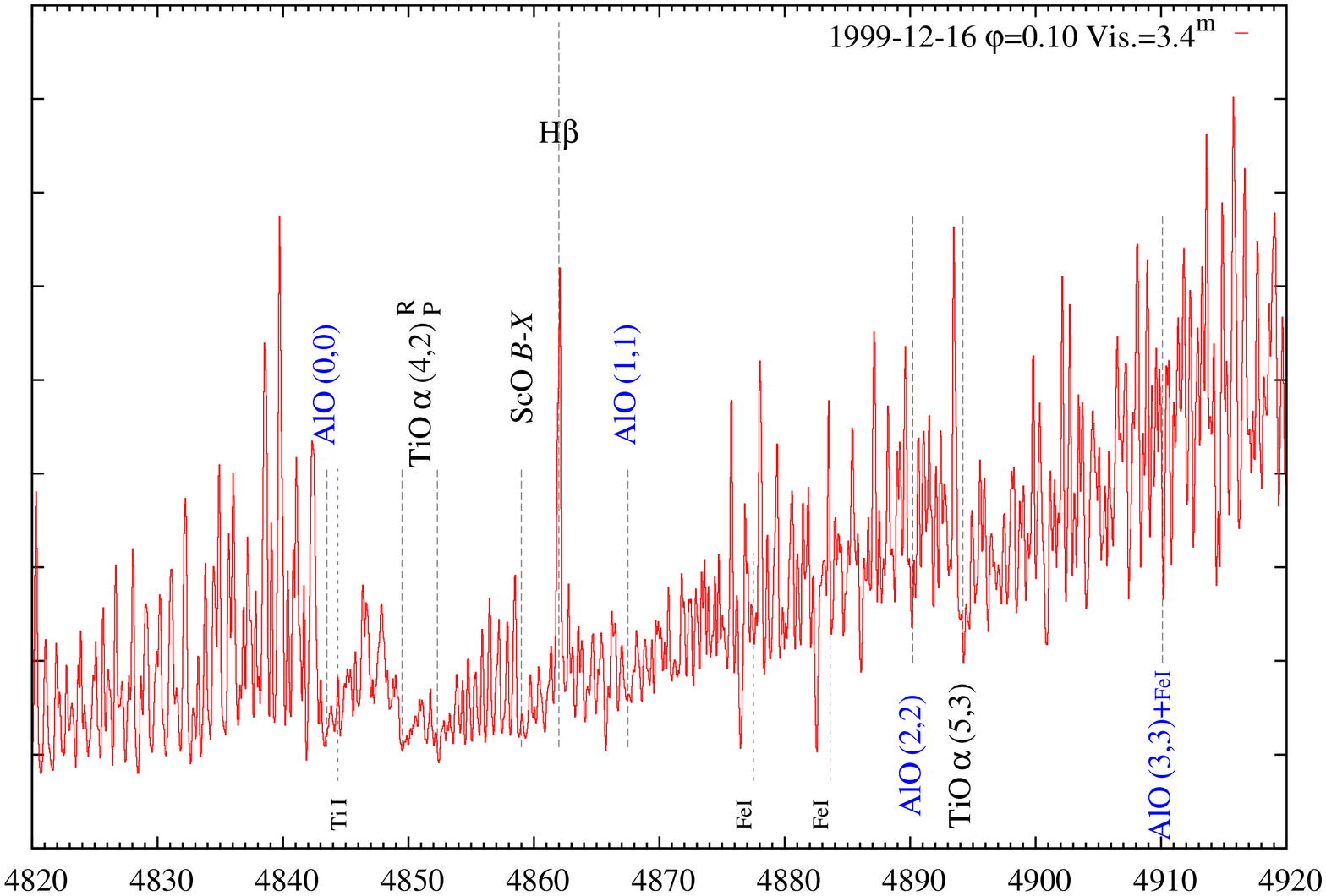}
\includegraphics[angle=270,width=0.85\textwidth]{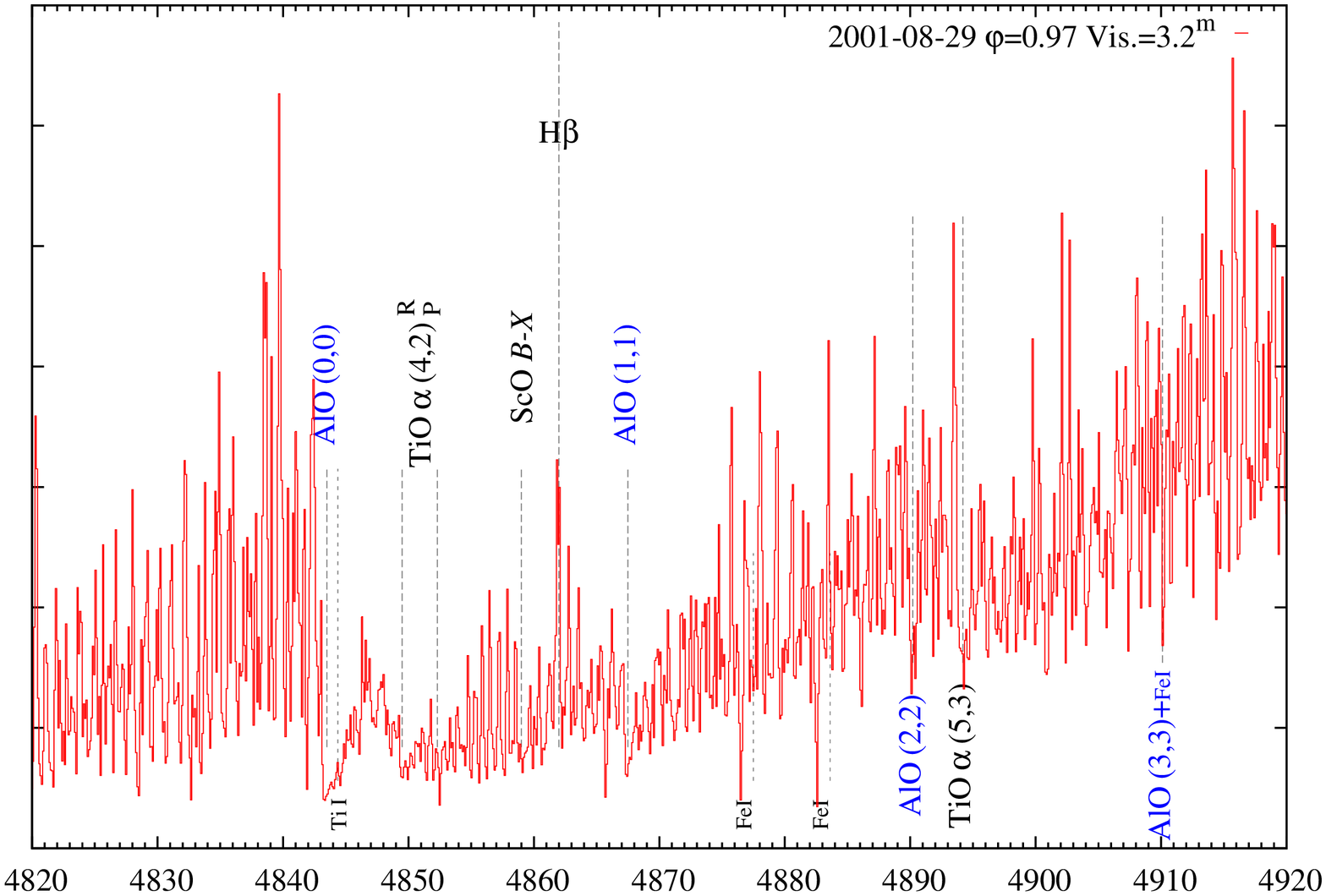}
\caption{Continued.}
\end{figure*}

  \setcounter{figure}{3}%

\begin{figure*} [tbh]
\centering
\includegraphics[angle=270,width=0.85\textwidth]{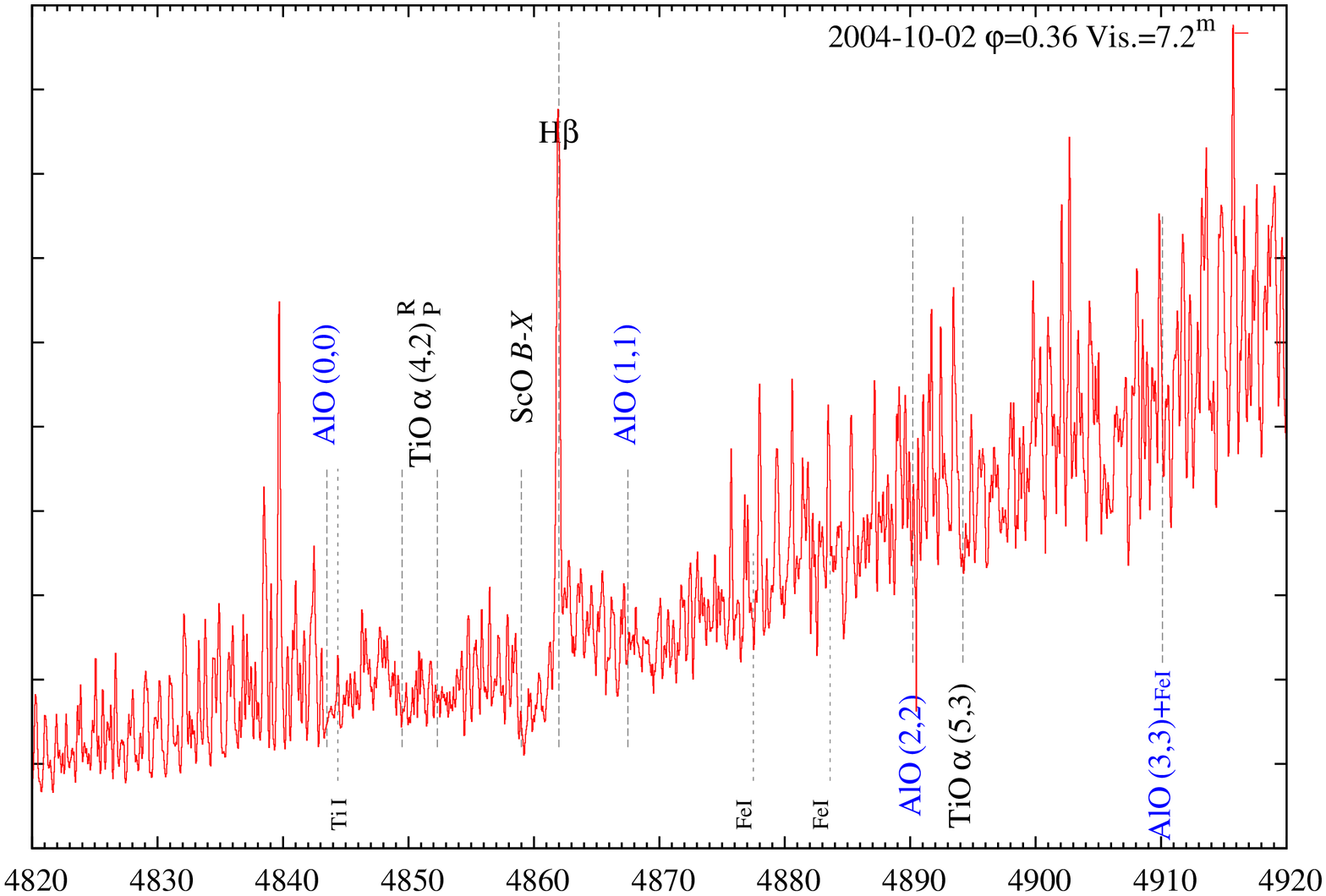}
\includegraphics[angle=270,width=0.85\textwidth]{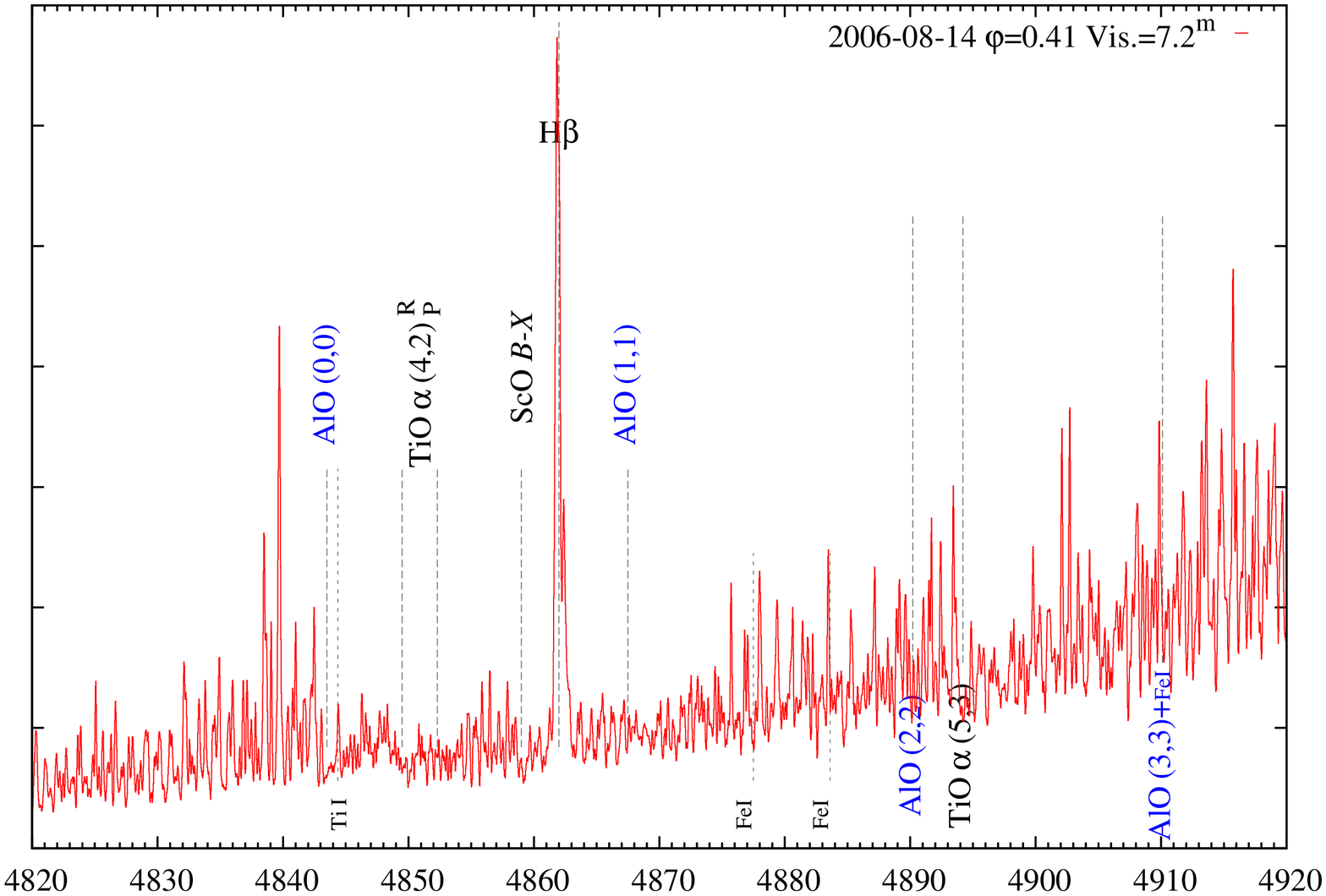}
\caption{Continued.}
\end{figure*}

  \setcounter{figure}{3}%

\begin{figure*} [tbh]
\centering
\includegraphics[angle=270,width=0.85\textwidth]{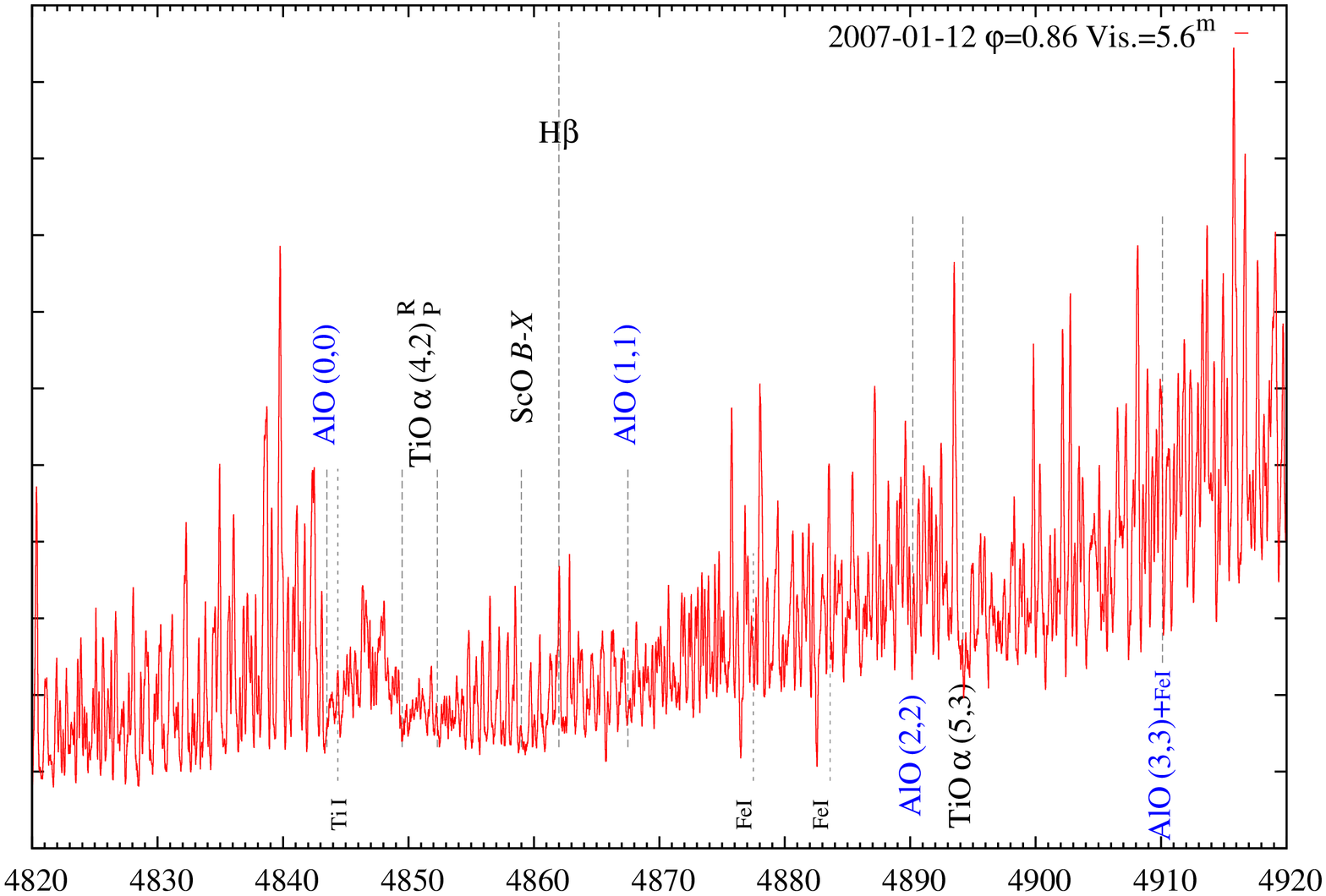}
\includegraphics[angle=270,width=0.85\textwidth]{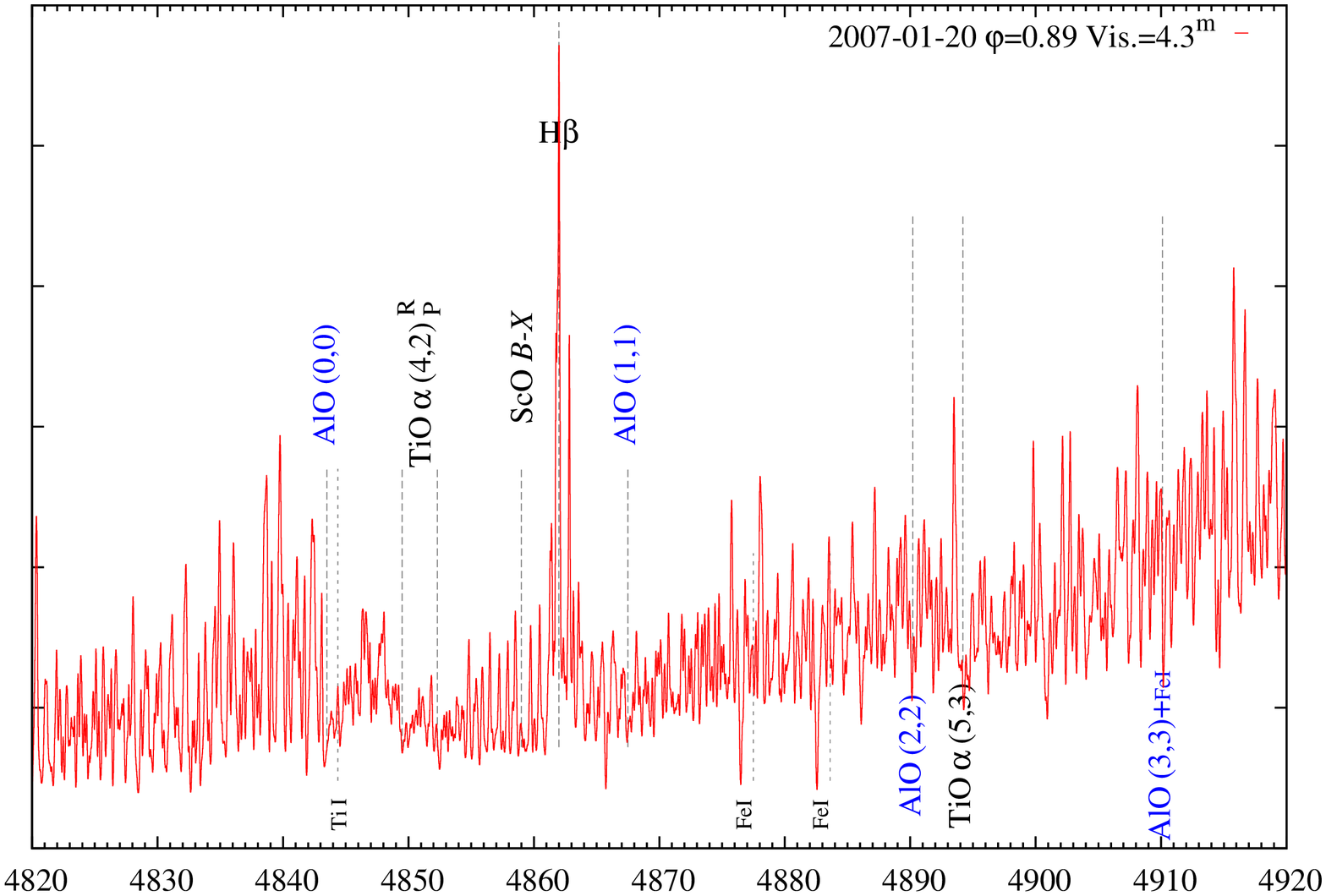}
\caption{Continued.}
\end{figure*}

  \setcounter{figure}{3}%

\begin{figure*} [tbh]
\centering
\includegraphics[angle=270,width=0.85\textwidth]{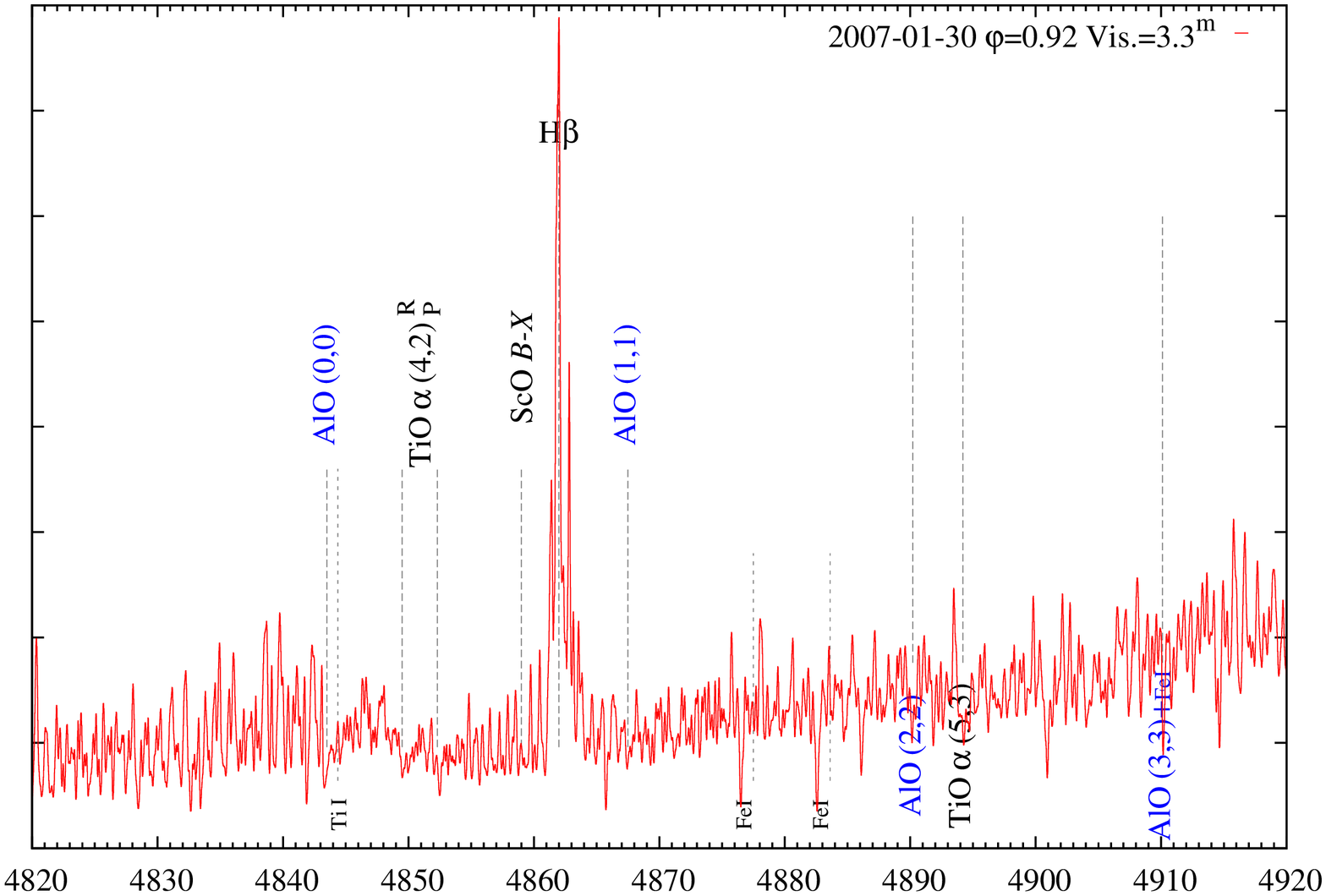}
\includegraphics[angle=270,width=0.85\textwidth]{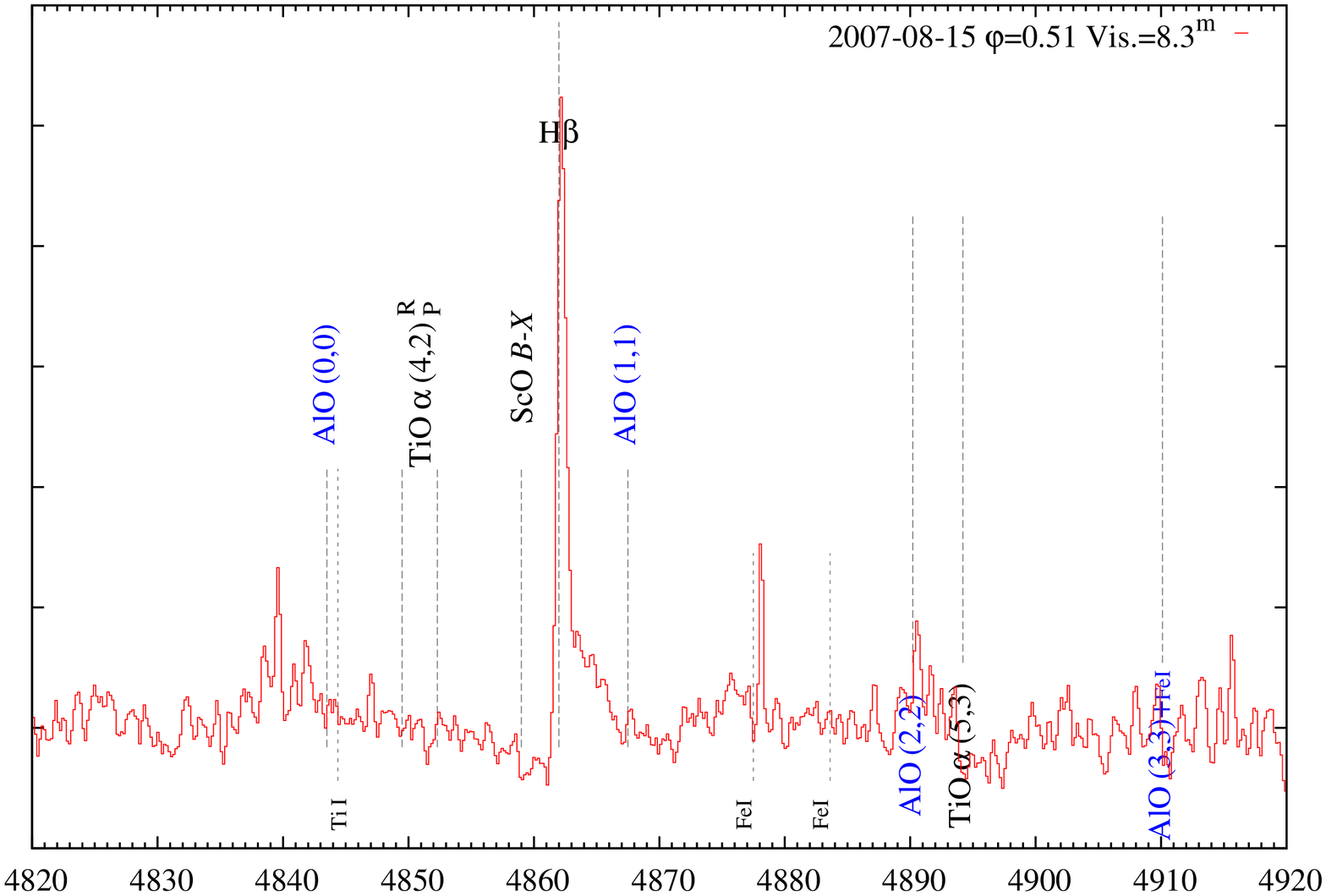}
\caption{Continued.}
\end{figure*}

  \setcounter{figure}{3}%

\begin{figure*} [tbh]
\centering
\includegraphics[angle=270,width=0.85\textwidth]{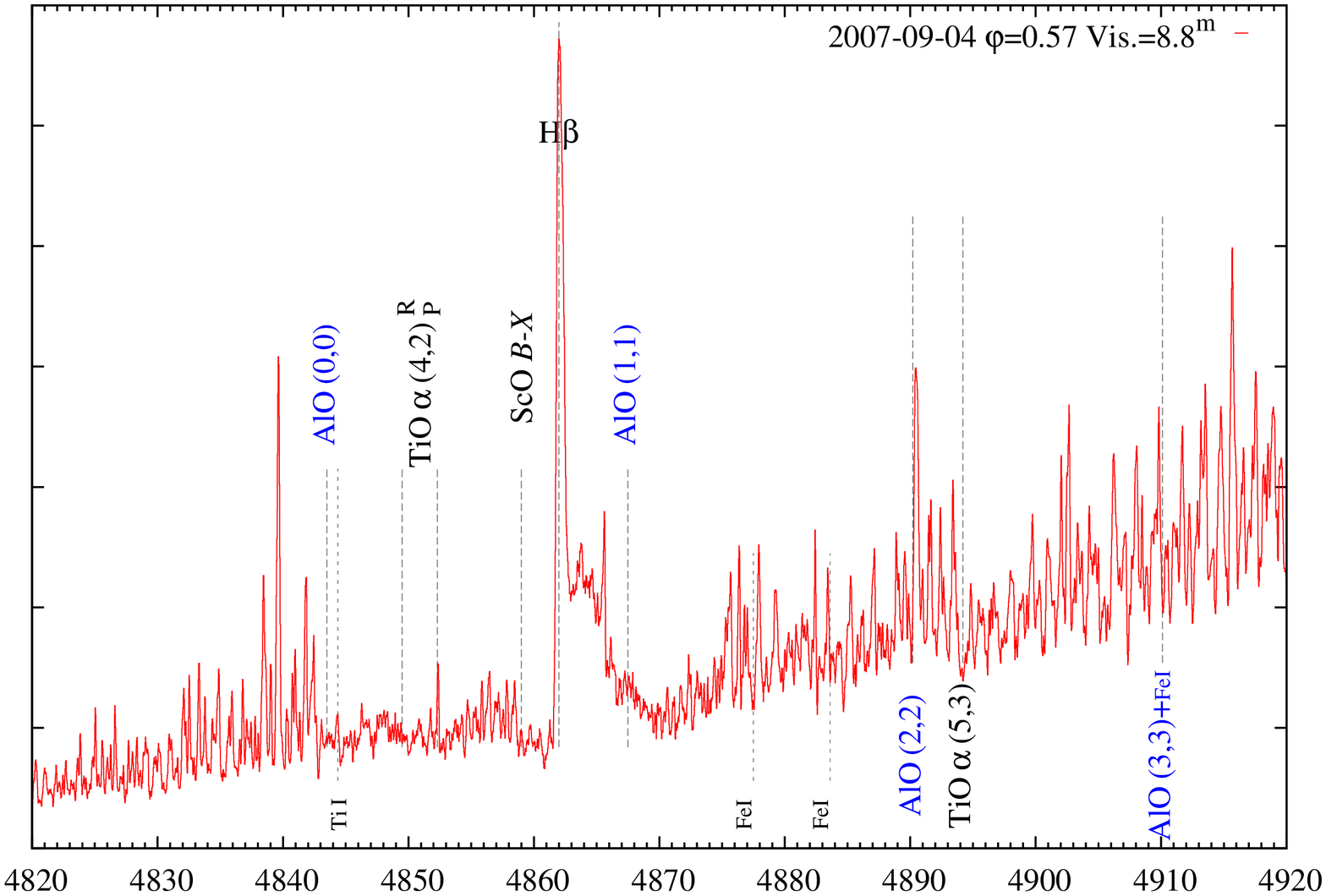}
\includegraphics[angle=270,width=0.85\textwidth]{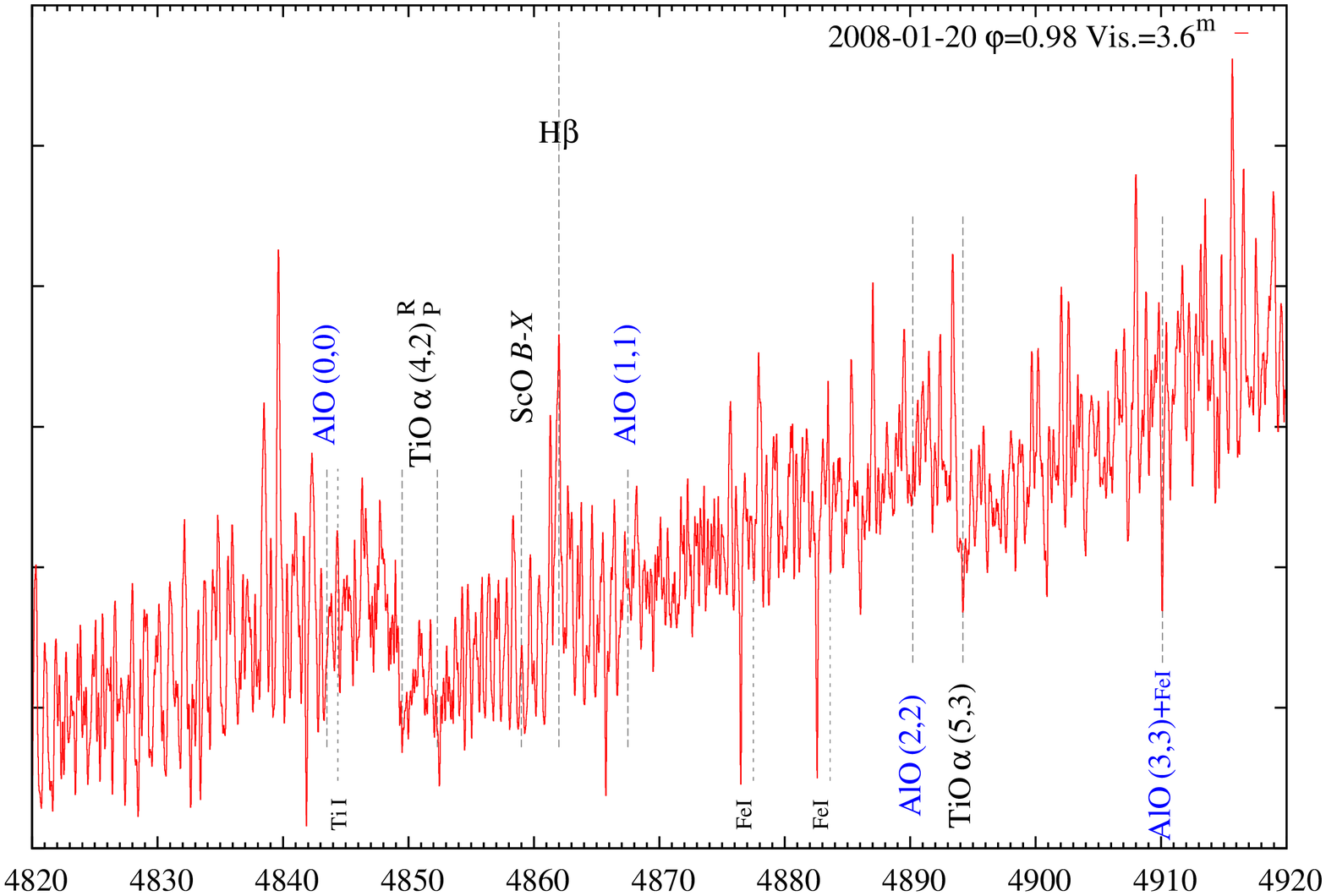}
\caption{Continued.}
\end{figure*}

  \setcounter{figure}{3}%

\begin{figure*} [tbh]
\centering
\includegraphics[angle=270,width=0.85\textwidth]{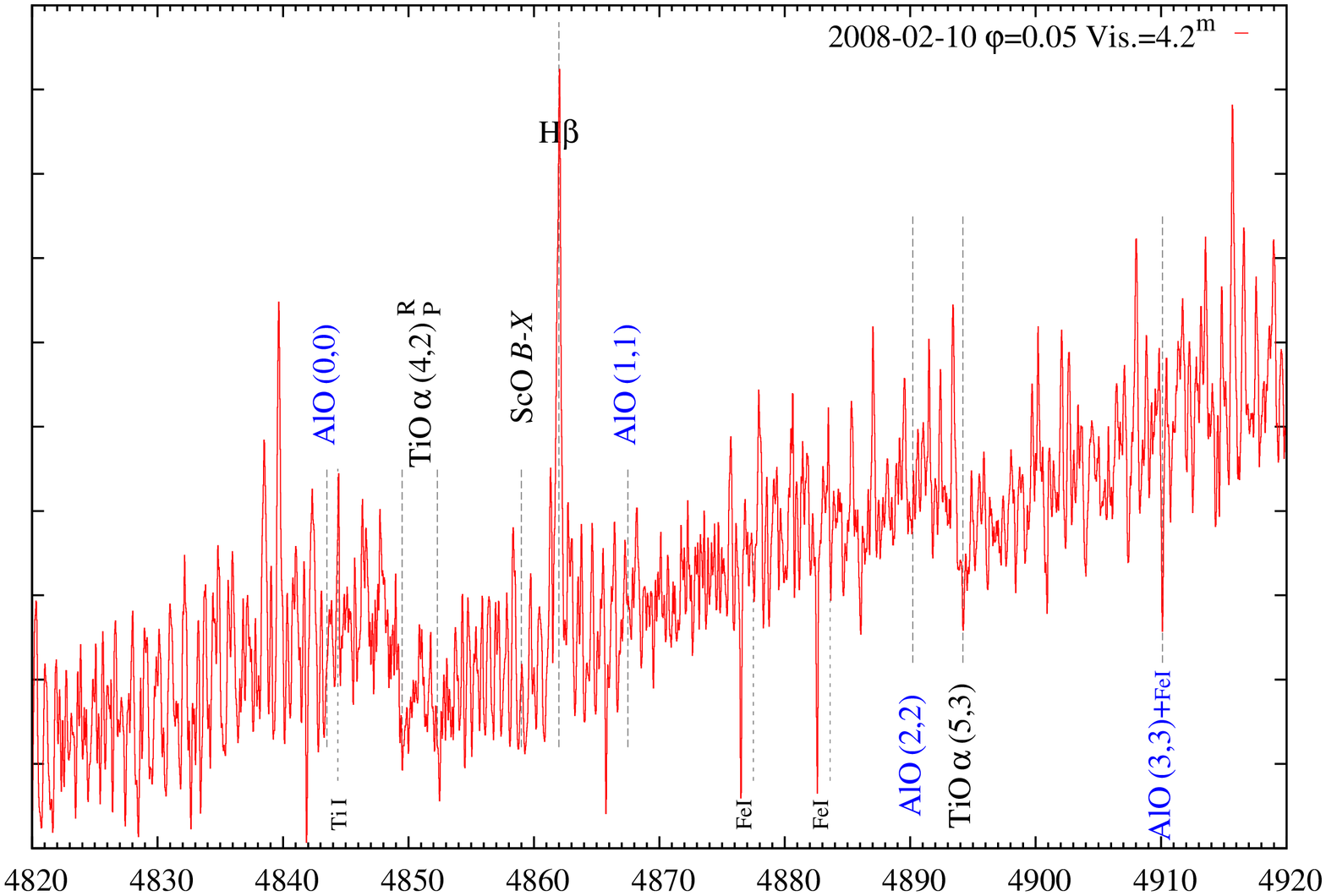}
\includegraphics[angle=270,width=0.85\textwidth]{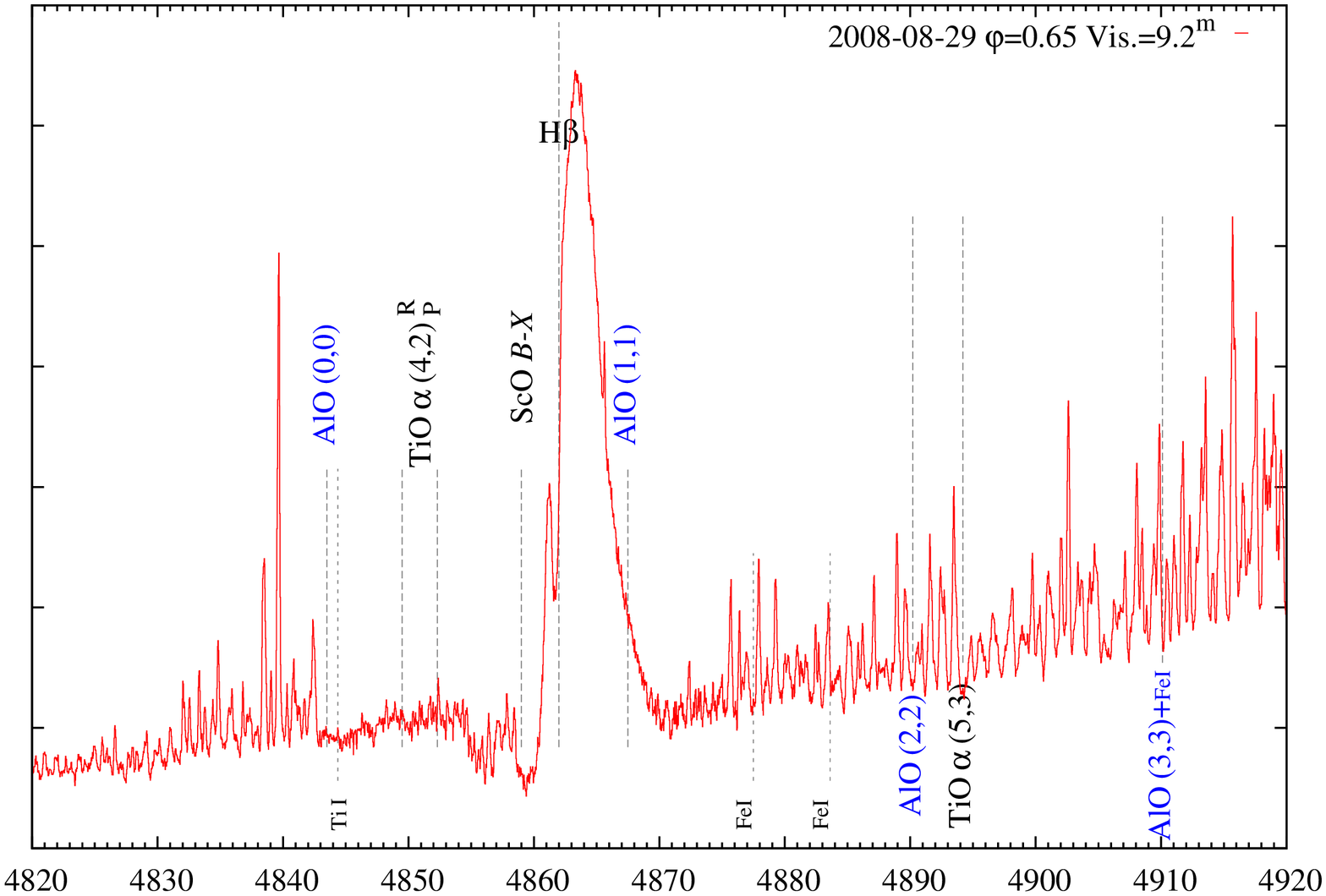}
\caption{Continued.}
\end{figure*}

  \setcounter{figure}{3}%

\begin{figure*} [tbh]
\centering
\includegraphics[angle=270,width=0.85\textwidth]{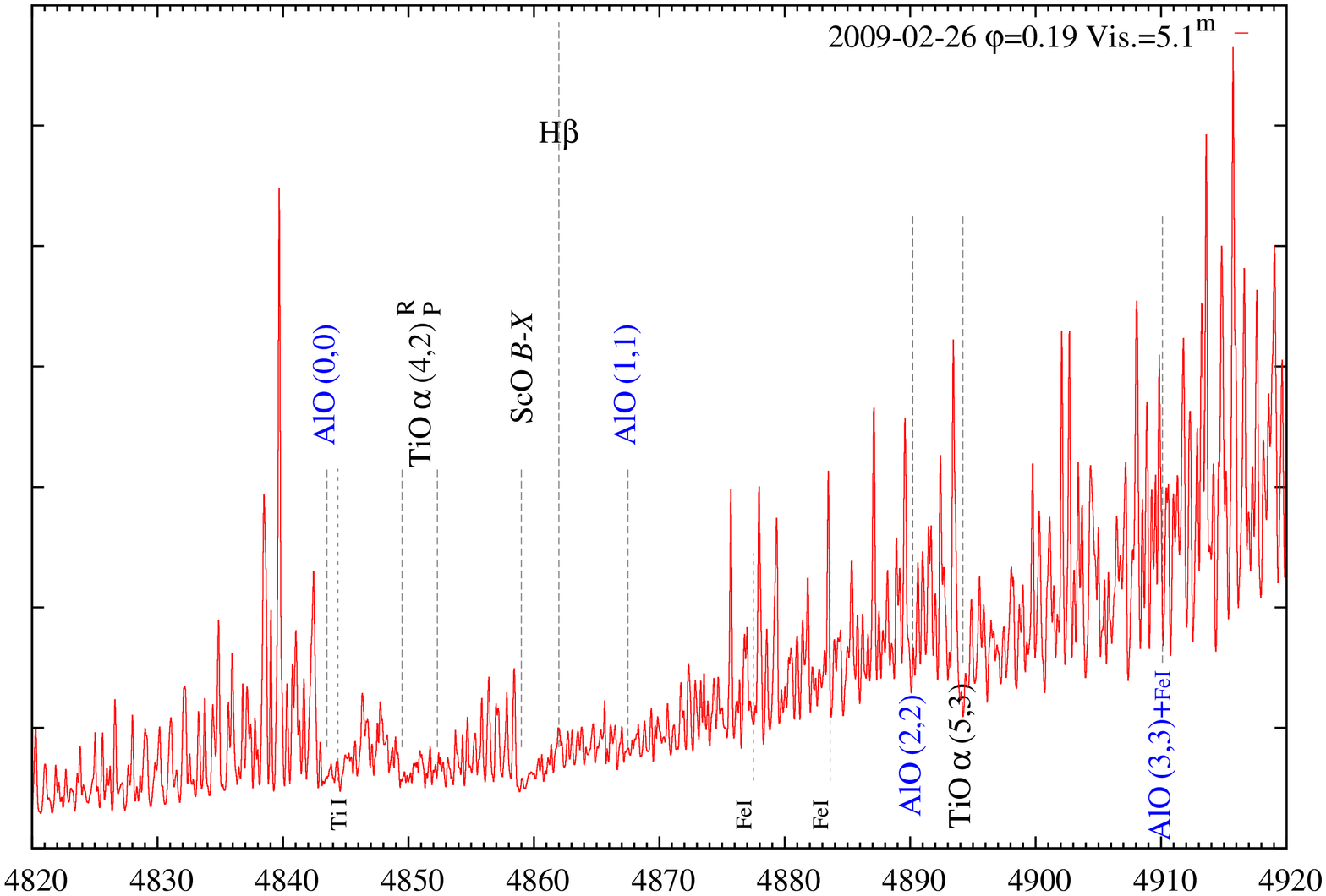}
\includegraphics[angle=270,width=0.85\textwidth]{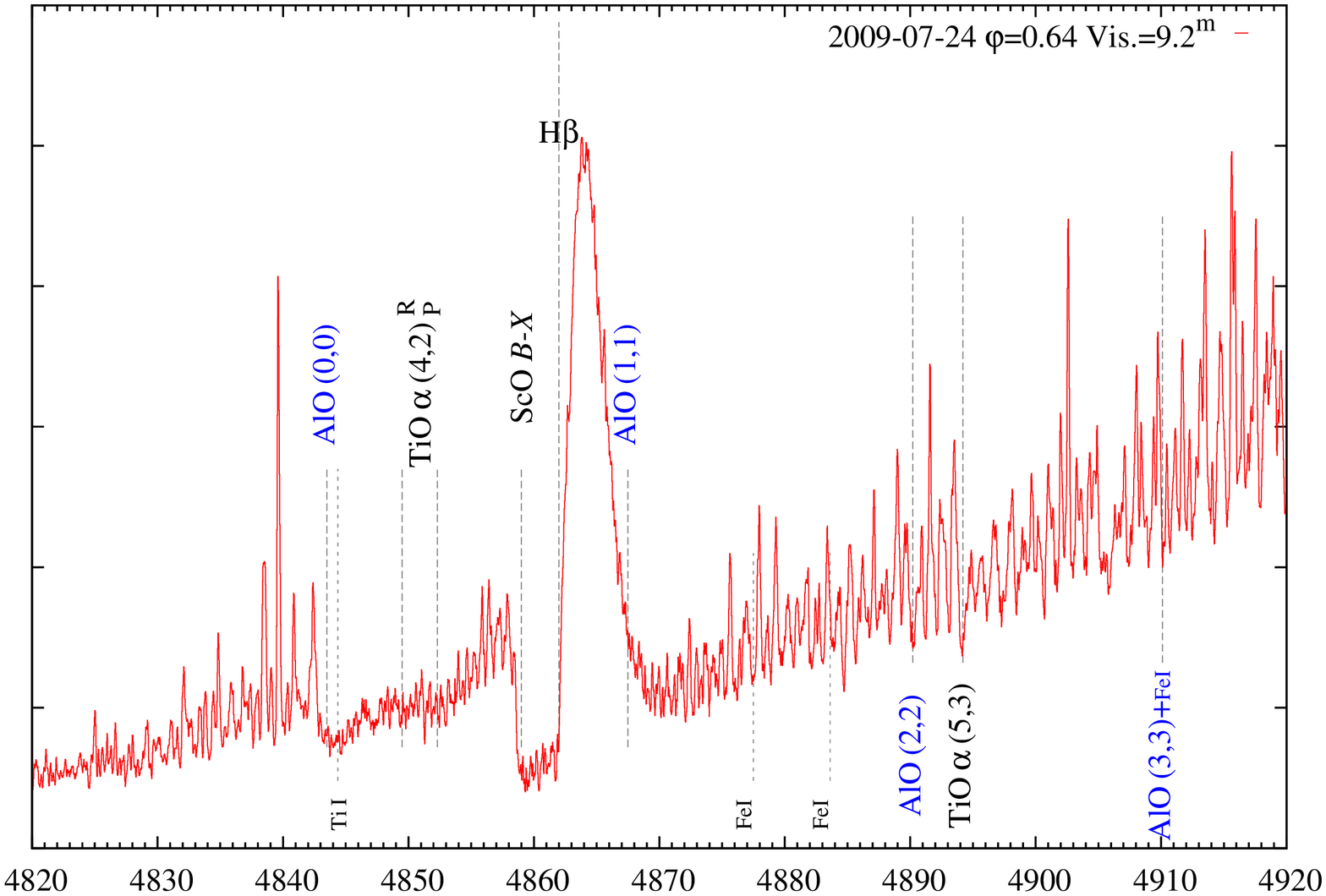}
\caption{Continued.}
\end{figure*}

  \setcounter{figure}{3}%

\begin{figure*} [tbh]
\centering
\includegraphics[angle=270,width=0.85\textwidth]{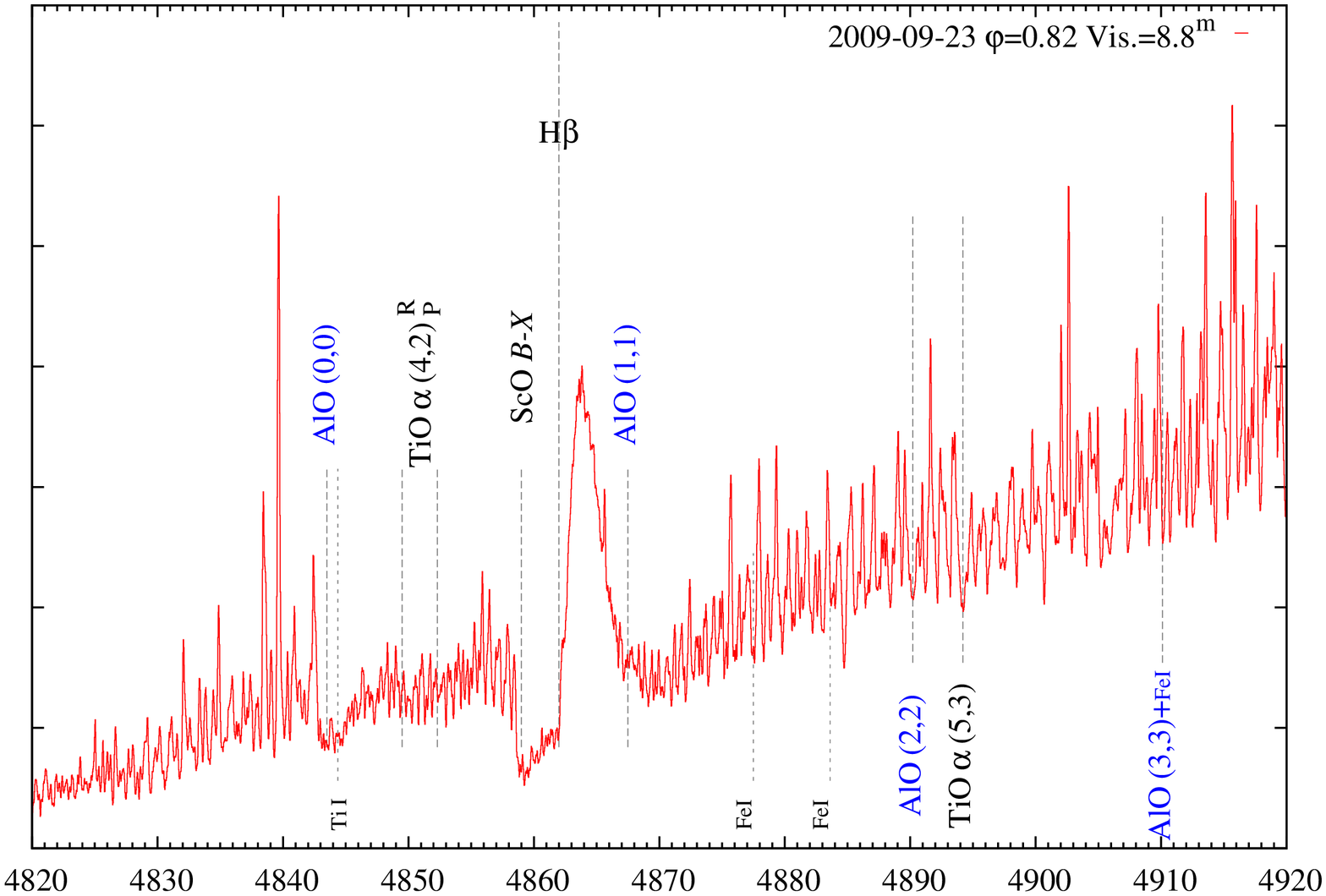}
\includegraphics[angle=270,width=0.85\textwidth]{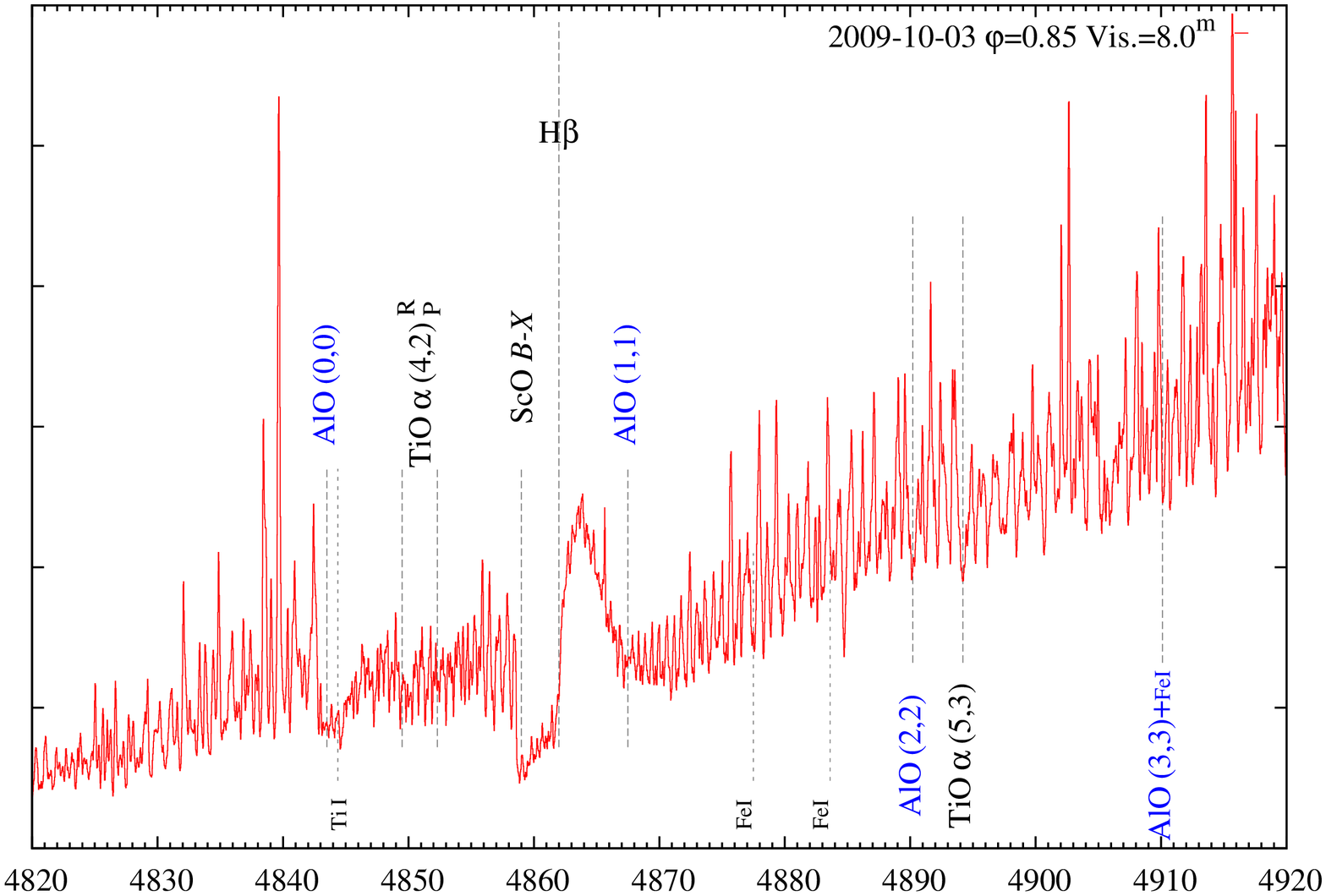}
\caption{Continued.}
\end{figure*}

  \setcounter{figure}{3}%

\begin{figure*} [tbh]
\centering
\includegraphics[angle=270,width=0.85\textwidth]{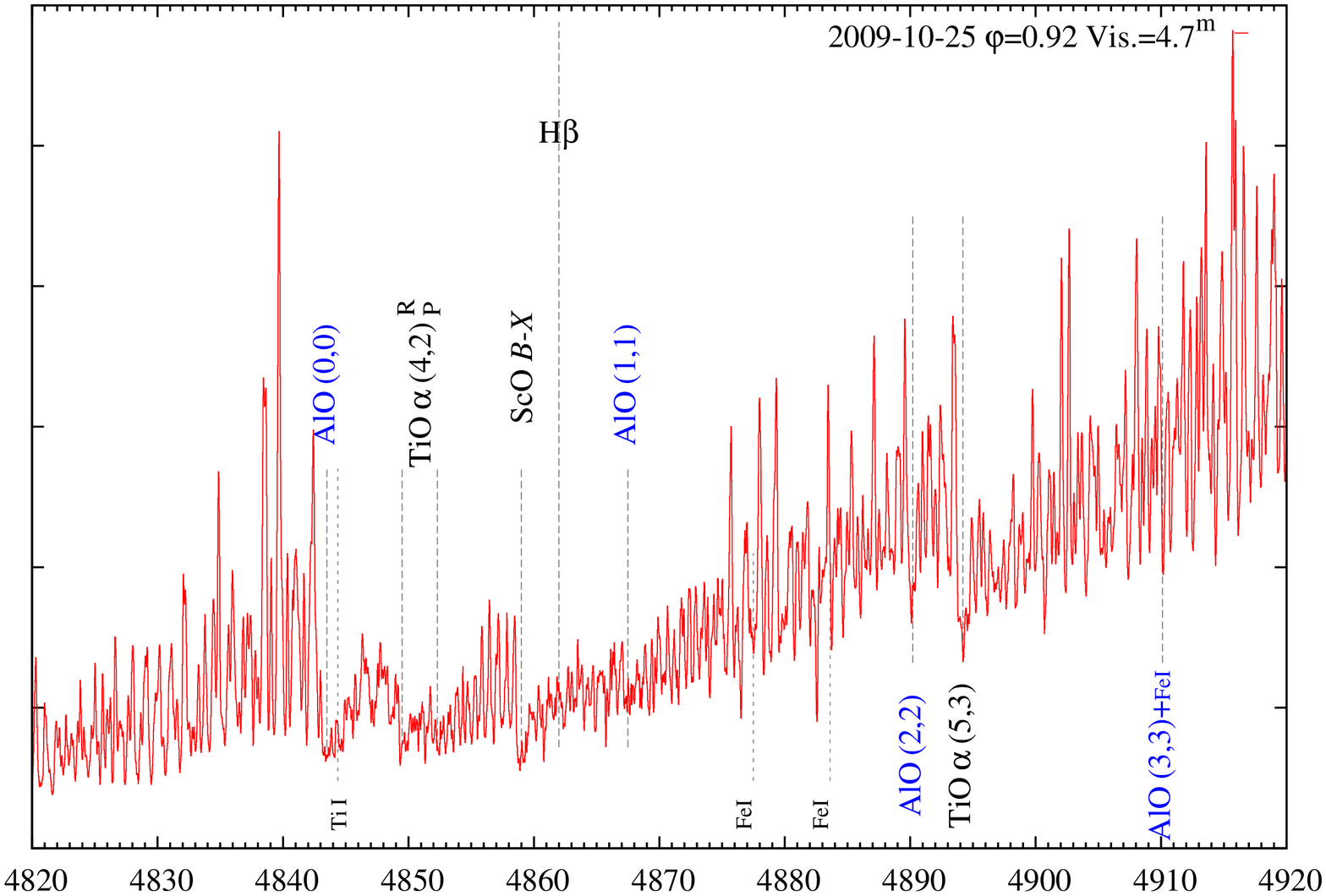}
\includegraphics[angle=270,width=0.85\textwidth]{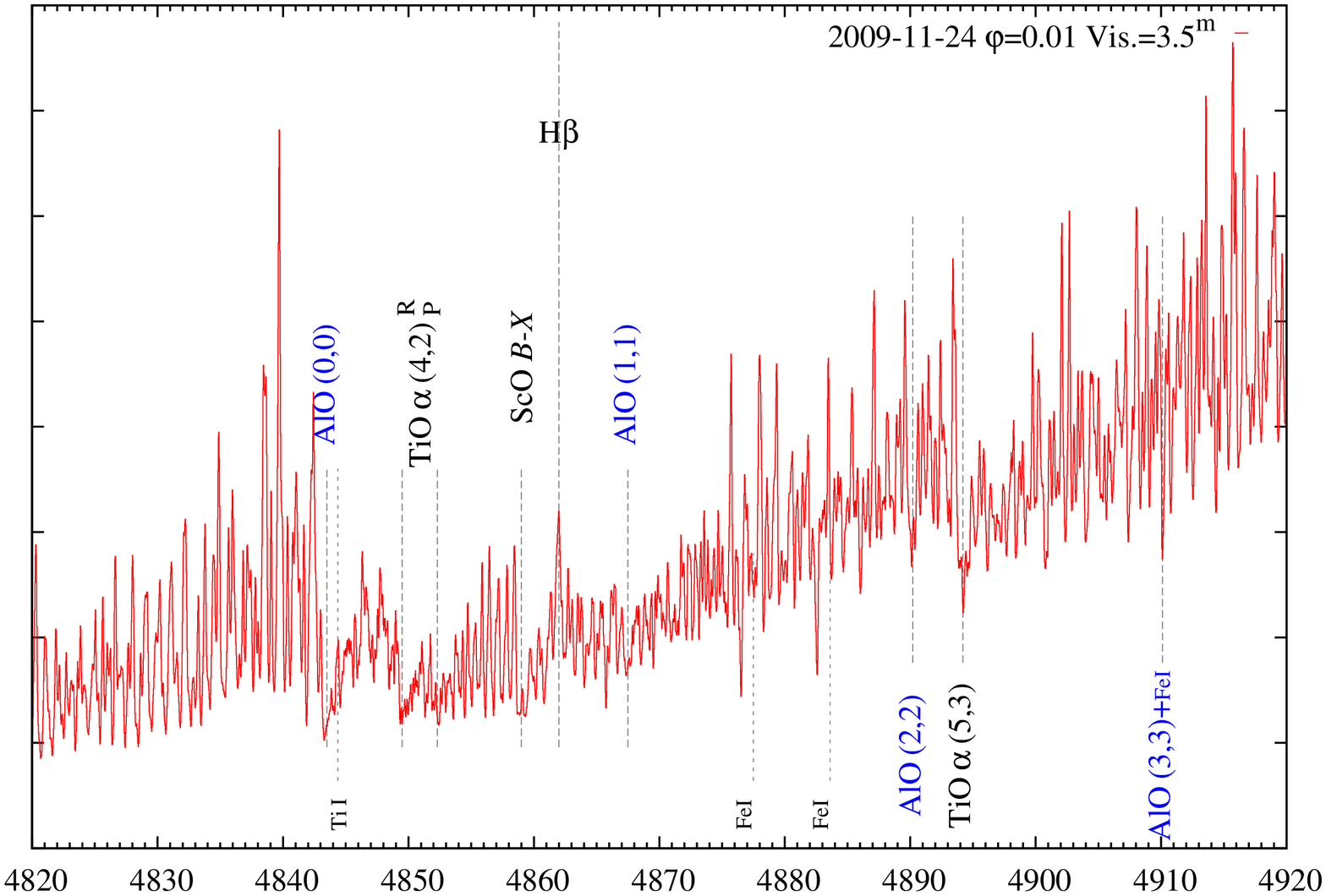}
\caption{Continued.}
\end{figure*}

  \setcounter{figure}{3}%

\begin{figure*} [tbh]
\centering
\includegraphics[angle=270,width=0.85\textwidth]{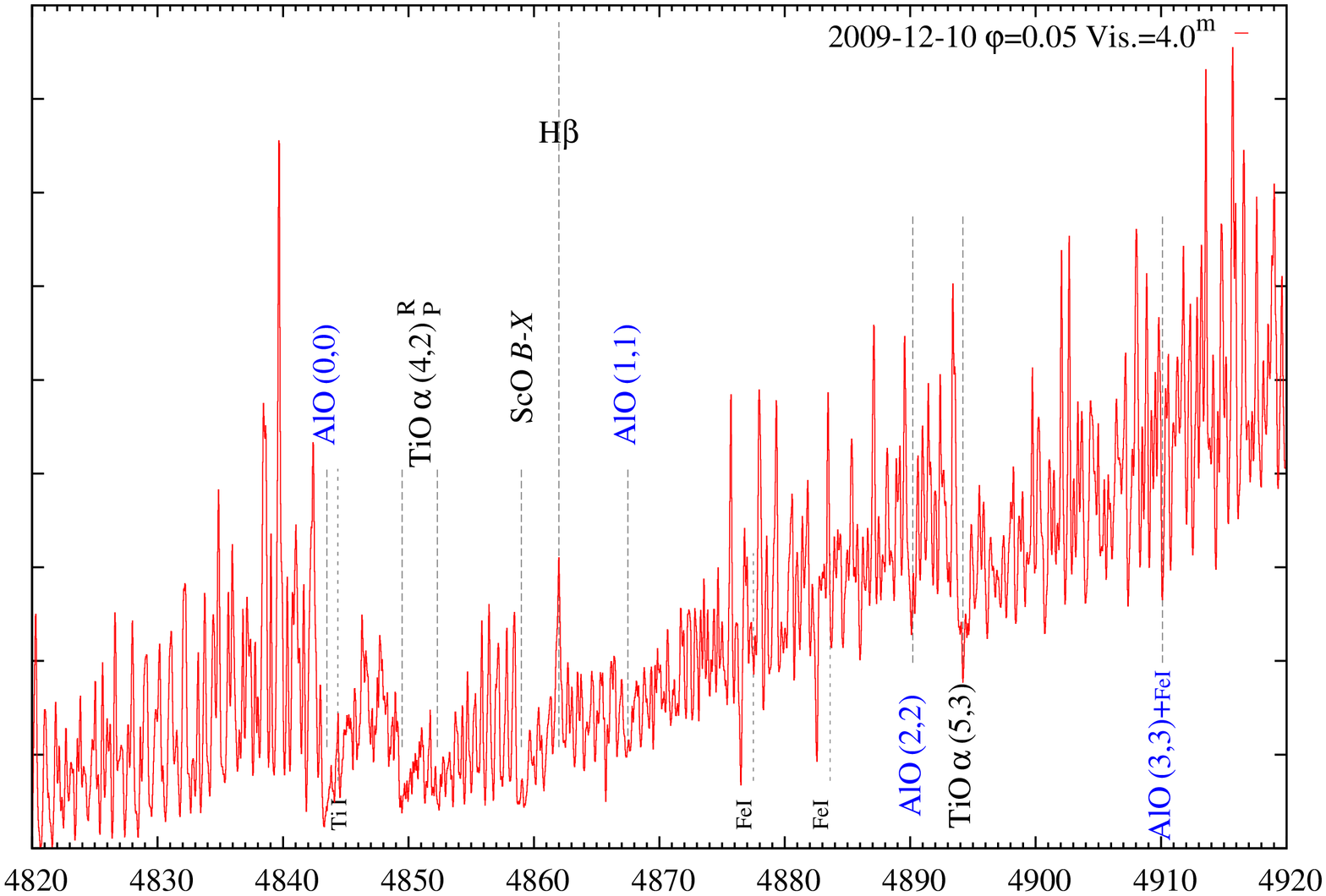}
\includegraphics[angle=270,width=0.85\textwidth]{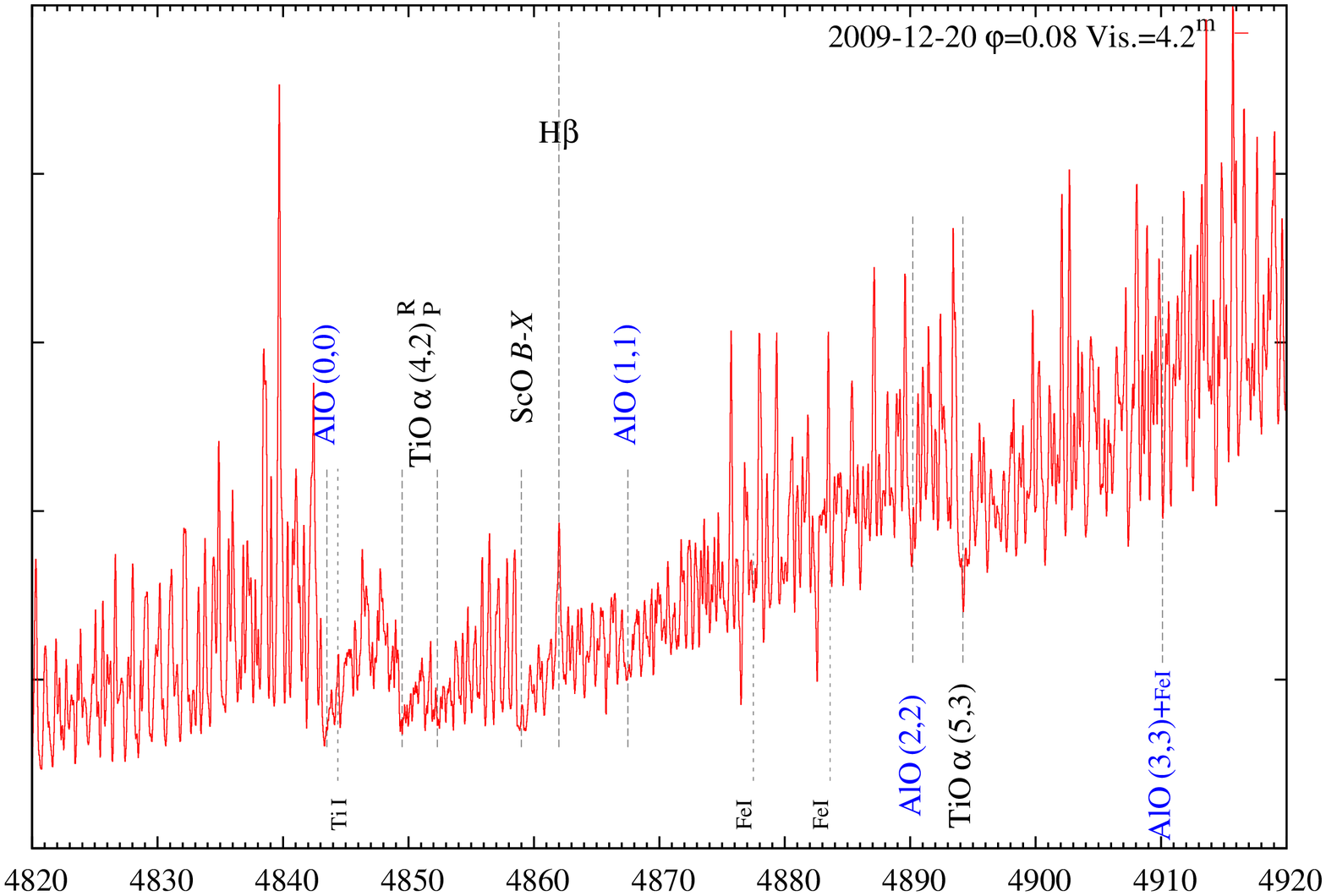}
\caption{Continued.}
\end{figure*}

  \setcounter{figure}{3}%

\begin{figure*} [tbh]
\centering
\includegraphics[angle=270,width=0.85\textwidth]{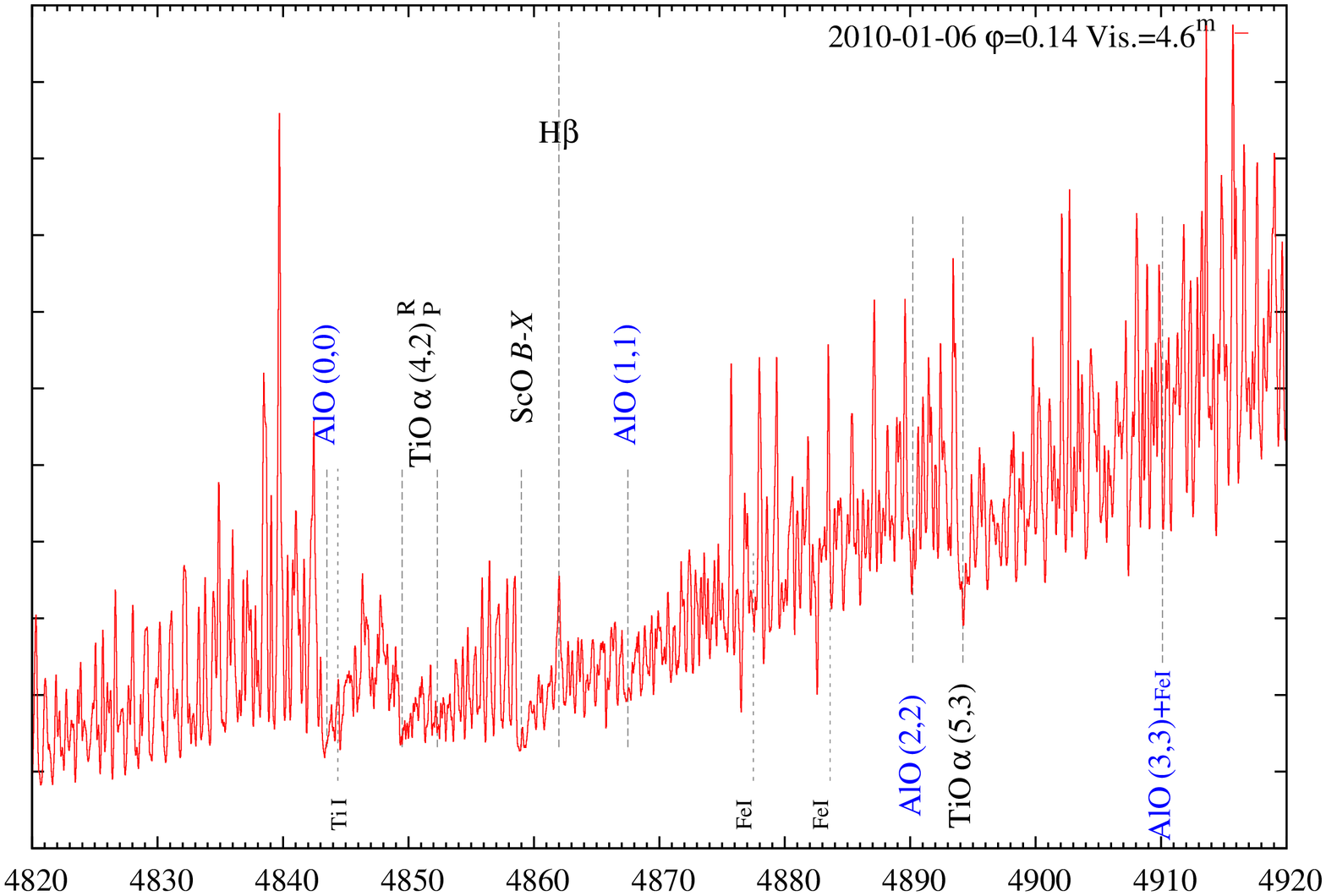}
\includegraphics[angle=270,width=0.85\textwidth]{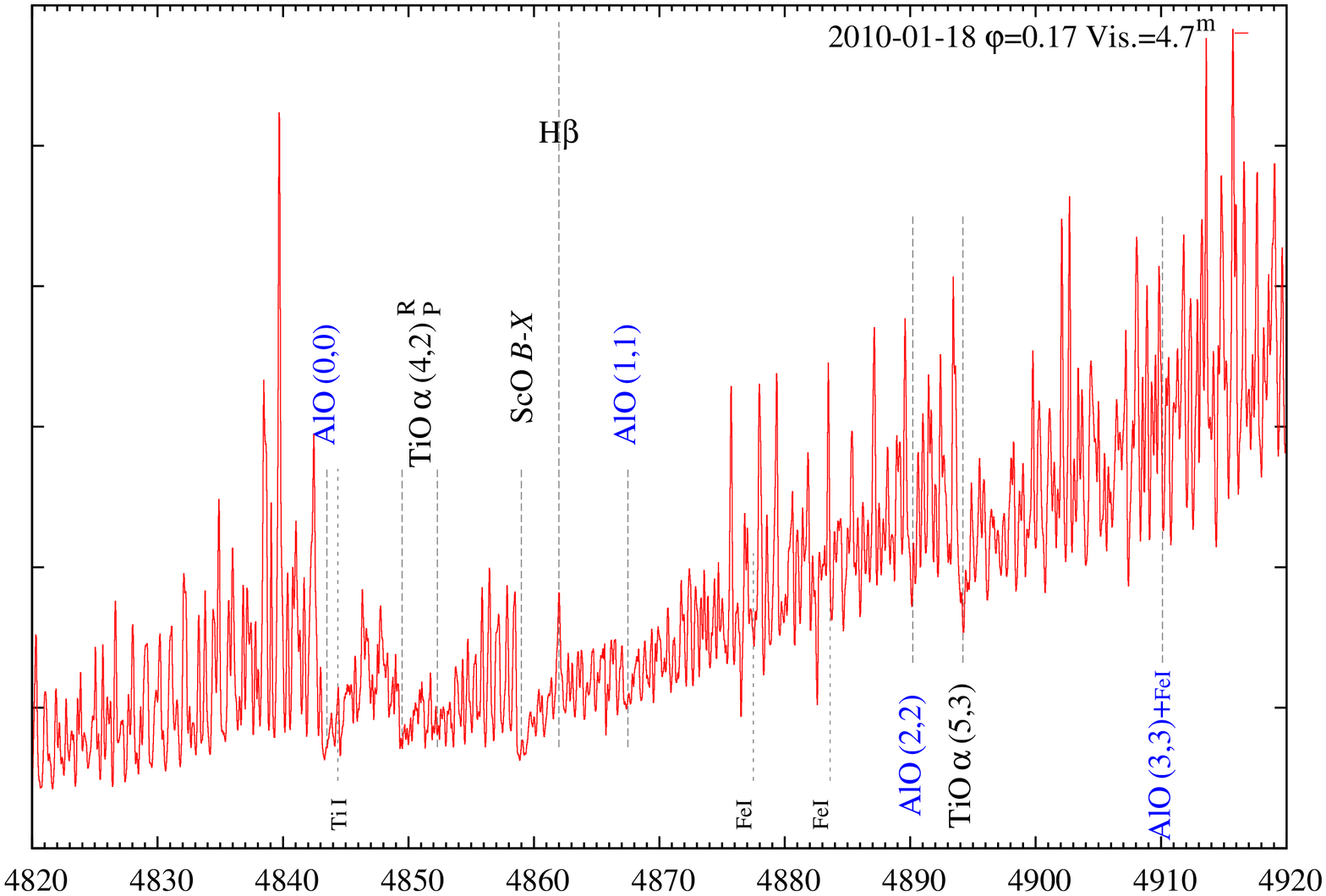}
\caption{Continued.}
\end{figure*}

  \setcounter{figure}{3}%

\begin{figure*} [tbh]
\centering
\includegraphics[angle=270,width=0.85\textwidth]{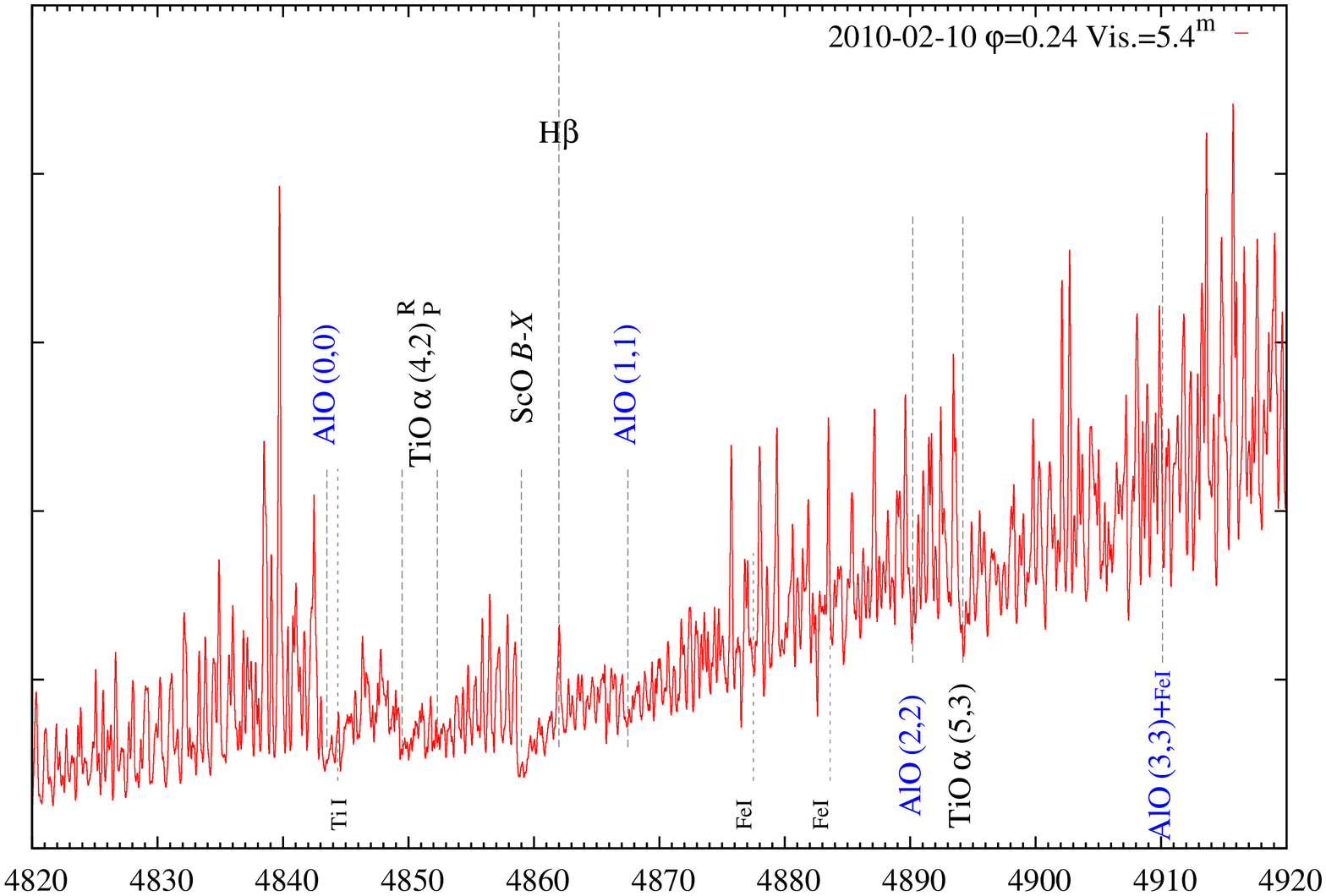}
\includegraphics[angle=270,width=0.85\textwidth]{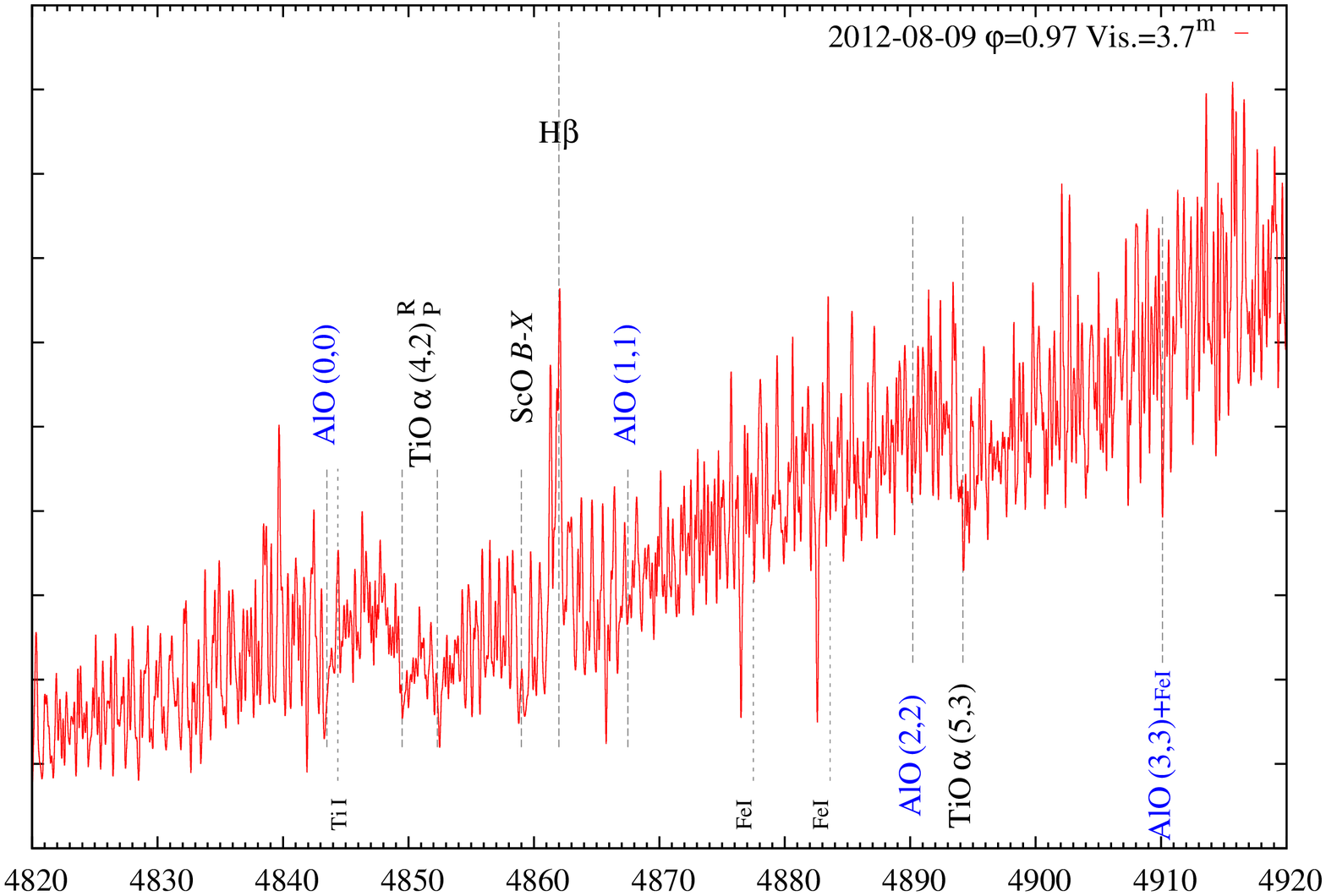}
\caption{Continued.}
\end{figure*}
\clearpage
\begin{figure*} [tbh]
\centering
\includegraphics[angle=270,width=0.85\textwidth]{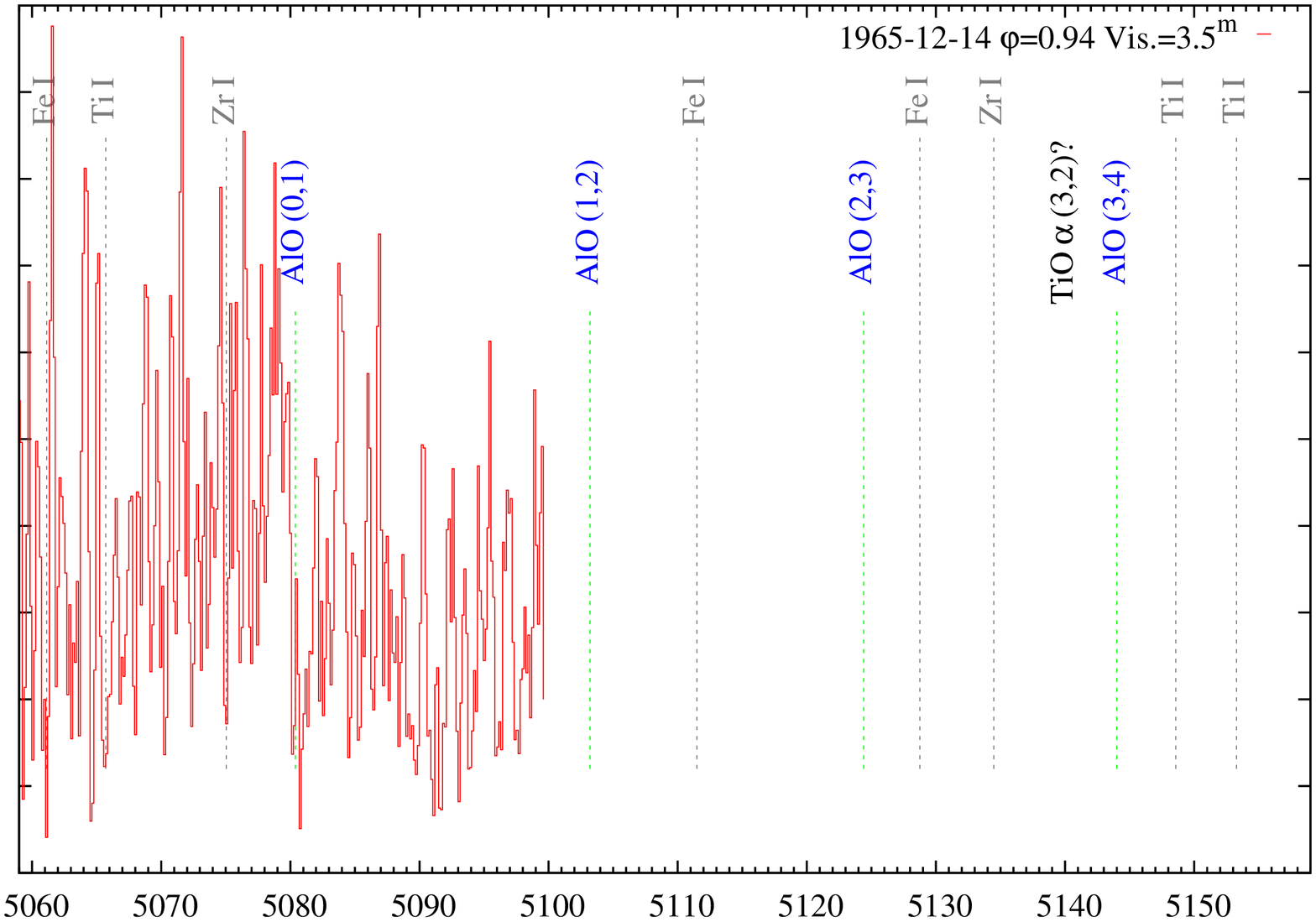}
\includegraphics[angle=270,width=0.85\textwidth]{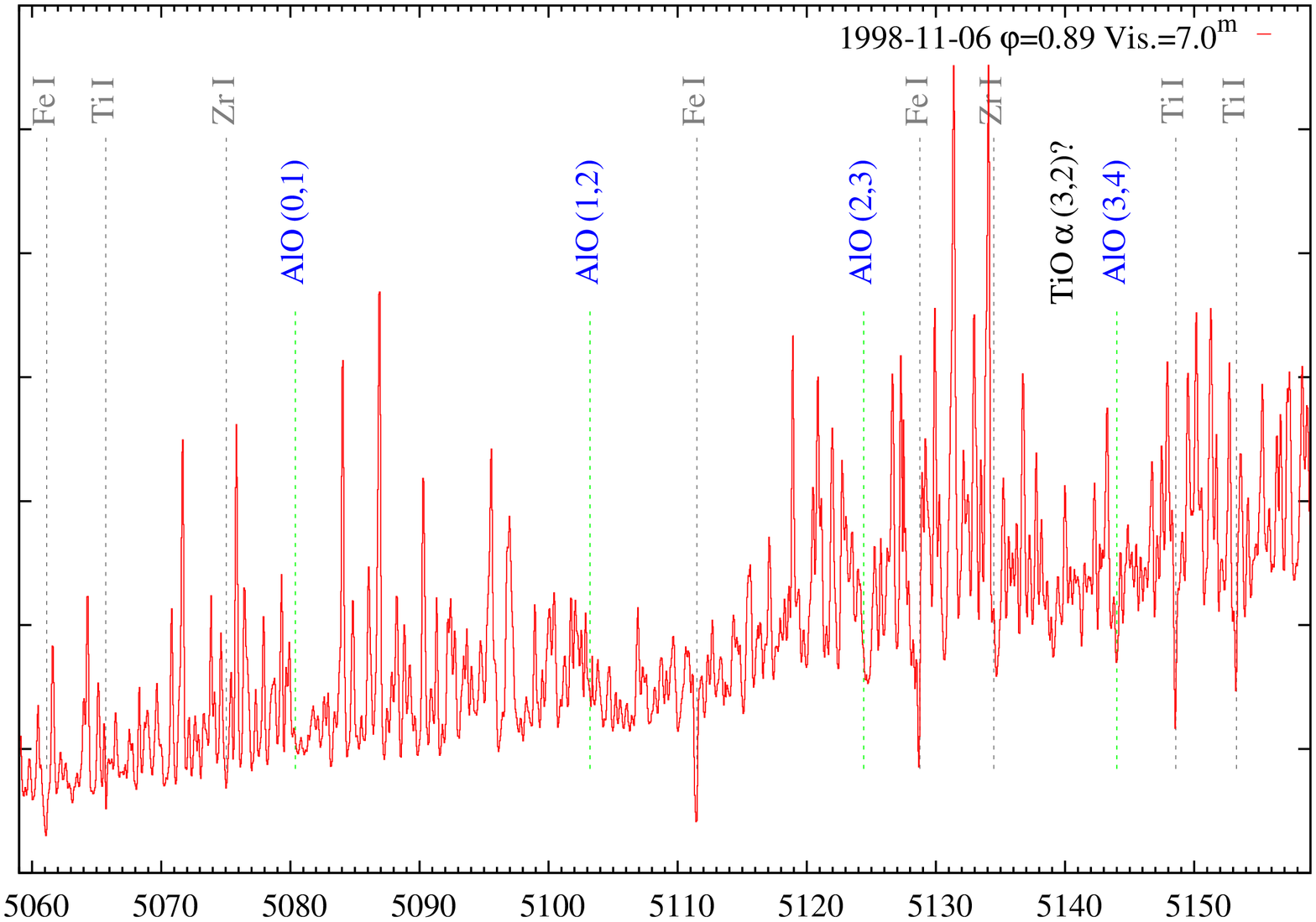}
\caption{The same as Fig.\,\ref{fig-AlOp2} but for the $\Delta\varv$=--1 sequence of AlO $B-X$. The spectrum from 2007-08-15 was diveded by a high-order polynomial.}\label{fig-AlOm1}
\end{figure*}

  \setcounter{figure}{4}%

\begin{figure*} [tbh]
\centering
\includegraphics[angle=270,width=0.85\textwidth]{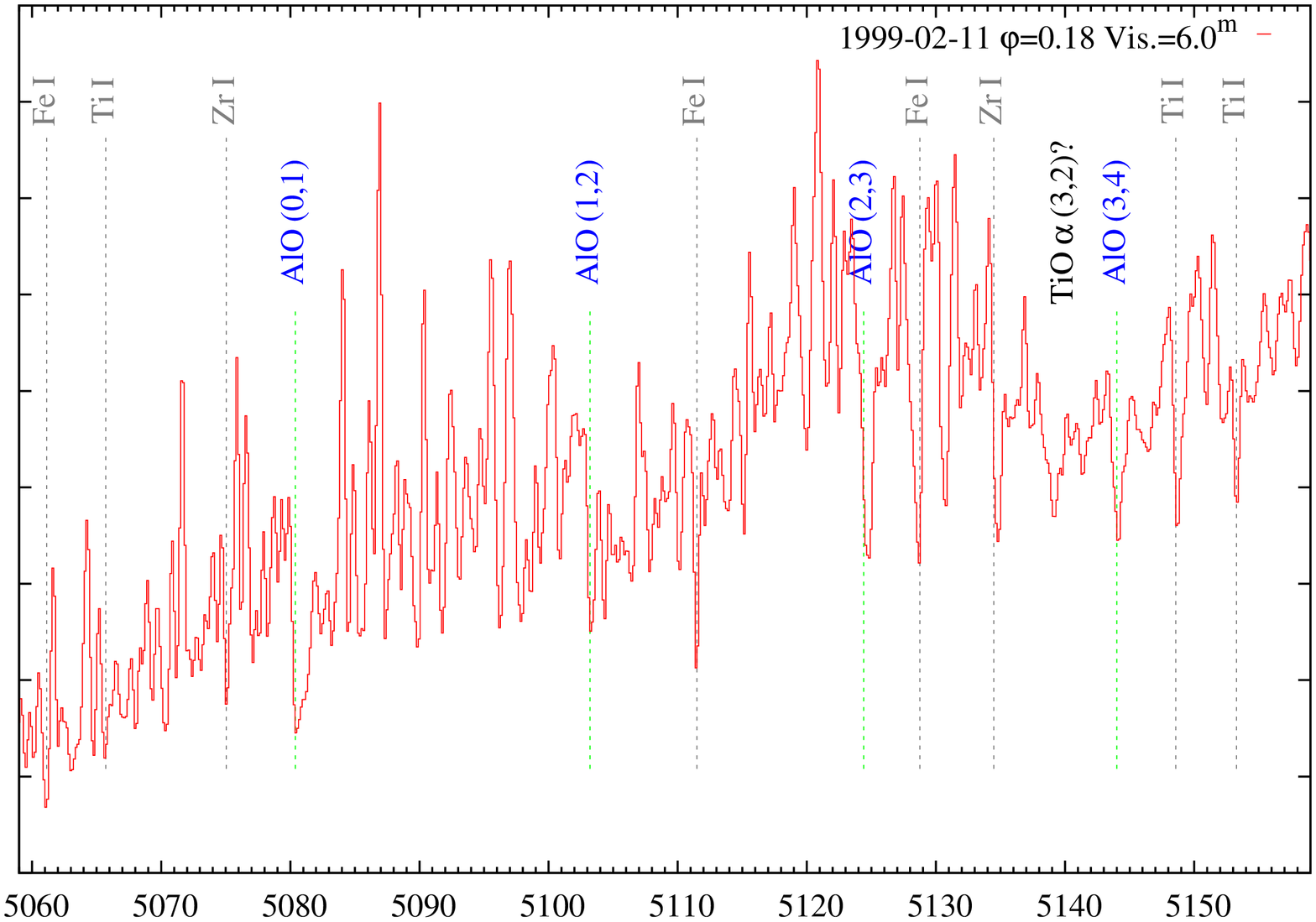}
\includegraphics[angle=270,width=0.85\textwidth]{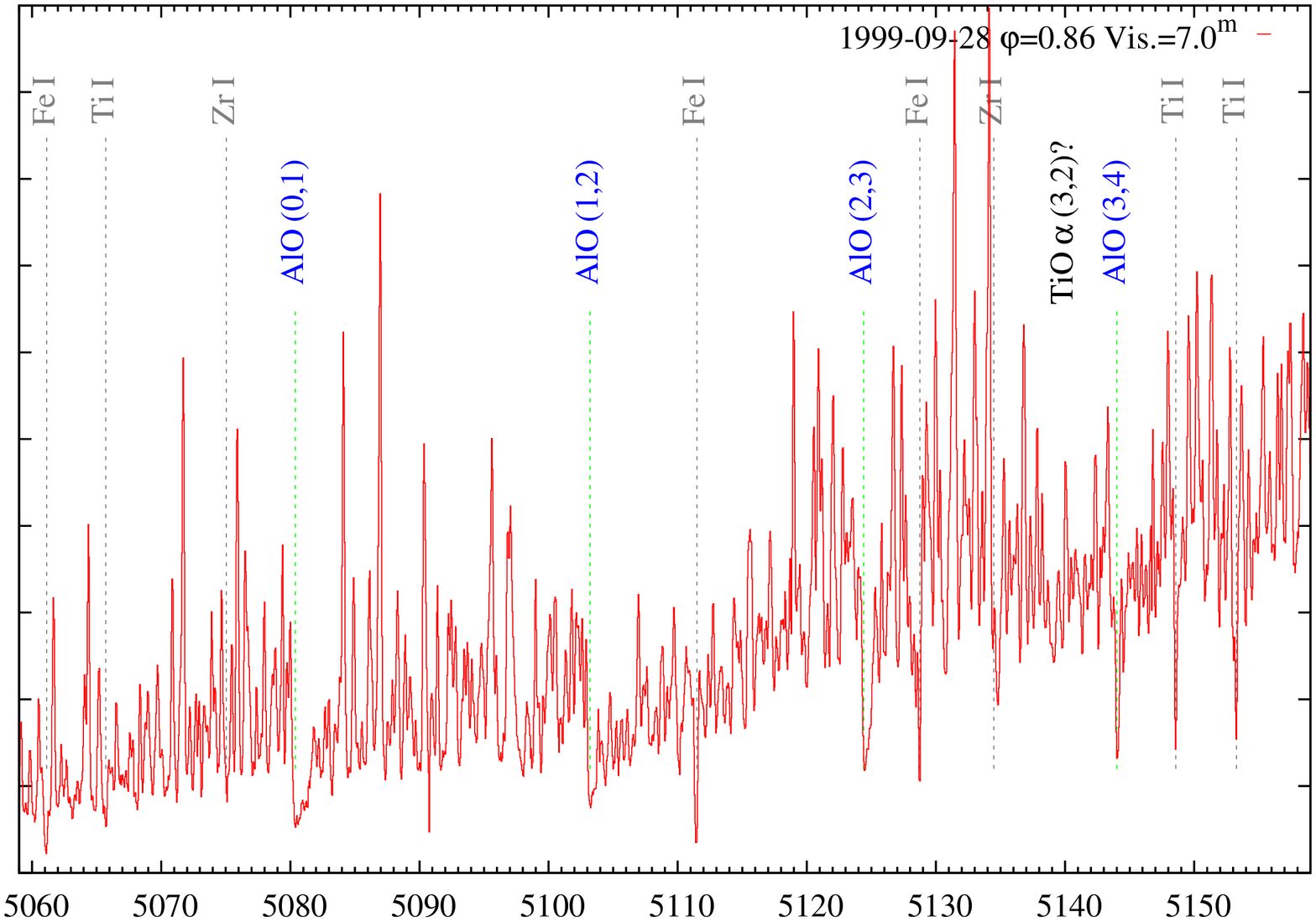}
\caption{Continued.}
\end{figure*}

  \setcounter{figure}{4}%

\begin{figure*} [tbh]
\centering
\includegraphics[angle=270,width=0.85\textwidth]{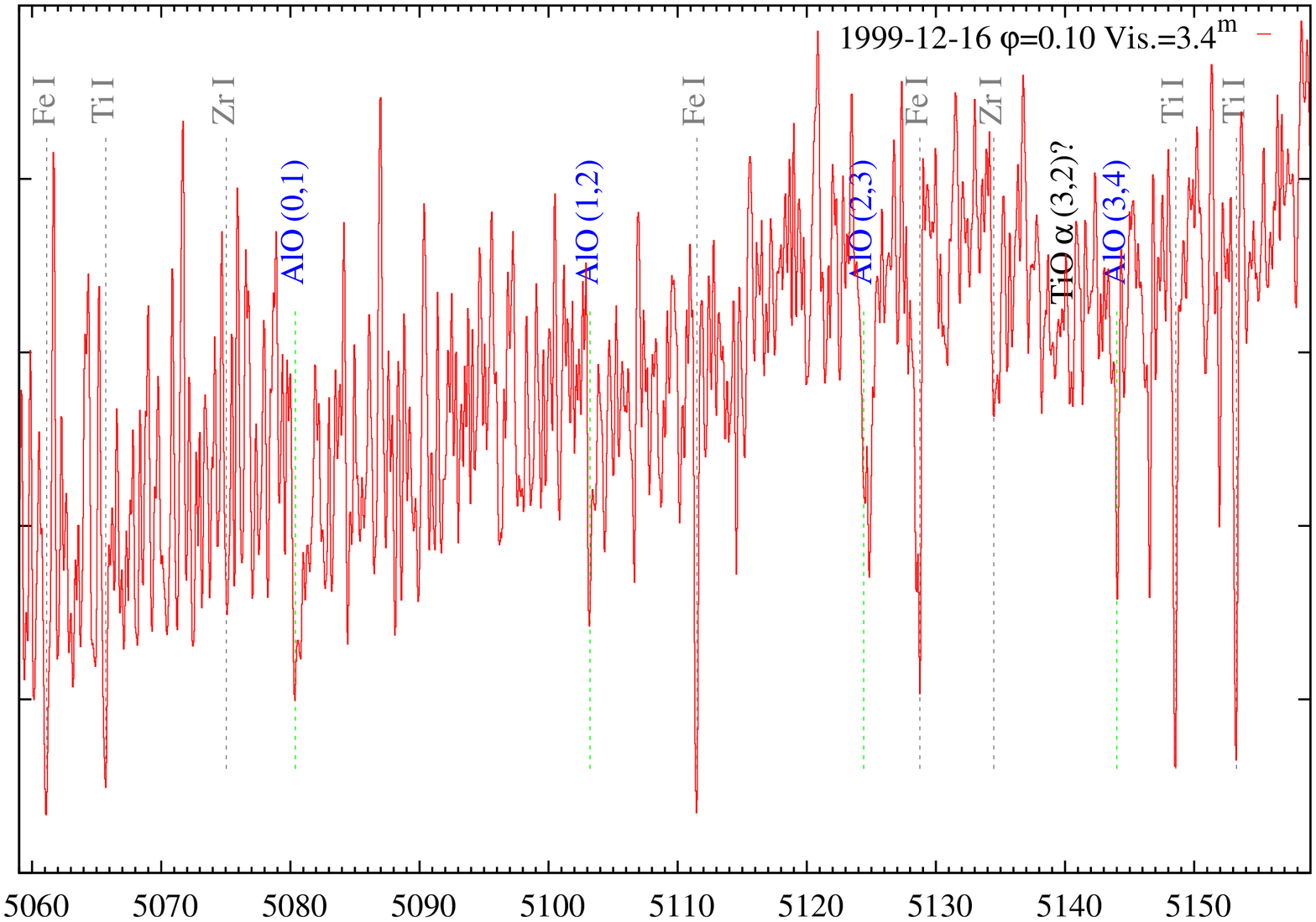}
\includegraphics[angle=270,width=0.85\textwidth]{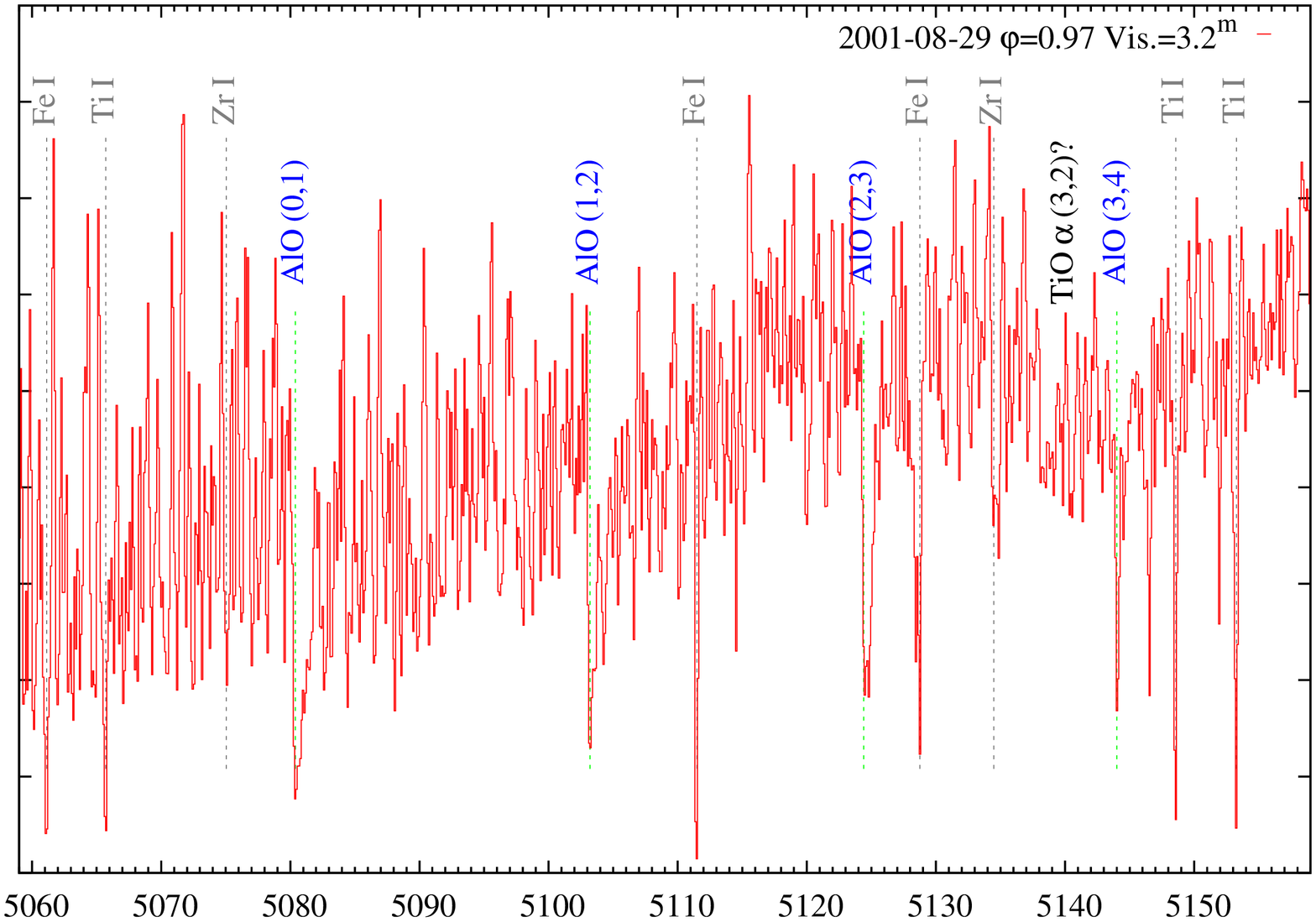}
\caption{Continued.}
\end{figure*}

  \setcounter{figure}{4}%

\begin{figure*} [tbh]
\centering
\includegraphics[angle=270,width=0.85\textwidth]{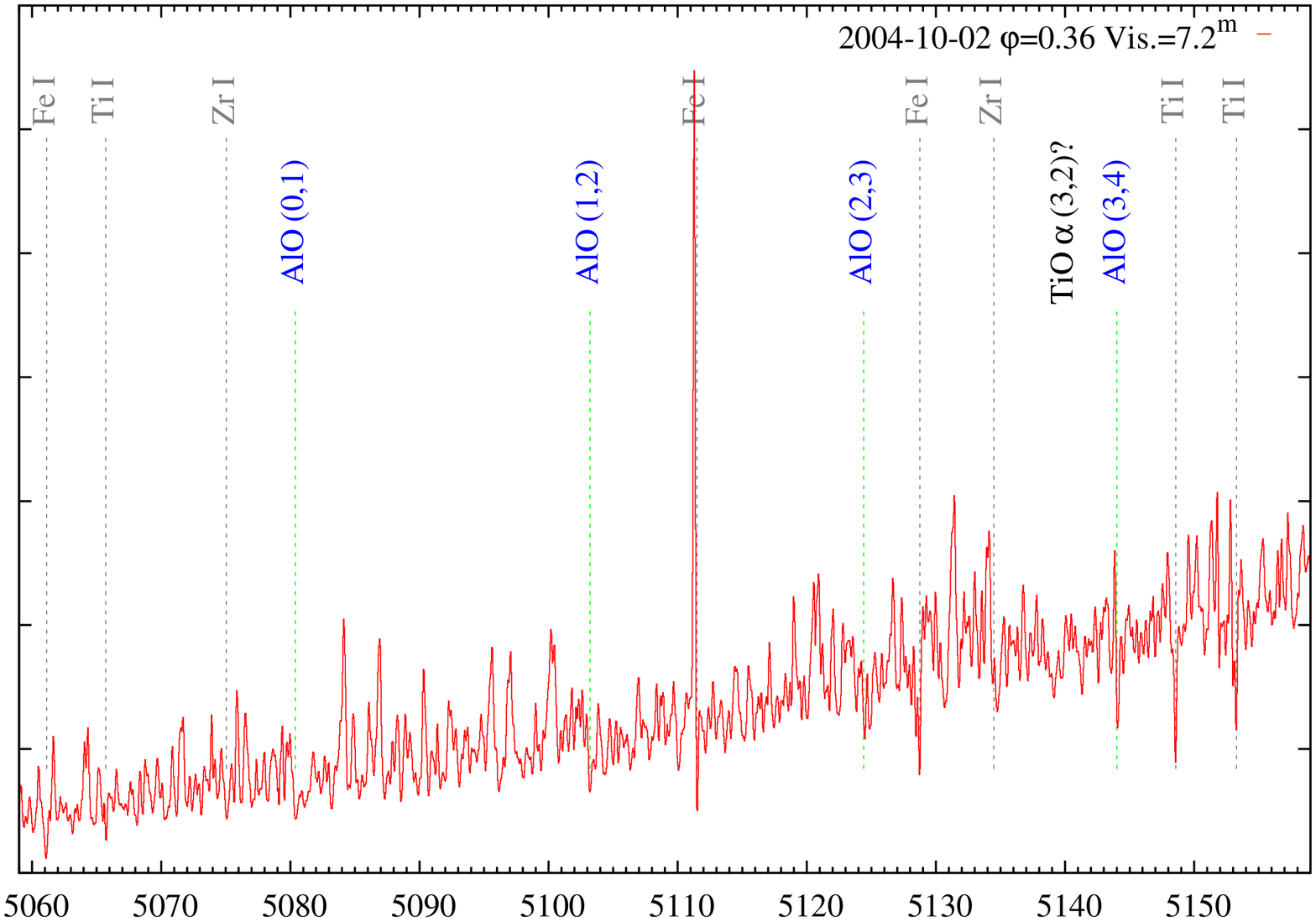}
\includegraphics[angle=270,width=0.85\textwidth]{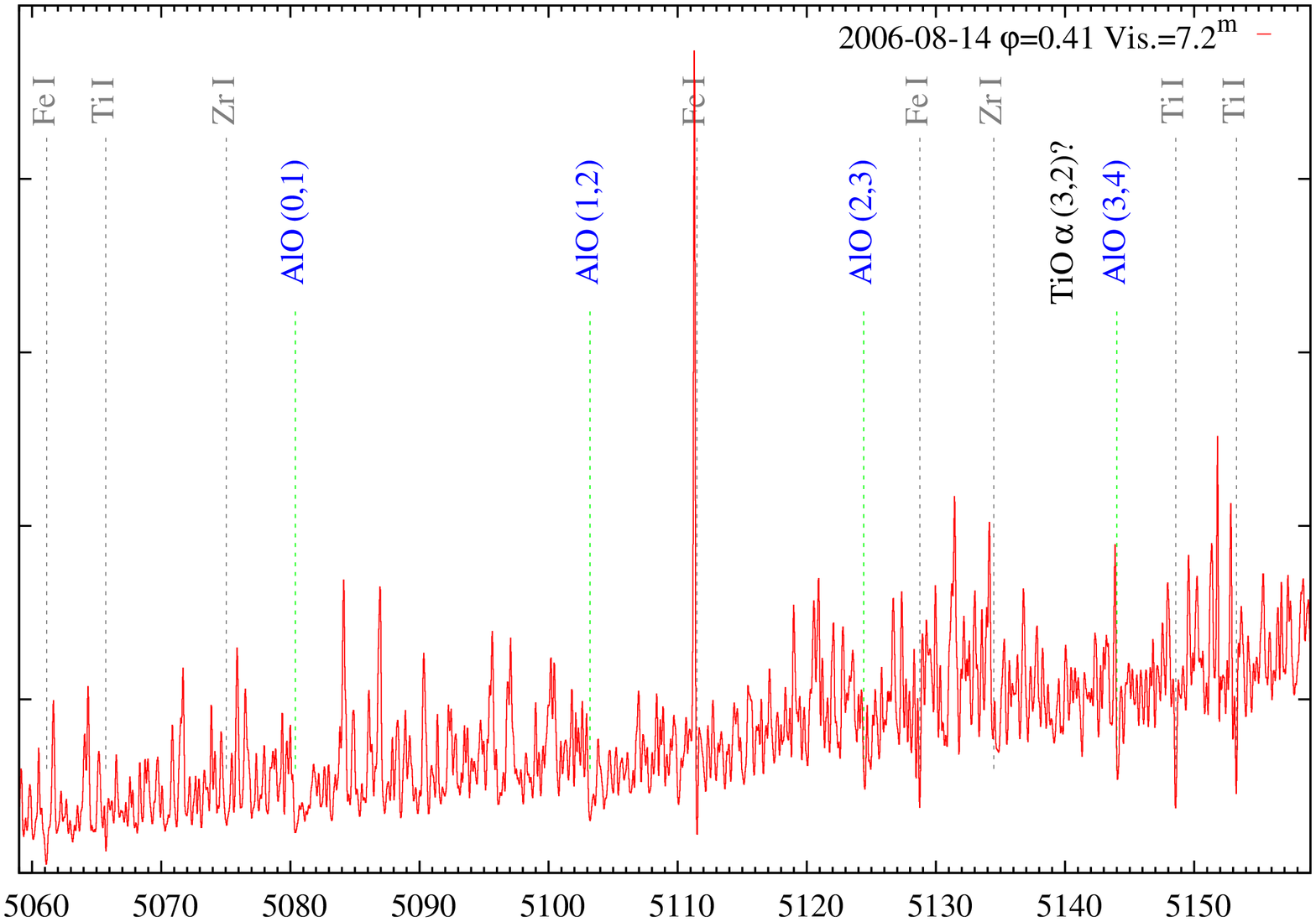}
\caption{Continued.}
\end{figure*}

  \setcounter{figure}{4}%

\begin{figure*} [tbh]
\centering
\includegraphics[angle=270,width=0.85\textwidth]{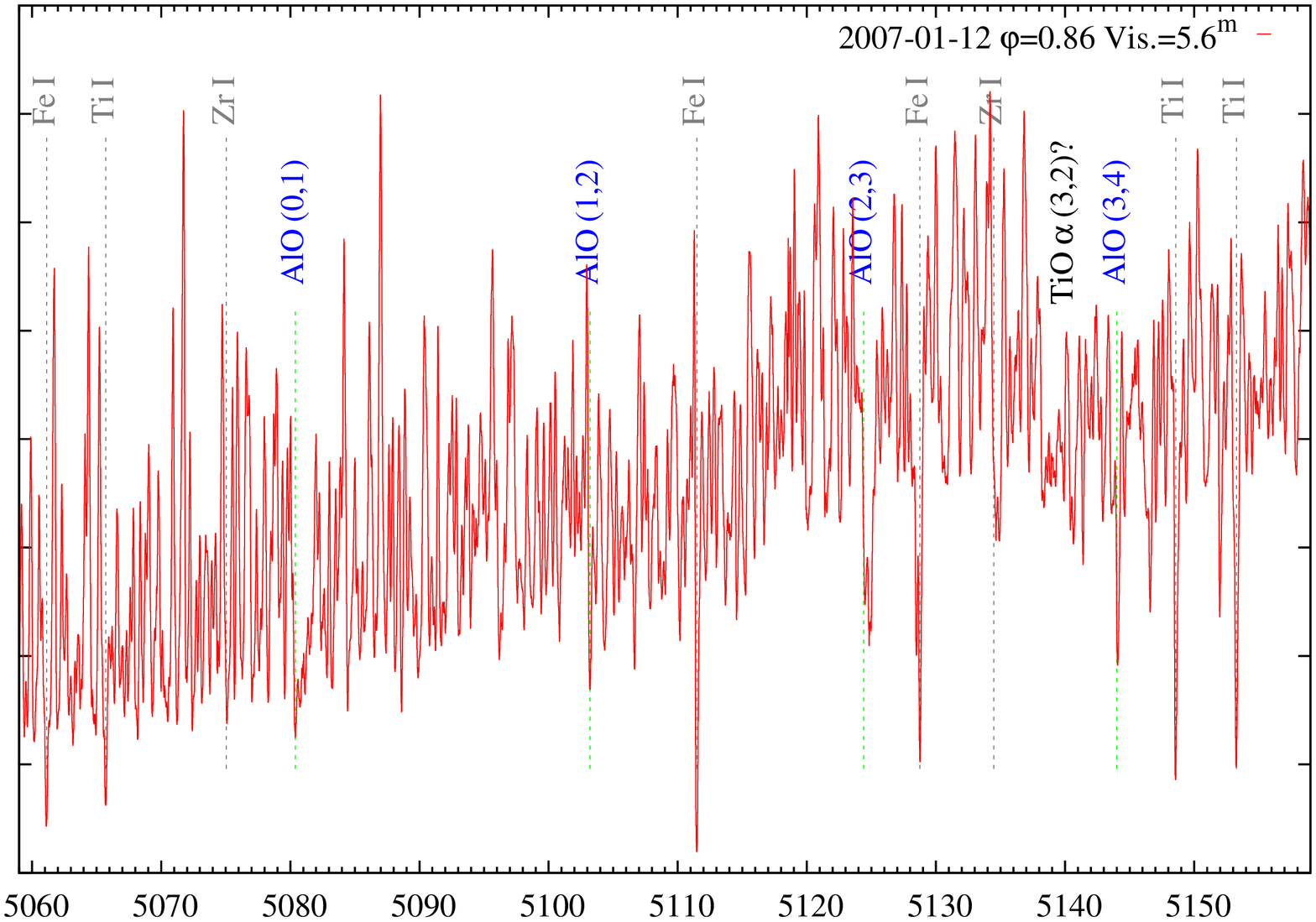}
\includegraphics[angle=270,width=0.85\textwidth]{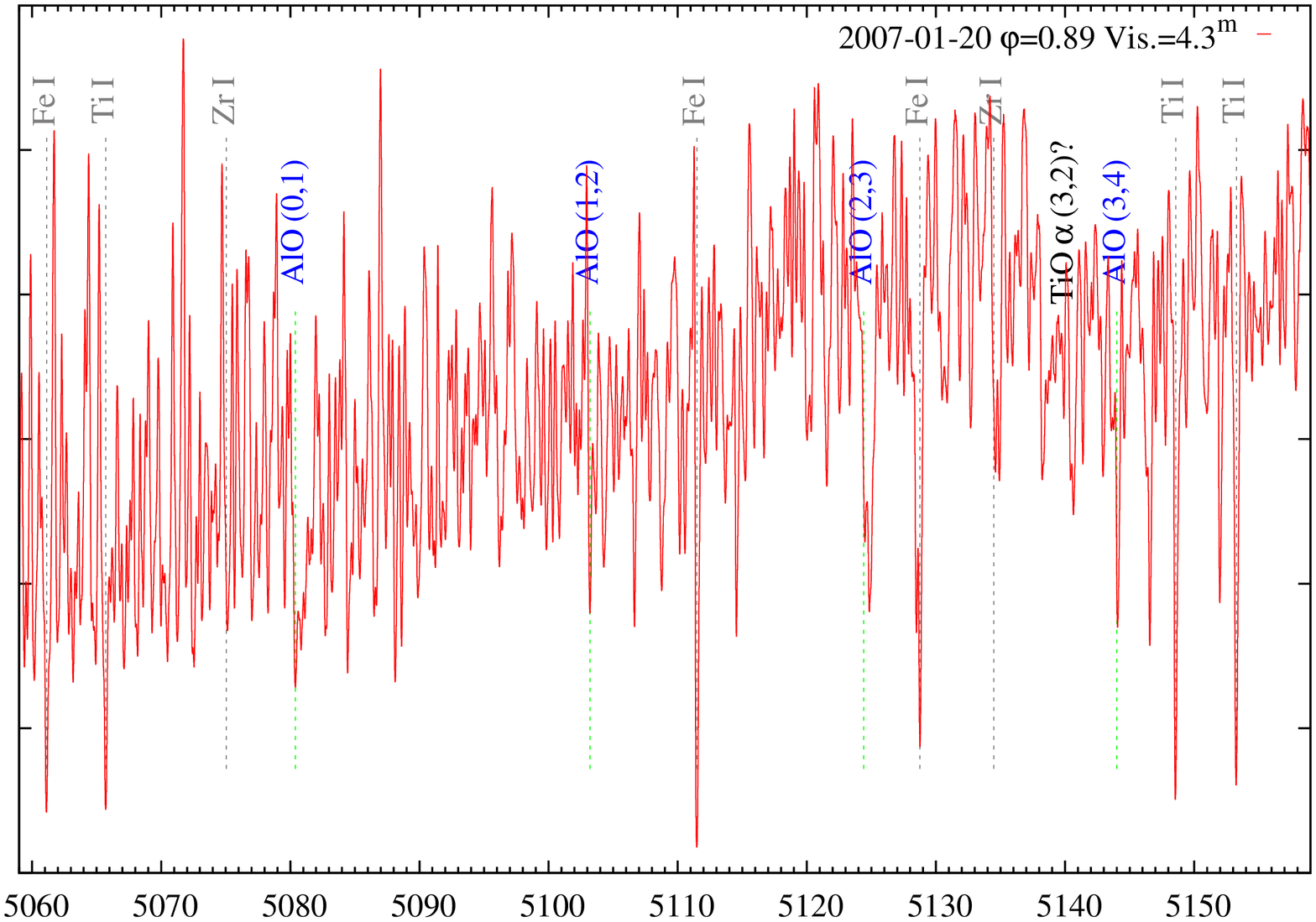}
\caption{Continued.}
\end{figure*}

  \setcounter{figure}{4}%

\begin{figure*} [tbh]
\centering
\includegraphics[angle=270,width=0.85\textwidth]{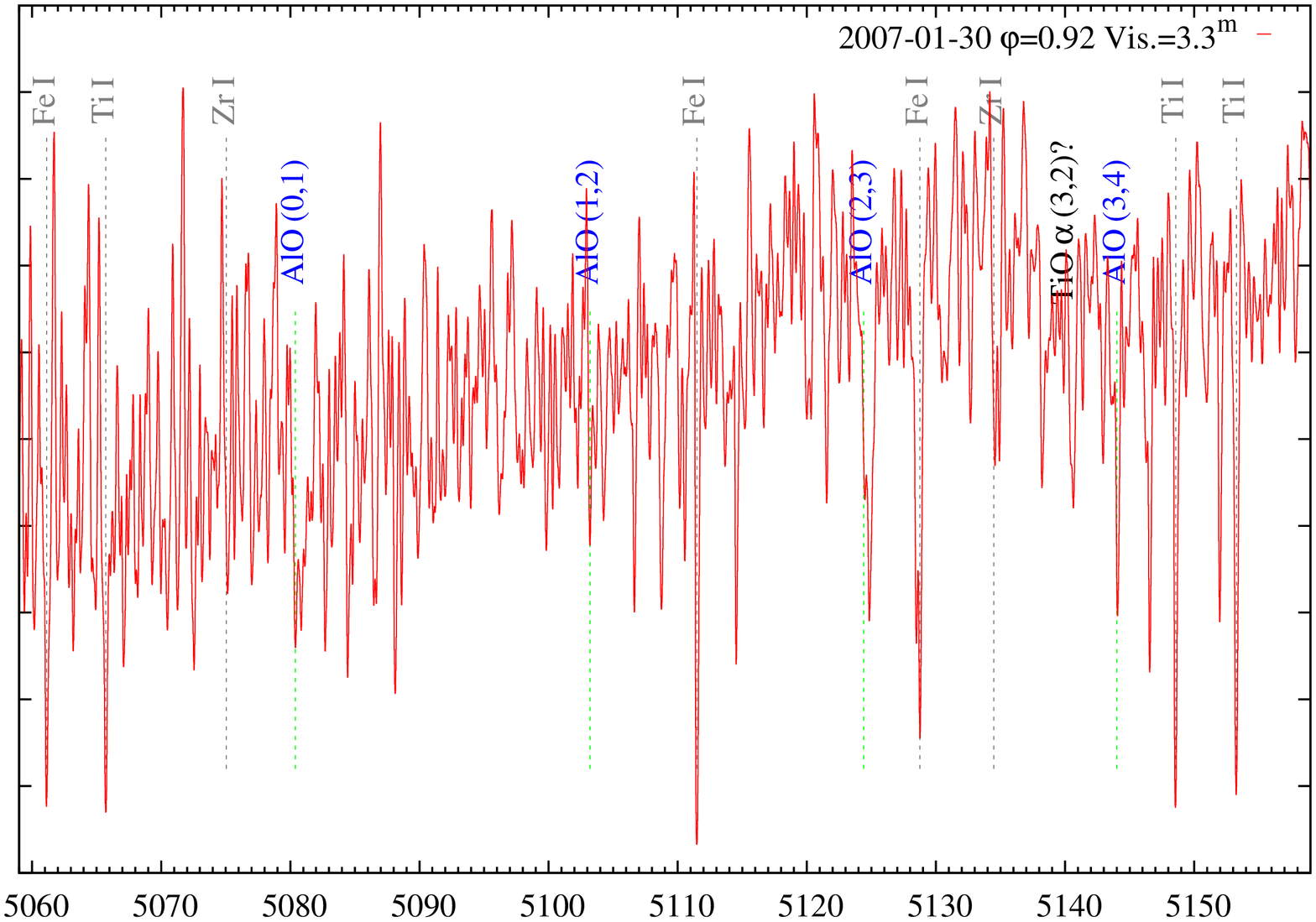}
\includegraphics[angle=270,width=0.85\textwidth]{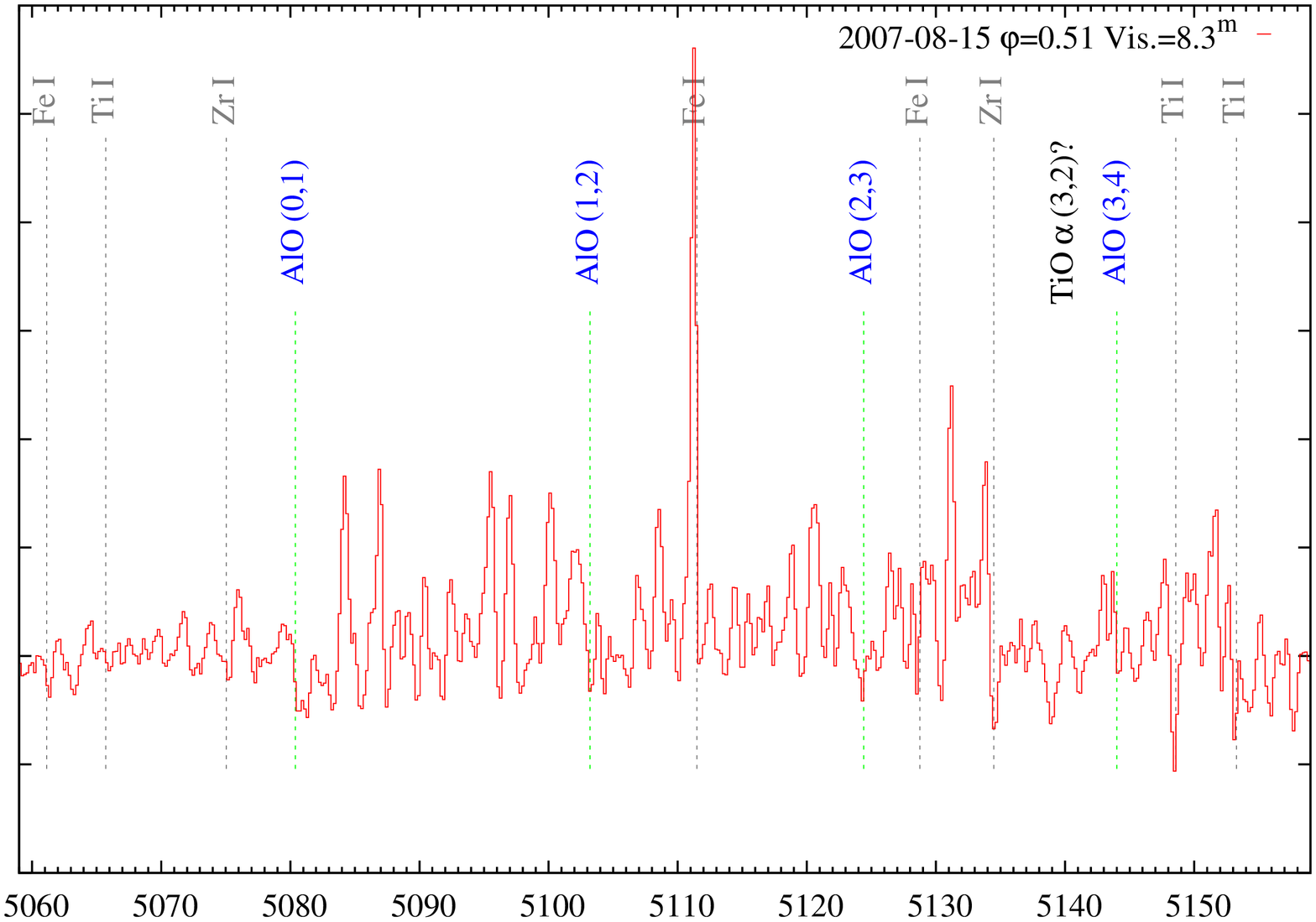}
\caption{Continued.}
\end{figure*}

  \setcounter{figure}{4}%

\begin{figure*} [tbh]
\centering
\includegraphics[angle=270,width=0.85\textwidth]{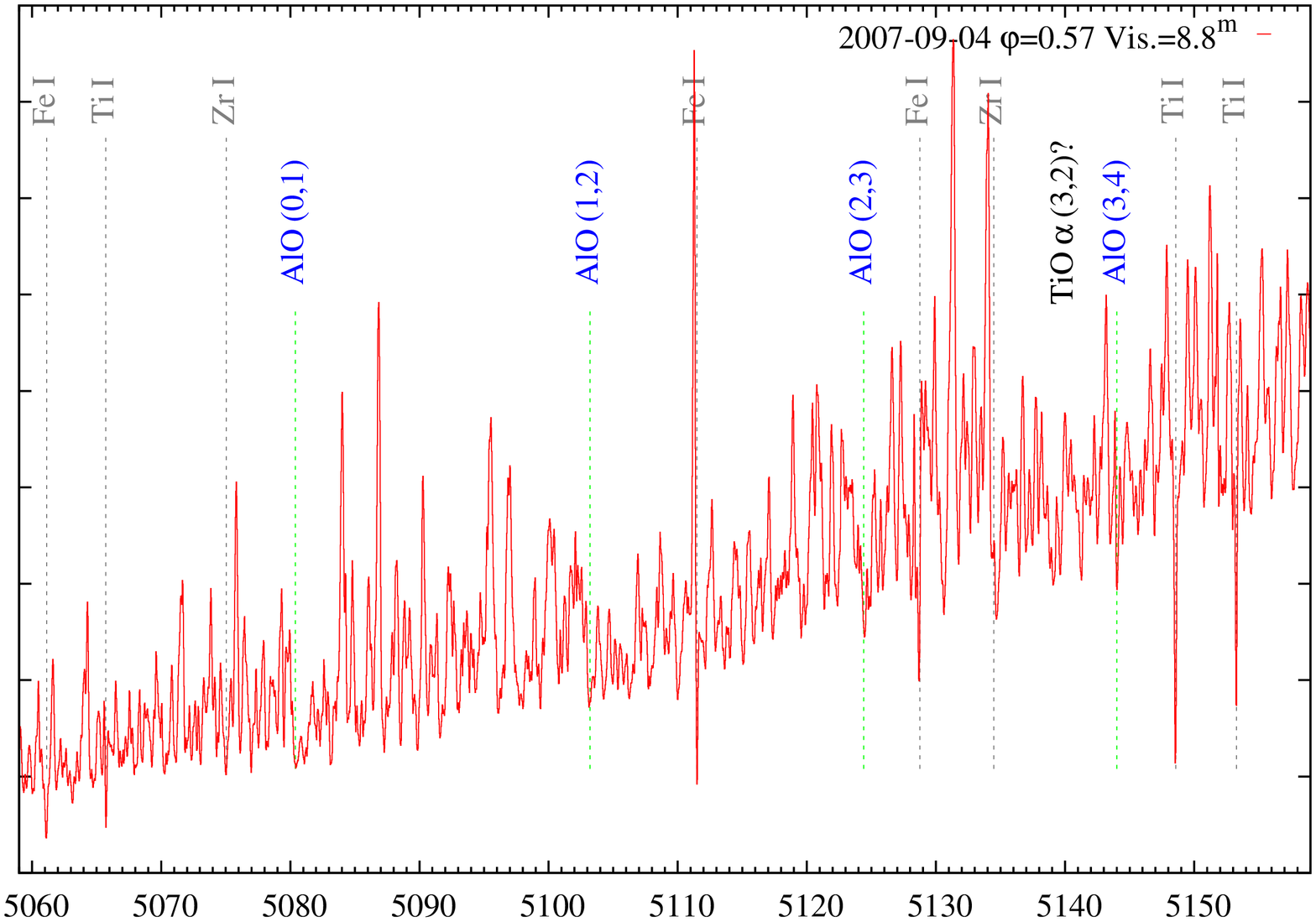}
\includegraphics[angle=270,width=0.85\textwidth]{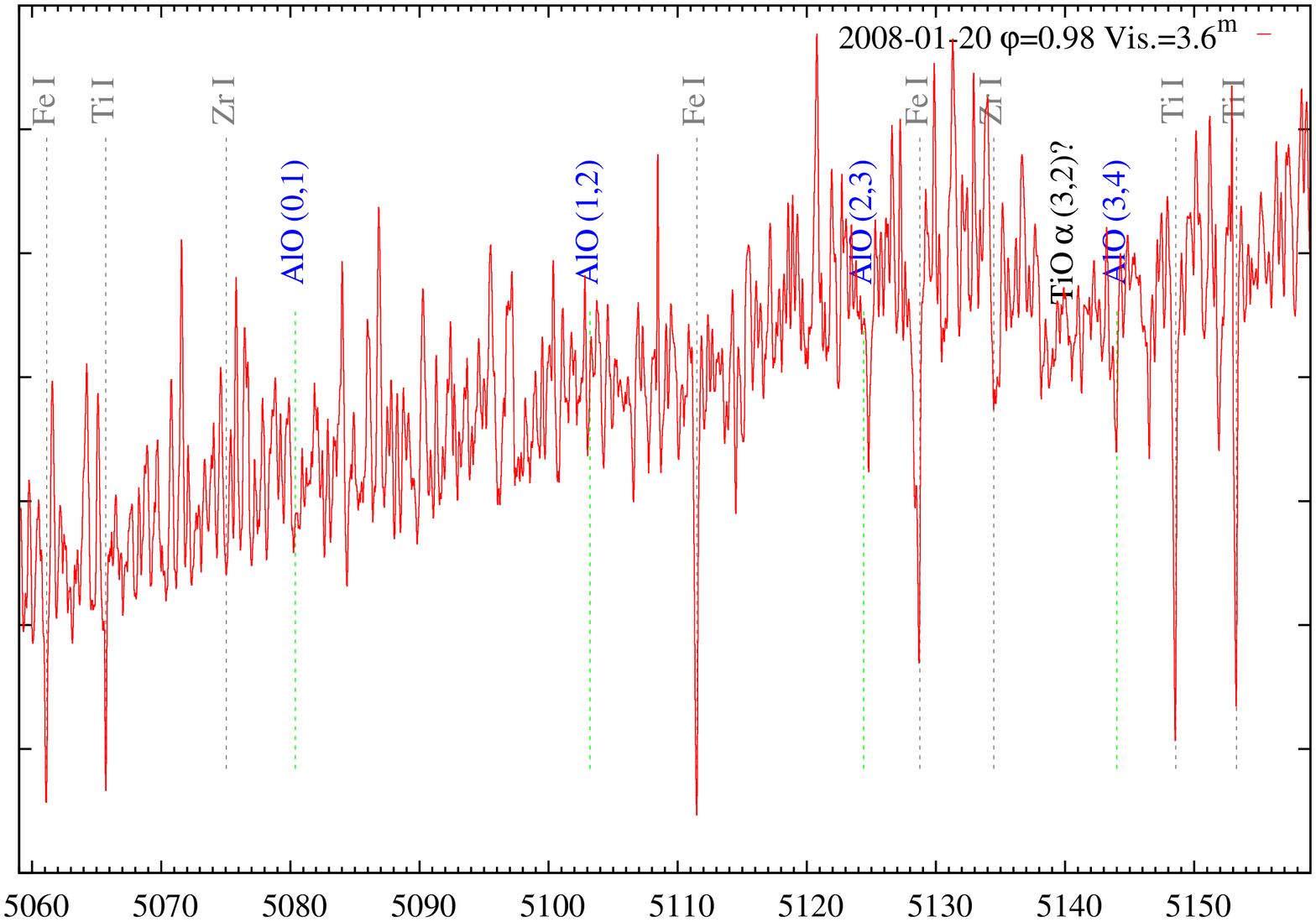}
\caption{Continued.}
\end{figure*}

  \setcounter{figure}{4}%

\begin{figure*} [tbh]
\centering
\includegraphics[angle=270,width=0.85\textwidth]{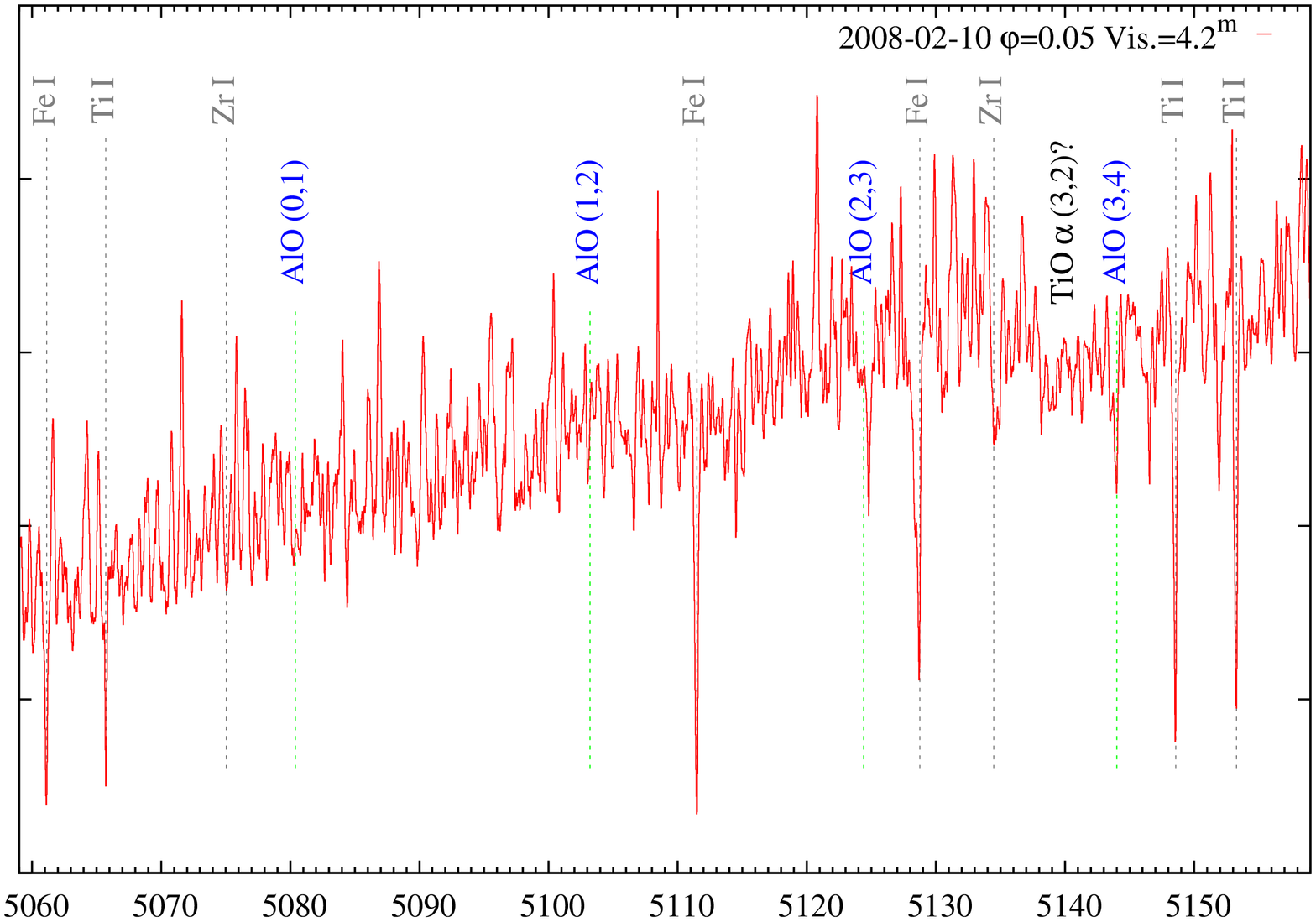}
\includegraphics[angle=270,width=0.85\textwidth]{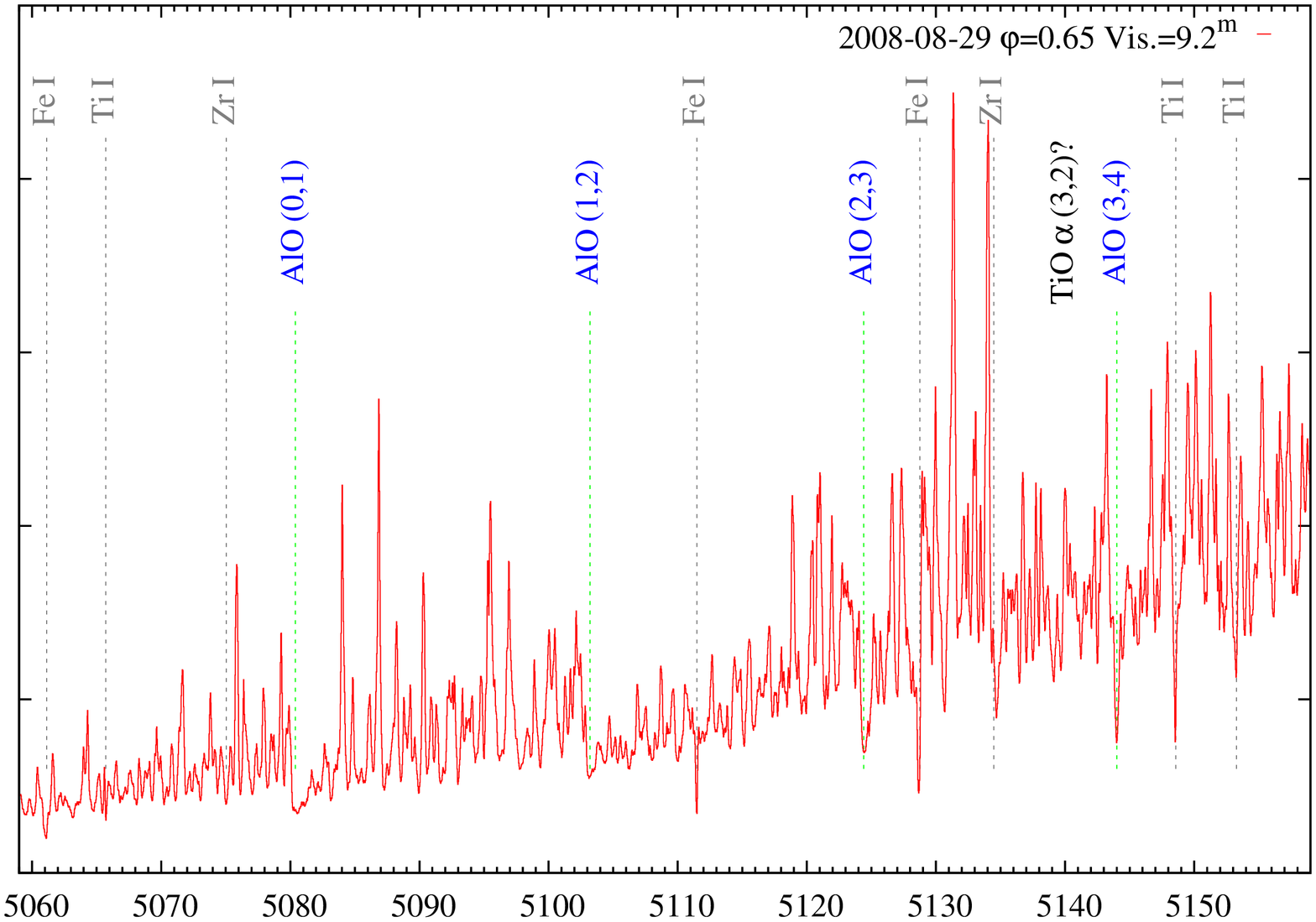}
\caption{Continued.}
\end{figure*}

  \setcounter{figure}{4}%

\begin{figure*} [tbh]
\centering
\includegraphics[angle=270,width=0.85\textwidth]{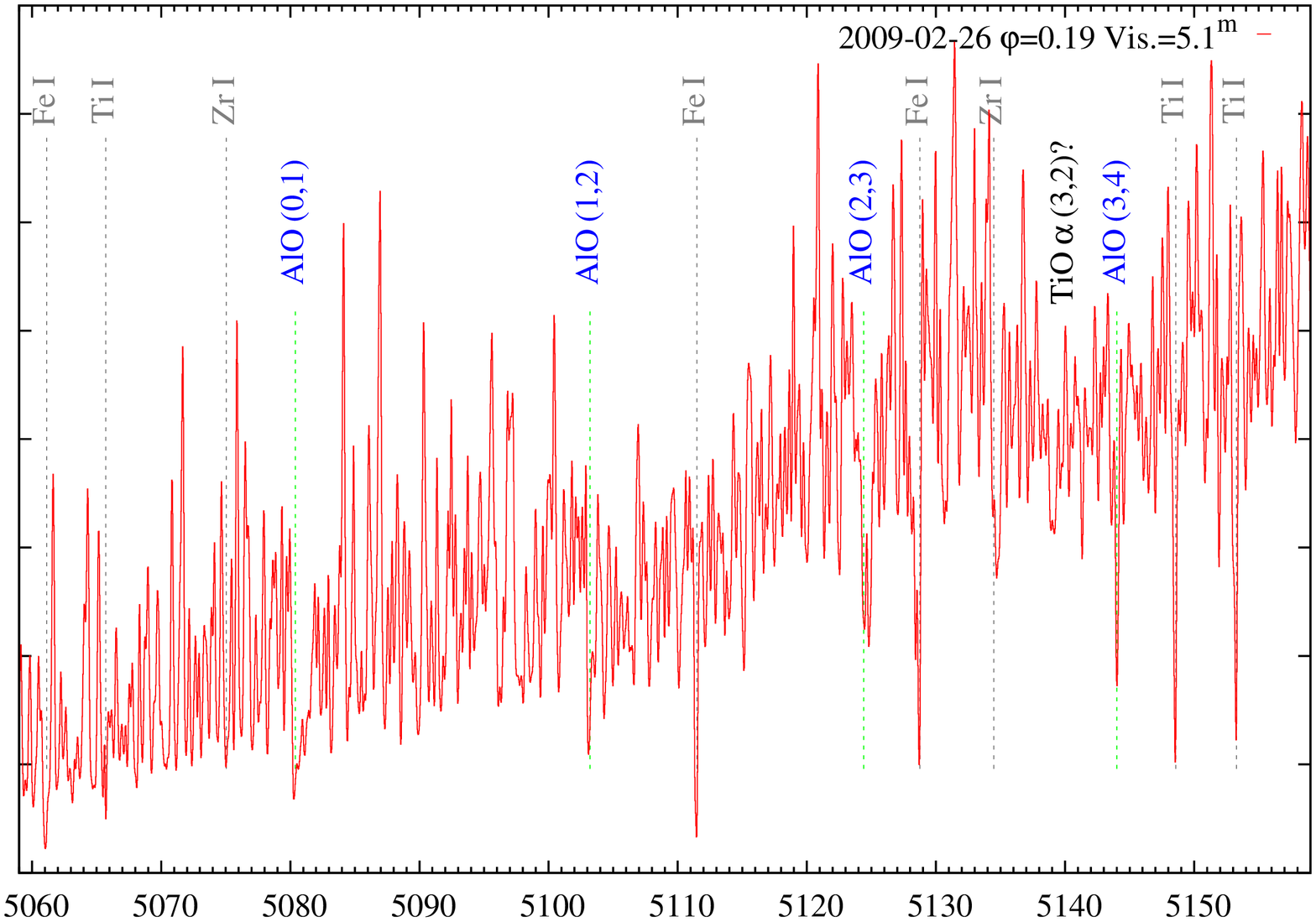}
\includegraphics[angle=270,width=0.85\textwidth]{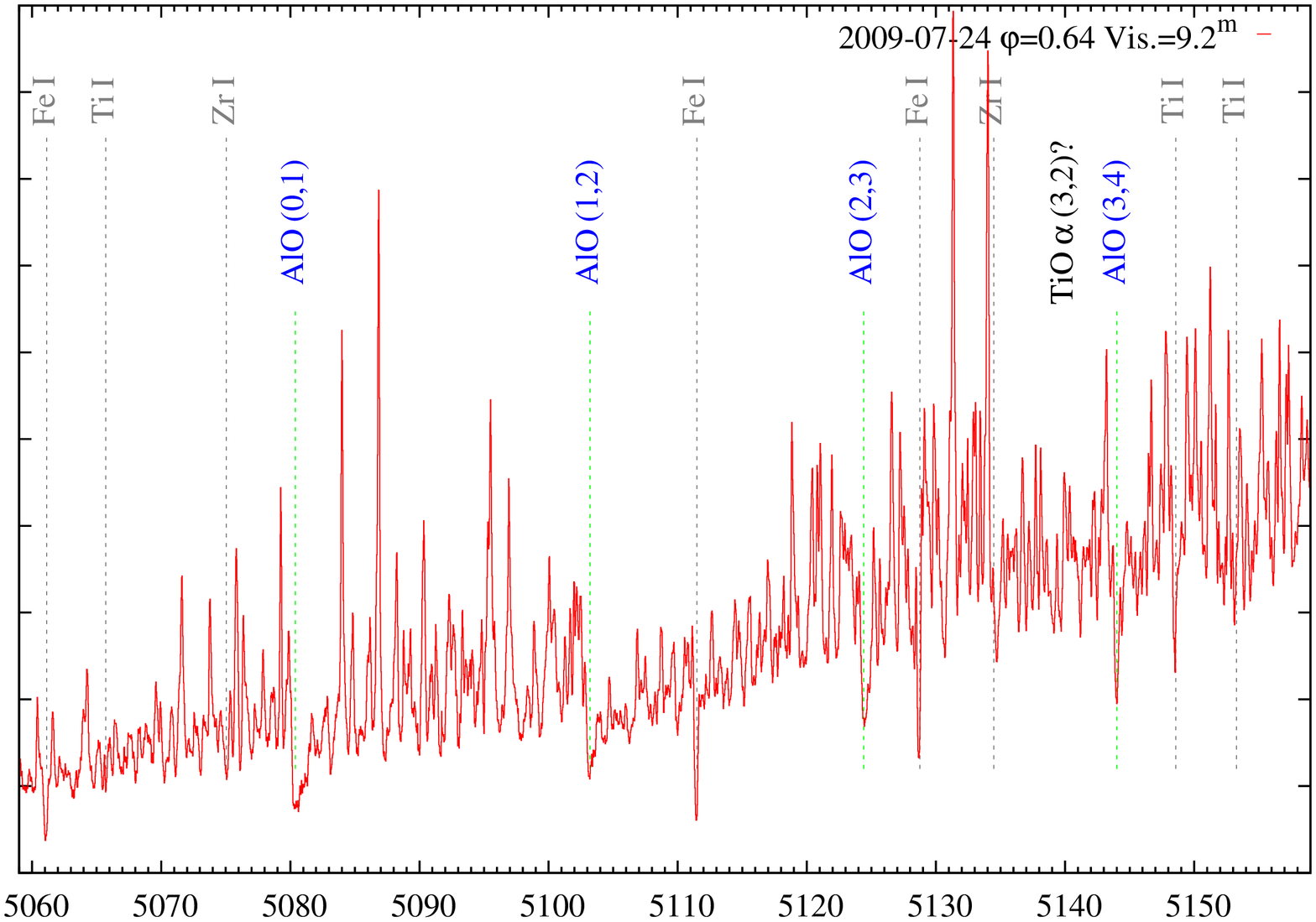}
\caption{Continued.}
\end{figure*}

  \setcounter{figure}{4}%

\begin{figure*} [tbh]
\centering
\includegraphics[angle=270,width=0.85\textwidth]{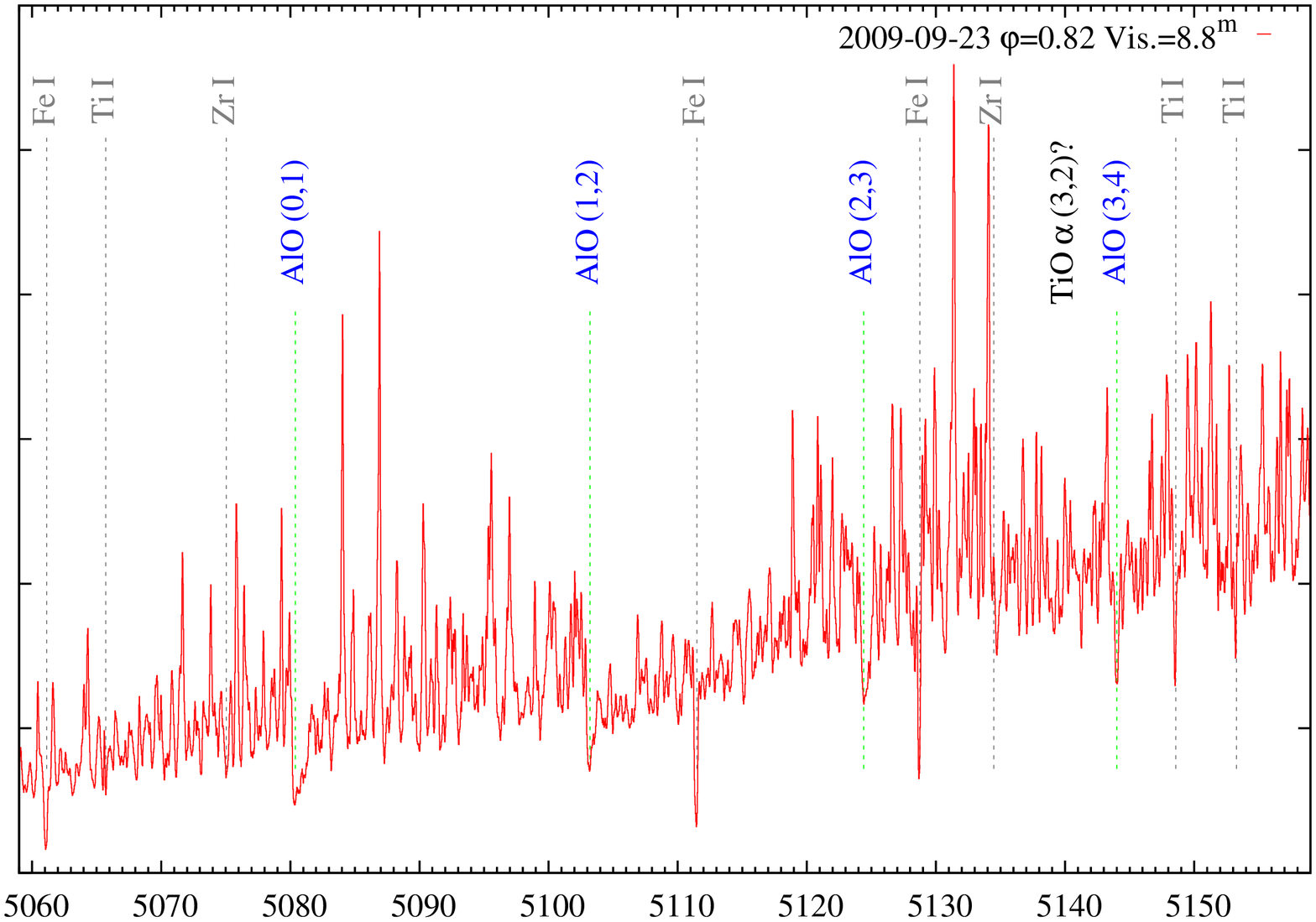}
\includegraphics[angle=270,width=0.85\textwidth]{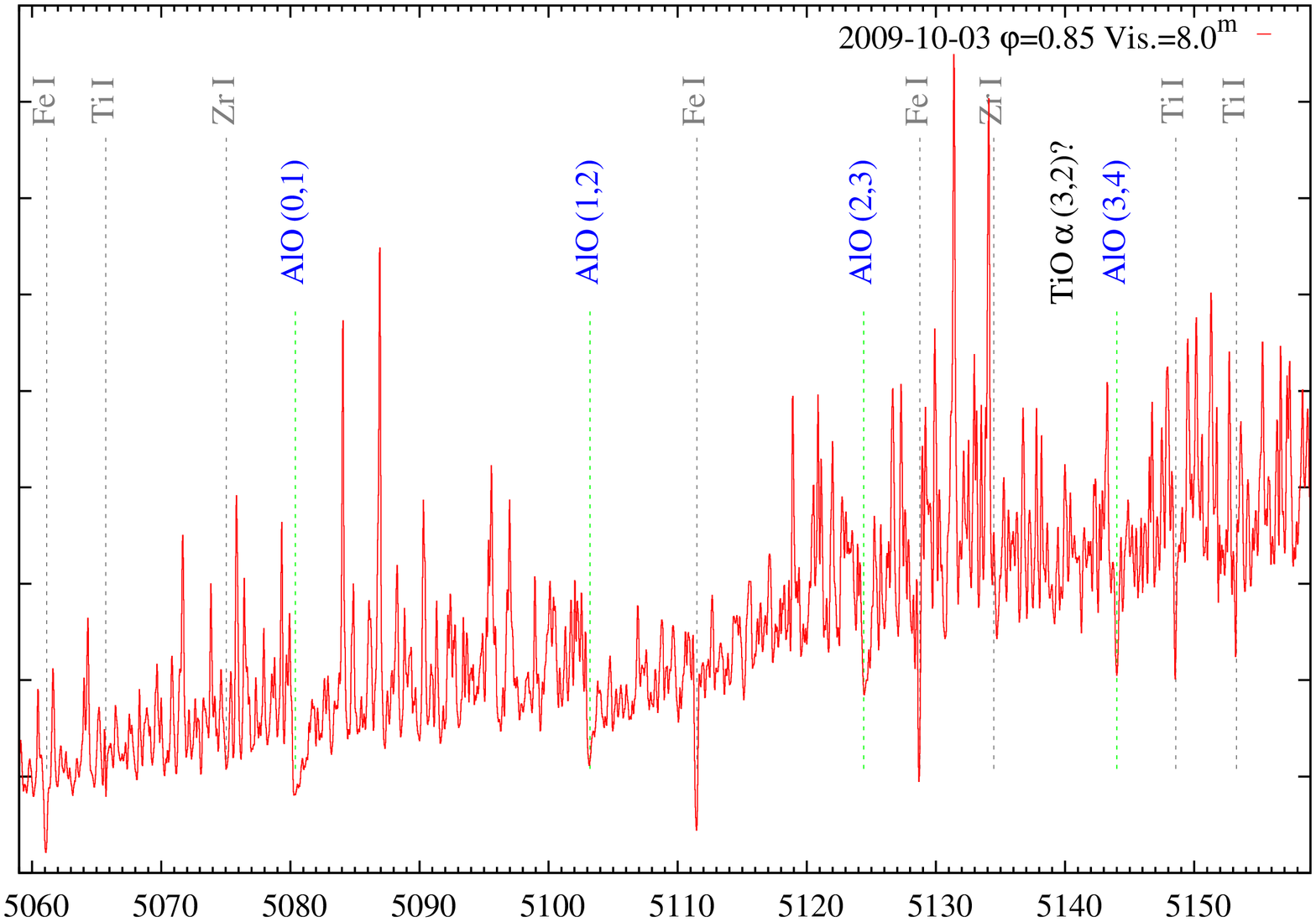}
\caption{Continued.}
\end{figure*}

  \setcounter{figure}{4}%

\begin{figure*} [tbh]
\centering
\includegraphics[angle=270,width=0.85\textwidth]{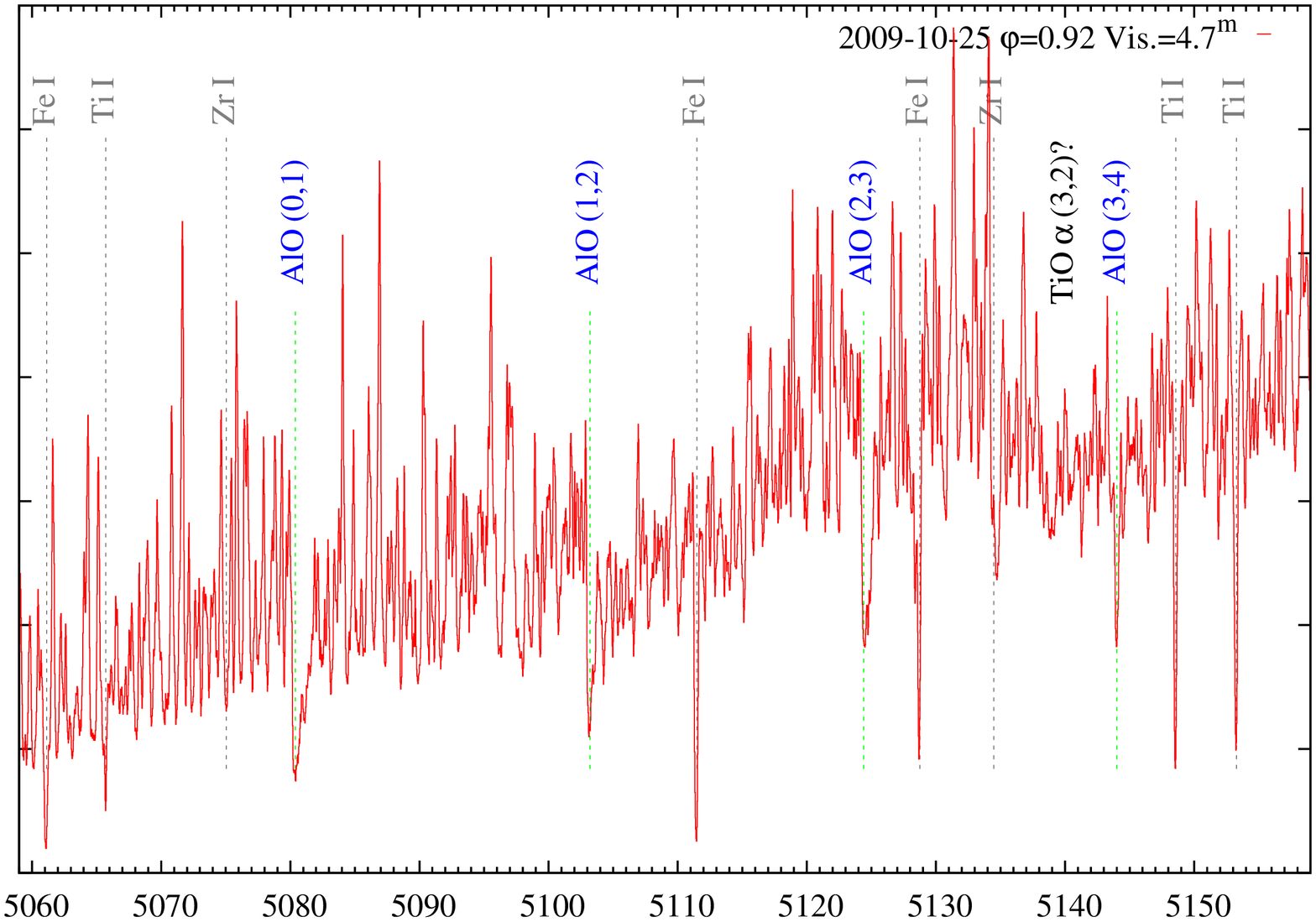}
\includegraphics[angle=270,width=0.85\textwidth]{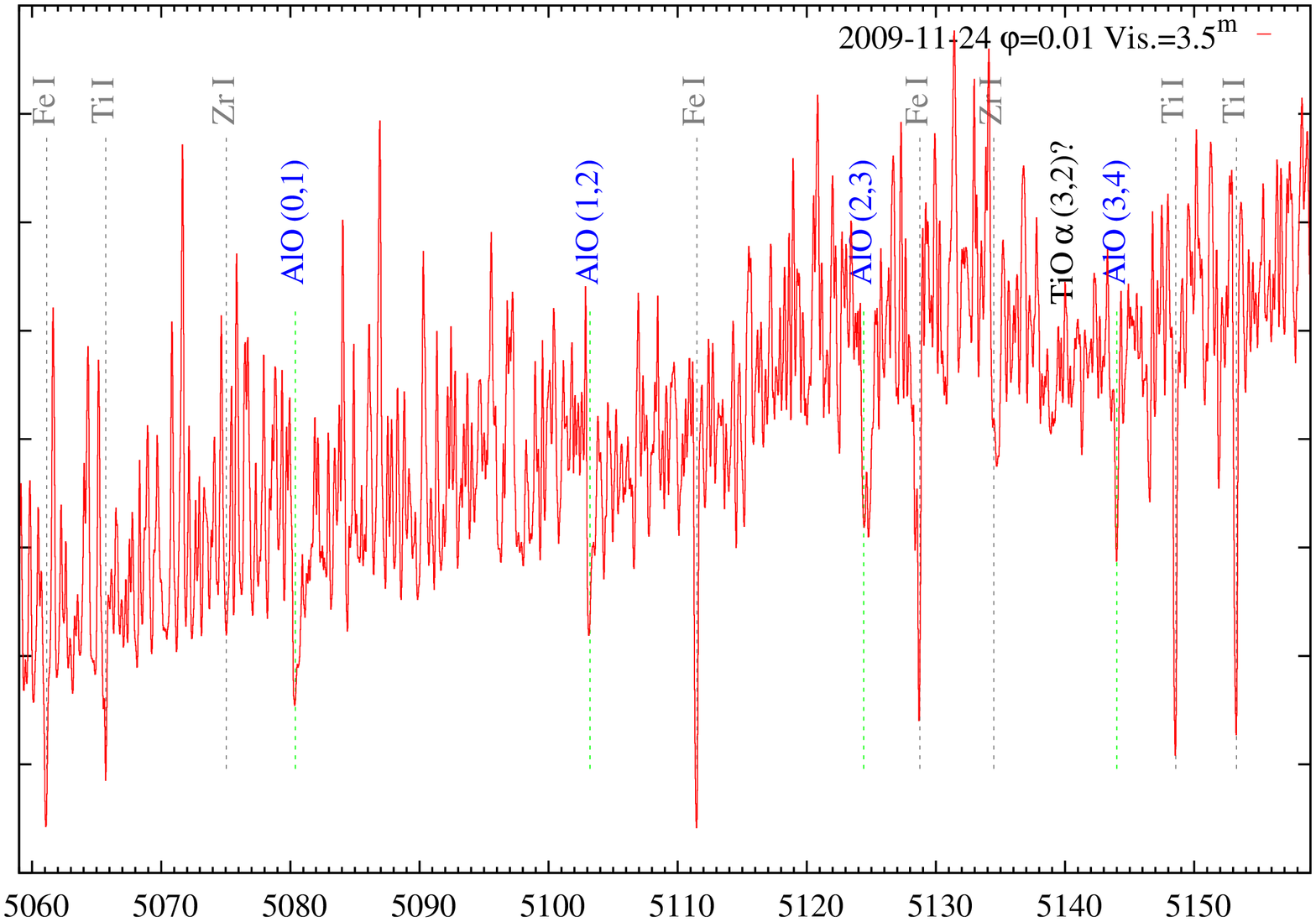}
\caption{Continued.}
\end{figure*}

  \setcounter{figure}{4}%

\begin{figure*} [tbh]
\centering
\includegraphics[angle=270,width=0.85\textwidth]{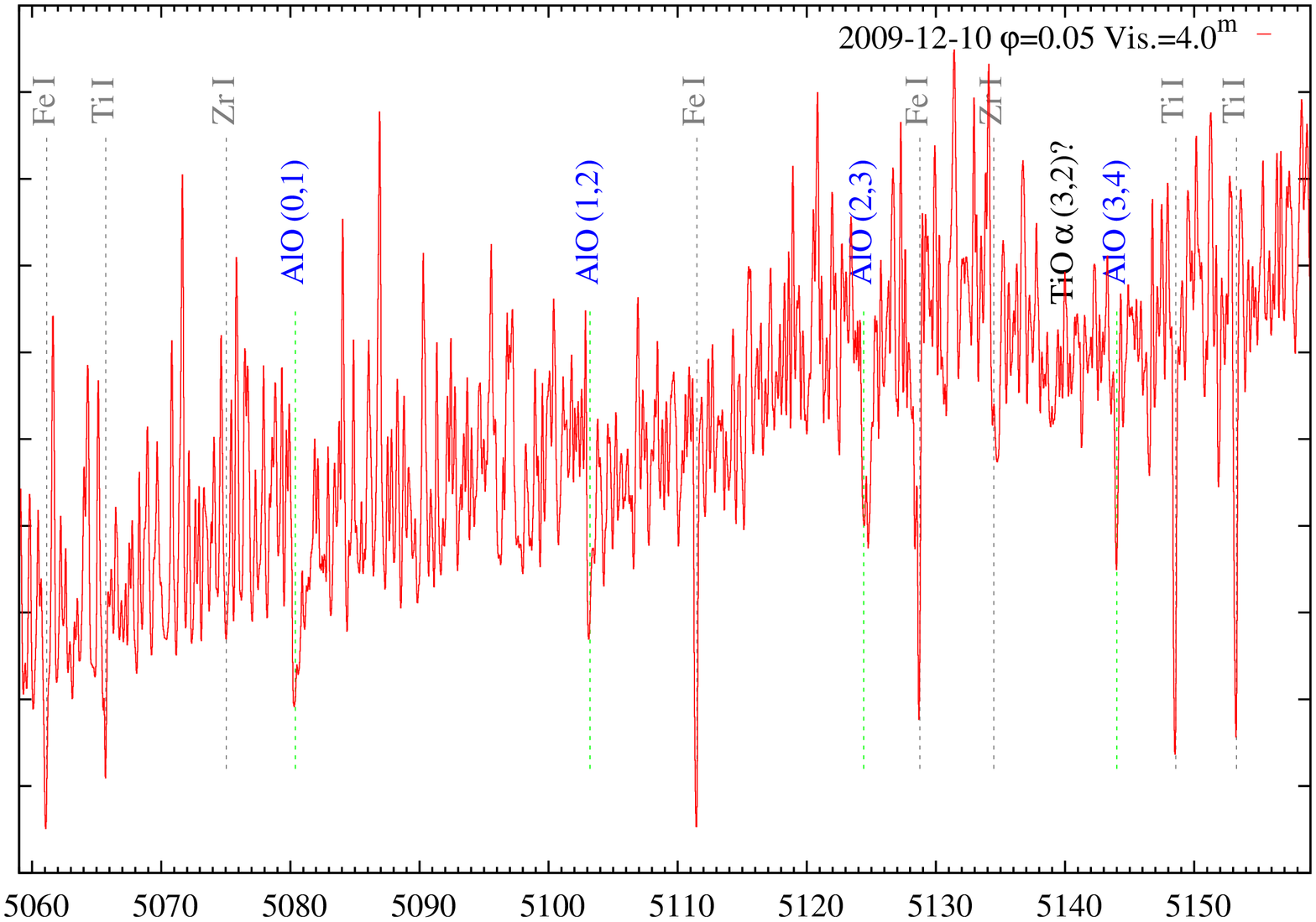}
\includegraphics[angle=270,width=0.85\textwidth]{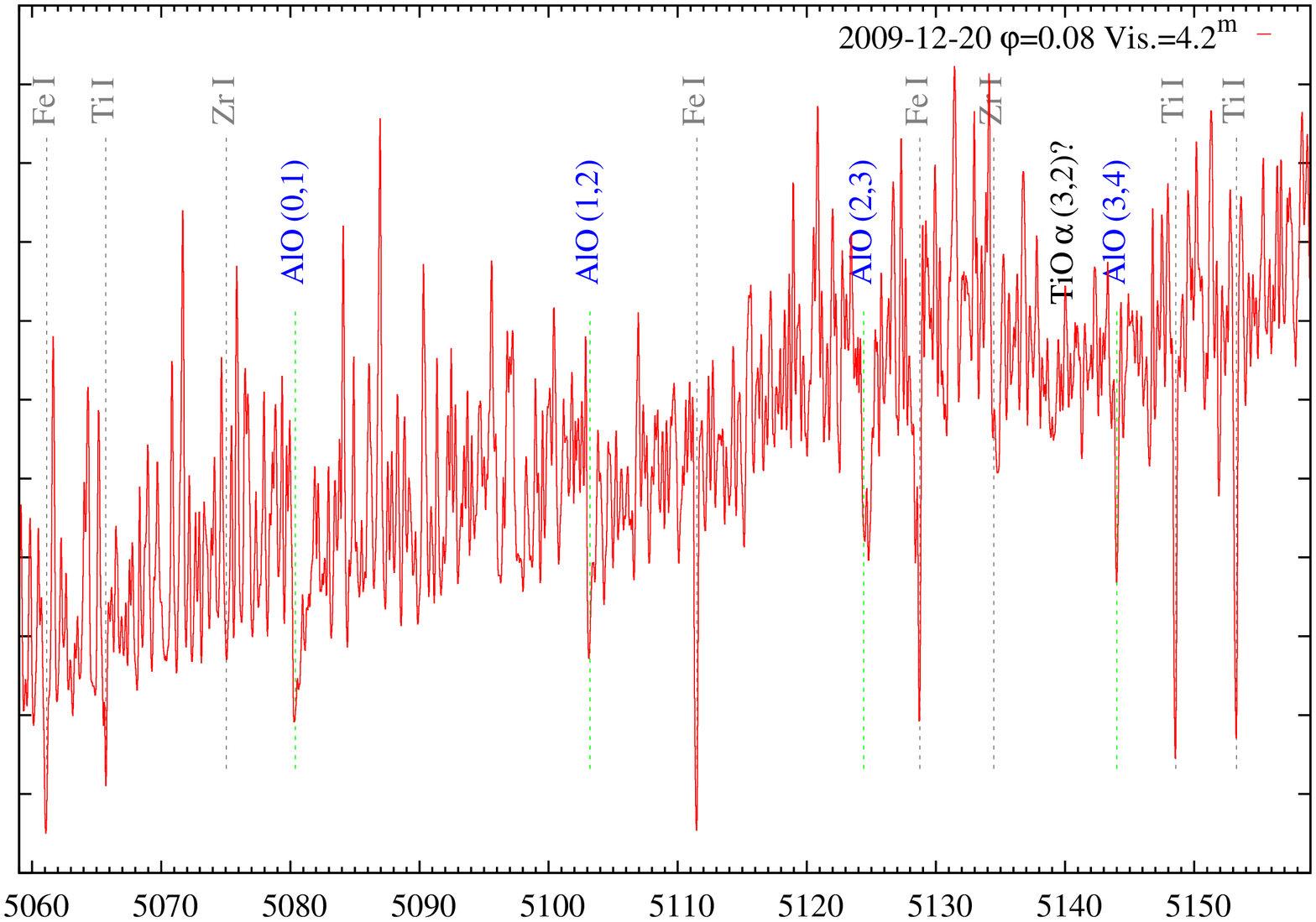}
\caption{Continued.}
\end{figure*}

  \setcounter{figure}{4}%

\begin{figure*} [tbh]
\centering
\includegraphics[angle=270,width=0.85\textwidth]{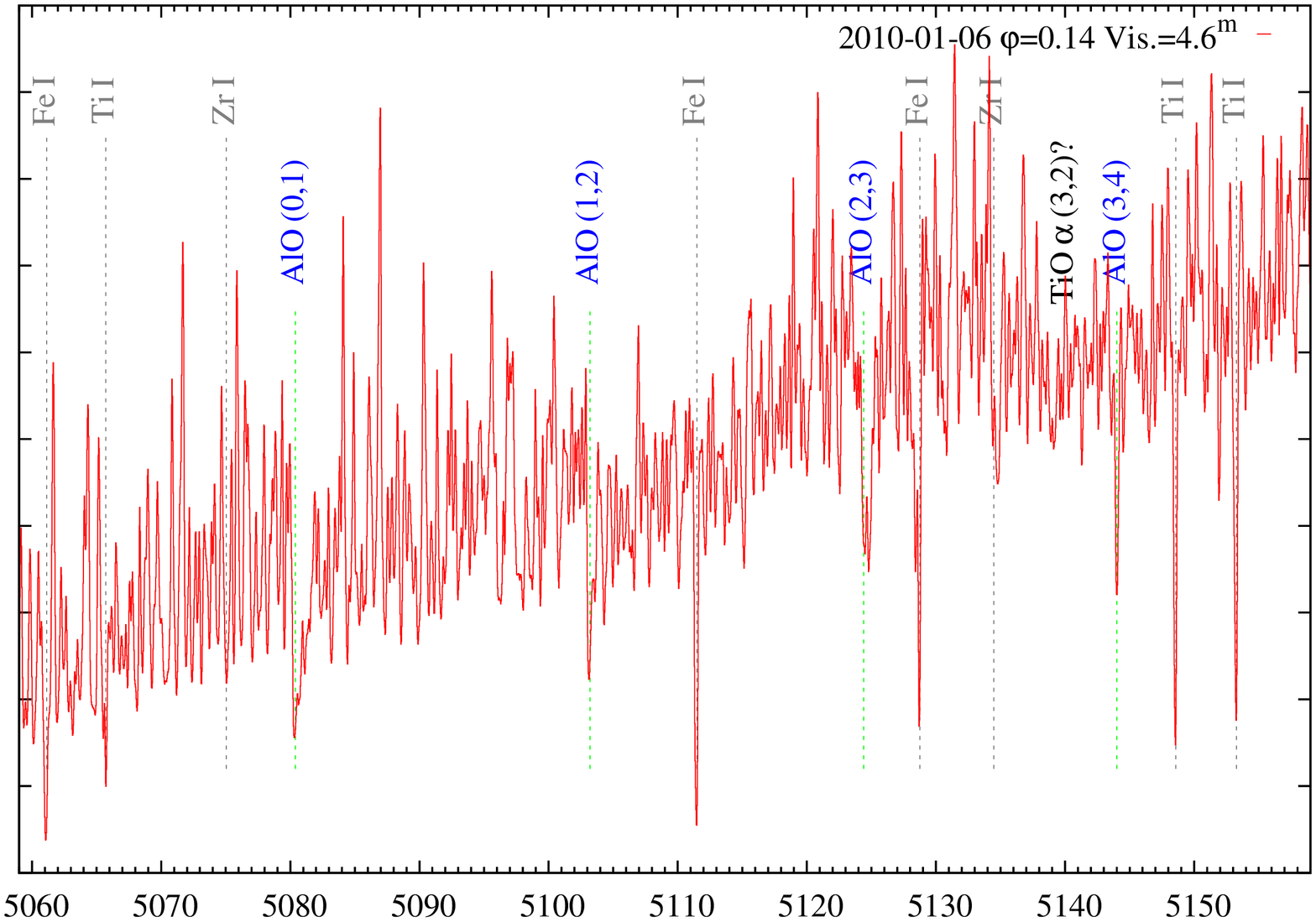}
\includegraphics[angle=270,width=0.85\textwidth]{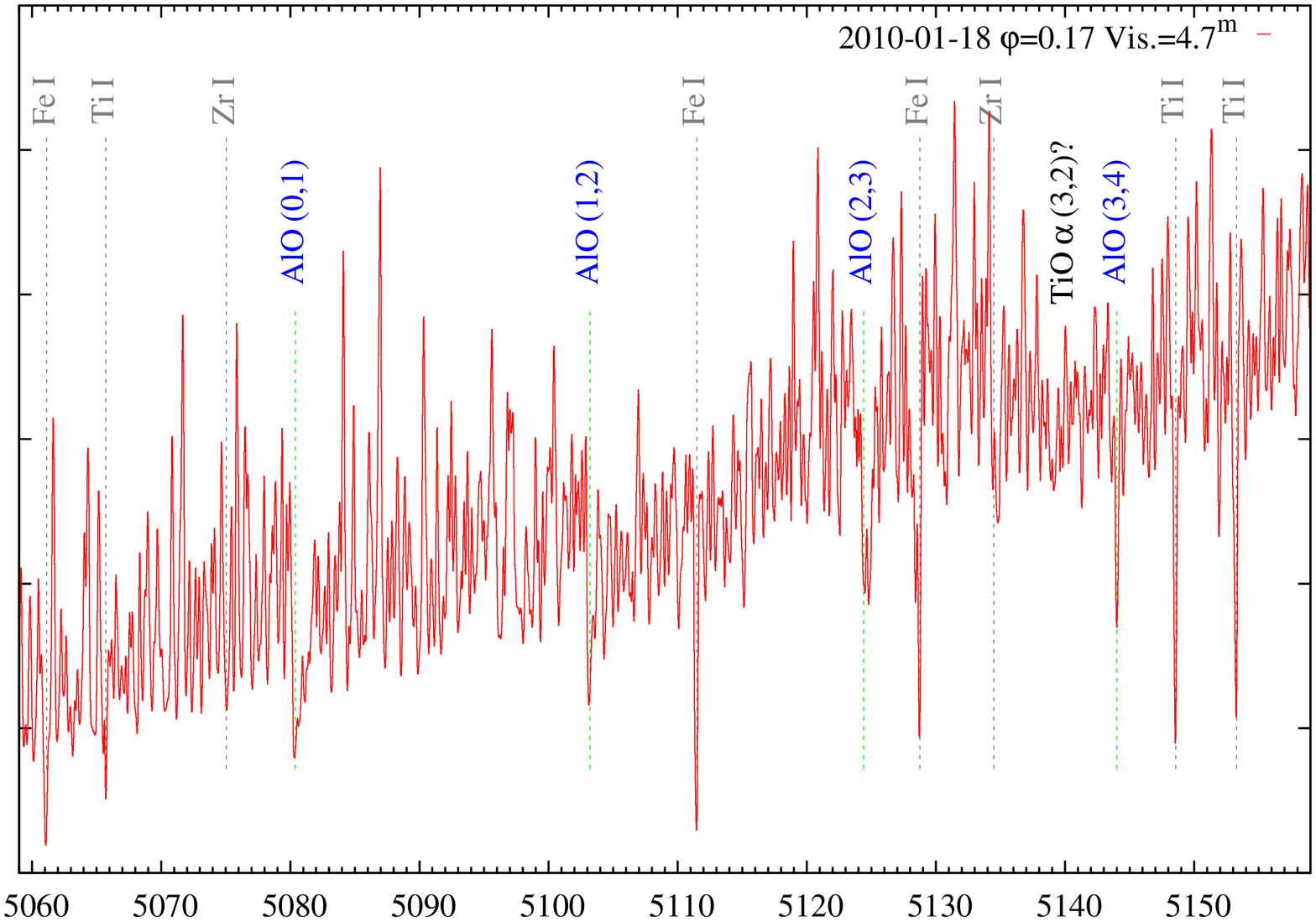}
\caption{Continued.}
\end{figure*}

  \setcounter{figure}{4}%

\begin{figure*} [tbh]
\centering
\includegraphics[angle=270,width=0.85\textwidth]{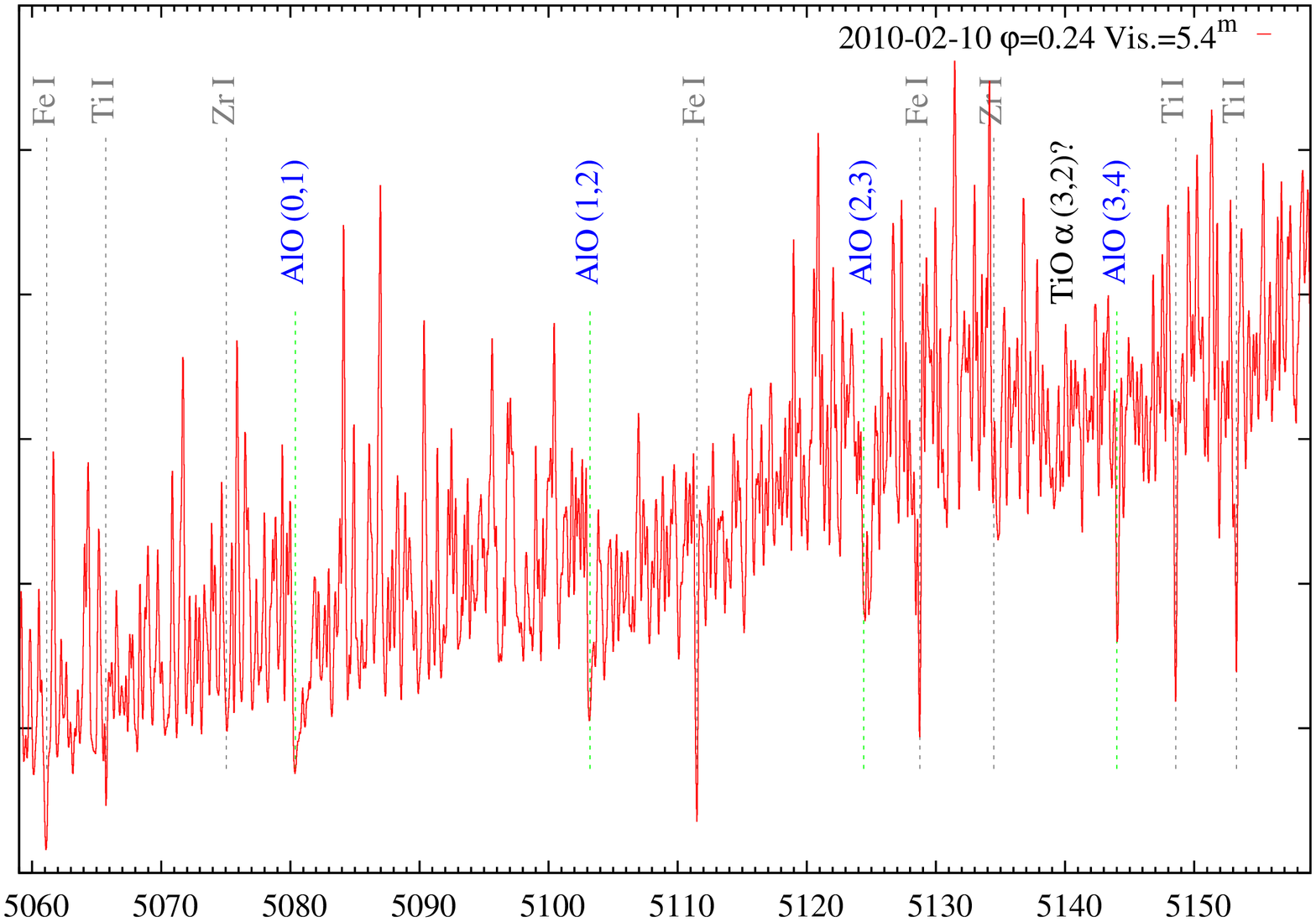}
\includegraphics[angle=270,width=0.85\textwidth]{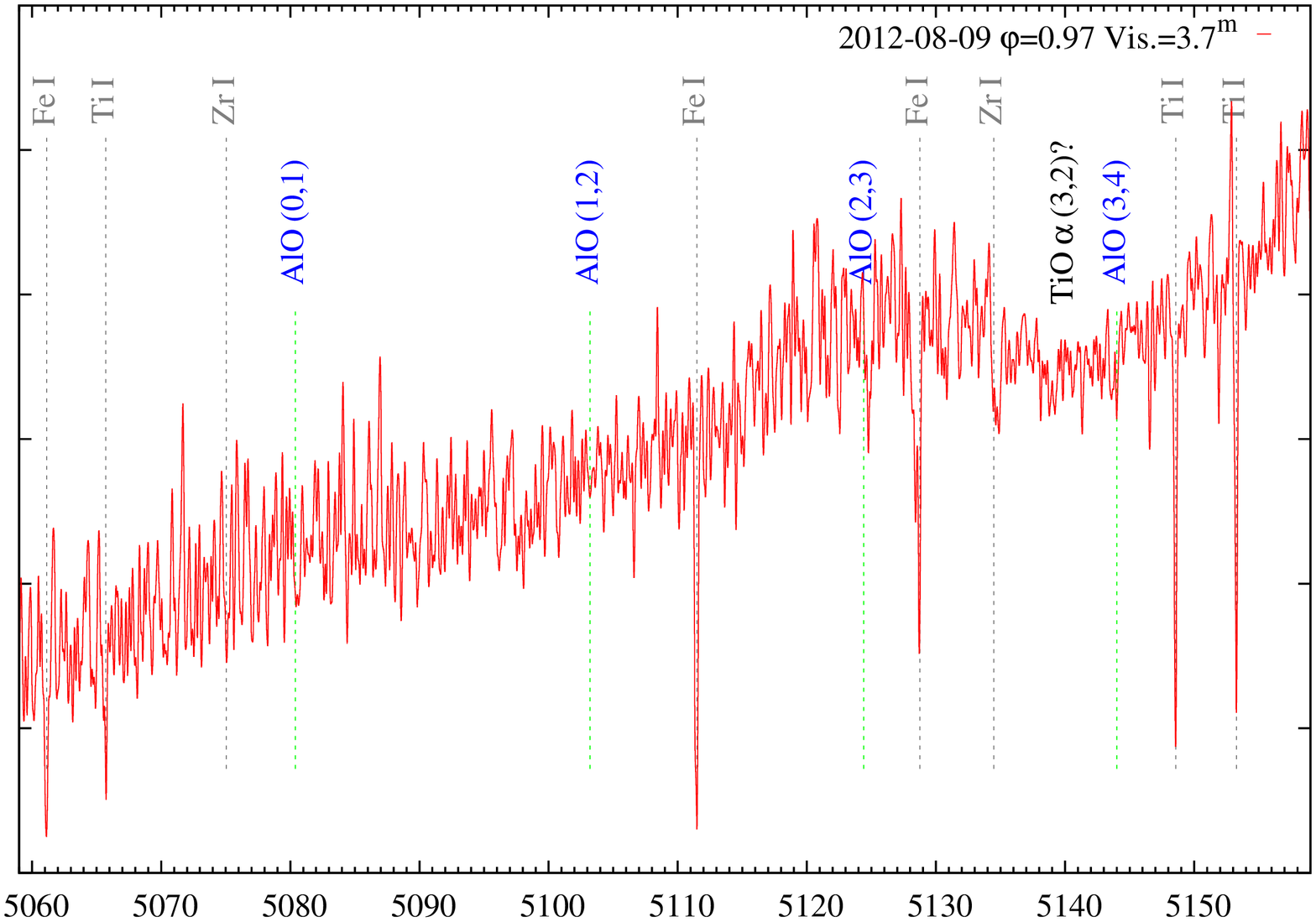}
\caption{Continued.}
\end{figure*}
\end{appendix}

\end{document}